\documentclass[12pt]{article}

\usepackage{amsmath, amssymb, amsthm, mathtools, cancel}
\newcommand{\aipw}{\textnormal{AIPW}}
\newcommand{\dr}{\textnormal{DR}}
\newcommand{\drs}{\textnormal{DR}\star}
\newcommand{\E}{\mathbb{E}}
\renewcommand{\P}{\mathbb{P}}
\newcommand{\Var}{\mathbb{V}\textnormal{ar}}
\newcommand{\N}{\mathcal{N}}

\usepackage{geometry}
\usepackage{setspace}
\makeatletter
\g@addto@macro\normalsize{%
  \setlength\abovedisplayskip{6pt}%
  \setlength\belowdisplayskip{6pt}%
  \setlength\abovedisplayshortskip{6pt}%
  \setlength\belowdisplayshortskip{6pt}%
}
\makeatother
\usepackage{amsthm}
\usepackage{appendix}
\usepackage{pifont}  % for \xmark or \ding
\usepackage{amssymb} % for \checkmark
\usepackage{enumitem}
\usepackage{titletoc}
\usepackage{etoc}
\usepackage[bottom]{footmisc}
\usepackage{titlesec}
\titleformat{\section}{\large\bfseries}{\thesection}{1em}{}
\titleformat{\subsection}{\normalsize\bfseries}{\thesubsection}{1em}{}
\titleformat{\subsubsection}{\small\bfseries}{\thesubsubsection}{1em}{}

\usepackage{graphicx, float, subcaption, caption}
\usepackage{booktabs, makecell, multirow, tabularx}
\usepackage{siunitx, threeparttable}
\usepackage{tikz}

\usepackage{algorithm, algpseudocode}

\usepackage[dvipsnames, table]{xcolor}
\usepackage[colorlinks=true, linkcolor={blue!50!black}, urlcolor=blue, citecolor={blue!50!black}, linktocpage=true]{hyperref}
\usepackage{natbib}

\newtheoremstyle{plain}{7pt}{7pt}{\itshape}{}{\rmfamily\bfseries}{:}{.5em}{}
\theoremstyle{plain}
\newtheorem{proposition}{Proposition}
\newtheorem{assumption}{Assumption}
\newtheorem{lemma}{Lemma}
\newtheorem{theorem}{Theorem}
\newtheorem{corollary}{Corollary}
\newtheorem{remark}{Remark}

\newcommand{\arefsub}[1]{\hyperref[#1]{Appendix~\ref*{#1}}}

\usepackage[color=yellow!20]{todonotes}
\presetkeys{todonotes}{inline}{}
\title{Debiased Conditional Independence Tests \\ for Causal Effects\footnote{Maria Nareklishvili is also affiliated with the University of Glasgow. E-mail: maria.nareklishvili@glasgow.ac.uk. We gratefully acknowledge financial support from the Golub Capital Social Impact Lab. We thank participants at the Stanford Econometrics Lunch, the 2025 International Conference on Computational Social Science (IC2S2), and the 2025 RSS International Conference for helpful comments and discussions.  } }

\author{
  Susan Athey  \\
  \texttt{athey@stanford.edu}
  \and
  Maria Nareklishvili \\
  \texttt{marnar@stanford.edu}
  \and
  Jann Spiess \\
  \texttt{jspiess@stanford.edu}
}
\date{First Draft: February 15, 2025}

\begin{document}

\maketitle

\begin{abstract}
\setstretch{1}
We propose robust tests for causal effects on potentially high-dimensional outcomes where there may be selection into treatment.
These tests measure conditional independence of outcomes from treatment assignment by asking whether treatment can be predicted from outcomes more accurately than from covariates alone.
We design three such tests of conditional independence.
First, we propose a conditional randomization test based on an inverse propensity weighted mean squared error test statistic.  
If the propensity score is known, the test controls size in finite samples.
Second, we propose a double-residualization approach, where we partial out covariate effects from both treatment and outcome-based treatment predictions, and test whether the correlation between the residuals is zero. Third, we propose a test based on an augmented inverse propensity weighted estimator built around doubly robust, Neyman-orthogonal population moments that are necessary and sufficient for conditional independence. Simulations show that all three tests improve power over a joint Wald test based on a seemingly unrelated regression across many non-linear settings, particularly under heavy tails, outliers, mixtures, heterogeneous effects, and effects with opposite signs.

\end{abstract}

\newpage

\section{Introduction}

Evaluating an intervention’s effectiveness with classical statistical methods involves at least three well-known difficulties.
First, an intervention may affect many outcomes, 
so researchers have to make a choice of which ones to focus on.
Second, researchers must choose which regions of the outcome distribution to test: many interventions shift only parts of the distribution (e.g., tails) rather than the mean, so power depends on where one looks.
Third, even after prespecifying outcomes and parts of their distribution, estimation requires committing to functional forms and covariate adjustments for each specification; these choices are model-dependent, and misspecification can lead to biased estimates of treatment effects.
These challenges become particularly severe when there is selection into treatment.
In a randomized controlled trial where treatment is assigned independently of pre-treatment covariates and potential outcomes, tests of treatment–outcome dependence can be performed unconditionally \citep*{ludwig2017machine}. In observational data, both treatment and outcomes generally vary with observed covariates \(X_i\). Unconditional association between treatment \(D_i\) and outcomes \(Y_i\) is therefore uninformative about the causal effect of treatment on any outcome. The relevant null hypothesis instead is that of conditional independence, 
\[
H_0:\; D_i \perp Y_i |X_i.
\]

Our main contribution is to propose a family of conditional independence tests for causal effects motivated by the estimators developed in the semiparametric statistics literature, which assume finite-dimensional parameters of interest and infinite-dimensional nuisance parameters \citep*{ichimura1993semiparametric, athey2018approximate}, and by the idea of the reverse prediction of \citet*{ludwig2017machine}.
A distinguishing feature of the proposed tests is their flexibility and computational efficiency: the tests we propose allow for generic machine learning and only require estimation of a small number of nuisance functions, even when both outcomes and covariates are high-dimensional.
Under the standard unconfoundedness (ignorability) assumption,%
\footnote{Unconfoundedness means that treatment assignment is independent of the full vector of potential outcomes conditional on observed covariates; see \citet*{rosenbaum1983central,rubin1990formal}. \citet*{rosenbaum1988permutation} refers to this as ``$x$-adjustable treatment assignment.''}
our main goal is to test whether potentially high-dimensional outcomes \(Y_i\) contain additional information about treatment assignment beyond a given set of covariates \(X_i\). Formally, we implement this null hypothesis by testing whether 
\[
H_0:\; \P(D_i=1|Y_i,X_i)=\P(D_i = 1|X_i).
\]

Intuitively, the null hypothesis is that the ``posterior'' treatment probability conditional on outcomes and covariates equals the ``prior'' treatment assignment probability based on covariates alone. Under the null hypothesis, the outcomes do not improve prediction of treatment conditional on covariates. Rejection occurs when some outcome shifts the conditional probability of treatment given \(X_i\), indicating a violation of distributional invariance of \(Y_i\) to treatment.%
\footnote{The hypothesis can also be applied to a single outcome at a time with multiple hypothesis testing corrections, such as those proposed by \citet*{holm1979simple, benjamini1995controlling,  romano2005stepwise, list2019multiple}.}
This null hypothesis is between two classical benchmarks: it is weaker than Fisher’s sharp null (no unit-level effect) yet stronger than a mean-zero effect null, because it requires invariance of the entire conditional distribution of \(Y_i\) to treatment rather than only of its mean.

The proposed procedures are designed to test for the presence of the joint distributional causal effect of treatment on outcomes under the unconfoundedness assumption, not to estimate its sign or magnitude. By determining whether outcomes are informative about treatment assignment, our methods serve as a crucial first-stage diagnostic for estimating treatment effects. They answer the question of whether an effect exists before a researcher proceeds to the more demanding task of estimating what that effect is. This approach can guide empirical strategy. For example, a significant finding for employment outcomes in a job training program evaluation would provide a solid statistical basis for analyzing on which outcomes and for which individuals there are treatment effects, while a null result would caution against it.

We formulate three tests of conditional independence between potentially high-dimensional outcomes \(Y_i\) and binary treatment \(D_i\), conditionally on high-dimensional covariates \(X_i\).
The first is a loss-based criterion defined as an inverse propensity weighted mean squared error between the treatment indicator and the predicted posterior treatment probability, under the assumption that the propensity scores are known to the researcher.
Inverse propensity weighting in the loss function constructs a balanced pseudo-population that mimics random assignment. 
Rather than relying on asymptotic inference, we develop a conditional randomization procedure for the loss-based test that delivers finite-sample control of the Type I error rate. The proposed loss-based measure of conditional independence is motivated by inverse probability weighted estimators dating to \citet*{horvitz1952generalization, hajek1964asymptotic, hajek1971comment}, followed by \citet*{hirano2003efficient}, who show that, within the semiparametric class of estimators, the inverse propensity weighted estimator with an estimated propensity score attains the semiparametric efficiency bound of \citet*{hahn1998role} and is asymptotically more efficient than the analogous estimator that treats the propensity score as known.%

Having considered the case where the propensity score is known, we then move to observational settings where selection into treatment instead has to be estimated.
The second test of conditional independence is based on a double-residualization approach \citep*{robinson1988root, chernozhukov2018double, nie2021quasi}. The test statistic measures the standardized covariance between treatment and the prediction of treatment based on outcomes, both residualized with respect to the available set of covariates.
Residualization enables measurement of the causal link between treatment and the treatment probability predicted from outcomes by removing the set of confounders. We develop two versions of this approach: one (DR) following the partialling-out approach of \citet{nie2021quasi}, and a debiased variant (\(\drs\)) in the classical spirit of \citet{robinson1988root} which is robust to limited overlap between treated and control units. Unlike the loss-based measure, this test does not assume that the propensity score is known, and instead may leverage flexible nonparametric methods, such as the regression forest, boosting, or neural networks to estimate it. 

The third measure of conditional independence between treatment and multiple outcomes is based on the augmented inverse propensity weighted (AIPW) estimator \citep*{robins1994estimation, robins1995semiparametric, bang2005doubly}. In analogy to AIPW, we treat the predicted treatment probability \(f_0(Y_i, X_i) = \P(D_i = 1 |Y_i, X_i)\) as a pseudo-outcome and design an augmented inverse propensity weighted measure of conditional independence.
That is, we ask whether an estimate of the function $f_0(Y_i, X_i)$ is systematically different between the treatment and control groups.
This approach combines an outcome regression component where we use the predicted treatment probability as a pseudo-outcome and inverse propensity weighted residuals that correct for selection bias.

Both the double-residualized and augmented inverse propensity weighted measures of conditional independence yield one-sided tests. When outcomes are informative about treatment status, the expected scores conditional on covariates of the population are strictly positive; otherwise, they are zero, guaranteeing that their population moments are both necessary and sufficient for testing conditional independence.
We show that the test statistics based on the augmented inverse propensity weighting as well as the classical double-residualization strategy also have a robustness property: they are Neyman-orthogonal with respect to the estimated propensity score (specifically, the test is robust to small estimation mistakes in the treatment probability as a function of the covariates). Consequently, valid asymptotic size based on cross-fitted estimation can be achieved by applying standard results from the framework of double/debiased machine learning \citep*{kennedy2016semiparametric, chernozhukov2017double, chernozhukov2018double,wager2018estimation}.
This allows for the use of flexible nonparametric learners, provided that the root mean squared error of these learners converges to zero at a rate faster than \(n^{-1/4}\) under the null hypothesis.

In addition to the asymptotic properties, we study the finite-sample performance of the suggested tests in a simulation study. Monte Carlo simulations, covering nine data-generating processes with varying treatment effect distributions, sample sizes, and outcomes, demonstrate that the proposed tests are markedly more powerful than the conventional joint Wald test based on the seemingly unrelated regression. In designs with heterogeneous or non-Gaussian treatment effects, improvements are substantial, often exceeding 100\%. Importantly, these performance gains are achieved without increasing false positive rates: the simulations and theoretical results confirm that all proposed procedures maintain control of the Type I error rate at the nominal significance level. We also show that the principal advantages of augmentation and residualization emerge in low-signal environments: residualization removes baseline-driven variation in outcomes and treatment assignment, thereby achieving significantly higher power when the signal-to-noise ratio is as low as 7\%.

\section{Related Literature}

Our approach brings together the literature on conditional independence testing, semiparametric estimation, and testing based on reverse predictions.

Early conditional independence tests discretize covariates or outcomes into strata and test independence within each stratum. The Cochran--Mantel--Haenszel procedure \citep*{mantel1959statistical} stacks \(K\) stratum-specific \(2\times 2\) tables and tests a common odds ratio of one, equivalently conditional independence within strata. \citet*{margaritis2005distribution} bins continuous variables into fixed \(2\times 2\) and \(3\times 3\) histograms and aggregates stratum-wise likelihood evidence into one statistic. \citet*{huang2010testing} partitions on the conditioning set, measures maximal correlation among scaled residualized variables locally, and aggregates into a global test. These designs inherit the curse of dimensionality: the partition cardinality grows exponentially in covariate/outcome dimension, driving exponential sample-size demands. Related approaches that
test the correlation between treatment residuals and indicators for
subsets of outcomes \citep*{angrist2011causal} likewise become computationally prohibitive as dimension and grid resolution increase.

The subsequent literature has developed kernel-based estimators for conditional independence testing that avoid explicit discretization. For instance, \citet*{su2007consistent} test for conditional independence between variables by comparing conditional characteristic functions, estimated by the Nadaraya–Watson kernel regression.  \citet*{su2008nonparametric} adopt a similar strategy, measuring (Hellinger) distance between conditional densities estimated via kernel methods, with similar limitations in high-dimensional settings. Other powerful tests that directly estimate conditional densities \citep*{su2010testing,su2014testing, mei2025dimension} share this sensitivity to bandwidth selection, which becomes increasingly problematic as the dimension of the conditioning variables grows.

Another related direction of conditional independence tests is based on mapping variables into a reproducing kernel Hilbert space (RKHS). For instance,  \cite*{fukumizu2004dimensionality, fukumizu2007kernel} test for conditional independence by orthogonalizing two sets of variables with respect to the conditioning variable and then evaluating whether the residual dependence remains---a kernel analogue of partial covariance. Recent literature has improved the efficiency of these statistics and developed better approximations to the null distribution \citep*{zhang2012kernel, doran2014permutation, pfister2018kernel, zhang2018large,strobl2019approximate}, including \citet*{zhang2024doubly} who propose a doubly robust test based on generative adversarial networks. Kernel methods are highly flexible and can handle high-dimensional covariates and mixed data types. However, they often face computational challenges: forming kernel matrices requires quadratic memory, and matrix operations require cubic time, with costs often multiplied by the number of resamples for permutation-based inference. Another related test statistic, the generalized covariance measure of \citet*{shah2020hardness} tests conditional independence by residualizing each variable with respect to the conditioning set and checking whether the covariance between the residuals deviates from zero. While the measure is well suited to univariate outcomes, residualization becomes computationally demanding with high-dimensional outcome variables.

Our tests most closely relate to \cite{ludwig2017machine}. \cite{ludwig2017machine} predict treatment assignment from outcomes with a flexible
machine learning model, and test for independence via a permutation-based reference distribution. \citet{zhang2025inverse} study a similar reverse-regression strategy for tests on many outcomes based on linear regression.
The proposed test statistics in our project share the same goal and approach. In this project, we generalize the reverse prediction approach of \cite{ludwig2017machine} to stratified experiments by proposing a test statistic based on the inverse propensity weighted loss function. For the loss-based test statistic, propensity scores are assumed to be known by the researcher.  We then propose novel measures of conditional independence for observational data that use the reverse prediction, $\P(D_i = 1 |Y_i, X_i)$, as a one-dimensional sufficient statistic within augmented inverse propensity weighting proposed by   \cite{robins1994estimation, chernozhukov2018double}, and double-residualization frameworks of \cite{robinson1988root, nie2021quasi}. The tests inherit desirable large-sample properties, such as Neyman orthogonality and double robustness under the null hypothesis.

The primary advantage of the proposed measures of conditional independence is their computational efficiency. For instance, if we estimate propensity scores with a regression forest with a fixed minimum leaf size and a fixed number of candidate covariates per split, the training complexity is on the order of $O(N \log N)$, with constants determined by the number of trees, average tree depth, and the number of candidate covariates considered per split. Estimating the posterior treatment probability $\P(D_i = 1 |Y_i, X_i)$ with a forest has the same order of complexity. Conditional on these two predictions, computing out-of-fold predictions, scores, and the test statistic requires a single pass through the data, i.e., $O(N)$ time and $O(N)$ memory. By contrast,  conditional independence tests based on reproducing kernel Hilbert spaces construct $N \times N$ Gram matrices, requiring $O(N^2)$ memory and typically $O(N^3)$ time, often repeated for bandwidth selection or resampling, making them substantially more costly at scale.

\section{Setup}

In this section, we formalize testing for causal effects on many outcomes within the Neyman--Rubin potential outcomes framework, state the corresponding identification assumptions, connect the resulting null hypothesis to conditional independence tests, and illustrate the connection between conditional independence tests and reverse prediction.  

\subsection{Potential Outcomes and Null Hypothesis}

We formalize testing for causal effects with potentially high-dimensional outcomes in the Neyman--Rubin potential outcomes framework \citep*{rubin1974estimating}, which we then connect to conditional independence testing. We consider $N$ individuals $i$ drawn independently and identically from a distribution
\[Y_i(1), Y_i(0), D_i, X_i \sim \P^\star\]
of potential outcomes $Y_i(1), Y_i(0)$, binary treatment $D_i \in \{0,1\}$, and covariates $X_i$. The potential outcomes $Y_i(1), Y_i(0)$ denote the outcomes of individual $i$ if $i$ was treated ($D_i = 1$) or not treated ($D_i = 0$), respectively.%
\footnote{
    We implicitly assume the Stable Unit Treatment Value Assumption (SUTVA), which requires that the potential outcomes of any individual are unaffected by the treatment assignments of other individuals.
}
We do not restrict the outcome space. Outcomes may be binary, continuous, categorical, or any combination thereof (including mixed discrete–continuous vectors), and can be scalar or multivariate.

Within this potential-outcomes framework, we state the null hypothesis of interest. Specifically, the null hypothesis is that the distributions of potential outcomes are equal under treated and control states,
\begin{align}\label{eq:null_distr}
    H_0: \big(Y_i(1) \stackrel{d}{=} Y_i(0)\big) |X_i
\end{align}
almost surely,
where $\stackrel{d}{=}$ denotes equality in distribution. The null hypothesis implies that treatment has no effect on the distribution of outcomes, not just on average, but for the entire distribution, without requiring that outcomes be exactly equal for every unit. The null hypothesis occupies an intermediate position between two classical benchmarks. It is weaker than Fisher’s sharp null, which requires exact equality of outcomes under treatment and control for each individual, \( Y_i(1) = Y_i(0) \) almost surely. At the same time, it is stronger than Neyman’s null hypothesis of zero average treatment effect, which only requires that the mean difference is zero in the population \(
\E[Y_i(1) - Y_i(0)] = 0.
\)
This null hypothesis is also stronger than the \emph{conditional} version of Neyman’s null hypothesis that $\E[Y_i(1) - Y_i(0)|X_i] = 0$ a.s., so our proposed tests may be sensitive to treatment effect heterogeneity even when there is no effect on the overall marginal distributions.

In practice, we do not observe the potential outcomes $Y_i(1), Y_i(0)$,
and instead only have access to the realized outcomes $Y_i$.
This observed outcome is a deterministic function of the treatment assignment and the potential outcomes,
\begin{align*}
    Y_i = Y_i(D_i)
    = \begin{cases} Y_i(1) & \text{if } D_i = 1 \\ Y_i(0) & \text{if } D_i = 0 \end{cases}.
\end{align*}
As a result, the observed data \(\{(Y_i, D_i, X_i)\}_{i=1}^{N}\) consist of \(N\) independent and identically distributed (i.i.d.) observations, drawn from the common joint probability measure 
\[
(Y_i, D_i, X_i) \sim \P, \quad i = 1, \dots, N
\]
implied by $\P^\star$.
Our goal is to test the null hypothesis \eqref{eq:null_distr} of no causal effects from this implied distribution $\P$ of the observed data.

\subsection{From Causal Effects to Conditional Independence}

We assume that treatment assignment is independent of potential outcomes after conditioning on observed covariates, commonly referred to as unconfoundedness or selection on observables.

\begin{assumption}[Unconfoundedness]\label{asm:unconfoundedness}
Treatment assignment is independent of potential outcomes given observed covariates:
\begin{align*}
    Y_i(0), Y_i(1) \perp D_i |X_i.
\end{align*}
\end{assumption}

Suppose that unconfoundedness holds and \(0 < e_0(X_i) < 1\) almost surely. Then the null hypothesis in \eqref{eq:null_distr} is \emph{equivalent} to the null hypothesis of conditional independence between treatment assignment \( D_i \) and the observed outcome \( Y_i \) conditional on covariates \( X_i \),
\begin{align} \label{eq:null_independence}
    H_0: D_i \perp Y_i |X_i
\end{align}
almost surely.
Unlike the original null hypothesis of no treatment effects, this null hypothesis is defined on the \emph{observed} distribution $\P$.
In principle, this null hypothesis of conditional independence may be of interest beyond the case of causal inference, and our approach applies for conditional independence tests more generally (as long as the variable \(D_i\) is binary).
It is the main null hypothesis we aim to test for the remainder of the paper.

\citet*{rosenbaum1983central} show that if \autoref{asm:unconfoundedness} holds, then this independence also holds conditional on the propensity score \( e_0(X_i) = \P(D_i = 1 |X_i) \).
Specifically, treatment is independent of potential outcomes, conditional on the propensity score
\(
Y_i(0), Y_i(1) \perp D_i |e_0(X_i).
\)
This result justifies the use of the propensity score as a sufficient statistic for selection: within strata defined by \( e_0(X_i) \), the distribution of observed covariates is the same for treated and control units.
By the same logic, the null hypothesis \eqref{eq:null_distr} implies independence conditional on the propensity score,
\begin{align*}
    H_0: D_i \perp Y_i |e_0(X_i)
\end{align*}
almost surely, as long as unconfoundedness holds.

For our analysis, we also impose strong overlap, that is, that the probability of treatment conditional on covariates is bounded away from zero and one:

\begin{assumption}[Strong overlap]\label{asm:overlap}
For all individuals \( i \) and all values of the covariates \( X_i \), the propensity score \( e_0(X_i) = \P(D_i = 1 | X_i) \) satisfies:
\[
\eta  \leq e_0(X_i) \leq 1 - \eta, \ \eta > 0.
\]
\end{assumption}

While not strictly necessary for creating valid tests of the null hypothesis in \eqref{eq:null_independence}, we rely on overlap in order to leverage standard estimators from the semiparametric statistics literature and to ensure power against the null.

\subsection{Equivalence to Reverse Predictions}

In order to test the null hypothesis \eqref{eq:null_independence} of conditional independence between treatment \(D_i\) and outcomes \(Y_i\) given covariates \(X_i\),
we note that it is equivalent to an inability to predict treatment \(D_i\) better from outcomes \(Y_i\) and covariates \(X_i\) than from covariates \(X_i\) alone:

\begin{proposition}[Equivalence to reverse prediction]\label{prop:reverse_prediction}
The null hypothesis \eqref{eq:null_independence} of conditional independence is equivalent to
\begin{align}
    \label{eq:null_predictability}
    H_0: \P(D_i = 1|Y_i, X_i) =  \P(D_i = 1|X_i)
\end{align}
almost surely. The proof is provided in \arefsub{app:proof_reverse_prediction}.
\end{proposition}

This result captures predictability since the null hypothesis \eqref{eq:null_predictability} is readily seen to be equivalent to
\(
    \label{eq:null_prediction}
    \E[(\P(D_i = 1|Y_i, X_i) - D_i)^2]
    =
    \E[(\P(D_i = 1| X_i) - D_i)^2],
\)
that is, we cannot predict $D_i$ better than using the propensity score $e_0(X_i) = \P(D_i = 1| X_i)$ (based on a suitable loss function like $L_2$ loss in this case).

Testing the null hypothesis in \eqref{eq:null_predictability} is especially useful when the goal is to confirm the presence of treatment effects across a collection of outcomes, without the need to estimate treatment effects for each. Predicting a single binary treatment variable is typically easier than specifying the full joint distribution of multiple, potentially correlated outcomes.
A rejection of the null hypothesis brings statistical evidence that treatment has shifted at least some region of the conditional outcome distribution (given \(X_i\)), without requiring prior specification of which outcomes or which regions of the outcome space are affected.

By moving from the null of no causal effect in \eqref{eq:null_independence} to the null of no predictability in \eqref{eq:null_predictability}, our methodological framework reverses the traditional causal question. Instead of testing whether treatment affects outcomes, we test whether outcomes predict treatment assignment after controlling for observed confounders. Under the null hypothesis of no treatment effect, outcomes should be uninformative about treatment assignment once we account for confounding through covariates. \autoref{fig:framework} illustrates the idea.

\begin{figure}[H]
    \centering
 \begin{tikzpicture}[
    node distance=1.2cm,
    box/.style={rectangle, draw, minimum width=2.5cm, minimum height=1cm, align=center, font=\small},
    arrow/.style={->, thick, >=stealth},
    label/.style={font=\footnotesize, color=gray!60!black}
]

% Horizontal separation
\def\xsep{4.5}

% Standard causal inference (left side)
\node[box, fill=gray!20, draw=gray!60] (treatment) at (-\xsep, 2) {Treatment \\ $D$};
\node[box, fill=gray!20, draw=gray!60] (outcome) at (-\xsep, 0) {Outcomes \\ $Y$};
\node[box, fill=gray!20, draw=gray!60] (covariates) at (-\xsep, -2) {Covariates \\ $X$};

% Staggered arrows for clarity
\draw[arrow, gray!60] (treatment) -- (outcome);
\draw[arrow, gray!60, bend left=40] (covariates) to (treatment);
\draw[arrow, gray!60, bend right=25] (covariates) to (outcome);

\node[label] at (-\xsep, 3.5) {\textbf{Forward Inference}};
\node[label] at (-\xsep, 3) {Test: $D \rightarrow Y$};

% Proposed approach (right side)
\node[box, fill=blue!20, draw=blue!60] (outcome2) at (\xsep, 2) {Outcomes \\ $Y$};
\node[box, fill=blue!20, draw=blue!60] (treatment2) at (\xsep, 0) {Treatment \\ $D$};
\node[box, fill=blue!20, draw=blue!60] (covariates2) at (\xsep, -2) {Covariates \\ $X$};

% Staggered arrows for clarity
\draw[arrow, blue!60] (outcome2) -- (treatment2);
\draw[arrow, blue!60, bend left=25] (covariates2) to (treatment2);
\draw[arrow, blue!60, bend right=40] (covariates2) to (outcome2);

\node[label] at (\xsep, 3.5) {\textbf{Reverse Inference}};
\node[label] at (\xsep, 3) {Test: $Y \rightarrow D$};

% Inversion arrow (longer and more central)
\draw[arrow, thick, color=red!70] (-2, 1) -- (2, 1);
\node[label, color=red!70] at (0, 1.5) {\textbf{Inversion}};

% Null hypothesis statement
\node[box, fill=gray!10, draw=gray!40, minimum width=8cm, minimum height=0.8cm] at (0, -4) {
    \textbf{Null Hypothesis:}  $Y$ should not predict $D$ after controlling for $X$
};

\end{tikzpicture}
\caption{Predictive inference in observational data.}
\label{fig:framework}
\end{figure}

This perspective is conceptually related to the framework of variational
autoencoders in machine learning. In our context, under the alternative hypothesis, the outcomes can be
viewed as noisy proxies for an underlying ``latent'' treatment variable. Since
outcomes are measured after the treatment has taken place, they inherently carry (encode) information
about treatment assignment. Our testing procedure then seeks to recover (decode) this
information: by evaluating whether treatment can be predicted by outcomes, we are asking whether the observed outcomes serve as proxy treatments that
can be mapped back to the original treatment assignment.

This approach generalizes the approach of \citet*{ludwig2017machine}, which examines unconditional independence \(\P(D_i  = 1|Y_i) = \P(D_i = 1)\) in randomized controlled trials in order to test the null of no treatment effects. The authors construct a measure of independence, namely, the mean squared error between observed treatment assignment and expected treatment conditional on outcomes. Under the randomization assumption where \(\P(D_i = 1|X_i) = p\), they generate the null distribution of the test statistic by permuting treatment assignments. In what follows, we generalize this framework to stratified randomized experiments in which propensity scores are known by design, and propose new measures for assessing conditional independence in observational data when propensity scores must be estimated.

\section{Rerandomization-Based Test with Known Propensity Score}

This section develops a formal test of conditional independence between treatment and outcomes given covariates when the propensity score is known, such as for a stratified experiment. We construct a test statistic based on inverse propensity weighted loss. Building on this statistic, we formulate a rerandomization test and derive its asymptotic behavior under the null hypothesis. The analysis motivates the loss-based criterion, clarifies the role of the propensity score, and specifies regularity conditions under which the statistic is asymptotically normally distributed and the resulting test attains correct size for a given sample.

\subsection{Motivation and Identification}

Throughout this section, we assume the propensity score \(e_0(X_i)=\P(D_i=1 |X_i)\) is known and formalize conditional independence tests for the inverse propensity weighted pseudo-population. This approach is motivated by the Horvitz--Thompson estimator, which serves as an unbiased estimator of the population mean when individuals have heterogeneous inclusion probabilities in the selected sample (or the experiment) from the universe  \citep*{horvitz1952generalization}.%
\footnote{Propensity scores may be known when treatment assignment is experimentally controlled or externally specified. By contrast, the probability of treatment given outcomes and covariates, $f_0(Y_i, X_i) = \P(D_i = 1 |Y_i, X_i)$, remains unknown and must be estimated from the available data.}  Specifically, with known propensities, inverse propensity weighting constructs a balanced pseudo-population in which the covariate distribution is equal across treatment arms, rendering \(D_i\) independent of \(X_i\) under the reweighted distribution. This balance permits unconditional tests of \(Y_i \perp D_i\) in the pseudo-population that are equivalent to tests of \(Y_i \perp D_i |X_i\) in the observed population, thereby enabling inference on distributional causal effects in the presence of observable confounders.

One common strategy for creating a balanced pseudo-population is to reweight the sample by inverse propensity score weights (IPW), defined by
\[
W_i = w(D_i, X_i)  = w\big(D_i, e_0(X_i)\big) :=  \frac{D_i}{e_0(X_i)} + \frac{1 - D_i}{1 - e_0(X_i)}.
\]
We use inverse propensity weights to construct a pseudo-randomized population in which each unit has an equal inclusion probability in the selected data set. \autoref{lem:pseudo_randomization} shows that, under the null hypothesis of conditional independence between treatment and outcomes given covariates, reweighting the joint likelihood of the data by the inverse of the propensity score introduces a pseudo-randomized experiment in which treatment is independent of outcomes and assigned with equal probability.

\begin{lemma}[Pseudo-randomization] \label{lem:pseudo_randomization}
Assume \autoref{asm:overlap} holds.
Write $\P_w$ for the distribution over $(Y_i,D_i,X_i)$ after weighting by $w$.%
\footnote{That is, $\displaystyle \P_w(E)=\mathbb E[W_i\,\mathbf 1\{(D_i,X_i,Y_i)\in E\}] / \mathbb E[W_i]$ for all measurable sets $E$.
\label{fn:pw}}
Then
\begin{align*}
    \P_w(D_i = 1|Y_i, X_i) = \tfrac{1}{2} \text{ a.s. if and only if } Y_i \perp D_i |X_i \text{ a.s.},
\end{align*}
that is, if and only if the null hypothesis \eqref{eq:null_independence} holds. Proof in \arefsub{app:proof_pseudo_randomization}.
\end{lemma}

\autoref{lem:pseudo_randomization} establishes that, when the propensity score is known, the inverse propensity weighted probability of treatment conditional on outcomes and covariates is 0.5 under the null hypothesis.

Correctly specifying the joint distribution of treatment, covariates, and outcomes can be difficult, especially with multiple outcomes and covariates.
Instead, the result from \autoref{lem:pseudo_randomization} suggests a more tractable strategy for testing the null hypothesis of conditional independence by asking whether we can predict treatment better than by $\tfrac{1}{2}$ in the weighted population.
Let \( f: \mathcal{Y} \times \mathcal{X} \to [0,1] \) denote a measurable function that predicts treatment assignment based on outcomes $Y \in \mathcal{Y}$ and covariates $X \in \mathcal{X}$. Define the weighted squared loss we wish to minimize as
\begin{align*}
L(f) := \E\left[ W_i \cdot (D_i - f(Y_i, X_i))^2  \right] = 2 \: \E_w\left[ (D_i - f(Y_i, X_i))^2  \right]
\end{align*}
\noindent The function \( f_w(\cdot)\) that minimizes this loss is the conditional treatment probability under the reweighted distribution
\begin{align}
f_w(y, x) := \E_w[D_i |Y_i{=}y, X_i{=}x] = \P_w(D_i{=}1 |Y_i{=}y, X_i{=}x)
= \frac{\mathbb E\!\left[W_iD_i|Y_i=y,X_i=x\right]}{\mathbb E\!\left[W_i|Y_i=y,X_i=x\right]},
\label{eq:fw}
\end{align}
which directly follows from Chapter 2.4 of \citet*{hastie2009elements} (shown in \arefsub{app:proof_loss_minimization}). Under the null (when treatment and outcomes are conditionally independent), \autoref{lem:pseudo_randomization} shows that the inverse propensity weighted squared loss is minimized at the constant function \( f_w(y, x) = \tfrac{1}{2} \).

\autoref{thm:loss_criterion} uses the globally minimized (oracle) loss $L(f_w)$ as a measure of conditional independence between treatment and outcomes, where inverse propensity weights are used as balancing weights and are known by design.

\begin{theorem}[One-sided loss-based criterion]\label{thm:loss_criterion}
Let \autoref{asm:overlap} hold.
Let $H_0$ denote the null hypothesis \eqref{eq:null_independence} of conditional independence and $H_1$ its alternative, that is, the presence of \emph{some} dependence.
Then $H_0$ and $H_1$ are equivalent to
\begin{align*}
    H_0:~& L(f_w) = \tfrac{1}{2}, &
    H_1:~& L(f_w) < \tfrac{1}{2}.
\end{align*}

Proof in \arefsub{app:proof_loss_criterion}.
\end{theorem}

The proof of \autoref{thm:loss_criterion} formally shows that the proposed measure of conditional independence represents the inverse propensity weighted variance of the realized treatment. Intuitively, this corresponds to the variance of the treatment assignment that would be observed after implementing a hypothetical randomized experiment, where the inverse propensity weights adjust for the fact that treatment assignment in the data is not random.

\subsection{Construction of the Test}

\autoref{thm:loss_criterion} motivates a one-sided test based on predictability of treatment $D_i$ in the weighted sample:
fit a function that predicts $D_i$ from $X_i$ and $Y_i$ after weighting by the (known) weights $W_i$, then check whether treatment is predicted better than the trivial benchmark of no predictability.
For constructing our test procedure, we begin by partitioning the sample into two subsets, \(\mathcal{I}\) and \(\mathcal{I}^c\). We train the model for the conditional treatment probability using observations in \(\mathcal{I}^c\), and evaluate it by the sample loss on \(i \in \mathcal{I}\) as our test statistic. To approximate the null distribution of this test statistic, we repeatedly generate new treatment assignments on \(\mathcal{I}\) by sampling them from the known propensity scores. In each case, we compute the resampled value of the statistic using the fixed predicted treatment probability and the resampled treatment assignments on \(\mathcal{I}\). If the observed test statistic falls sufficiently below the resampled distribution, we reject the null hypothesis, concluding that the outcomes contain predictive information about treatment assignment. \autoref{alg:crt} details the implementation.

\begin{algorithm}[htbp]
\caption{Conditional Randomization Test}\label{alg:crt}
\begin{algorithmic}[1]
\setlength{\itemsep}{0.3pt}
\setlength{\parskip}{0.3pt}

\State Input: Data $\{(Y_i, D_i, X_i)\}_{i=1}^N$, known propensity scores $e_0(X_i)$, the number of resampling draws $B$.

\State Randomly split the data into two non-overlapping samples
\(\mathcal{I}\) and \(\mathcal{I}^c\), containing \(n\) and \(N-n\)
observations, respectively. 

\State Compute
\begin{align*}
S_n := \frac{1}{n}\sum_{i \in \mathcal{I}}\left[\bigg(\frac{D_i}{e_0(X_i)} + \frac{1 - D_i}{1 - e_0(X_i)}\bigg)\big(D_i - \hat{f}^{\mathcal{I}^c}_w(Y_i, X_i)\big)^2\right],
\end{align*}
where \(\hat{f}^{\mathcal{I}^c}_w(Y_i, X_i)\) is trained on \(\mathcal{I}^c\) using (known) inverse propensity weights.

\State  For $b = 1, \ldots, B$:
\begin{itemize}[noitemsep, topsep=0pt]
    \item[(a)] Sample $D_i^{(b)} \sim \text{Bernoulli}(e_0(X_i))$ for $i \in \mathcal{I}$.
    \item[(b)] Compute the resampled statistic:
\begin{align*}
S_n^{(b)} := \frac{1}{n}\sum_{i \in \mathcal{I}}\left[\bigg(\frac{D_i^{(b)}}{e_0(X_i)} + \frac{1 - D_i^{(b)}}{1 - e_0(X_i)}\bigg)\big(D_i^{(b)} - \hat{f}^{\mathcal{I}^c}_w(Y_i, X_i)\big)^2\right].
\end{align*}

\end{itemize}

\State Obtain the p-value $p = \frac{1 + \sum_{b=1}^B \mathbf{1}\{S_n^{(b)} \leq S_n\}}{1 + B}$, where \(\mathbf{1}\{.\}\) is a binary indicator.

\State  Reject the null hypothesis of conditional independence if \(p \leq \alpha\) for a prespecified significance level \(\alpha\).
\end{algorithmic}
\end{algorithm}

\subsection{Properties of the Test}

We now analyze the test in two ways.
First, we establish validity.
Indeed, \autoref{prop:valid_test} shows that the test procedure described in \autoref{alg:crt} achieves finite-sample control of the rejection probability under the null hypothesis.

\begin{proposition}[Finite-sample validity]\label{prop:valid_test}
Let \autoref{asm:overlap} hold, and the sample split \(\mathcal{S} = (\mathcal{I}, \mathcal{I}^c)\) be independent of the data \((Y_i, D_i, X_i)_{i=1}^N\). If the null hypothesis of conditional independence holds, $D_i \perp Y_i |X_i$ a.s., then the test in \autoref{alg:crt} controls size. Specifically, for any sample size $n$ of the evaluation set \(\mathcal{I}\) and any number \(B\) of resampling draws,
\[
\P(p \leq \alpha) \leq \alpha \quad \text{for all } \alpha \in [0,1].
\]
Proof in \arefsub{app:proof_valid_test}.
\end{proposition}

Second, we study the asymptotic distribution of the test statistic.
Our above testing procedure does \emph{not} depend on its large-sample properties since validity is guaranteed in finite samples based on resampling.
Note also that this validity does not depend on the specific test statistic we use.
However, it may be interesting to describe the statistical properties of this loss function in order to understand how it behaves and to see how this approach could or could not be extended to cases where the propensity score is not known.

The test statistic of the loss-based test is \begin{align}\label{eq:extracted_loss}
S_n = \frac{1}{n}\sum_{i \in \mathcal{I}}\left[\bigg(\frac{D_i}{e_0(X_i)} + \frac{1 - D_i}{1 - e_0(X_i)}\bigg)\big(D_i - \hat{f}^{\mathcal{I}^c}_w(Y_i, X_i)\big)^2\right].
\end{align}
Here, \(\hat{f}^{\mathcal{I}^c}_w(Y_i, X_i)\) is a conditional treatment probability estimated with a flexible learner on the auxiliary sample \(\mathcal{I}^c\). 
This test statistic is a sample analogue of the minimized loss.
Since \(\hat{f}^{\mathcal{I}^c}_w(Y_i, X_i)\) is estimated, the test statistic may not exactly reach its theoretical minimum even under the null hypothesis.
In order to derive its asymptotic distribution, we therefore assume that, under the null hypothesis, the estimated inverse propensity weighted predicted treatment probability converges to the true value sufficiently fast.

\begin{assumption}[Convergence of the conditional treatment probability]\label{asm:ipw_rate}
When treatment is independent of outcomes \(D_i \perp Y_i |X_i\), we assume
\begin{align*}
\left\|\hat{f}^{\mathcal{I}^c}_w - \tfrac{1}{2}\right\|_{L_2} = o_p\big(n^{-\frac{1}{4}}\big),
\end{align*}
where
\(
\left\|\hat{f}^{\mathcal{I}^c}_w - \tfrac{1}{2}\right\|_{L_2} = \sqrt{\E \left[\left(\hat{f}^{\mathcal{I}^c}_w(Y_i, X_i) - \frac{1}{2}\right)^2 \middle| \mathcal{I}^c \right]}
\)
is the population $L_2$-norm of the estimated inverse propensity weighted conditional treatment probability under the null hypothesis.%
\footnote{The expectation is taken over a new, independent draw of the test point \((Y_i, X_i)\) for a given realization of \(\mathcal{I}^c.\)}
\end{assumption}

\autoref{asm:ipw_rate} states that the distance between the inverse propensity weighted conditional treatment probability (estimated on an auxiliary sample) and the oracle value under the null hypothesis must converge to zero at a rate strictly faster than \(n^{-\frac{1}{4}}\). Here,  \(o_p(.)\) denotes the convergence of the sequence in probability. Let the estimator of the test statistic, evaluated at the true function \(f_w(Y_i, X_i)\), be denoted as:
\begin{align*}
    S^\star = \frac{1}{n}\sum_{i \in \mathcal{I}}\left[\bigg(\frac{D_i}{e_0(X_i)} + \frac{1 - D_i}{1 - e_0(X_i)}\bigg)\big(D_i - f_w(Y_i, X_i)\big)^2\right].
\end{align*}
Then \autoref{prop:asymptotics_ipw} shows that the sample analogue of the test statistic, centered at its true value, is asymptotically normally distributed.

\begin{proposition}[Asymptotic normality]\label{prop:asymptotics_ipw}
Let \autoref{asm:overlap} and \autoref{asm:ipw_rate} hold. Then under the null hypothesis of conditional independence, i.e. \(D_i \perp Y_i |X_i\), we have that
\[
\sqrt{n}\!\left(S_n - \tfrac{1}{2}\right) \xrightarrow{d} \N\!\left(0, \tfrac{1}{16}\,\Var(W_i)\right).
\]
Proof in \arefsub{app:proof_asymptotics_ipw}.
\end{proposition}

The proof of \autoref{prop:asymptotics_ipw} also shows that the loss-based test is Neyman orthogonal with respect to \(f_w(Y_i, X_i)\) under the null hypothesis when treatment is independent of outcomes conditional on covariates, but not with respect to the propensity score.
Therefore, \autoref{asm:ipw_rate} is sufficient when the propensity scores are known, and we may use a flexible nonparametric learner to estimate this conditional treatment probability with known inverse propensity weights. Under the alternative hypothesis, when treatment is conditionally dependent on (some) outcomes given covariates, any additional information from outcomes helps us reject the null hypothesis, as long as this information dominates the bias and variance of the estimated \(f_w(Y_i, X_i)\) in \eqref{eq:extracted_loss}. The proof of \autoref{prop:asymptotics_ipw} also shows that if the propensity score is unknown and must be estimated from the data, \autoref{asm:ipw_rate} is typically not sufficient, as the test statistic is not Neyman orthogonal with respect to the propensity score.
As a consequence, we cannot directly extend the test based on the loss-function statistic to the case of estimated propensity scores.

\section{Robust Tests Based on the Estimated Propensity Score}

We now consider the case of observational data where the propensity score is not known and instead has to be estimated.
In this case, simply replacing known propensity scores with estimated propensities and using the same test as in the previous section does not guarantee a valid test.
Indeed, the previous results cease to apply even under relatively small mistakes in the estimation of the propensity score.
We therefore instead construct three test statistics based on efficient, Neyman-orthogonal, doubly robust scores that measure the predictability of treatment \(D_i\) from outcomes \(Y_i\) conditional on \(X_i\), and thus translate into a test of conditional independence.

\subsection{Motivation and Identification}

Our approach to testing for conditional independence is based on the following idea: \autoref{prop:reverse_prediction} shows that the null hypothesis is equivalent to our inability to predict treatment from outcomes better than by the controls alone.
The true prediction $f_0(Y_i,X_i) = \P(D_i=1|Y_i,X_i)$ of treatment from outcomes $Y_i$ and controls $X_i$ is therefore uncorrelated with treatment $D_i$ conditional on $X_i$ if and only if conditional independence holds.
Any evidence of $f_0(Y_i,X_i)$ being correlated with $D_i$ after controlling for $X_i$ is therefore an indication of a violation of the null hypothesis.

\paragraph{Augmented inverse propensity weighting (AIPW).}

Our first measure of conditional independence between treatment and a set of outcomes is based on the augmented inverse propensity weighted estimator (AIPW), proposed by \citet*{robins1994estimation, robins1995semiparametric}.
Our strategy is to estimate the effect of treatment $D_i$ on the pseudo-outcome $f_0(Y_i,X_i)$, which plays the role of the $\aipw$ outcome variable.
It is based on the following debiased representation of the pseudo-treatment effect:
\begin{align}
    &\theta_0
    =
    \E\left[\E[f_0(Y_i, X_i) |X_i, D_i{=}1] - \E[f_0(Y_i, X_i) |X_i, D_i{=}0]\right]
    \label{eq:pseudote}
    \\
    &=
    \E\left[\mu_1(X_i) - \mu_0(X_i) + \frac{D_i(f_0(Y_i, X_i) - \mu_1(X_i))}{e_0(X_i)} - \frac{(1-D_i)(f_0(Y_i, X_i) - \mu_0(X_i))}{1 - e_0(X_i)} \right].
    \nonumber
\end{align}
Here, $\mu_1(X_i) = \E[f_0(Y_i, X_i) |D_i{=}1, X_i]$ and $\mu_0(X_i) = \E[f_0(Y_i, X_i) |D_i{=}0, X_i]$ are predictions of the pseudo-outcome $f_0(Y_i,X_i)$ for the treated and control groups, respectively.
Hence, we can write $\theta_0$
as a solution to the moment equation $\E[\psi_i^{\aipw}(\theta_0; e_0, \mu_0, \mu_1)] = 0$
using the $\aipw$ score
\begin{equation}\label{eq:score_aipw}
\begin{aligned}
\psi_i^{\aipw}(\theta_0; e_0, \mu_0, \mu_1) &= \mu_1(X_i) {-} \mu_0(X_i)
\\
&\phantom{=}
+ \frac{D_i(f_0(Y_i, X_i) {-} \mu_1(X_i))}{e_0(X_i)} {-} \frac{(1{-}D_i)(f_0(Y_i, X_i) {-} \mu_0(X_i))}{1 {-} e_0(X_i)} - \theta_0.
\end{aligned}
\end{equation}

The score consists of two components: the outcome model \(\mu_d(X_i)\) for \(d \in \{0, 1\}\), and the inverse propensity weighted residual for each treatment group.  \citet*{cassel1976some} propose a similar adjustment term when the outcome model is linearly additive in parameters. \citet*{kang2007demystifying} summarize the development of inverse propensity weighted estimators. Intuitively, if the outcome model \(\mu_d(X_i)\) is correct, then the second term is mean zero noise, and if the outcome model is misspecified, the second term corrects  for the bias present in the outcome model. Therefore, the augmented inverse propensity weighted estimator is doubly robust.

We can now reformulate the null hypothesis as
\begin{align*}
H_0:\
\theta_0 = \E\left[\mu_1(X_i) {-} \mu_0(X_i) {+} \frac{D_i(f_0(Y_i, X_i) {-} \mu_1(X_i))}{e_0(X_i)} {-} \frac{(1{-}D_i)(f_0(Y_i, X_i) {-} \mu_0(X_i))}{1 {-} e_0(X_i)}\right] = 0.
\end{align*}
Under the null hypothesis, \(f_0(Y_i, X_i) = e_0(X_i) = \mu_0(X_i) = \mu_1(X_i)\) pointwise, and the score is degenerate.

\begin{theorem}[One-sided effect criterion, $\aipw$]\label{thm:one_sided_aipw}
Let \autoref{asm:overlap} hold. Then
\begin{align*}
&\E\!\left[\mu_1(X_i) {-} \mu_0(X_i) {+} \frac{D_i(f_0(Y_i, X_i) {-} \mu_1(X_i))}{e_0(X_i)} {-} \frac{(1{-}D_i)(f_0(Y_i, X_i) {-} \mu_0(X_i))}{1 {-} e_0(X_i)} \;\middle|\; X_i\right]
\\
&=\frac{\Var\!\bigl(f_0(Y_i, X_i)|X_i\bigr)}{e_0(X_i)(1-e_0(X_i))}
\;\ge 0
\end{align*}
a.s., with equality almost everywhere if and only if the null hypothesis of conditional independence holds. In particular, $\theta_0 \ge 0$, so the alternative hypothesis is one-sided, $H_1: \theta_0 > 0$.
Proof in \arefsub{app:proof_one_sided_aipw}.
\end{theorem}

\paragraph{Double residualization (DR).}

Our second measure of conditional independence between treatment assignment and outcome variables is based on the double-residualization approach, motivated by the semiparametric statistics literature \citep*{robinson1988root, robins1995semiparametric, rotnitzky1998semiparametric, chernozhukov2018double, chernozhukov2022locally, nie2021quasi}.  The double-residualized score arises naturally within the R-learner framework, which traces back to Robinson’s partialling-out approach \citep*{robinson1988root} and is generalized by \citet*{nie2021quasi} to nonparametric treatment effect estimation. In this formulation, we test whether $f_0(Y_i,X_i)$ is correlated with $D_i$ conditional on $X_i$ by residualizing both components with respect to $X_i$ and testing for correlation in the remainders.

Recall that $e_0(X_i)=\P(D_i=1|X_i)$ is the propensity score, and let $m_0(X_i) := \E[f_0(Y_i, X_i)|X_i]$ be the conditional expectation of the predicted treatment $f_0(Y_i,X_i)$ given $X_i$
(we have that $m_0(X_i) = e_0(X_i)$; we keep $m_0(X_i)$ separate from $e_0(X_i)$ both because the construction still applies for different, potentially suboptimal choices of prediction function and because we will later estimate $e_0(X_i)$, $f_0(Y_i, X_i)$, and $m_0(X_i)$ separately).
We treat $f_0(Y_i,X_i)$ as a pseudo-outcome and estimate the effect of $D_i$ on it by regressing pseudo-outcome residuals on treatment residuals.
As \citet*{nie2021quasi} show, the effect conditional on $X_i$ can be written in terms of the R-loss as
\begin{align*}
    \theta_0(X_i)
    &:=
    \E[f_0(Y_i, X_i) |X_i, D_i{=}1] - \E[f_0(Y_i, X_i) |X_i, D_i{=}0]
    \\
    &=
    \underset{\theta(X_i)}{\arg \min} \  \E\left[\left(f_0(Y_i, X_i) - m_0(X_i) - \theta(X_i) (D_i - e_0(X_i))\right)^2\middle|X_i\right].
\end{align*}
\citet*{foster2023orthogonal} show that the R-loss is Neyman orthogonal: the expected derivative of the loss with respect to the nuisance parameters $(e_0(X_i), m_0(X_i))$, evaluated at their true values, is zero, so estimation errors in the nuisance parameters have only a second-order effect.
By taking the first-order condition with respect to $\theta(X_i)$, we obtain the representation
\begin{align}\label{eq:conditional_pseudo_te}
    \theta_0(X_i)
    &=
    \E\left[\frac{D_i - e_0(X_i)}{e_0(X_i) (1-e_0(X_i))}(f_0(Y_i, X_i) - m_0(X_i)) \middle| X_i\right]
\end{align}
for the conditional effect of $D_i$ on $f_0(Y_i,X_i)$.
Under the null hypothesis $D_i \perp Y_i |X_i$, we have $f_0(Y_i, X_i) = e_0(X_i) = m_0(X_i)$ a.s., so the pseudo-outcome residual is zero and $\theta_0(X_i) = 0$ almost surely.
Analogously to \eqref{eq:pseudote}, this representation of conditional average treatment effects suggests an expression for the overall effect $\theta_0 = \E[\theta_0(X_i)]$.
Specifically, we have that
\begin{align*}
    \theta_0 = \E\left[\frac{D_i - e_0(X_i)}{e_0(X_i) (1-e_0(X_i))}(f_0(Y_i, X_i) - m_0(X_i))\right].
\end{align*}
This expression corresponds to the solution to the moment condition \(\E\big[\psi_i^{\dr}(\theta_0; e_0, m_0)\big] = 0\)
with the $\dr$-based score
\begin{align}\label{eq:score_function}
    \psi_i^{\dr}(\theta_0; e_0, m_0) = \frac{D_i - e_0(X_i)}{e_0(X_i)(1 - e_0(X_i))} (f_0(Y_i, X_i) - m_0(X_i)) - \theta_0.
\end{align}

This score function \eqref{eq:score_function} has the advantage over the $\aipw$ score function \eqref{eq:score_aipw} that it only requires the simple outcome prediction $m_0(X_i)$ as a nuisance parameter (alongside the propensity score $e_0(X_i)$), rather than separate predictions $\mu_1(X_i)$ and $\mu_0(X_i)$.
However, unlike the $\aipw$ score function this score function is not Neyman orthogonal with respect to the propensity score.
Specifically, its expected derivative with respect to the propensity score is generally only zero under the null hypothesis, in which case the two scores agree.
Indeed, the scores are equivalent for
\[m_0(X_i) = (1-e_0(X_i))\,\mu_1(X_i) + e_0(X_i)\,\mu_0(X_i),\]
which is not generally the same as $e_0(X_i)$.
That said, \autoref{thm:one_sided_dr} shows that the double-residualization approach yields a valid test because of the orthogonality under the null hypothesis.

\begin{theorem}[One-sided effect criterion, $\dr$]\label{thm:one_sided_dr}
Let \autoref{asm:overlap} hold. Then
\begin{align*}
&\frac{\E[(D_i {-} e_0(X_i))(f_0(Y_i, X_i) {-} m_0(X_i))|X_i]}{e_0(X_i)(1 {-} e_0(X_i))} = \frac{\Var\!\bigl(f_0(Y_i, X_i)|X_i\bigr)}{e_0(X_i)(1-e_0(X_i))}
\;\ge 0
\end{align*}
a.s., with equality almost everywhere if and only if the null hypothesis of conditional independence holds. In particular, $\theta_0 \ge 0$, so the alternative hypothesis is one-sided, $H_1: \theta_0 > 0$.
Proof in \arefsub{app:proof_one_sided_aipw}.
\end{theorem}

\paragraph{Direct implementation of double residualization ($\drs$).}

In addition to the non-orthogonal score in \eqref{eq:score_function},
we also directly apply the double-residualization approach to obtain an overall measure of the conditional relationship between $f_0(Y_i,X_i)$ and $D_i$ by the residual-on-residual regression
\begin{align*}
    \theta^\star_0 &:= \underset{\theta}{\arg \min} \  \E\left[\left(f_0(Y_i, X_i) - m_0(X_i) - \theta (D_i - e_0(X_i))\right)^2\right]
    \\
    &= \frac{\E\left[(D_i - e_0(X_i))(f_0(Y_i, X_i) - m_0(X_i))\right]}{\E\left[e_0(X_i) \: (1 - e_0(X_i))\right]},
\end{align*}
which is the unconditional equivalent of the R-loss \eqref{eq:conditional_pseudo_te} from \cite{nie2021quasi}.
By taking the first-order condition, we see that $\theta^\star_0$
solves the moment condition
\(
    \E\big[\psi_i^{\drs}(\theta_0^\star; e_0, m_0)\big] = 0
\)
with the Neyman-orthogonal score
\begin{align}\label{eq:score_drs}
    \psi_i^{\drs}(\theta_0^\star; e_0, m_0) =
    (D_i - e_0(X_i))
    (f_0(Y_i, X_i) - m_0(X_i)) - \theta_0^\star \: (D_i - e_0(X_i))^2.
\end{align}
This score is now debiased/orthogonalized.

Using $\theta_0^\star$ as a measure of conditional independence, we note that we can reformulate our null hypothesis of conditional independence as
\begin{equation*}
\begin{aligned}
H_0:\ &
\theta_0^\star = \E\left[
  \frac{D_i - e_0(X_i)}{\E\!\left[e_0(X_i)(1-e_0(X_i))\right]}
  ( f_0(Y_i, X_i) - m_0(X_i) )
\right] = 0.
\end{aligned}
\end{equation*}
\autoref{thm:one_sided_drs} shows that testing based on this formulation of the null hypothesis is naturally one-sided and that the score is pointwise degenerate under the null hypothesis of conditional independence.

\begin{theorem}[One-sided effect criterion, \(\drs\)]\label{thm:one_sided_drs}
Let \(\P(e_0(X_i) \in (0,1)) > 0\) hold. Then 
\[
\E\left[
  \frac{D_i - e_0(X_i)}{\E\!\left[e_0(X_i)(1-e_0(X_i))\right]}
  ( f_0(Y_i, X_i) - m_0(X_i) )
\middle| X_i \right]
=\frac{\Var\!\bigl(f_0(Y_i, X_i)|X_i\bigr)}{\E\!\left[e_0(X_i)(1-e_0(X_i))\right]}
\;\ge 0,
\]
with a.s.\ equality if and only if the null hypothesis of conditional independence holds. In particular,
$\theta_0^\star \geq 0$, and
the alternative hypothesis is one-sided, $H_1: \theta_0^\star > 0$.
Proof in \arefsub{app:proof_one_sided_aipw}.
\end{theorem}

Note that $\theta_0^\star$ is not generally the same as the (pseudo-)treatment effect $\theta_0$ of $D_i$ on $f_0(Y_i,X_i)$ since the residual-on-residual regression leads instead to a \emph{weighted} effect
\begin{equation*}
    \theta_0^\star = \E\left[\frac{e_0(X_i)(1-e_0(X_i))}{\E\left[e_0(X_i)(1-e_0(X_i))\right]} \, \theta_0(X_i)\right].
\end{equation*}
In the context of testing, this may have two advantages: First, testing based on this estimate avoids extreme weights in parts with limited overlap.
Second and relatedly, this approach allows us to relax the assumption of strong overlap (\autoref{asm:overlap}).
At the same time, a test based on this weighting may be underpowered if the difference in distributions concentrates in the part where propensity scores are close to zero or one.

% -----------------------------------------------------------------------
\subsection{Construction of the Tests}
% -----------------------------------------------------------------------

We now describe the sample-splitting and estimation procedure used to construct the $\dr$, $\aipw$, and $\drs$ test statistics. The key challenge is that, unlike in the previous section, the propensity score and all other nuisance functions must be estimated from the data. We address this through sample splitting: nuisance functions are estimated on auxiliary samples and the test statistic is evaluated on a held-out sample, ensuring that estimation error in the nuisance functions is independent of the evaluation data.

Specifically, we partition the full sample into three non-overlapping subsamples: a test sample
$\mathcal{I}$ with $n$ observations, as before, and two auxiliary samples
$\mathcal{I}^{c,1}$ and $\mathcal{I}^{c,2}$ containing the remaining $N-n$
observations.%
\footnote{Triple sample splitting is not strictly required, but it may reduce overfitting bias relative to double sample
splitting. Cross-fitting, as recommended by \citet*{chernozhukov2018double},
\citet*{newey2018cross}, and \citet*{wager2024causal}, is often more
statistically efficient. In the implementation below, all nuisance functions
are estimated using honest regression forests as implemented in the
\texttt{grf} package in \texttt{R} \citep*{tibshirani2018package,
athey2019generalized}.}
All nuisance functions are estimated on the auxiliary samples and evaluated on
$\mathcal{I}$, so that the estimation error in the nuisance functions is
independent of the sampling variability in the evaluation data set. The sample counterparts of observation-level score contributions for the augmented inverse propensity weighted (AIPW) and double
residualized (DR) test statistics
are%
\footnote{The conditional treatment probability
$\hat{f}^{\mathcal{I}^{c,1}}(Y_i, X_i)$ is estimated without inverse
propensity weighting.}
\begin{equation*}
\resizebox{\textwidth}{!}{$
\begin{aligned}
\hat{\psi}_i^{\aipw}
&:= \hat{\mu}^{\mathcal{I}^{c,2}}_{1}(X_i) - \hat{\mu}^{\mathcal{I}^{c,2}}_{0}(X_i)
 + \frac{D_i\bigl(\hat{f}^{\mathcal{I}^{c,1}}(Y_i, X_i)
                    - \hat{\mu}^{\mathcal{I}^{c,2}}_{1}(X_i)\bigr)}
        {\hat{e}^{\mathcal{I}^{c,2}}(X_i)}
 - \frac{(1-D_i)\bigl(\hat{f}^{\mathcal{I}^{c,1}}(Y_i, X_i)
                       - \hat{\mu}^{\mathcal{I}^{c,2}}_{0}(X_i)\bigr)}
        {1 - \hat{e}^{\mathcal{I}^{c,2}}(X_i)},\\
\hat{\psi}_i^{\dr}
&:= \frac{D_i - \hat{e}^{\mathcal{I}^{c,2}}(X_i)}
         {\hat{e}^{\mathcal{I}^{c,2}}(X_i)(1 - \hat{e}^{\mathcal{I}^{c,2}}(X_i))}
   \bigl(\hat{f}^{\mathcal{I}^{c,1}}(Y_i, X_i)
         - \hat{m}^{\mathcal{I}^{c,2}}(X_i)\bigr),
\end{aligned}
$}
\end{equation*}
leading to the test statistics
\[
\hat{\theta}^{\aipw} := \frac{1}{n}\sum_{i \in \mathcal{I}} \hat{\psi}_i^{\aipw}
\quad\text{and}\quad
\hat{\theta}^{\dr} := \frac{1}{n}\sum_{i \in \mathcal{I}} \hat{\psi}_i^{\dr},
\]
with associated asymptotic variance estimators
\begin{align*}
\hat{\sigma}^2_{\aipw}
  &:= \frac{1}{n}\sum_{i \in \mathcal{I}}(\hat{\psi}_i^{\aipw} - \hat{\theta}^{\aipw})^2,
&
\hat{\sigma}^2_{\dr}
  &:= \frac{1}{n}\sum_{i \in \mathcal{I}}(\hat{\psi}_i^{\dr} - \hat{\theta}^{\dr})^2.
\end{align*}
The alternative implementation of double residualization ($\drs$) uses the test statistic
\begin{align*}
    \hat{\theta}^{\drs}
    &:= \frac{\frac{1}{n}\sum_{i \in \mathcal{I}} (D_i - \hat{e}^{\mathcal{I}^{c,2}}(X_i))(\hat{f}^{\mathcal{I}^{c,1}}(Y_i, X_i) - \hat{m}^{\mathcal{I}^{c,2}}(X_i))}{\frac{1}{n}\sum_{i \in \mathcal{I}} (D_i - \hat{e}^{\mathcal{I}^{c,2}}(X_i))^2},
\end{align*}
with asymptotic variance estimate
\begin{align*}
    \hat{\sigma}^2_{\drs}
    &:= \frac{\frac{1}{n}\sum_{i \in \mathcal{I}} \bigl((D_i - \hat{e}^{\mathcal{I}^{c,2}}(X_i))(\hat{f}^{\mathcal{I}^{c,1}}(Y_i, X_i) - \hat{m}^{\mathcal{I}^{c,2}}(X_i)) - \hat{\theta}^{\drs} (D_i - \hat{e}^{\mathcal{I}^{c,2}}(X_i))^2\bigr)^2}{\bigl(\frac{1}{n}\sum_{i \in \mathcal{I}} (D_i - \hat{e}^{\mathcal{I}^{c,2}}(X_i))^2\bigr)^2}.
\end{align*}

\noindent The underlying nuisance functions are given by
\begin{alignat*}{3}
\hat{e}^{\mathcal{I}^{c,2}}(X_i)
  &:= \hat{\E}^{\mathcal{I}^{c,2}}[D_i |X_i],
  \quad&&\text{(propensity score),} \\
\hat{f}^{\mathcal{I}^{c,1}}(Y_i, X_i)
  &:= \hat{\E}^{\mathcal{I}^{c,1}}[D_i |Y_i, X_i],
  \quad&&\text{(predicted treatment probability),} \\
\hat{m}^{\mathcal{I}^{c,2}}(X_i)
  &:= \hat{\E}^{\mathcal{I}^{c,2}}
     \bigl[\hat{f}^{\mathcal{I}^{c,1}}(Y_i, X_i) |X_i\bigr],
  \quad&&\text{(marginal mean),} \\
\hat{\mu}^{\mathcal{I}^{c,2}}_{d}(X_i)
  &:= \hat{\E}^{\mathcal{I}^{c,2}}
     \bigl[\hat{f}^{\mathcal{I}^{c,1}}(Y_i, X_i) |D_i = d,\, X_i\bigr],
  \quad&&\text{(conditional means for } d \in \{0,1\}\text{).}
\end{alignat*}

\begin{algorithm}[p]
\caption{$\aipw$, $\dr$, and $\drs$ Tests }
\label{alg:aipw_dr}
\begin{algorithmic}[1]
\Require Data $\{(Y_i, X_i, D_i)\}_{i=1}^{N}$; a random split
$\mathcal{S}=(\mathcal{I},\mathcal{I}^{c,1},\mathcal{I}^{c,2})$
into three non-overlapping samples;
  significance level $0 < \alpha < 1$; jitter constant $\delta > 0$.
\State On $\mathcal{I}^{c,1}$, estimate a pseudo-outcome \(
  \hat{f}^{\mathcal{I}^{c,1}}(Y_i, X_i)
    = \hat{\E}^{\mathcal{I}^{c,1}}[D_i |Y_i, X_i].
\)

\noindent On $\mathcal{I}^{c,2}$, estimate nuisance functions
\begin{align*}
  \hat{e}^{\mathcal{I}^{c,2}}(X_i)
    &= \hat{\E}^{\mathcal{I}^{c,2}}[D_i |X_i], \\
  \hat{m}^{\mathcal{I}^{c,2}}(X_i)
    &= \hat{\E}^{\mathcal{I}^{c,2}}
       \bigl[\hat{f}^{\mathcal{I}^{c,1}}(Y_i, X_i) |X_i\bigr], \\
  \hat{\mu}^{\mathcal{I}^{c,2}}_{d}(X_i)
    &= \hat{\E}^{\mathcal{I}^{c,2}}
       \bigl[\hat{f}^{\mathcal{I}^{c,1}}(Y_i, X_i) |D_i = d,\, X_i\bigr],
       \quad d \in \{0,1\}.
\end{align*}
\State For each $i \in \mathcal{I}$, compute unit score contributions
\begin{equation*}
\resizebox{0.97\textwidth}{!}{$
\begin{aligned}
\hat{\psi}_i^{\aipw}
  &= \hat{\mu}^{\mathcal{I}^{c,2}}_{1}(X_i)
   - \hat{\mu}^{\mathcal{I}^{c,2}}_{0}(X_i)
   + \frac{D_i\,\bigl(
       \hat{f}^{\mathcal{I}^{c,1}}(Y_i, X_i)
       - \hat{\mu}^{\mathcal{I}^{c,2}}_{1}(X_i)\bigr)}
          {\hat{e}^{\mathcal{I}^{c,2}}(X_i)}
   - \frac{(1-D_i)\,\bigl(
       \hat{f}^{\mathcal{I}^{c,1}}(Y_i, X_i)
       - \hat{\mu}^{\mathcal{I}^{c,2}}_{0}(X_i)\bigr)}
          {1 - \hat{e}^{\mathcal{I}^{c,2}}(X_i)}, \\[4pt]
\hat{\psi}_i^{\dr}
  &= \frac{D_i - \hat{e}^{\mathcal{I}^{c,2}}(X_i)}
          {\hat{e}^{\mathcal{I}^{c,2}}(X_i)
           \bigl(1 - \hat{e}^{\mathcal{I}^{c,2}}(X_i)\bigr)}
    \cdot \Bigl(
      \hat{f}^{\mathcal{I}^{c,1}}(Y_i, X_i)
      - \hat{m}^{\mathcal{I}^{c,2}}(X_i)
    \Bigr),
    \\
    \hat{\psi}_i^{\drs,1}
    &= (D_i - \hat{e}^{\mathcal{I}^{c,2}}(X_i))
       (\hat{f}^{\mathcal{I}^{c,1}}(Y_i, X_i) - \hat{m}^{\mathcal{I}^{c,2}}(X_i)),
    \qquad
    \hat{\psi}_i^{\drs,2}
    = (D_i - \hat{e}^{\mathcal{I}^{c,2}}(X_i))^2.
\end{aligned}
$}
\end{equation*}
\State Form the test statistics and variance estimates:
\begin{align*}
  \hat{\theta}^{\aipw} &= \frac{1}{n}\sum_{i\in\mathcal{I}}\hat{\psi}_i^{\aipw},
  &
  \hat{\sigma}^2_{\aipw} &= \frac{1}{n}\sum_{i\in\mathcal{I}}(\hat{\psi}_i^{\aipw} - \hat{\theta}^{\aipw})^2,
  \\
  \hat{\theta}^{\dr} &= \frac{1}{n}\sum_{i\in\mathcal{I}}\hat{\psi}_i^{\dr},
  &
  \hat{\sigma}^2_{\dr} &= \frac{1}{n}\sum_{i\in\mathcal{I}}(\hat{\psi}_i^{\dr} - \hat{\theta}^{\dr})^2,
  \\
  \hat{\theta}^{\drs} &= \frac{\frac{1}{n}\sum_{i\in\mathcal{I}} \hat{\psi}_i^{\drs,1}}{\frac{1}{n}\sum_{i\in\mathcal{I}} \hat{\psi}_i^{\drs,2}},
  &
  \hat{\sigma}^2_{\drs} &= \frac{\frac{1}{n}\sum_{i\in\mathcal{I}} (\hat{\psi}_i^{\drs,1} - \hat{\theta}^{\drs} \hat{\psi}_i^{\drs,2})^2}{(\frac{1}{n}\sum_{i\in\mathcal{I}} \hat{\psi}_i^{\drs,2})^2}.
\end{align*}

\State For each $k \in \{\aipw, \dr, \drs\}$, draw $Z_k \overset{\mathrm{iid}}{\sim} \N(0,1)$
independently of the data and compute jittered test statistics:
\[
  T^k_n
    = \frac{\sqrt{n}\,\hat{\theta}^k + \delta\,Z_k}{\sqrt{\hat{\sigma}^2_k + \delta^2}}.
\]
\State Let $z_{1-\alpha}$ be the $(1-\alpha)$-quantile of
  $\N(0,1)$. Reject $H_0$ in favor of the one-sided alternative if
  $T^k_n > z_{1-\alpha}$ for a given $k \in \{\aipw, \dr, \drs\}$.
\end{algorithmic}
\end{algorithm}

As established by \autoref{thm:one_sided_aipw}, \autoref{thm:one_sided_dr}, and \autoref{thm:one_sided_drs}, under the null hypothesis the oracle $\dr$/$\drs$ and $\aipw$ scores are pointwise zero and their population variances are degenerate. Consequently, if the pseudo-outcome is consistent for the propensity score then the associated variance estimators satisfy $\hat{\sigma}^2_{\dr} \xrightarrow{p} 0$, \(\hat{\sigma}^2_{\drs} \xrightarrow{p} 0\), and $\hat{\sigma}^2_{\aipw} \xrightarrow{p} 0$. This violates the non-degenerate variance condition required for standard asymptotic inference (part (d) of Assumption 3.2 in \citealp*{chernozhukov2018double}), and precludes a direct application of the usual central limit theorem.

To address this degeneracy, let $\delta > 0$ denote a small constant and let
$Z_{\aipw},\, Z_{\dr},\, Z_{\drs} \overset{\mathrm{iid}}{\sim} \N(0,1)$
be standard normal random variables independent of the data. We then define the jittered and studentized test statistics as
\[
T^{\aipw}_n
:= \frac{\sqrt{n}\, \hat{\theta}^{\aipw} {+} \delta\,Z_{\aipw}}
        {\sqrt{\hat{\sigma}^2_{\aipw} {+} \delta^2}},
\quad
T^{\dr}_n
:= \frac{\sqrt{n}\, \hat{\theta}^{\dr} {+} \delta\,Z_{\dr}}
        {\sqrt{\hat{\sigma}^2_{\dr} {+} \delta^2}},
\quad
T^{\drs}_n
:= \frac{\sqrt{n}\, \hat{\theta}^{\drs} {+} \delta\,Z_{\drs}}
        {\sqrt{\hat{\sigma}^2_{\drs} {+} \delta^2}}.
\]

Under the null hypothesis, the tests will be asymptotically centered at zero by construction since the limiting distributions of the $\hat{\theta}^k$ and the jitter variables $Z_k$ have mean zero.
Based on \autoref{thm:one_sided_aipw}, \autoref{thm:one_sided_dr}, and \autoref{thm:one_sided_drs}, we consider one-sided tests.
\autoref{alg:aipw_dr} summarizes the procedure and construction of the tests.

\subsection{Properties of the Tests}

We now establish the asymptotic properties of the $\aipw$, $\dr$, and $\drs$ tests. All three scores are doubly robust under the null hypothesis, and the $\aipw$ also under the alternative: the test statistic is consistent for $\theta_0$ if either the outcome models $\mu_d(X_i)$ or the propensity score $e_0(X_i)$ is correctly specified \citep*{robins1994estimation}. We impose the following assumptions to establish asymptotic normality of the jittered test statistics.

\begin{assumption}[Nuisance convergence rates]\label{asm:nuisance_rates}
Each nuisance parameter estimator converges to its target at a rate strictly faster than $n^{-1/4}$ in root mean squared error.%
\footnote{Here, we define the estimation targets as the conditional means of the fixed prediction function, $\mu_{d, \hat f}(X_i) := \E[\hat{f}^{\mathcal{I}^{c,1}}(Y_i, X_i) \,|\, \mathcal{I}^{c,1}, X_i, D_i {=} d]$ for $d \in \{0, 1\}$, and $m_{\hat f}(X_i) := \E[\hat{f}^{\mathcal{I}^{c,1}}(Y_i, X_i) \,|\, \mathcal{I}^{c,1}, X_i]$. See \arefsub{app:proof_asymptotics} for details.} That is, for each nuisance function $g \in \{e_0, 1/e_0, 1/(1-e_0), \mu_{1, \hat f}, \mu_{0, \hat f}, m_{\hat f}\}$ and its estimator $\hat{g}$,
\(
\bigl\|\hat{g} - g\bigr\|_{L_2} = o_p(n^{-1/4}).
\)
In addition, \(\|\hat{f}^{\mathcal{I}^{c,1}} - f_0\|_{L_2} = o_p(1)\) and $\|\hat{f}^{\mathcal{I}^{c,1}}\|_{L_\infty}$ is a.s.\ bounded, uniformly in $n$.%
\footnote{Here, the norms are defined conditional on the training data $\mathcal{I}^{c,1}, \mathcal{I}^{c,2}$, analogously to \autoref{asm:ipw_rate}. Moreover, under overlap, if the estimated propensity score is also bounded
away from zero and one, for example by trimming, then the convergence rates
for $1/\hat e^{\mathcal I^{c, 2}}$ and $1/(1-\hat e^{\mathcal I^{c, 2}})$ are implied by the corresponding rate for
$\hat e^{\mathcal I^{c, 2}}$.}
\end{assumption}

Under the proposed assumptions, we can establish the asymptotic distribution of the test statistics.

\begin{theorem}[Asymptotic distribution of test statistics]\label{thm:asymptotics_orthogonal}

    Let \autoref{asm:nuisance_rates} hold, and write
    \[
        \theta_{\hat{f}}(X_i) := \E[\hat{f}^{\mathcal{I}^{c,1}}(Y_i, X_i) |\mathcal{I}^{c,1}, X_i, D_i{=}1] - \E[\hat{f}^{\mathcal{I}^{c,1}}(Y_i, X_i) |\mathcal{I}^{c,1}, X_i, D_i{=}0]
    \]
    for the conditional pseudo-treatment effect for a given function $\hat{f} = \hat{f}^{\mathcal{I}^{c,1}}$.
    \begin{enumerate}
        \item If \autoref{asm:overlap} also holds, then
        \[
            \sqrt{n}\bigl(\hat{\theta}^{\aipw} - \theta_{\hat{f}}\bigr) \xrightarrow{d} \N(0, \sigma^2_{\aipw})
        \]
        where $\theta_{\hat{f}} := \E_{X_i}[\theta_{\hat{f}}(X_i)]$ and $\sigma^2_{\aipw}$ is the variance of the oracle $\aipw$ score.

        \item If \(\P(e_0(X_i) \in (0,1)) > 0\), then
        \[
            \sqrt{n}\bigl(\hat{\theta}^{\drs} - \theta_{\hat{f}}^\star\bigr) \xrightarrow{d} \N(0, \sigma^2_{\drs}),
        \]
        where $\theta_{\hat{f}}^\star = \E_{X_i}\!\left[\frac{e_0(X_i)(1-e_0(X_i))}{\E[e_0(X_i)(1-e_0(X_i))]} \, \theta_{\hat{f}}(X_i)\right]$ and $\sigma^2_{\drs}$ is the variance of the oracle $\drs$ score normalized by $(\E[e_0(X_i)(1-e_0(X_i))])^2$.

        \item If also \autoref{asm:overlap} holds and the null hypothesis $D_i \perp Y_i |X_i$ is true, then
        \[
            \sqrt{n}\bigl(\hat{\theta}^{\dr} - \theta_{\hat{f}}\bigr) \xrightarrow{p} 0.
        \]
    \end{enumerate}
Proof in \arefsub{app:proof_asymptotics}.
\end{theorem}

In the theorem, the estimand itself can depend on the data through $\hat{f}^{\mathcal{I}^{c,1}}(Y_i, X_i)$, since the scores are not generally orthogonal with respect to this prediction function.
Since the $\aipw$ and $\drs$ scores are otherwise Neyman orthogonal, and the $\dr$ score is Neyman orthogonal under the null hypothesis, we can establish asymptotic normality based on standard results from the debiased machine-learning literature.
\autoref{thm:asymptotics_orthogonal} follows from Theorems~3.2 and 4.2 of \citet*{wager2024causal} and Theorems~4.1 and 5.1 of \citet*{chernozhukov2018double} (conditional on $\mathcal{I}^{c,1}$).
For all three statistics, the case where the null holds requires special treatment, since the asymptotic variance is then degenerate when the prediction function is consistent for $f_0$; for misspecified prediction functions, the variance is generally positive even under the null (see the discussion following \autoref{cor:asymptotics_test}).

\begin{corollary}[Asymptotic distribution, validity, and consistency of the test]\label{cor:asymptotics_test}

Let \autoref{asm:nuisance_rates} hold, with also $\|\hat{e}^{\mathcal{I}^{c,2}}\|_{L_\infty}, \|\hat{\mu}^{\mathcal{I}^{c,2}}_1\|_{L_\infty}, \|\hat{\mu}^{\mathcal{I}^{c,2}}_{0}\|_{L_\infty}, \|\hat{m}^{\mathcal{I}^{c,2}}\|_{L_\infty}$ a.s.\ bounded, uniformly across $n$, for the nuisance estimators from \autoref{alg:aipw_dr}.
Also assume \autoref{asm:overlap} for the $\aipw$ and $\dr$ tests and $\P(e_0(X_i) \in (0,1)) > 0$ for the $\drs$ test.
Then, the tests are valid and consistent:
Under the null hypothesis of conditional independence,
we have $\theta_{\hat{f}} = 0 = \theta^\star_{\hat{f}}$ a.s.\ and $T_n^{k} \xrightarrow{d} \N(0,1)$, so the size of the tests from \autoref{alg:aipw_dr} converges to $\alpha$.
Under the alternative that there is conditional dependence with positive probability,
we have $\theta_{\hat{f}} \xrightarrow{p} \theta_0 > 0, \theta^\star_{\hat{f}} \xrightarrow{p} \theta^\star_0 > 0$,
so the tests reject with probability approaching one.
Furthermore,
\begin{align}
    T_n^{\aipw} - \sqrt{n} \frac{\theta_{\hat{f}}}{\sqrt{\hat{\sigma}^2_{\aipw} + \delta^2}}
    &\xlongrightarrow{d} \N(0,1),
    &
    T_n^{\drs} - \sqrt{n} \frac{\theta_{\hat{f}}^\star}{\sqrt{\hat{\sigma}^2_{\drs} + \delta^2}}
    &\xlongrightarrow{d} \N(0,1).
    \label{eq:jittered_clt}
\end{align}
Proof in \arefsub{app:proof_corollary}.
\end{corollary}

For any fixed significance level \(0 < \alpha < 1\), the procedure that rejects \(H_0\) when \(T^k_n > z_{1-\alpha}\) for \(k \in \{\aipw, \dr, \drs\}\) attains asymptotic size \(\alpha\). The choice of $\delta$ does not affect the asymptotic size of the resulting tests; under fixed alternatives, asymptotic power still tends to one. It can, however, affect size and power of the test statistics in finite samples or under a local alternative, and $\delta$ should therefore be chosen to be small. 

The statements of \autoref{thm:asymptotics_orthogonal} and \autoref{cor:asymptotics_test} assume that the prediction $\hat{f}^{\mathcal{I}^{c,1}}$ of $D_i$ given $Y_i$ and $X_i$ is consistent for the true
$f_0(Y_i, X_i) = \P(D_i {=} 1 | Y_i, X_i)$, in the sense that $\|\hat{f}^{\mathcal{I}^{c,1}} - f_0\|_{L_2} = o_p(1)$.
In practice, the prediction function may be misspecified, and only converge to a (potentially simpler) function $\tilde{f}_0$.
The proofs show that if we replace $f_0$ by $\tilde{f}_0$ in \autoref{thm:asymptotics_orthogonal} and in \autoref{cor:asymptotics_test}, and assume that $\|\hat{f}^{\mathcal{I}^{c,1}} - \tilde{f}_0\|_{L_2} = o_p(1)$, then the results in \autoref{thm:asymptotics_orthogonal} still go through, with asymptotic variances depending on $\tilde{f}_0$.
For the corollary, \eqref{eq:jittered_clt} still applies in general, and the test still has correct size.
Under the alternative, the tests only reject with probability approaching one if the effects of $D_i$ on $\tilde{f}_0(Y_i, X_i)$,
\begin{align*}
    \theta_{\tilde{f}_0}
    &= \E\Bigl[\E[\tilde{f}_0(Y_i, X_i) \,|\, X_i, D_i {=} 1] - \E[\tilde{f}_0(Y_i, X_i) \,|\, X_i, D_i {=} 0]\Bigr],\\
    \theta^\star_{\tilde{f}_0}
    &= \E\biggl[\frac{e_0(X_i)(1{-}e_0(X_i))}{\E[e_0(X_i)(1{-}e_0(X_i))]}
        \Bigl(\E[\tilde{f}_0(Y_i, X_i) \,|\, X_i, D_i {=} 1] - \E[\tilde{f}_0(Y_i, X_i) \,|\, X_i, D_i {=} 0]\Bigr)\biggr]
\end{align*}
are strictly positive for the $\aipw$/$\dr$ and $\drs$ tests, respectively, which is not generally guaranteed for arbitrary $\tilde{f}_0$.
Hence, misspecification affects power, but not size.
Indeed, under the null hypothesis of conditional independence, $\E[\tilde{f}_0(Y_i, X_i)|X_i, D_i{=}d]$ does not depend on $d$, so that $\theta_{\tilde{f}_0} = 0 = \theta^\star_{\tilde{f}_0}$ for any $\tilde{f}_0$; since the jittered test statistics are asymptotically standard normal whether or not the asymptotic variance is degenerate, the tests retain asymptotic size $\alpha$ under misspecification.

\section{Monte Carlo Simulations}\label{sec:simulations}

We conduct Monte Carlo experiments to examine finite-sample properties of the
proposed tests. In each experiment, we draw samples of sizes
$\{500,\,1000,\,2000\}$. Each observation is characterized by a vector of covariates:
\[
\begin{aligned}
X_{i1} &\sim \text{Discrete Uniform}\{18,19,\dots,65\} \\
X_{i2} &\sim \text{Bernoulli}(0.5) \\
X_{i3} &\sim \text{Discrete Uniform}\{1,2,3,4\} \\
X_{i4} &\sim \text{Bernoulli}(0.8)
\end{aligned}
\]
These covariates are then standardized to have mean zero and standard deviation
one. Binary treatment is assigned probabilistically according to the
propensity score, winsorized to $[0.1,0.9]$,\footnote{While the implementation winsorizes the propensity score to $[0.1, 0.9]$ to ensure overlap under an arbitrary DGP, the operation does not bind in this particular design: since the covariates are bounded, $e_0(X_i) \in [0.28, 0.72]$, so the winsorization leaves every propensity score unchanged.}
\begin{align*}
e_0(X_i) &= \min\!\left(\max\!\left(0.5 + 0.1\,X_{i1} + 0.05\,X_{i2},\,0.1\right),\,0.9\right),\\
D_i &\sim \mathrm{Bernoulli}\!\left(e_0(X_i)\right).
\end{align*}

\subsection{Treatment Effect Decay}\label{subsection_decay}

An $M$-dimensional vector of outcomes $Y_i$ with $M \in \{5, 10, 30\}$
is generated according to the data-generating process
\begin{align}\label{eq:dgp_decay}
Y_{im} &= \mu(X_i) + D_i\,\tau(X_i)\,\frac{\kappa}{\sqrt{m}} + \varepsilon_{im},\\
\varepsilon_i &\sim \N(0,\Sigma),\nonumber
\end{align}
where \(\mu(X_i)\) denotes the baseline, common across outcomes, and \(\tau(X_i)\) denotes the conditional treatment effect. \(\varepsilon_{im}\) represents the error term.\footnote{The error terms \(\varepsilon_{im}\) share a fixed $M\times M$ covariance matrix $\Sigma$ with
unit variances and off-diagonal correlations that decrease with the distance
between outcome indices:
\[
\Sigma_{jk}=
\begin{cases}
1, & j=k,\\[4pt]
\max\!\bigl(0,\ \rho_0-(|j-k|-1)\,\delta\bigr), & j\neq k,
\end{cases}
\qquad j,k=1,\dots,M,
\]
with $\rho_0=0.5$ and $\delta=0.1$. By construction $\Sigma$ is symmetric, and
under this parameterization positive-definite for every $M$ used in the
simulations. The errors are correlated across outcomes, with correlation
decaying linearly to zero once $|j-k|\ge 6$.} The factor $m^{-1/2}$ introduces
treatment effect decay across outcomes and thereby heterogeneity across outcome dimensions, and
$\tau(X_i)$ governs covariate-driven heterogeneity in the conditional
treatment effect.
We consider the following models for $\mu$ and $\tau$:

\begin{equation}
\label{eq:designs}
\begin{aligned}
\mu(X_i) &= 0.5 X_{i1} + 0.3 X_{i2} + 0.2 X_{i3} + 0.1 X_{i4}, \\
\tau(X_i) &=  X_{i4}^2
&& \text{(Full Heterogeneity)} \\[0.8em]
\mu(X_i) &= 0, \\
\tau(X_i) &=  \bigl(0.3 X_{i1} + 0.1 X_{i3}\bigr)
&& \text{(No Baseline)} \\[0.8em]
\mu(X_i) &= 0.5 X_{i1} + 0.3 X_{i2} + 0.2 X_{i3} + 0.1 X_{i4}, \\
\tau(X_i) &= 1.5 \cdot \mathbf{1}\{X_{i3} \le 0\} - 1.5 \cdot \mathbf{1}\{X_{i3} > 0\}
&& \text{(Mixtures)} \\[0.8em]
\mu(X_i) &= 0.5 X_{i1} + 0.3 X_{i2} + 0.2 X_{i3} + 0.1 X_{i4}, \\
\tau(X_i) &= 0.3\,(X_{i1}X_{i2})^2
&& \text{(Interactions)}
\end{aligned}
\end{equation}

\noindent  In these designs the baseline $\mu(X_i)$ and the
conditional treatment effect $\tau(X_i)$ are held fixed across outcomes, so that
the only source of variation across the outcome dimension is the decay factor
$m^{-1/2}$ in \eqref{eq:dgp_decay}.%
\footnote{We approximate the null distribution of the loss-based (IPW) test statistic by resampling treatment assignments from (known) propensity scores 300 times.  Moreover, the IPW test statistic uses known propensity scores in the loss function. The pseudo-outcome and the nuisance parameters are estimated with the generalized random forest \citep*{tibshirani2018package} using honest sample splitting, 1000 trees, and a minimum leaf size of 10 observations. As a parametric benchmark, we also report a joint Wald test based on a seemingly unrelated regression. Implementation details of the joint Wald test are provided in \autoref{app:wald}.}

The scalar $\kappa\ge 0$ controls the overall strength of treatment
effects and is calibrated, rather than fixed, so that the comparison across
designs is not confounded by differences in the strength of the signal.
Specifically, we define the share of the total additive variance
attributable to the treatment component at the reference outcome $m=1$ as
\begin{align}\label{eq:snr}
\rho
\;:=\;
\frac{\Var\!\bigl(D_i\,\tau(X_i)\,\kappa\bigr)}
     {\Var\!\bigl(D_i\,\tau(X_i)\,\kappa\bigr)+\Var(\mu(X_i))+\bar\sigma^{2}_{\varepsilon}},
\qquad
\bar\sigma^{2}_{\varepsilon}\;:=\;\frac{1}{M}\sum_{m=1}^{M}\Var(\varepsilon_{im}).
\end{align}
Here, \(\bar\sigma^{2}_{\varepsilon} = 1\) since the error variance of each outcome
equals one. We refer to $\rho\in[0,1)$ as the signal-to-noise ratio (SNR). As the
numerator scales as $\kappa^{2}\,\Var(D_i\,\tau(X_i))$, solving \eqref{eq:snr} for
$\kappa$ gives
\begin{equation}\label{eq:kappa}
\kappa(\rho)
\;=\;
\sqrt{\,\frac{\rho}{1-\rho}\cdot
      \frac{\Var(\mu(X_i))+\bar\sigma^{2}_{\varepsilon}}{\Var\!\bigl(D_i\,\tau(X_i)\bigr)}\,},
\qquad \rho\in[0,1).
\end{equation}
We evaluate $\Var(\mu(X_i))$ from the simulated baseline,
$\Var(\varepsilon_{im})$ from the diagonal of $\Sigma$, and
$\Var(D_i\,\tau(X_i))$ from the realized product in the simulated sample. By
construction, at the reference outcome the treatment component accounts for a
share $\rho$ of the total additive variance. For subsequent outcomes $m\ge 2$ the
effective coefficient is $\kappa(\rho)/\sqrt{m}$, so the per-outcome share
$\rho_m$ decreases in $m$.

\subsection{Mixed Heterogeneity}\label{subsection_mixed}

We also consider the Mixed Heterogeneity design
in which the conditional treatment effect and the baseline are themselves allowed
to differ across outcomes:
\begin{align}\label{eq:dgp_mixed}
Y_{im} &= \mu_m(X_i) + D_i\,\tau_m(X_i)\,\kappa_m + \varepsilon_{im}.
\end{align}
Both the baseline $\mu_m(X_i)$ and the conditional treatment effect
$\tau_m(X_i)$ cycle through five functional forms, each governing one-fifth of the
outcomes. The error term follows \(\varepsilon_i \sim \N(0,\Sigma)\) as described in \autoref{subsection_decay}.  Under the null hypothesis, the treatment effect is identically zero,
$\tau_m(X_i)=0$, so heterogeneity enters through the baseline alone.
\begin{equation}
\label{eq:designs_mixed}
\begin{aligned}
\tau_m(X_i) &= \begin{cases}
\text{ (a) } X_{i4}^2 \\[2pt]
\text{ (b) } 0.3 X_{i1} + 0.1 X_{i3} \\[2pt]
\text{ (c) } 1.5\cdot\mathbf{1}\{X_{i3}\le 0\} - 1.5\cdot\mathbf{1}\{X_{i3} > 0\} \\[2pt]
\text{ (d) } 0.3\,(X_{i1}X_{i2})^2 \\[2pt]
\text{ (e) } \sin(1.5\,X_{i1})
\end{cases}
&& \text{(treatment effect)} \\[4pt]
\mu_m(X_i) &= \begin{cases}
\text{ (a) } 0.5 X_{i1} + 0.3 X_{i2} + 0.2 X_{i3} + 0.1 X_{i4} \\[2pt]
\text{ (b) } 0.6\,X_{i3} \\[2pt]
\text{ (c) } 0.4\,X_{i1}^2 \\[2pt]
\text{ (d) } 0.5\,X_{i1} X_{i2} \\[2pt]
\text{ (e) } 0.8\cdot\mathbf{1}\{X_{i3}\le 0\} - 0.8\cdot\mathbf{1}\{X_{i3} > 0\}
\end{cases}
&& \text{(baseline)}.
\end{aligned}
\end{equation}
In the mixed design the scaling is outcome-specific and the $m^{-1/2}$ decay is
suppressed. $\kappa_m$ is calibrated per outcome to attain the target
signal-to-noise ratio $\rho$. Using the per-outcome, total-variance counterpart
of \eqref{eq:kappa} with the outcome-specific baseline $\mu_m(X_i)$,
\begin{equation*}
\kappa_m(\rho)=\sqrt{\,\frac{\rho}{1-\rho}\cdot
\frac{\Var(\mu_m(X_i))+\bar\sigma^{2}_{\varepsilon}}
{\Var(D_i\,\tau_m(X_i))}\,},
\qquad \rho\in[0,1).
\end{equation*}
Additional simulated experiments with varying intercepts,
treatment effects, and error distributions can be found in \autoref{app:sims_additional}.

\subsection{Power under Weak and No Effects}

We simulate two regimes: an absence of treatment effects when
$\rho=0$ (equivalently $\kappa=0$) and weak effects with
$\rho = 0.2$. The value $\rho=0.2$ corresponds to a weak signal-to-noise ratio.\footnote{We use sample splitting to keep estimation independent of inference.
For the debiased tests (AIPW and DR/DR$^\star$), we partition the data randomly into three
disjoint folds of roughly $33\%$ each: the first estimates the propensity score
$\hat e(X_i)$ and pseudo-outcome $\hat f(Y_i,X_i)$, the second trains the outcome
regressions ($\hat\mu_1,\hat\mu_0$ for AIPW, or $\hat m(X_i)$ for DR/DR$^\star$), and
the third, held out from all nuisance estimation, forms the test statistic. For
the loss-based IPW test, where the propensity score is known, an even random $50/50$
split suffices: $50\%$ trains the pseudo-outcome model, and $50\%$ evaluates the
inverse propensity weighted loss against its resampled null. All splits are random and
seeded for reproducibility.}

\autoref{tab:strong_effects_main} reports rejection probabilities under the alternative hypothesis for five tests across multiple designs, each aggregated across 1,000 Monte Carlo replications. \autoref{tab:strong_effects_main} yields three main findings. First, the proposed
conditional independence tests significantly outperform the joint Wald test in
designs where the unconditional treatment effect is small or zero but
covariate-driven heterogeneity remains. Under the Mixtures and No Baseline
specifications, both of which satisfy $\mathbb{E}[\tau(X_i)] \approx 0$, the
Wald test rejects at approximately the nominal level across all sample sizes
and numbers of outcomes (between 4\% and 10\%), whereas the $\aipw$, $\dr$, and
$\drs$ tests retain non-trivial power. At $N=500$, for instance, the rejection probabilities of these tests range from
roughly 12\% to 63\%, whereas the Wald test rejects at approximately its nominal level.
Power rises substantially with the sample size, exceeding 80\% at $N=2000$ across all
specifications except the Mixtures design with $M=30$, in which averaging over a large
number of outcomes attenuates the signal and reduces power to about 40\%.

\begin{table}[H]
\centering
\caption{\label{tab:strong_effects_main} Power of the conditional independence tests under weak effects. The table reports rejection probabilities under the alternative hypothesis for test statistics based on inverse propensity weighting (IPW, \autoref{alg:crt}), augmented inverse propensity weighting (AIPW, \autoref{alg:aipw_dr}), double residualization (DR / \(\drs\), \autoref{alg:aipw_dr}), and a joint Wald test (\autoref{app:wald}). Each cell reports the average over 1,000 Monte Carlo replications.   Signal-to-noise ratio $\rho=0.2$ at the reference outcome $m=1$, held fixed across outcomes in the Mixed Heterogeneity design and decaying in the others.
}
\centering
\resizebox{\ifdim\width>\linewidth\linewidth\else\width\fi}{!}{
\fontsize{9}{11}\selectfont
\begin{tabular}[t]{r>{}rrr>{}rrr>{}rrr>{}rrr>{}rrr}
\toprule
\multicolumn{1}{c}{ } & \multicolumn{3}{c}{IPW} & \multicolumn{3}{c}{$\aipw$} & \multicolumn{3}{c}{$\dr$} & \multicolumn{3}{c}{\(\drs\)} & \multicolumn{3}{c}{Wald} \\
\cmidrule(l{3pt}r{3pt}){2-4} \cmidrule(l{3pt}r{3pt}){5-7} \cmidrule(l{3pt}r{3pt}){8-10} \cmidrule(l{3pt}r{3pt}){11-13} \cmidrule(l{3pt}r{3pt}){14-16}
\multicolumn{1}{c}{ } & \multicolumn{15}{c}{Number of outcomes} \\
\cmidrule(l{3pt}r{3pt}){2-16}
\multirow{2}{*}{\raisebox{2ex}[0pt][0pt]{\makecell[l]{Sample size}}} & 5 & 10 & 30 & 5 & 10 & 30 & 5 & 10 & 30 & 5 & 10 & 30 & 5 & 10 & 30\\
\midrule
\addlinespace[0.8em]
\multicolumn{16}{l}{Full Heterogeneity}\\
\hspace{1em}500 & 0.544 & 0.436 & 0.302 & 0.826 & 0.811 & 0.802 & 0.856 & 0.850 & 0.832 & 0.860 & 0.854 & 0.836 & 0.987 & 0.986 & 0.973\\
\hspace{1em}1000 & 0.949 & 0.847 & 0.641 & 0.995 & 0.977 & 0.976 & 0.998 & 0.987 & 0.983 & 0.998 & 0.989 & 0.981 & 1.000 & 1.000 & 1.000\\
\hspace{1em}2000 & 1.000 & 1.000 & 0.979 & 1.000 & 1.000 & 1.000 & 1.000 & 1.000 & 1.000 & 1.000 & 1.000 & 1.000 & 1.000 & 1.000 & 1.000\\
\midrule
\addlinespace[0.8em]
\multicolumn{16}{l}{No Baseline}\\
\hspace{1em}500 & 0.505 & 0.363 & 0.170 & 0.561 & 0.439 & 0.229 & 0.625 & 0.476 & 0.236 & 0.625 & 0.480 & 0.235 & 0.049 & 0.061 & 0.099\\
\hspace{1em}1000 & 0.957 & 0.887 & 0.542 & 0.984 & 0.954 & 0.801 & 0.987 & 0.961 & 0.817 & 0.985 & 0.954 & 0.811 & 0.055 & 0.046 & 0.062\\
\hspace{1em}2000 & 1.000 & 1.000 & 0.980 & 1.000 & 1.000 & 1.000 & 1.000 & 1.000 & 1.000 & 1.000 & 1.000 & 1.000 & 0.044 & 0.045 & 0.059\\
\midrule
\addlinespace[0.8em]
\multicolumn{16}{l}{Mixtures}\\
\hspace{1em}500 & 0.263 & 0.139 & 0.053 & 0.278 & 0.219 & 0.160 & 0.277 & 0.192 & 0.119 & 0.281 & 0.194 & 0.120 & 0.055 & 0.055 & 0.096\\
\hspace{1em}1000 & 0.793 & 0.460 & 0.115 & 0.687 & 0.475 & 0.211 & 0.715 & 0.470 & 0.172 & 0.727 & 0.476 & 0.177 & 0.046 & 0.058 & 0.055\\
\hspace{1em}2000 & 1.000 & 0.977 & 0.418 & 0.993 & 0.930 & 0.467 & 0.997 & 0.937 & 0.463 & 0.996 & 0.940 & 0.465 & 0.064 & 0.052 & 0.057\\
\midrule
\addlinespace[0.8em]
\multicolumn{16}{l}{Interactions}\\
\hspace{1em}500 & 0.937 & 0.864 & 0.653 & 0.971 & 0.959 & 0.942 & 0.980 & 0.976 & 0.966 & 0.977 & 0.973 & 0.963 & 1.000 & 1.000 & 1.000\\
\hspace{1em}1000 & 0.999 & 1.000 & 0.992 & 1.000 & 1.000 & 1.000 & 1.000 & 1.000 & 1.000 & 1.000 & 1.000 & 1.000 & 1.000 & 1.000 & 1.000\\
\hspace{1em}2000 & 1.000 & 1.000 & 1.000 & 1.000 & 1.000 & 1.000 & 1.000 & 1.000 & 1.000 & 1.000 & 1.000 & 1.000 & 1.000 & 1.000 & \vphantom{1} 1.000\\
\midrule
\addlinespace[0.8em]
\multicolumn{16}{l}{Mixed Heterogeneity}\\
\hspace{1em}500 & 0.997 & 0.999 & 1.000 & 0.991 & 0.999 & 1.000 & 0.997 & 1.000 & 1.000 & 0.998 & 1.000 & 1.000 & 1.000 & 1.000 & 1.000\\
\hspace{1em}1000 & 1.000 & 1.000 & 1.000 & 1.000 & 1.000 & 1.000 & 1.000 & 1.000 & 1.000 & 1.000 & 1.000 & 1.000 & 1.000 & 1.000 & 1.000\\
\hspace{1em}2000 & 1.000 & 1.000 & 1.000 & 1.000 & 1.000 & 1.000 & 1.000 & 1.000 & 1.000 & 1.000 & 1.000 & 1.000 & 1.000 & 1.000 & 1.000\\
\bottomrule
\end{tabular}}
\end{table}

Second, under the small-sample, weak-signal design considered here, the Wald test performs better than the proposed conditional independence tests in the Full Heterogeneity and Interactions designs, in which $\E[\tau(X_i)]\neq 0$ and the marginal mean shift is directly detectable. This reflects the finite-sample variance cost of nonparametric nuisance parameter estimation: when both the sample size and the signal-to-noise ratio are small, slower nuisance function convergence rates translate into noisier score functions and reduced power. Moreover, all methods perform comparably well in the Mixed Heterogeneity design, rejecting with
probability close to one across sample sizes and numbers of outcomes. This reflects the
design's calibration: the signal-to-noise ratio is held fixed across outcomes (the
$m^{-1/2}$ decay is suppressed), so each outcome carries the full target signal rather
than a diminishing share. Combined with the nonzero mean effects, this yields high power for the proposed tests and the joint Wald test alike.

\begin{table}[H]
\centering
\caption{\label{tab:null_effects_main}Size of the conditional independence tests. The table reports rejection probabilities under the null hypothesis for test statistics based on inverse propensity weighting (IPW, \autoref{alg:crt}), augmented inverse propensity weighting (AIPW, \autoref{alg:aipw_dr}), double residualization (DR / \(\drs\), \autoref{alg:aipw_dr}), and a joint Wald test (\autoref{app:wald}). Each cell reports the average over 1,000 Monte Carlo replications.   }
\centering
\resizebox{\ifdim\width>\linewidth\linewidth\else\width\fi}{!}{
\fontsize{9}{11}\selectfont
\begin{tabular}[t]{r>{}rrr>{}rrr>{}rrr>{}rrr>{}rrr}
\toprule
\multicolumn{1}{c}{ } & \multicolumn{3}{c}{IPW} & \multicolumn{3}{c}{$\aipw$} & \multicolumn{3}{c}{$\dr$} & \multicolumn{3}{c}{\(\drs\)} & \multicolumn{3}{c}{Wald} \\
\cmidrule(l{3pt}r{3pt}){2-4} \cmidrule(l{3pt}r{3pt}){5-7} \cmidrule(l{3pt}r{3pt}){8-10} \cmidrule(l{3pt}r{3pt}){11-13} \cmidrule(l{3pt}r{3pt}){14-16}
\multicolumn{1}{c}{ } & \multicolumn{15}{c}{Number of outcomes} \\
\cmidrule(l{3pt}r{3pt}){2-16}
\multirow{2}{*}{\raisebox{2ex}[0pt][0pt]{\makecell[l]{Sample size}}} & 5 & 10 & 30 & 5 & 10 & 30 & 5 & 10 & 30 & 5 & 10 & 30 & 5 & 10 & 30\\
\midrule
\addlinespace[0.8em]
\multicolumn{16}{l}{Full Heterogeneity}\\
\hspace{1em}500 & 0.054 & 0.041 & 0.046 & 0.076 & 0.099 & 0.148 & 0.065 & 0.075 & 0.103 & 0.067 & 0.075 & 0.102 & 0.052 & 0.056 & 0.088\\
\hspace{1em}1000 & 0.064 & 0.041 & 0.044 & 0.079 & 0.073 & 0.120 & 0.061 & 0.062 & 0.082 & 0.065 & 0.064 & 0.084 & 0.053 & 0.055 & 0.075\\
\hspace{1em}2000 & 0.048 & 0.045 & 0.048 & 0.061 & 0.047 & 0.080 & 0.054 & 0.043 & 0.059 & 0.050 & 0.040 & 0.062 & 0.048 & 0.049 & 0.054\\
\midrule
\addlinespace[0.8em]
\multicolumn{16}{l}{No Baseline}\\
\hspace{1em}500 & 0.061 & 0.051 & 0.050 & 0.076 & 0.076 & 0.085 & 0.066 & 0.072 & 0.085 & 0.066 & 0.075 & 0.079 & 0.053 & 0.061 & 0.101\\
\hspace{1em}1000 & 0.050 & 0.043 & 0.055 & 0.058 & 0.053 & 0.055 & 0.048 & 0.054 & 0.062 & 0.051 & 0.056 & 0.064 & 0.063 & 0.053 & 0.062\\
\hspace{1em}2000 & 0.040 & 0.047 & 0.047 & 0.050 & 0.045 & 0.049 & 0.057 & 0.045 & 0.059 & 0.062 & 0.049 & 0.061 & 0.047 & 0.046 & 0.061\\
\midrule
\addlinespace[0.8em]
\multicolumn{16}{l}{Mixtures}\\
\hspace{1em}500 & 0.039 & 0.052 & 0.040 & 0.060 & 0.085 & 0.065 & 0.053 & 0.070 & 0.064 & 0.052 & 0.069 & 0.064 & 0.056 & 0.061 & 0.093\\
\hspace{1em}1000 & 0.057 & 0.040 & 0.053 & 0.054 & 0.052 & 0.066 & 0.056 & 0.055 & 0.070 & 0.055 & 0.055 & 0.073 & 0.049 & 0.058 & 0.053\\
\hspace{1em}2000 & 0.053 & 0.045 & 0.061 & 0.050 & 0.064 & 0.068 & 0.054 & 0.070 & 0.066 & 0.057 & 0.072 & 0.065 & 0.064 & 0.046 & 0.055\\
\midrule
\addlinespace[0.8em]
\multicolumn{16}{l}{Interactions}\\
\hspace{1em}500 & 0.041 & 0.055 & 0.053 & 0.089 & 0.119 & 0.135 & 0.065 & 0.095 & 0.096 & 0.063 & 0.093 & 0.101 & 0.036 & 0.061 & 0.099\\
\hspace{1em}1000 & 0.037 & 0.036 & 0.048 & 0.074 & 0.068 & 0.125 & 0.059 & 0.064 & 0.088 & 0.058 & 0.064 & 0.094 & 0.053 & 0.069 & 0.074\\
\hspace{1em}2000 & 0.033 & 0.043 & 0.053 & 0.061 & 0.063 & 0.075 & 0.058 & 0.057 & 0.068 & 0.056 & 0.057 & 0.069 & 0.061 & 0.058 & 0.053\\
\midrule
\addlinespace[0.8em]
\multicolumn{16}{l}{Mixed Heterogeneity}\\
\hspace{1em}500 & 0.057 & 0.060 & 0.062 & 0.063 & 0.079 & 0.083 & 0.071 & 0.070 & 0.068 & 0.069 & 0.075 & 0.074 & 0.054 & 0.064 & 0.117\\
\hspace{1em}1000 & 0.052 & 0.055 & 0.059 & 0.061 & 0.068 & 0.084 & 0.054 & 0.060 & 0.070 & 0.054 & 0.059 & 0.068 & 0.057 & 0.056 & 0.057\\
\hspace{1em}2000 & 0.043 & 0.053 & 0.052 & 0.048 & 0.049 & 0.067 & 0.052 & 0.046 & 0.069 & 0.053 & 0.044 & 0.069 & 0.038 & 0.043 & 0.052\\
\bottomrule
\end{tabular}}
\end{table}

Third, the $\aipw$, $\dr$, and $\drs$ tests are systematically more powerful and
considerably more robust to the dimension $M$ than the IPW-based test, despite
the deliberately weak signal $\rho=0.2$.  This suggests that the augmentation and double
residualization mitigate the power loss induced by averaging
over a growing number of outcomes. The advantage of the IPW test statistic, however, is apparent under the null hypothesis, as reported in \autoref{tab:null_effects_main}. By \autoref{prop:valid_test}, the IPW statistic controls size in small samples. In contrast, the $\aipw$, $\dr$, and $\drs$ tests can over-reject under the null hypothesis in small samples, because their validity is established only asymptotically. \autoref{tab:strong_effects_app} of \autoref{app:sims_additional} shows that under strong treatment effects, the proposed
conditional independence tests substantially outperform the joint Wald test
across a range of nonlinear specifications of $\tau(X_i)$.

\subsection{Calibrating Signal Strength}
Here, we define the outcome equation without effect decay,
\begin{align*}
Y_{im} \;=\; \mu(X_i) \;+\; D_i\,\tau_m(X_i)\,\kappa_m \;+\; \varepsilon_{im},
\end{align*}
where $\tau_m(X_i)=\tau(X_i)$ and $\kappa_m=\kappa$ for all designs except the
mixed design, in which the conditional treatment effect varies across outcomes.
The baseline $\mu(X_i)$ is held fixed across outcomes in every design, as
specified in \eqref{eq:designs}.
In the mixed design it likewise remains the
common linear form $\mu(X_i) = 0.5\,X_{i1} + 0.3\,X_{i2} + 0.2\,X_{i3} +
0.1\,X_{i4}$, so that outcome-level heterogeneity enters through the treatment
effect $\tau_m(X_i)$ alone.

\begin{figure}[H]
\centering
\begin{subfigure}{0.47\textwidth}
    \centering
    \includegraphics[width=\textwidth]{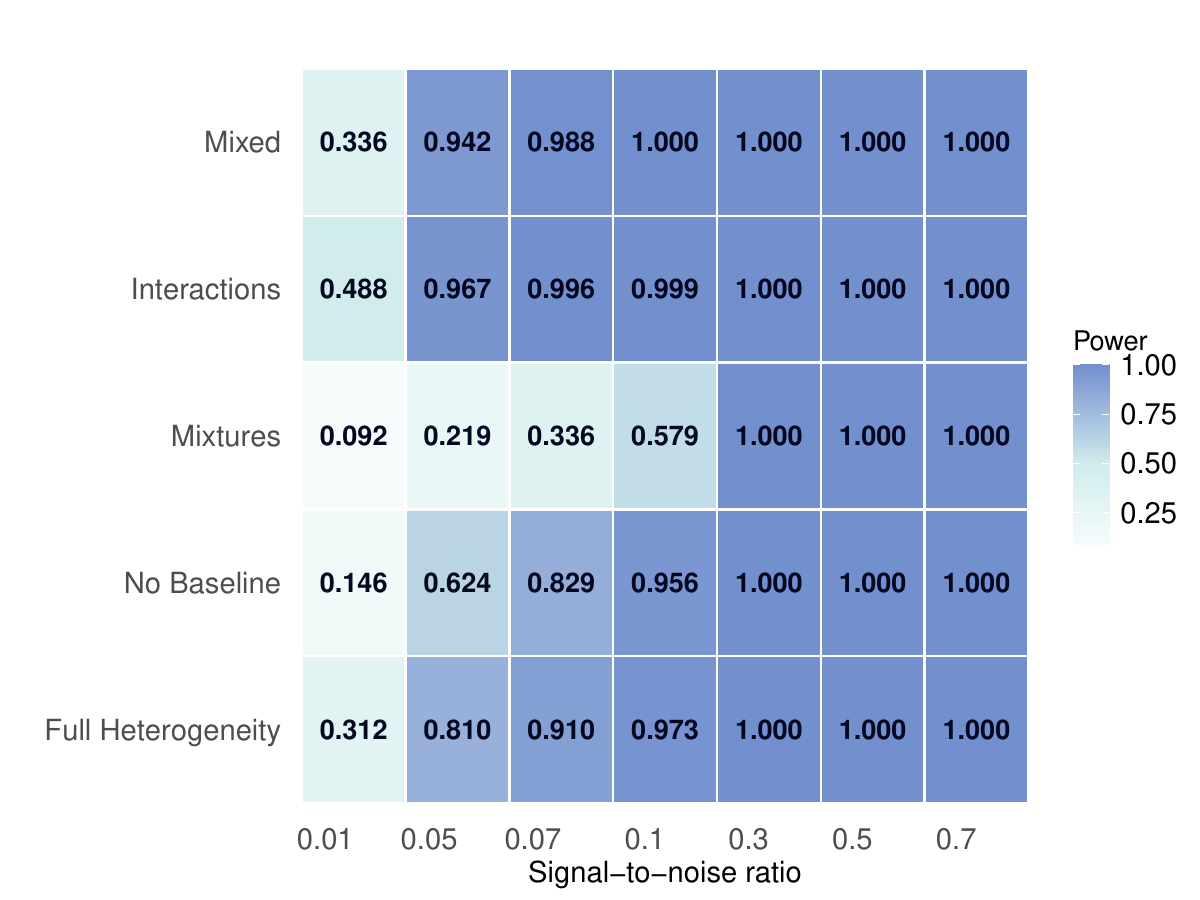}
    \caption{$\aipw$}
    \label{fig:aipw}
\end{subfigure}
\hfill
\begin{subfigure}{0.47\textwidth}
    \centering
    \includegraphics[width=\textwidth]{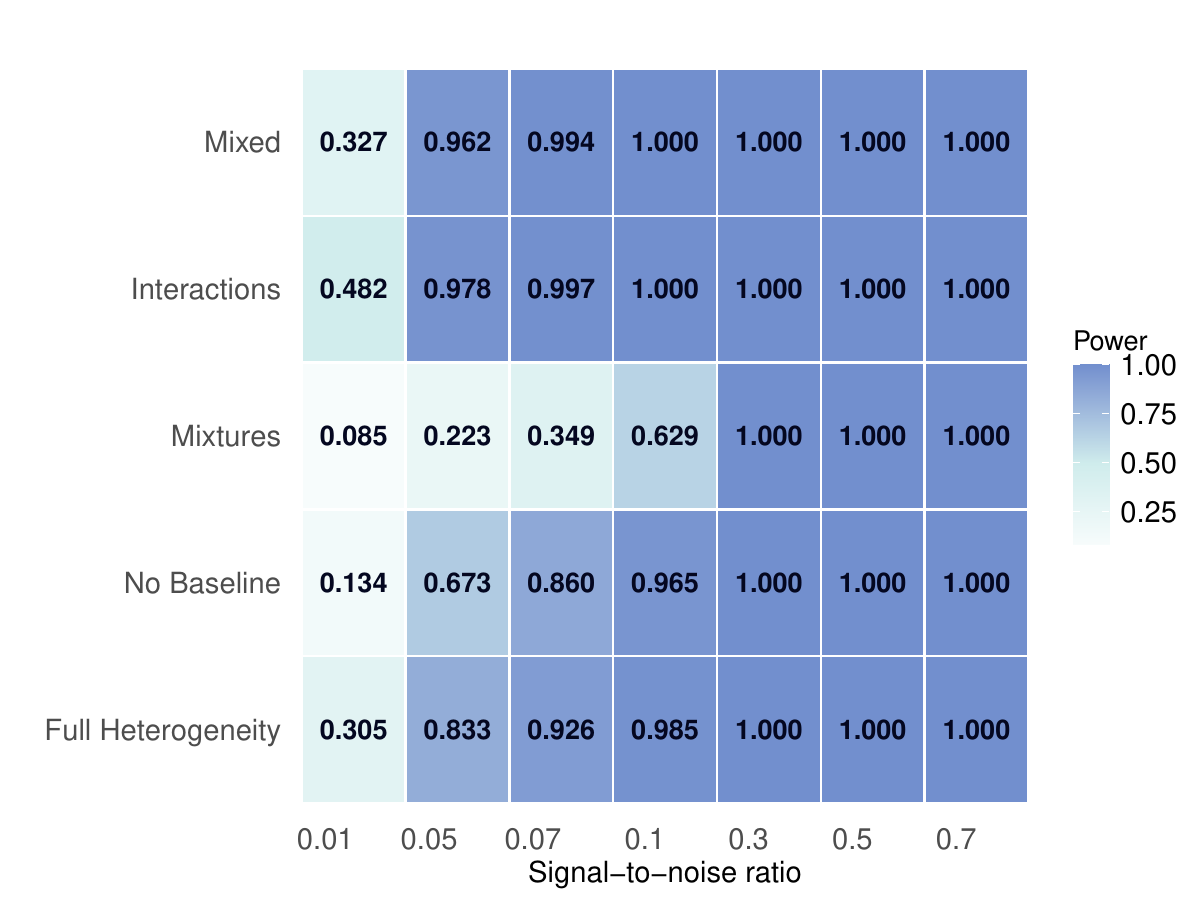}
    \caption{DR}
    \label{fig:rlearner}
\end{subfigure}

\vspace{0.8em}

\begin{subfigure}{0.47\textwidth}
    \centering
    \includegraphics[width=\textwidth]{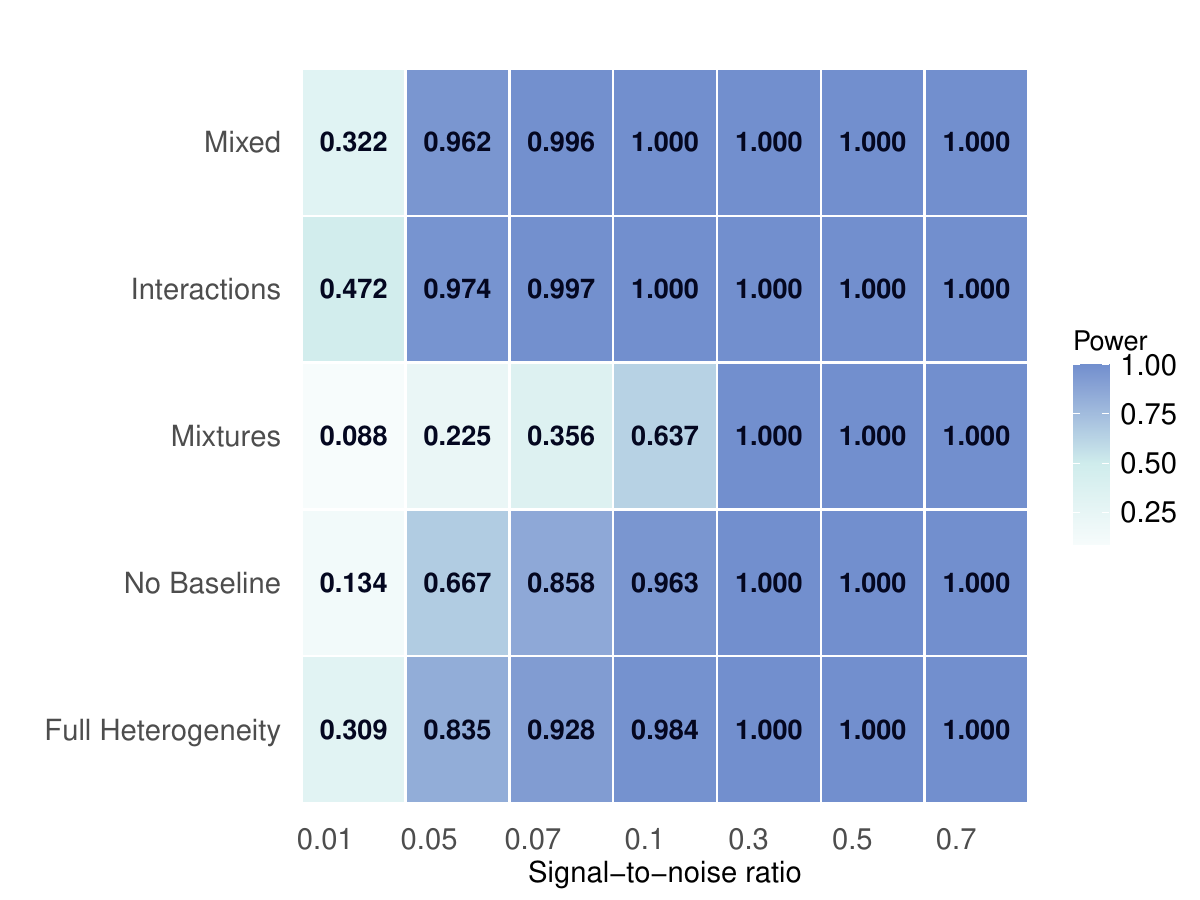}
    \caption{$\drs$}
    \label{fig:drstar}
\end{subfigure}
\hfill
\begin{subfigure}{0.47\textwidth}
    \centering
    \includegraphics[width=\textwidth]{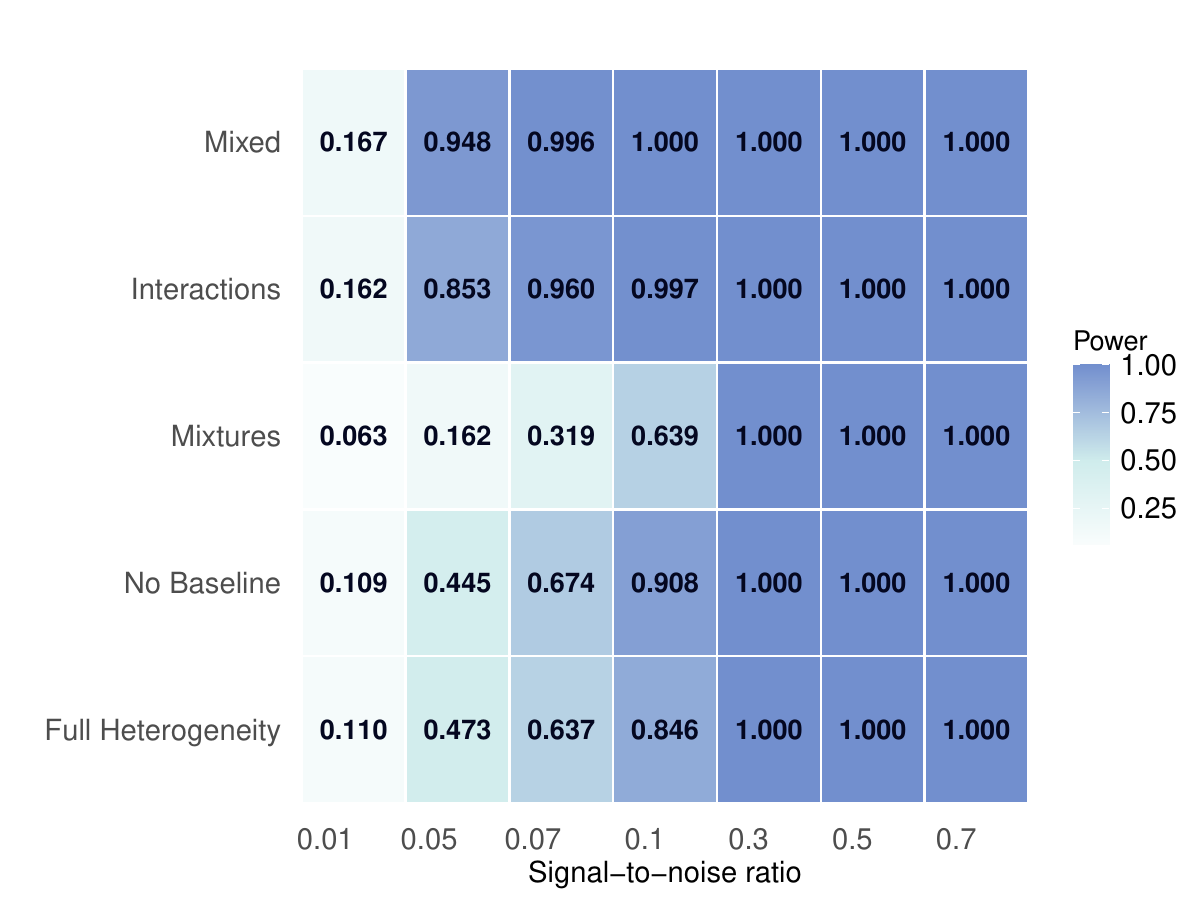}
    \caption{IPW}
    \label{fig:ipw}
\end{subfigure}
\caption{Power heatmaps for $\aipw$, $\dr$, $\drs$, and IPW test statistics across different signal-to-noise ratios (\(\rho\)) and design classes. Empirical power (rejection probability under the alternative hypothesis) is estimated from 1{,}000 Monte Carlo replications per cell, with sample size \(1000\) and five outcomes in each replication. Higher values indicate more detectable treatment effects. For instance, \(\rho=0.05\) means that the treatment
component is calibrated to account for 5\% of the additive variance components
used in the simulation design.}
\label{fig:all_heatmaps}
\end{figure}

The calibration of $\rho$ follows the signal-to-noise construction of
\autoref{subsection_decay} (see \eqref{eq:snr}--\eqref{eq:kappa}): $\rho$ is the share of
an outcome's total additive variance due to the treatment component, and
$\kappa_m(\rho)$ is set to hit that target from the realized variances of the
baseline, errors, and treatment interaction. Two features differ here. First, we suppress the $m^{-1/2}$ decay, so every
outcome carries the full signal. Second, we calibrate to the same target $\rho$
outcome by outcome.\footnote{Since $\tau_m(X_i)$ is constant across outcomes in every
design except Mixed Heterogeneity, a single $\kappa$ achieves this target for all
outcomes there. In the mixed design, where $\tau_m(X_i)$ varies, each outcome
requires its own $\kappa_m$ to reach the common $\rho$.}  Larger values of $\rho$ correspond to stronger, more
easily detectable treatment effects, while smaller values represent low-signal
regimes. Because the same value of $\rho$ is applied to every design and every
outcome, differences in power across rows of the heatmaps in \autoref{fig:all_heatmaps} reflect how the shape of
$\tau_m(X_i)$ interacts with each test, rather than differences in raw effect
magnitude.

\autoref{fig:all_heatmaps} reports empirical power as a function of the
variance share $\rho$. Power rises sharply with the SNR: at $\rho=0.10$ the
$\aipw$, $\dr$, and $\drs$ tests reject with high probability in the Full
Heterogeneity, No Baseline, and Interactions designs, with IPW close
behind; by $\rho=0.30$ empirical power reaches one across all tests
and designs. The Mixtures design saturates more slowly because
$\mathbb{E}[\tau(X_i)]=0$ by construction. The advantage of the augmented
and residualized tests over IPW is most pronounced in the low-signal regime
($\rho\le 0.05$), where residualizing out covariate-explained variation in the outcome and the treatment
raises the effective signal, making departures from the null hypothesis easier to detect.

\autoref{fig:ipw_loss_null} (\autoref{app:sims_distributions}) shows that, under the null hypothesis, the IPW statistic is centered at its theoretical mean of 0.5. Under the alternative hypothesis (\autoref{fig:ipw_loss_alternative}), the realized IPW statistic shifts downward relative to its null distribution, indicating predictive content of outcomes for treatment beyond chance. \autoref{fig:aipw_density} further shows that the $\aipw$ statistic is significantly greater than zero under the alternative hypothesis, consistent with \autoref{thm:one_sided_aipw}. Under the null hypothesis (\autoref{fig:aipw_density_null}), the $\aipw$ statistic is sharply centered at zero and closely follows its normal approximation, even in small samples.  The same conclusion applies to the tests based on double residualization in \autoref{fig:r_learner_density}, \autoref{fig:r_learner_density_null}, \autoref{fig:dr_star_density_alternative}, and \autoref{fig:dr_star_density_null}. Additional simulations in \autoref{app:sims_additional} confirm the size and superior power across various nonlinear specifications of the treatment effects relative to the joint Wald test. Furthermore, \autoref{app:small_sample} compares the proposed tests to two existing conditional independence tests: the generalized covariance measure (GCM; \citealp{shah2020hardness}) and the kernel conditional independence test (KCI; \citealp{zhang2012kernel}). In small samples with weak treatment effects, the two families of conditional independence tests are complementary: $\dr$, $\aipw$, and $\drs$ are sensitive to any departure
from conditional independence and dominate when the treatment effect averages to
zero, while KCI, GCM, and the joint Wald tests dominate when the mean treatment effect is nonzero.

\section{Conclusion}

We introduce a family of conditional independence tests for causal effects in the presence of multiple outcomes and covariates in observational data. The construction centers on a pseudo-outcome, the ``posterior'' treatment probability after conditioning on outcomes and covariates, and evaluates the association between treatment assignment and the pseudo-outcome. We propose three measures of conditional independence: an inverse propensity weighted loss-based test statistic; a test statistic based on double residualization that partials out covariates from both the treatment and the posterior treatment probability; and an augmented inverse propensity weighted test statistic. The inverse propensity weighted test statistic requires a known propensity score, while the tests based on double residualization and augmented inverse propensity weighting accommodate observational settings in which the propensity score is unknown and must be estimated. The population moments underlying the tests based on double residualization and augmented inverse propensity weighting are Neyman orthogonal and necessary and sufficient for conditional independence. Simulations demonstrate high power under complex treatment heterogeneity and in low-signal environments. Future work includes extending the framework to continuous and mixed treatment assignments, as well as to settings with multiple treatments.

\bibliographystyle{apalikefull3}
\bibliography{refs}

\clearpage

\begin{appendix}

\appendix

\section{Motivating Application}
\label{app:application}

This section illustrates how modern field experiments routinely involve multiple outcomes and subgroups, using the New Jersey minimum wage study and the Job Corps evaluation as motivating examples, and thereby motivates the conditional independence tests proposed in this paper.

\citet*{list2019multiple} report that from 2004 to 2006, 84\% of experiments published in leading social science journals analyzed at least five outcomes, while 61\% examined ten or more. In addition, 76\% of field experiments in top economics journals considered multiple subgroups, with 29\% analyzing ten or more. For instance, \citet*{card1993minimum} study the 1992 increase in New Jersey's minimum wage from \$4.25 to \$5.05 using a difference-in-differences design, comparing fast-food restaurants in New Jersey (treated) to restaurants in eastern Pennsylvania (control), and also comparing low-wage to already-high-wage restaurants within New Jersey. Using survey data from before and after the policy change, they find that the minimum wage increase did not reduce employment; to the contrary, employment in New Jersey fast-food restaurants rose relative to Pennsylvania. The authors also find no strong evidence that firms offset higher wages by cutting non-wage benefits or sharply changing promotion and raise policies, while meal prices increased in New Jersey relative to Pennsylvania, consistent with partial cost pass-through to consumers.

Another example is Job Corps, a federally run residential training program for disadvantaged youth that began in 1964. \citet*{schochet2008does} evaluate cohorts of applicants who were randomly assigned between late 1994 and early 1996. The treatment is being offered a Job Corps slot (program group) versus being denied access for three years but remaining free to participate in other education or training programs (control group). The authors study multiple outcomes: education and training participation and credentials, earnings and employment over nine years, and crime (arrests and convictions). They find that Job Corps substantially increases time in education and training, raises General Educational Development (GED) and especially vocational certificate attainment, reduces arrests and convictions, and generates meaningful short-run earnings gains (roughly a 12\% increase in year-four earnings), but these earnings gains largely fade out after about five years, remaining persistent only for the oldest applicants (ages 20--24).

The proposed conditional independence tests allow us to evaluate whether treatment assignment (denoted \(D_i\), for example, fast-food restaurants located in New Jersey) is associated with outcomes (denoted \(Y_i\))---such as full-time workers per store, employment growth rates, hours and capacity measures, wage and non-wage compensation, and growth in outlets and new openings across states---after conditioning on covariates (denoted \(X_i\)), such as region indicators and store-level characteristics. The approach constructs the posterior treatment probability \(\mathbb{P}(D_i=1|Y_i,X_i)\), which summarizes the information in \((Y_i,X_i)\) relevant for predicting treatment assignment. One may then test the null hypothesis by evaluating this association using either a loss-based test statistic (when propensity scores are known) or a standardized covariance measure between residualized treatment and the posterior treatment probability.

\section{Proofs}\label{app:proofs}

\subsection{Proof of \autoref{prop:reverse_prediction}}\label{app:proof_reverse_prediction}

\begin{proof}[Proof of \autoref{prop:reverse_prediction}]
Assume first that \eqref{eq:null_predictability} holds.
    Let $A$ be a measurable set and $d \in \{0,1\}$.
    Then
    \begin{align*}
        \P(D_i=d, Y_i\in A |X_i)
        &=
        \P(D_i=d |Y_i\in A , X_i) \: \P(Y_i\in A |X_i) \\
        &\stackrel{\mathclap{\eqref{eq:null_predictability}}}{=}
        \P(D_i=d |X_i)
        \: \P(Y_i\in A |X_i)
    \end{align*}
    and hence \eqref{eq:null_independence}.
    Conversely, if \eqref{eq:null_independence} and $\P(Y_i\in A |X_i) > 0$, then
    \begin{align*}
        \P(D_i=d |Y_i\in A , X_i)
        &=
        \frac{\P(D_i=d, Y_i\in A |X_i)}{\P(Y_i\in A |X_i)}
        \\
        &\stackrel{\mathclap{\eqref{eq:null_independence}}}{=}
        \frac{\P(D_i=d |X_i) \: \P(Y_i\in A |X_i)}{\P(Y_i\in A |X_i)}
        =
        \P(D_i=d |X_i)
    \end{align*}
    and thus \eqref{eq:null_predictability} holds.
\end{proof}

\subsection{Proof of \autoref{lem:pseudo_randomization}} \label{app:proof_pseudo_randomization}

\begin{proof}[Proof of \autoref{lem:pseudo_randomization}]
For both parts, note first that $\P_w$ (defined in \autoref{fn:pw}) is a probability measure with
    \begin{align*}
        \P_w(D_i = 1 |X_i)
        \;=\; \frac{\E[W_i D_i |X_i]}{\E[W_i |X_i]}
        \;=\; \frac{1}{2} \quad \text{a.s.} 
    \end{align*}
since \(\mathbb{E}[W_iD_i\mid X_i] = \frac{\mathbb E[D_i\mid X_i]}{e_0(X_i)} = \frac{e_0(X_i)}{e_0(X_i)} = 1\), and \(\mathbb{E}[W_i\mid X_i] = \frac{e_0(X_i)}{e_0(X_i)} + \frac{1 - e_0(X_i)}{ 1 - e_0(X_i)} = 2\). Moreover,
    \begin{align*}
        \P_w(Y_i \in A |X_i, D_i = d)
        &\;=\;  \frac{\mathbb{P}_w\left( Y_i \in A, D_i = d \mid X_i\right)}{\mathbb{P}_w\left(  D_i = d \mid X_i\right)} = \frac{\E[W_i \mathbf 1\{Y_i \in A, D_i = d\} |X_i]}{\E[W_i \mathbf 1\{D_i = d\} |X_i]}\\
        &\;=\; \frac{w(d, X_i)\,\P(Y_i \in A, D_i = d |X_i)}{w(d, X_i)\,\P(D_i = d |X_i)}
        \;=\; \P(Y_i \in A |X_i, D_i = d) \quad \text{a.s.},
    \end{align*}
    for all $d \in \{0,1\}$ and measurable sets $A$. That is, once we fix the treatment arm, the inverse propensity weighting does not distort the outcomes inside either group. 

    For the first part, by Bayes' rule for all measurable sets $A$,
    \begin{align}
        \P_w(D_i{=}1|Y_i \in A, X_i)
        &=
        \frac{\P_w(Y_i \in A | D_i{=}1, X_i)\,\P_w(D_i{=}1 | X_i)}{\P_w(Y_i \in A | D_i{=}1, X_i)\,\P_w(D_i{=}1 | X_i) + \P_w(Y_i \in A | D_i{=}0, X_i)\,\P_w(D_i{=}0 | X_i)} \nonumber
        \\
        &=
        \frac{\P(Y_i \in A | D_i{=}1, X_i)\,\tfrac{1}{2}}{\P(Y_i \in A | D_i{=}1, X_i)\,\tfrac{1}{2} + \P(Y_i \in A | D_i{=}0, X_i)\,\tfrac{1}{2}} \nonumber
        \\
        &=
        \frac{\P(Y_i \in A | D_i{=}1, X_i)}{\P(Y_i \in A | D_i{=}1, X_i) + \P(Y_i \in A | D_i{=}0, X_i)}. \label{eq:pw_bayes_set}
    \end{align}
    In the particular case when $Y_i$ conditional on $D_i$ and $X_i$ admits a probability density $f$ with respect to Lebesgue measure under $\P$, we analogously obtain
    \begin{align*}
        \P_w(D_i{=}1|Y_i =y,X_i =x)
        =
        \frac{f(y |D_i{=}1, X_i = x)}{f(y |D_i{=}1, X_i = x) + f(y |D_i{=}0, X_i = x)}.
    \end{align*}

    For the second part,
    it follows from \eqref{eq:pw_bayes_set}
    that $\P_w(D_i{=}1|Y_i \in A, X_i) = \frac{1}{2}$ a.s.\
    if and only if
    \begin{align*}
        \P(Y_i \in A | D_i{=}d, X_i) = \P(Y_i \in A | X_i)
    \end{align*}
    a.s.\ for $d \in \{0,1\}$.

    Assume first that
    $\P_w(D_i{=}d|Y_i \in A, X_i) = \frac{1}{2}$ a.s.\ for all $d \in \{0,1\}$ and measurable sets $A$.
     Since 
    \begin{align*}
        \P(D_i{=}d,Y_i \in A | X_i)
        &= \P(Y_i \in A | D_i{=}d, X_i)\,\P(D_i{=}d| X_i)\\
        &= \P(Y_i \in A | X_i)\,\P(D_i{=}d| X_i)
        = \P(D_i{=}d| X_i)\,\P(Y_i \in A | X_i)
    \end{align*}
    a.s.\ for all $d \in \{0,1\}$ and measurable sets $A$,
    $Y_i \perp D_i | X_i$ a.s.\ and \eqref{eq:null_independence} holds.
    Assume now that $Y_i \perp D_i | X_i$ a.s.\ so that
    \(
        \P(Y_i \in A | D_i{=}d, X_i) = \P(Y_i \in A | X_i)
    \)
    a.s.\ for all $d \in \{0,1\}$ and measurable sets $A$.
    Then also, whenever $\P(Y_i \in A | X_i) > 0$,
    \begin{align*}
        \P_w(D_i{=}1 | Y_i \in A, X_i)
        = \frac{\P(Y_i \in A | D_i{=}1, X_i)}{\P(Y_i \in A | D_i{=}1, X_i) + \P(Y_i \in A | D_i{=}0, X_i)}
        = \frac{\P(Y_i \in A | X_i)}{2\,\P(Y_i \in A | X_i)}
        = \tfrac{1}{2}.
    \end{align*}
\end{proof}

\subsection{Proof of Loss Minimization}\label{app:proof_loss_minimization}

\begin{proof}[Proof of loss minimization]
Take any measurable function $f$ of $Y_i$ and $X_i$. Since $\P_w$ is a probability measure, we have
    \begin{align}
        L(f)
        &= \E\!\left[W_i\,\big(D_i - f(Y_i, X_i)\big)^2\right]
        =
        2 \: \E_w\!\left[\big(D_i - f(Y_i, X_i)\big)^2\right] \nonumber
        \\
        &= 2\,\E_w\!\left[\big((D_i - \E_w[D_i |Y_i, X_i]) + (\E_w[D_i |Y_i, X_i] - f(Y_i, X_i))\big)^2\right] \nonumber
        \\
        &= 2\,\E_w\!\left[\big(D_i - \E_w[D_i |Y_i, X_i]\big)^2\right]
        + 2\,\E_w\!\left[\big(\E_w[D_i |Y_i, X_i] - f(Y_i, X_i)\big)^2\right], \label{eq:loss_decomp}
    \end{align}
    where we have used the tower property to argue that
    \begin{align*}
        &\E_w\!\big[
            \big(\E_w[D_i |Y_i, X_i] - f(Y_i, X_i)\big)
            \big(D_i - \E_w[D_i |Y_i, X_i]\big)
        \big]
        \\
        &= \E_w\!\big[\big(\E_w[D_i |Y_i, X_i] - f(Y_i, X_i)\big)\,\underbrace{\E_w\!\left[D_i - \E_w[D_i |Y_i, X_i] \,\middle|\, Y_i, X_i\right]}_{=0}\big]
        = 0.
    \end{align*}
    Since $\E_w\!\left[\big(D_i - \E_w[D_i |Y_i, X_i]\big)^2\right]$ does not depend on $f$
    and $\E_w\!\left[\big(\E_w[D_i |Y_i, X_i] - f(Y_i, X_i)\big)^2\right]$ is non-negative and zero for $f(Y_i, X_i) = f_w(Y_i, X_i) = \E_w[D_i |Y_i, X_i]$,
    it follows from \eqref{eq:loss_decomp} that $f_w$ is a minimizer of $L(f)$.
\end{proof}

\subsection{Proof of \autoref{thm:loss_criterion}} \label{app:proof_loss_criterion}

\begin{proof}[Proof of \autoref{thm:loss_criterion}]
By \eqref{eq:fw},
$f_w(Y_i,X_i) = \E_w[D_i|Y_i,X_i]$,
so
\begin{align*}
    L(f_w)
    &= 2 \: \E_w[(D_i - \E_w[D_i|Y_i,X_i])^2]
    =
    2 \: \E_w[\E_w[(D_i - \E_w[D_i|Y_i,X_i])^2|Y_i,X_i]]
    \\
    &=
    2 \: \E_w[\Var_w(D_i |Y_i,X_i)].
\end{align*}
Since
\begin{equation*}
    \Var_w(D_i|Y_i, X_i) = \P_w(D_i{=}1|Y_i,X_i) \: (1{-}\P_w(D_i{=}1|Y_i,X_i)) \leq \tfrac{1}{4},
\end{equation*}
with equality if and only if $\P_w(D_i{=}1|Y_i,X_i) = \tfrac{1}{2}$,
we have that $L(f_w) \leq \tfrac{1}{2}$ in general,
and $L(f_w) = \tfrac{1}{2}$ if and only if $\P_w(D_i{=}1|Y_i,X_i) = \tfrac{1}{2}$ a.s.
It follows by \autoref{lem:pseudo_randomization} that $L(f_w) = \tfrac{1}{2}$ if and only if $H_0$ holds, as well as $L(f_w) < \tfrac{1}{2}$ if and only if it does not.
\end{proof}

\begin{remark}[Weighted prediction under the unweighted loss]\label{rem:ndr_weights}
If the inverse propensity weights are removed from the test statistic while the inverse propensity weighted conditional treatment probability $f_w(Y_i,X_i)$ is retained, 
then $f_w$ does not generally minimize $\E\left[(D_i-f(Y_i,X_i))^2\right]$ over $f$.
% and
% (ii) $\E\left[(D_i-f_w(Y_i,X_i))^2\right]$ can be smaller or larger than $\E\left[(D_i-\tfrac{1}{2})^2\right]$.
\end{remark}

\begin{proof}

\(f_w(Y_i, X_i)\) minimizes the weighted risk
\(\E[W_i(D_i-f(Y_i, X_i))^2]\), not the unweighted risk
\(\E[(D_i-f(Y_i, X_i))^2]\).
Writing $f_0(Y_i,X_i) = \P(D_i{=}1|Y_i,X_i)$ for the minimizer of $\E[(D_i-f(Y_i, X_i))^2]$,
we have the familiar decomposition
\begin{align*}
    \E\!\left[(D_i-f_w(Y_i, X_i))^2\right]
    =
    \E\!\left[(f_0(Y_i, X_i)-f_w(Y_i, X_i))^2\right]
    + \E\!\left[\Var(D_i|Y_i,X_i)\right].
\end{align*}
This expression is only minimal when \(f_w(Y_i, X_i) = f_0(Y_i, X_i)\) a.s.; since this is not true in general, we have the claim.

\end{proof}

\begin{remark}[Non-robustness with respect to the unweighted prediction]\label{rem:ndr_unweighted}
If the inverse propensity weighted prediction $f_w(Y_i,X_i)$ is replaced by the unweighted conditional probability $f_0(Y_i,X_i) = \E[D_i\mid Y_i,X_i]$, then under $H_0$ the weighted loss $L(f_0)$ is equal to $\tfrac{1}{2}$ \emph{only if} $e_0(X_i)=\tfrac12$ a.s., and otherwise larger. Outside balanced designs the test that compares to the fixed value $\tfrac12$ is therefore miscalibrated.
\end{remark}

\begin{proof}
If we use the unweighted conditional expectation \(f_0(Y_i, X_i) = \P(D_i = 1|Y_i, X_i)\), then since \(f_0(Y_i, X_i) = \P(D_i = 1|Y_i, X_i) = \P(D_i = 1|X_i) = e_0(X_i)\) under the null hypothesis, the oracle loss simplifies to
\begin{align*}
\E\!\left[W_i\big(D_i - e_0(X_i)\big)^2 \,\middle|\, X_i\right]
&= e_0(X_i) \cdot \frac{\big(1 - e_0(X_i)\big)^2}{e_0(X_i)}
   + \big(1-e_0(X_i)\big) \cdot \frac{\big(e_0(X_i)\big)^2}{1-e_0(X_i)} \\
&= \big(1-e_0(X_i)\big)^2 + \big(e_0(X_i)\big)^2.
\end{align*}
Hence, the unconditional loss

\[\mathbb{E}\bigg\{\E\!\left[W_i\big(D_i - e_0(X_i)\big)^2 \,\middle|\, X_i\right]\bigg\} = \E\bigg[(1-e_0(X_i))^2 + (e_0(X_i))^2\bigg]\geq \tfrac{1}{2}
\]

\noindent is only minimized at $\tfrac{1}{2}$ when \(e_0(X_i) = \tfrac{1}{2}\) a.s., so that minimum is only attained under balanced data (e.g. randomized controlled trials). For unbalanced designs, the test that compares to $\tfrac{1}{2}$ would be miscalibrated.
\end{proof}

\subsection{Proof of \autoref{prop:asymptotics_ipw}}\label{app:proof_asymptotics_ipw}

\begin{proof}[Proof of \autoref{prop:asymptotics_ipw}]
This proof establishes that the inverse propensity weighted test statistic \( S_n \) is asymptotically normally distributed when the null hypothesis holds, i.e. treatment is independent of outcomes conditional on covariates, \(D_i \perp Y_i |X_i\). 
Let
\begin{align*}
\Psi_i & := W_i\,\big(D_i - \hat{f}^{\mathcal{I}^c}_w(Y_i, X_i)\big)^2, &
\Psi_i^\star & := W_i\,\big(D_i - f_w(Y_i, X_i)\big)^2
\end{align*}
denote the observation-specific loss contributions with estimated and oracle prediction functions, respectively.
Write
\begin{align*}
    S_n &= \frac{1}{n}\sum_{i \in \mathcal{I}} \Psi_i = \frac{1}{n}\sum_{i \in \mathcal{I}}W_i\,\big(D_i - \hat{f}^{\mathcal{I}^c}_w(Y_i, X_i)\big)^2, \\
    S^\star &= \frac{1}{n}\sum_{i \in \mathcal{I}} \Psi_i^\star = \frac{1}{n}\sum_{i \in \mathcal{I}}W_i\,\big(D_i - f_w(Y_i, X_i)\big)^2,
\end{align*}
where \(n\) is the number of observations in a sample \(\mathcal{I}\) and all nuisance parameters are estimated on an auxiliary sample \(\mathcal{I}^c\) with \(N - n\) observations, and
\(
    W_i = \frac{D_i}{e_0(X_i)} + \frac{1 - D_i}{1 - e_0(X_i)}.
\)

Under the null, we have that
\begin{align*}
    \E[\Psi_i^\star] &= \E\big[W_i\,\big(D_i - f_w(Y_i, X_i)\big)^2\big] = \E\Big[W_i\,\big(D_i - \tfrac{1}{2}\big)^2\Big] = \tfrac{1}{4}\,\E[W_i] = \tfrac{1}{2},
    \\
    \Var(\Psi_i^\star) &= \Var\Big(W_i\,\big(D_i - \tfrac{1}{2}\big)^2\Big) = \Var\big(\tfrac{1}{4} W_i\big) = \tfrac{1}{16} \Var(W_i),
\end{align*}
so the goal is to establish asymptotic normality of the test statistic centered at its oracle expectation,
\begin{align}
\sqrt{n}(S_n - \E[\Psi_i^\star]) \xrightarrow{d} \N(0, \Var(\Psi_i^\star)).
\label{eq:simplegoal}
\end{align}
We decompose the statistic by adding and subtracting its oracle counterpart:
\begin{align}
\sqrt{n}(S_n - \E[\Psi_i^\star]) =
\underbrace{\sqrt{n}(S^\star - \E[\Psi_i^\star])}_{\text{asymptotically normal by CLT}}
\;+\;
\underbrace{\sqrt{n}(S_n - S^\star)}_{\text{vanishes as } n \to \infty}.
\label{eq:simpledecomp}
\end{align}
For the first term,
\begin{align}
    \sqrt{n}(S^\star - \E[\Psi_i^\star])
    \xrightarrow{d}
    \N(0, \Var(\Psi_i^\star))
    \label{eq:simpleclt}
\end{align}
by the Central Limit Theorem (CLT).
For the second term,
\begin{align*}
    &\sqrt{n}(S_n - S^\star)
    =
    \frac{1}{\sqrt{n}}\sum_{i \in \mathcal{I}} W_i\,\Big(\big(D_i {-} \hat{f}^{\mathcal{I}^c}_w(Y_i, X_i)\big)^2 - \big(D_i {-} f_w(Y_i, X_i)\big)^2\Big)
    \\
    &=
    \frac{1}{\sqrt{n}}\sum_{i \in \mathcal{I}} W_i\,\big(f_w(Y_i, X_i) {-} \hat{f}^{\mathcal{I}^c}_w(Y_i, X_i)\big)\,\big(2 (D_i {-} f_w(Y_i, X_i)) +  (f_w(Y_i, X_i) {-} \hat{f}^{\mathcal{I}^c}_w(Y_i, X_i))\big)
    \\
    &=
    - 2 \frac{1}{\sqrt{n}}\sum_{i \in \mathcal{I}} W_i\,\big(D_i {-} f_w(Y_i, X_i)\big)\,\delta_{f,i}
    +
    \frac{1}{\sqrt{n}}\sum_{i \in \mathcal{I}} W_i\,\delta_{f,i}^2,
\end{align*}
where we write $\delta_{f,i} = \hat{f}^{\mathcal{I}^c}_w(Y_i, X_i) - f_w(Y_i, X_i)$.
Note that $|W_i| \leq \tfrac{1}{\eta}$ by \autoref{asm:overlap}
and $|D_i {-} f_w(Y_i, X_i)| \leq 1$, which we will use below.
Conditional on $\mathcal{I}^{c}$,
the terms
$W_i\,\big(D_i {-} f_w(Y_i, X_i)\big)\,\delta_{f,i}$ are mean-zero and independent across $i$, so
\begin{align*}
    &\E\left[
        \left(
            \frac{1}{\sqrt{n}}\sum_{i \in \mathcal{I}} W_i\,\big(D_i {-} f_w(Y_i, X_i)\big)\,\delta_{f,i}
        \right)^2
    \,\middle|\, \mathcal{I}^{c}
    \right]
    \\
    &=
    \frac{1}{n} \E\Big[\sum_{i \in \mathcal{I}} W_i^2\,\big(D_i {-} f_w(Y_i, X_i)\big)^2\,\delta_{f,i}^2 \,\Big|\, \mathcal{I}^{c}\Big]
    \leq
    \frac{1}{\eta^2} \E\big[\delta_{f,i}^2 \,\big|\, \mathcal{I}^{c}\big]
    \xrightarrow{p} 0
\end{align*}
by \autoref{asm:ipw_rate}.
Also,
\begin{align*}
    \E\left[\frac{1}{\sqrt{n}}\sum_{i \in \mathcal{I}} W_i\,\delta_{f,i}^2 \,\middle|\, \mathcal{I}^{c}\right]
    \leq \frac{\sqrt{n}}{\eta} \E\big[\delta_{f,i}^2 \,\big|\, \mathcal{I}^{c}\big] \xrightarrow{p} 0,
\end{align*}
again by \autoref{asm:ipw_rate}.
In particular, by Markov's inequality conditional on $\mathcal{I}^{c}$ along with dominated convergence,
we obtain, unconditionally,
\begin{align*}
    \sqrt{n}(S_n - S^\star)
    =
    - 2 \frac{1}{\sqrt{n}}\sum_{i \in \mathcal{I}} W_i\,\big(D_i {-} f_w(Y_i, X_i)\big)\,\delta_{f,i}
    +
    \frac{1}{\sqrt{n}}\sum_{i \in \mathcal{I}} W_i\,\delta_{f,i}^2
    \xrightarrow{p} 0.
\end{align*}
Together with \eqref{eq:simpledecomp} and \eqref{eq:simpleclt}, Slutsky yields \eqref{eq:simplegoal}. This establishes the proposition.
\end{proof}

\begin{remark}[Non-orthogonality with respect to the propensity score]\label{rem:non_orthogonality_e}
Assume the propensity score $e_0$ is estimated by $\hat{e}^{\mathcal{I}^c}$, rather than known.
Consider the test statistic
\begin{align*}
    \widehat{S}_n = \frac{1}{n}\sum_{i \in \mathcal{I}} \widehat{W}_i\,\big(D_i - \hat{f}^{\mathcal{I}^c}_w(Y_i, X_i)\big)^2,
    \qquad
    \widehat{W}_i = \frac{D_i}{\hat{e}^{\mathcal{I}^c}(X_i)} + \frac{1 - D_i}{1 - \hat{e}^{\mathcal{I}^c}(X_i)}.
\end{align*}
Then the corresponding score is not Neyman orthogonal with respect to the propensity score: in general the derivative of the expected score with respect to the propensity score does not vanish, even under $H_0$. Consequently, if $e_0$ is estimated at an $o_p(n^{-1/4})$ rate rather than known, the asymptotic distribution of $\sqrt{n} (\widehat{S}_n - \E[\Psi_i^\star])$ may retain a first-order contribution from the propensity-score estimation error.
\end{remark}

\begin{proof}
    For realizations $Y,X,D$ and conformable functions $f, e$, define
    \begin{align}
        \Psi_0(Y,X,D; f,e) = \left(\frac{D}{e(X)} + \frac{1 - D}{1 - e(X)}\right) (D - f(Y,X))^2.
        \label{eq:psi0}
    \end{align}
    Then we have that
    \begin{align*}
        \widehat{S}_n = \frac{1}{n} \sum_{i \in \mathcal{I}} \widehat{\Psi}_i,
        \qquad
        \widehat{\Psi}_i = \Psi_0(Y_i,X_i,D_i; \hat{f}^{\mathcal{I}^c}_w,\hat{e}^{\mathcal{I}^c}).
    \end{align*}
    (Similarly, in the proof of \autoref{prop:asymptotics_ipw}, we have that $\Psi_i = \Psi_0(Y_i,X_i,D_i; \hat{f}^{\mathcal{I}^c}_w, e_0)$ and $\Psi_i^\star = \Psi_0(Y_i,X_i,D_i; f_w, e_0)$ to obtain $S_n$ and $S^\star$, respectively.)
    $\widehat{S}_n$ estimates the parameter
    \begin{align*}
        \phi_0 = \E[\Psi_0(Y_i,X_i,D_i; f_w,e_0)]
    \end{align*}
    for the true conditional expectation $f_w(Y_i, X_i) = \E_w[D_i \,|\, Y_i, X_i]$ and propensity score $e_0(X_i) = \P(D_i = 1 \,|\, X_i)$.
    Neyman orthogonality of the corresponding score with respect to the propensity score is equivalent to the Gateaux derivative of $\E[\Psi_0(Y_i,X_i,D_i; f_w,e)]$ with respect to $e$ at $e_0$ being zero, that is,
    \begin{align}
        \frac{d}{dt}\,\E\bigl[\Psi_0(Y_i, X_i, D_i;\, f_w,\, e_0 + t\,h)\bigr]\Big|_{t=0} = 0
        \label{eq:gateaux}
    \end{align}
    for all directions $h = e - e_0$ (which are functions) with $e$ a candidate propensity score.

    For our case, checking orthogonality simplifies because the score depends on the nuisance parameters only through the fitted values,
    \begin{align}
        \Psi_0(Y,X,D; f,e) = \Psi(D; f(Y,X), e(X)),
        \qquad
        \Psi(D; F, E)
        =
        \left(\frac{D}{E} + \frac{1 - D}{1 - E}\right) (D - F)^2.
        \label{eq:psi_fitted}
    \end{align}
    By the chain rule and dominated convergence,
    \begin{align*}
        \frac{d}{dt}\,\E\bigl[\Psi_0(Y_i, X_i, D_i;\, f_w,\, e_0 + t\,h)\bigr]\Big|_{t=0}
        =
        \E\left[\frac{\partial \Psi(D_i; f_w(Y_i, X_i), E)}{\partial E} \Bigg|_{E = e_0(X_i)} \cdot h(X_i)\right].
    \end{align*}

\noindent   Hence,
    orthogonality as in \eqref{eq:gateaux} is equivalent to
    \begin{align*}
        \E\left[\frac{\partial \Psi(D_i; f_w(Y_i, X_i), E)}{\partial E} \Bigg|_{E = e_0(X_i)}\,\middle|\, X_i\right] = 0
    \end{align*}
    almost surely.
    Here, the partial derivative with respect to the propensity score is
    \begin{align*}
        \frac{\partial \Psi(D; F, E)}{\partial E} = \left( -\frac{D}{E^2} + \frac{1 - D}{(1 - E)^2} \right) (D - F)^2.
    \end{align*}
    Taking the conditional expectation and applying the law of total expectation:
    \begin{align*}
        &\E\left[\frac{\partial \Psi(D_i; f_w(Y_i, X_i), E)}{\partial E} \Bigg|_{E = e_0(X_i)}\,\middle|\, X_i\right]
        \\
        &=
        \E\left[\left( -\frac{D_i}{e_0(X_i)^2} + \frac{1 - D_i}{(1 - e_0(X_i))^2} \right) \big(D_i - f_w(Y_i, X_i)\big)^2 \,\middle|\, X_i\right]
        \\
        &=
        -\frac{1}{e_0(X_i)} \E\Big[\big(1 - f_w(Y_i, X_i)\big)^2 \,\Big|\, X_i, D_i {=} 1\Big]
         + \frac{1}{1 - e_0(X_i)}\, \E\Big[f_w(Y_i, X_i)^2 \,\Big|\, X_i, D_i {=} 0\Big],
    \end{align*}
    where we have used that $e_0(X_i) = \P(D_i = 1 \,|\, X_i)$.
    Under the null hypothesis, $f_w(Y_i, X_i) = \tfrac{1}{2}$ by \autoref{lem:pseudo_randomization}. Hence,
    \begin{align*}
        \E\left[\frac{\partial \Psi(D_i; f_w(Y_i, X_i), E)}{\partial E} \Bigg|_{E = e_0(X_i)}\,\middle|\, X_i\right]
        = \frac{1}{4}\left(\frac{1}{1 - e_0(X_i)} - \frac{1}{e_0(X_i)}\right)
        = \frac{2e_0(X_i) - 1}{4 e_0(X_i)(1 - e_0(X_i))}.
    \end{align*}
    This expression is generally non-zero; it is zero only when $e_0(X_i) = \tfrac{1}{2}$.
    Beyond the inverse propensity weighted loss, Example 2 of \citet*{ahrens2025introduction} verifies that, in general, the moments underlying the inverse propensity weighted estimators are not Neyman orthogonal. To ensure that the contribution of this estimation error is asymptotically negligible, one would require the estimation error in the propensity score to converge sufficiently fast. Specifically, a sufficient condition would be
    \begin{align*}
        \sqrt{\E\Big[\big(\hat{e}^{\mathcal{I}^c}(X_i) - e_0(X_i)\big)^2 \,\Big|\, \mathcal{I}^c\Big]} = o_p(n^{-1/2}),
    \end{align*}
    where \(o_p(n^{-1/2})\) denotes convergence of the root-mean squared error of the propensity score to zero in probability at a rate faster than \(1/\sqrt{n}\). This condition is rarely satisfied when the propensity scores are estimated with flexible machine-learning models (\citealp*{wager2024causal}, pp.~33--34). Hence, the assumption of a known propensity score cannot be easily dropped for this test statistic.
\end{proof}

\begin{remark}[Neyman orthogonality of the loss-based score with respect to the prediction function under the null hypothesis]\label{rem:orthogonality_f}
Assume that the propensity score is known, and consider the test statistic $S_n$ from \autoref{prop:asymptotics_ipw}.
Under the null hypothesis, the corresponding score is Neyman orthogonal with respect to the predicted treatment probability.
Hence $o_p(n^{-1/4})$ errors from estimating $f_w$ on the auxiliary sample $\mathcal I^c$ do not affect the asymptotics of $S_n$ at the $\sqrt{n}$ rate.
\end{remark}

\begin{proof}

With $\Psi_0$ from \eqref{eq:psi0} in the proof of \autoref{rem:non_orthogonality_e}, we have that
\begin{align*}
    S_n = \frac{1}{n} \sum_{i \in \mathcal{I}} \Psi_i,
    \qquad
    \Psi_i = \Psi_0(Y_i,X_i,D_i; \hat{f}^{\mathcal{I}^c}_w, e_0).
\end{align*}
$S_n$ estimates the parameter $\phi_0 = \E[\Psi_0(Y_i,X_i,D_i; f_w,e_0)]$.
As above, Neyman orthogonality with respect to the predicted treatment probability corresponds to the Gateaux derivative of $\E[\Psi_0(Y_i,X_i,D_i; f,e_0)]$ with respect to $f$ at $f_w$ being zero.
Here, since the score only depends on the fitted values by \eqref{eq:psi_fitted}, Neyman orthogonality requires that, a.s.,
\begin{align*}
    \E\left[\frac{\partial \Psi(D_i; F, e_0(X_i))}{\partial F} \Bigg|_{F = f_w(Y_i, X_i)} \,\middle|\, Y_i, X_i\right] = 0.
\end{align*}
The derivative evaluates as
\begin{align*}
    \frac{\partial \Psi(D; F, E)}{\partial F} = - 2 \left(\frac{D}{E} + \frac{1 - D}{1 - E}\right) (D - F),
\end{align*}
so
\begin{align*}
    &\E\left[\frac{\partial \Psi(D_i; F, e_0(X_i))}{\partial F} \Bigg|_{F = f_w(Y_i, X_i)} \,\middle|\, Y_i, X_i\right]
    \\
    &=
    -2\, \E\left[\left(\frac{D_i}{e_0(X_i)} + \frac{1 - D_i}{1 - e_0(X_i)}\right) \big(D_i - f_w(Y_i, X_i)\big) \,\middle|\, Y_i, X_i\right]
    \\
    &=
    -2 \left( \P(D_i {=} 1 \,|\, Y_i, X_i) \, \frac{1 - f_w(Y_i, X_i)}{e_0(X_i)} - \P(D_i {=} 0 \,|\, Y_i, X_i) \, \frac{f_w(Y_i, X_i)}{1 - e_0(X_i)} \right),
\end{align*}
where we again write the conditional expectation as a sum over the two treatment values.
Under the null hypothesis, \( f_w(Y_i, X_i) = \tfrac{1}{2} \) and \(\P(D_i {=} 1 \,|\, Y_i, X_i) = \P(D_i {=} 1 \,|\, X_i) = e_0(X_i)\) by conditional independence,
so
\begin{align*}
    \E\left[\frac{\partial \Psi(D_i; F, e_0(X_i))}{\partial F} \Bigg|_{F = f_w(Y_i, X_i)} \,\middle|\, Y_i, X_i\right]
    =
    -2 \left( e_0(X_i) \, \frac{1 - \tfrac{1}{2}}{e_0(X_i)} - (1 - e_0(X_i)) \, \frac{\tfrac{1}{2}}{1 {-} e_0(X_i)} \right)
    = 0.
\end{align*}
This establishes Neyman orthogonality, and the proof of \autoref{prop:asymptotics_ipw} confirms that convergence of $\hat{f}^{\mathcal{I}^c}_w$ at the $o_p(n^{-1/4})$ rate is indeed sufficient for asymptotic normality of $S_n$.
\end{proof}

\begin{remark}[Closed-form null variance]\label{rem:closed_form_variance}
Under the null hypothesis $H_0$, the oracle loss contributions $\Psi_i^\star$ in the proof of \autoref{prop:asymptotics_ipw} reduce to $\Psi_i^\star = \tfrac14 W_i$, with conditional variance
\[
\Var(\Psi_i^\star \mid X_i) = \tfrac{1}{16} \Var(W_i \mid X_i) = \frac{(2e_0(X_i)-1)^2}{16\,e_0(X_i)\,(1-e_0(X_i))}.
\]
In particular, the asymptotic variance $\Var(\Psi_i^\star) = \tfrac{1}{16}\,\Var(W_i) = \tfrac{1}{16}\,\E\left[\frac{(2e_0(X_i)-1)^2}{e_0(X_i)\,(1-e_0(X_i))}\right]$ vanishes if and only if $e_0(X_i)=\tfrac12$ almost surely.
\end{remark}

\begin{proof}

Since $f_w(Y_i, X_i) = \tfrac{1}{2}$ under $H_0$ by \autoref{lem:pseudo_randomization}, the oracle loss contribution under the null hypothesis is given as
\begin{align*}
    \Psi_i^\star = W_i\big(D_i - \tfrac{1}{2}\big)^2 = \tfrac{1}{4}W_i,
\end{align*}
where \(W_i = D_i/e_0(X_i) + (1-D_i)/(1-e_0(X_i))\) as before.
Conditional on $X_i$, $W_i$ takes the value $1/e_0(X_i)$ with probability $e_0(X_i)$ and $1/(1{-}e_0(X_i))$ otherwise, so the conditional variance is
\begin{align*}
    \Var(\Psi_i^\star \mid X_i) = \frac{\Var(W_i \mid X_i)}{16}
    =
    \frac{e_0(X_i) (1{-}e_0(X_i))}{16}  \left(\frac{1}{e_0(X_i)} {-} \frac{1}{1 {-} e_0(X_i)}\right)^2
    =
    \frac{(2e_0(X_i){-}1)^2}{16\,e_0(X_i)\,(1{-}e_0(X_i))}.
\end{align*}
Since $\E[W_i \mid X_i] = 2$ is a.s.\ constant, $\Var(\Psi_i^\star) = \E[\Var(\Psi_i^\star \mid X_i)]$ by the law of total variance.
As the conditional variance is non-negative and equals zero if and only if $e_0(X_i) = \tfrac{1}{2}$, the asymptotic variance $\Var(\Psi_i^\star)$ vanishes if and only if $e_0(X_i) = \tfrac{1}{2}$ almost surely.
\end{proof}

The degeneracy at \(e_0(X_i) = \tfrac{1}{2}\) may affect the choice of inference procedure. If the propensity scores are constant, as in many randomized controlled trials, we may directly apply the permutation procedure proposed by \cite*{ludwig2017machine}.
When the propensity scores are known, but vary by covariates, a researcher may instead use our resampling procedure from \autoref{alg:crt}
or also use asymptotic inference. In the latter case, since the denominator of the conditional variance depends on \(e_0(X_i)(1 -e_0(X_i))\), when the propensity scores are closer to zero or one, the denominator approaches zero, and the variance approaches infinity. We may wish to weight or trim the sample in that case, as recommended by \citet*{crump2009dealing}.

\subsection{Proof of \autoref{prop:valid_test}}\label{app:proof_valid_test}

\begin{proof}[Proof of \autoref{prop:valid_test}]
We fix the auxiliary sample with index set \(\mathcal{I}^c\) as well as the covariates $X_i$ and outcomes $Y_i$ in the evaluation sample \(\mathcal{I}\)
to form the $\sigma$-algebra
\[
\mathcal{F} = \sigma\left( \mathcal{S},
    (X_i, Y_i)_{i \in \mathcal{I}},
    (X_i, Y_i, D_i)_{i \in \mathcal{I}^c},
    \hat{f}^{\mathcal{I}^c}_w
\right).
\]

Treatment assignments are from a Bernoulli distribution $D_i |X_i \sim \mathrm{Bernoulli}(e_0(X_i))$.
Under the null, we also have $D_i \perp Y_i |X_i$.
Since the $(X_i, Y_i, D_i)$ are drawn iid,
under the null we have that, conditional on $\mathcal{F}$,
$D_i \sim \mathrm{Bernoulli}(e_0(X_i))$, independently across $i \in \mathcal{I}$ (since the sample split \(\mathcal{S}\) is independent of the data).
Step~4 of \autoref{alg:crt} replicates this random assignment by drawing $B$ independent treatment vectors 
\[
\mathbf{D}^{(b)} = (D_i^{(b)})_{i \in \mathcal{I}}
\]
from the known probabilities $e_0(X_i)$ on the evaluation sample \(\mathcal{I}\).
These draws are also independent of the observed treatments given $\mathcal{F}$.
Stacking these alongside the observed vector
\[
\mathbf{D}^{(0)} = (D_i^{(0)})_{i \in \mathcal{I}} = (D_i)_{i \in \mathcal{I}}
\]
produces the matrix
\[
\mathbb{D}
= \big(D_i^{(b)}\big)_{b \in \{0, 1, \dots, B\},\, i \in \mathcal{I}},
\]
with rows that are independent and identically distributed conditional on $\mathcal{F}$. This implies that the rows are exchangeable given $\mathcal{F}$ under the null: any permutation of them has the same joint distribution.

Because both $\hat{f}^{\mathcal{I}^c}_w(Y_i, X_i)$ and $e_0(X_i)$ are $\mathcal{F}$-measurable, 
for all $b \in \{0,\ldots, B\}$ the test statistic
\[
S^{(b)} = \frac{1}{n} \sum_{i \in \mathcal{I}} w(D_i^{(b)}, X_i) \left(D_i^{(b)} - \hat{f}^{\mathcal{I}^c}_w(Y_i, X_i)\right)^2
\]
(so that $S^{(0)} = S_n$ and $S^{(b)} = S_n^{(b)}$ for $b \geq 1$ in \autoref{alg:crt}) is a function only of the row $\mathbf{D}^{(b)}$ conditional on $\mathcal{F}$, for a common function across $b$. Thus, the collection of statistics $\{S^{(0)}, S^{(1)}, \dots, S^{(B)}\}$ is conditionally exchangeable given $\mathcal{F}$ under the null.

Write $R$ for the ascending rank of $S^{(0)}$ among $\{S^{(0)}, S^{(1)}, \dots, S^{(B)}\}$, with ties broken uniformly at random and independently of the data and of the resampling draws.
Then $R$ is uniform on $\{1,\ldots, B + 1\}$ under the null, conditional on $\mathcal{F}$.
Also write $\tilde{R}$ for the number of entries in $\{S^{(0)}, S^{(1)}, \dots, S^{(B)}\}$ that are at most $S^{(0)}$.
Then $\tilde{R} \geq R$.
As a consequence, for the $p$-value
\[
p = \frac{1 + \sum_{b = 1}^{B} \mathbf{1}\!\left(S^{(b)} \leq S^{(0)}\right)}{B+1} = \frac{\tilde{R}}{B + 1},
\]
we have that
\begin{align*}
    \P(p \leq \alpha \,|\, \mathcal{F})
    &= \P\big(\tilde{R} \leq \lfloor \alpha (B+1) \rfloor \,\big|\, \mathcal{F}\big)
    \\
    &\leq \P\big(R \leq \lfloor \alpha (B+1) \rfloor \,\big|\, \mathcal{F}\big)
    = \frac{\lfloor \alpha (B+1) \rfloor}{B+1}
    \leq \alpha
    \qquad \text{for all } \alpha \in [0,1].
\end{align*}
Hence,
\begin{equation*}
     \P(p \leq \alpha)
     =
     \E[ \P(p \leq \alpha \,|\, \mathcal{F})]
         \leq \alpha
    \qquad \text{for all } \alpha \in [0,1].
    \qedhere
\end{equation*}
\end{proof}

\subsection{Proof of \autoref{thm:one_sided_aipw}, \autoref{thm:one_sided_dr}, and \autoref{thm:one_sided_drs}}\label{app:proof_one_sided_aipw}

\begin{proof}[Proof of \autoref{thm:one_sided_aipw}, \autoref{thm:one_sided_dr}, and \autoref{thm:one_sided_drs}]

Let the uncentered scores be defined as
\begin{align*}
\psi_i^{\aipw} &:= \psi_i^{\aipw}(0; e_0, \mu_0, \mu_1), \\[4pt]
\psi_i^{\dr} &:= \psi_i^{\dr}(0; e_0, m_0), \\[4pt]
\psi_i^{\drs} &:= \frac{\psi_i^{\drs}(0; e_0, m_0)}{\E\!\left[e_0(X_i)(1-e_0(X_i))\right]}
  = \frac{(D_i - e_0(X_i))\bigl( f_0(Y_i, X_i) - m_0(X_i) \bigr)}
         {\E\!\left[e_0(X_i)(1-e_0(X_i))\right]},
\end{align*}
for the $\aipw$ score of \eqref{eq:score_aipw}, the $\dr$ score of
\eqref{eq:score_function}, and the $\drs$ score component of \eqref{eq:score_drs}.
Using $e_0(X_i) = \E[D_i|X_i]$ and the tower property we find that
\begin{align*}
&\E[\psi_i^{\aipw}\mid X_i]
\\
&= \mu_1(X_i) {-} \mu_0(X_i)
{+} \frac{\E[D_i (f_0(Y_i, X_i) {-} \mu_1(X_i))|X_i]}{e_0(X_i)}
{-} \frac{\E[(1 {-} D_i)(f_0(Y_i, X_i) {-} \mu_0(X_i))|X_i]}{1 {-} e_0(X_i)} \\
&= \frac{\E[D_i f_0(Y_i, X_i)|X_i]}{e_0(X_i)} {-} \frac{\E[(1 {-} D_i) f_0(Y_i, X_i)|X_i]}{1 {-} e_0(X_i)}
= \frac{\E[(D_i {-} e_0(X_i)) f_0(Y_i, X_i)|X_i]}{e_0(X_i)(1 {-} e_0(X_i))},
\\[4pt]
& \E[\psi_i^{\dr} \mid X_i]
= \frac{\E[(D_i {-} e_0(X_i))(f_0(Y_i, X_i) {-} m_0(X_i))|X_i]}{e_0(X_i)(1 {-} e_0(X_i))} \\
&= \frac{\E[(D_i {-} e_0(X_i)) f_0(Y_i, X_i)|X_i] {-} m_0(X_i)\,\E[D_i {-} e_0(X_i)|X_i]}{e_0(X_i)(1 {-} e_0(X_i))}
=
\frac{\E[(D_i {-} e_0(X_i)) f_0(Y_i, X_i)|X_i]}{e_0(X_i)(1 {-} e_0(X_i))}, \\
& \E[\psi_i^{\drs}\mid X_i] = \frac{\E[(D_i {-} e_0(X_i))(f_0(Y_i, X_i) {-} m_0(X_i))|X_i]}{\E[e_0(X_i)(1 {-} e_0(X_i))]} = \frac{\E[(D_i {-} e_0(X_i)) f_0(Y_i, X_i)|X_i]}{\E[e_0(X_i)(1 {-} e_0(X_i))]}.
\end{align*}

In particular, $\mu_1(X_i), \mu_0(X_i), m_0(X_i)$ cancel out and the result does not depend on their choice.
By the tower property,
\begin{align}
&\E[(D_i {-} e_0(X_i)) f_0(Y_i, X_i)|X_i]
= \E\bigl[\E[(D_i {-} e_0(X_i)) f_0(Y_i, X_i)|X_i, Y_i] \,\big|\, X_i\bigr]
\notag
\\
&= \E\bigl[f_0(Y_i, X_i)\,(\E[D_i|X_i, Y_i] {-} e_0(X_i)) \,\big|\, X_i\bigr]
\notag
\\
&= \E\bigl[\P(D_i{=}1|Y_i, X_i)\,(\P(D_i{=}1|Y_i, X_i) {-} \P(D_i{=}1|X_i)) \,\big|\, X_i\bigr]
\notag
\\
&= \E\bigl[(\P(D_i{=}1|Y_i, X_i) {-} \P(D_i{=}1|X_i))^2 \,\big|\, X_i\bigr]
\notag
\\
&= \Var(\P(D_i{=}1|Y_i, X_i)|X_i) = \Var(f_0(Y_i, X_i)|X_i)
\label{eq:varequiv}
\end{align}
almost surely.
Hence,
\begin{align*}
\E[\psi_i^{\aipw} \mid X_i]
& = \E[\psi_i^{\dr}|X_i]  =  \frac{\Var(f_0(Y_i, X_i)|X_i)}{e_0(X_i)(1 {-} e_0(X_i))}
\geq 0, \\
\E[\psi_i^{\drs} \mid X_i]
& = \frac{\Var(f_0(Y_i, X_i)|X_i)}{\E[e_0(X_i)(1 {-} e_0(X_i))]} \geq 0,
\end{align*}
with a.s.\ equality if and only if $\Var(f_0(Y_i, X_i)|X_i) = 0$ a.s., i.e., $f_0(Y_i, X_i) = e_0(X_i)$ a.s. This is equivalent to $\P(D_i{=}1|X_i, Y_i) = \P(D_i{=}1|X_i)$ a.s., i.e., $D_i \perp Y_i|X_i$ a.s.\ by \autoref{prop:reverse_prediction}.
Hence, also $\theta_0 = \E[\psi_i^{\aipw}] = \E[\psi_i^{\dr}] \geq 0$ and $\theta_0^\star = \E[\psi_i^{\drs}] \geq 0$, with equality if and only if the null hypothesis holds.
\end{proof}

\subsection{Proof of \autoref{thm:asymptotics_orthogonal}}\label{app:proof_asymptotics}

\begin{proof}[Proof of \autoref{thm:asymptotics_orthogonal}]

We show the result of the theorem for the more general case where there is a bounded function $\tilde{f}_0$ with $\|\hat{f}^{\mathcal{I}^{c,1}} - \tilde{f}_0\|_{L_2} = o_p(1)$.
(The main statement of the theorem assumes that $\tilde{f}_0$ is well-specified, that is, $\tilde{f}_0 = f_0$, but the result in \autoref{thm:asymptotics_orthogonal} holds for more general $\tilde{f}_0$.)
Throughout, write $\hat{f} = \hat{f}^{\mathcal{I}^{c,1}}$.

\paragraph{Part 1: separation into components.}
We first separate out the estimators into their oracle components and the deviations from them.
Specifically, following \autoref{alg:aipw_dr},
write
\begin{align*}
    \hat{\theta}^{\aipw} &= \frac{1}{n}\sum_{i \in \mathcal{I}} \hat{\psi}_i^{\aipw}, &
    \hat{\theta}^{\dr} &= \frac{1}{n}\sum_{i \in \mathcal{I}} \hat{\psi}_i^{\dr}, \\
    \hat{\theta}^{\drs} &= \frac{\hat{\theta}^{\drs,1}}{\hat{\theta}^{\drs,2}}, &
    \hat{\theta}^{\drs,1} &= \frac{1}{n}\sum_{i \in \mathcal{I}} \hat{\psi}_i^{\drs,1},
    \qquad
    \hat{\theta}^{\drs,2} = \frac{1}{n}\sum_{i \in \mathcal{I}} \hat{\psi}_i^{\drs,2}
\end{align*}
for the estimators of interest.
To later prove asymptotic normality of
\begin{align*}
    \sqrt n\bigl(\hat\theta^{\aipw}{-}\theta_{\hat f}\bigr),
    \qquad
    \sqrt n\bigl(\hat\theta^{\dr}{-}\theta_{\hat f}\bigr),
    \qquad
    \sqrt n\bigl(\hat\theta^{\drs}{-}\theta_{\hat f}^\star\bigr)
    =
    (\hat\theta^{\drs,2})^{-1}
    \sqrt n 
    \bigl(\hat\theta^{\drs,1}{-} \hat\theta^{\drs,2} \: \theta_{\hat f}^\star\bigr),
\end{align*}
we add and subtract the conditional oracle estimators $\hat\theta^{\aipw}_{\hat f}$, $\hat\theta^{\dr}_{\hat f}$, $\hat\theta^{\drs,1}_{\hat f}$, $\hat\theta^{\drs,2}_{\hat f}$
to obtain
\begin{align}
    \sqrt n\bigl(\hat\theta^{\aipw}-\theta_{\hat f}\bigr)
    &= \sqrt n\bigl(\hat\theta^{\aipw}_{\hat f}{-}\theta_{\hat f}\bigr)
     + \sqrt n\bigl(\hat\theta^{\aipw}{-}\hat\theta^{\aipw}_{\hat f}\bigr),
    \notag
    \\
    \sqrt n\bigl(\hat\theta^{\dr}{-}\theta_{\hat f}\bigr)
    &= \sqrt n\bigl(\hat\theta^{\dr}_{\hat f}{-}\theta_{\hat f}\bigr)
     + \sqrt n\bigl(\hat\theta^{\dr}{-}\hat\theta^{\dr}_{\hat f}\bigr),
    \label{eq:tripledecomp} 
    \\
    \sqrt n\bigl(\hat\theta^{\drs,1}{-}\hat\theta^{\drs,2}\,\theta_{\hat f}^\star\bigr)
    &= \sqrt n\bigl(\hat\theta^{\drs,1}_{\hat f}{-}\hat\theta^{\drs,2}_{\hat f} \,\theta_{\hat f}^\star\bigr)
     + \sqrt n\bigl(\hat\theta^{\drs,1}{-}\hat\theta^{\drs,1}_{\hat f}\bigr)
     - \theta_{\hat f}^\star\,\sqrt n\bigl(\hat\theta^{\drs,2}{-}\hat\theta^{\drs,2}_{\hat f}\bigr),
     \notag
\end{align}
where
\(\hat\theta^{k}_{\hat f}
:= \frac{1}{n}\sum_{i \in \mathcal{I}} \psi^{k}_{i,\hat f}\)
for
\(k \in \{\aipw,\, \dr,\, (\drs,1),\, (\drs,2)\}\)
with
\begin{align*}
    \psi^{\aipw}_{i,\hat f}
    &:= \mu_{1,\hat f}(X_i) - \mu_{0,\hat f}(X_i)
     + \frac{D_i\bigl(\hat f(Y_i,X_i)-\mu_{1,\hat f}(X_i)\bigr)}{e_0(X_i)}
     - \frac{(1-D_i)\bigl(\hat f(Y_i,X_i)-\mu_{0,\hat f}(X_i)\bigr)}{1-e_0(X_i)}, \\
    \psi^{\dr}_{i,\hat f}
    &:= \frac{D_i - e_0(X_i)}{e_0(X_i)(1-e_0(X_i))}\bigl(\hat f(Y_i,X_i) - m_{\hat f}(X_i)\bigr), \\
    \psi^{\drs,1}_{i,\hat f}
    &:= (D_i - e_0(X_i))\bigl(\hat f(Y_i,X_i) - m_{\hat f}(X_i)\bigr),
    \qquad
    \psi^{\drs,2}_{i,\hat f}
    := (D_i - e_0(X_i))^2
\end{align*}

\noindent for
$\mu_{d,\hat f}(X_i) := \E[\hat f(Y_i, X_i) |\mathcal{I}^{c,1}, X_i, D_i {=} d]$
and
$m_{\hat f}(X_i) := \E[\hat f(Y_i, X_i) |\mathcal{I}^{c,1}, X_i]$.
(Note that $\psi^{\drs,2}_{i,\hat f}$ and $\hat\theta^{\drs,2}_{\hat f}$ do not depend on $\hat{f}$.)

\paragraph{Part 2: asymptotic normality of the oracle part.}
Second, we establish asymptotic normality with the desired limit for the first components in \eqref{eq:tripledecomp}.
Write
\begin{align*}
    \Delta^{\aipw}_{i,\hat{f}}
    := \psi^{\aipw}_{i,\hat{f}} - \theta_{\hat{f}},
    \qquad
    \Delta^{\dr}_{i,\hat{f}}
    := \psi^{\dr}_{i,\hat{f}} - \theta_{\hat{f}},
    \qquad
    \Delta^{\drs}_{i,\hat{f}}
    := \psi^{\drs,1}_{i,\hat{f}} - \theta_{\hat{f}}^\star\,\psi^{\drs,2}_{i,\hat{f}}.
\end{align*}
Then each of the three components can be expressed as $\frac{1}{\sqrt n}\sum_{i \in \mathcal{I}} \Delta^{k}_{i,\hat{f}}$ for $k \in \{\aipw,\, \dr,\, \drs\}$.
By definition of the estimands,
the $\Delta^{k}_{i,\hat{f}}$ are mean-zero and independent conditional on $\mathcal{I}^{c,1}$.

For the fixed target $\tilde{f}_0$ for which $\hat{f}$ is consistent, similarly define the $\Delta^{k}_{i,\tilde{f}_0}$ by plugging in $\tilde{f}_0$ instead of $\hat{f}$.
Then the
$\Delta^{k}_{i,\hat{f}} {-} \Delta^{k}_{i,\tilde{f}_0}$ are also mean-zero and independent conditional on $\mathcal{I}^{c,1}$,
with
\begin{align}
    &\E\left[
        \left(
            \frac{1}{\sqrt n}\sum_{i \in \mathcal{I}}
            \bigl(\Delta^{k}_{i,\hat{f}} {-} \Delta^{k}_{i,\tilde{f}_0}\bigr)
        \right)^2
        \middle|\mathcal{I}^{c,1}
    \right]
    =
    \Var\left(
        \frac{1}{\sqrt n}\sum_{i \in \mathcal{I}}
            \bigl(\Delta^{k}_{i,\hat{f}} {-} \Delta^{k}_{i,\tilde{f}_0}\bigr)
    \middle|\mathcal{I}^{c,1}\right)
    =
    \frac{1}{n}
    \sum_{i \in \mathcal{I}}
    \E\left[\left(
            \Delta^{k}_{i,\hat{f}} {-} \Delta^{k}_{i,\tilde{f}_0}
        \right)^2
        \middle|\mathcal{I}^{c,1}
        \right]
        \notag
    \\
    &=
    \E\left[\left(
            \Delta^{k}_{i,\hat{f}} {-} \Delta^{k}_{i,\tilde{f}_0}
        \right)^2
        \middle|\mathcal{I}^{c,1}
    \right]
    \leq
    C_k \,
    \E\left[\bigl(
            \hat{f}(Y_i, X_i) {-} \tilde{f}_0(Y_i, X_i)
        \bigr)^2
        \middle|\mathcal{I}^{c,1}
    \right]
    =
    C_k \, \bigl\|\hat{f} - \tilde{f}_0\bigr\|^2_{L_2}
    \label{eq:Ckbound}
\end{align}
with
\begin{align*}
    C_{\aipw} := \frac{36}{\eta^3},
    \qquad
    C_{\dr} := \frac{4}{\eta^2},
    \qquad
    C_{\drs} := 4\left(1 + \frac{1}{\E[e_0(X_i)(1-e_0(X_i))]}\right)^2.
\end{align*}

To derive these constants, write $u := \hat{f} - \tilde{f}_0$, $u_d := \mu_{d,\hat f} - \mu_{d,\tilde f_0}$, and $u_m := m_{\hat f} - m_{\tilde f_0}$, where
$\|u_d\|_{L_2} \leq \eta^{-1/2} \|u\|_{L_2}$ and $\|u_m\|_{L_2} \leq \|u\|_{L_2}$
by Jensen's inequality and \autoref{asm:overlap}.

To see this, note that, by linearity,
\begin{align*}
    u_m(X_i) = \E[u(Y_i,X_i) \mid \mathcal{I}^{c,1}, X_i],
    \qquad
    u_d(X_i) = \E[u(Y_i,X_i) \mid \mathcal{I}^{c,1}, X_i, D_i {=} d].
\end{align*}
Both $u_m$ and $u_d$ are thus conditional expectations of $u$ given
$\mathcal{I}^{c,1}$, and Jensen's inequality applied to $x \mapsto x^2$ gives
\begin{align*}
    u_m(X_i)^2 \leq \E\bigl[u(Y_i,X_i)^2 \mid \mathcal{I}^{c,1}, X_i\bigr],
    \qquad
    u_d(X_i)^2 \leq \E\bigl[u(Y_i,X_i)^2 \mid \mathcal{I}^{c,1}, X_i, D_i {=} d\bigr].
\end{align*}
For $u_m$, taking expectations over $X_i$ conditional on $\mathcal{I}^{c,1}$ and
applying the tower property yields
\begin{align*}
    \|u_m\|_{L_2}^2
    = \E\bigl[u_m(X_i)^2 \mid \mathcal{I}^{c,1}\bigr]
    \leq \E\Bigl[\E\bigl[u(Y_i,X_i)^2 \mid \mathcal{I}^{c,1}, X_i\bigr]
        \Bigm| \mathcal{I}^{c,1}\Bigr]
    = \E\bigl[u(Y_i,X_i)^2 \mid \mathcal{I}^{c,1}\bigr]
    = \|u\|_{L_2}^2 .
\end{align*}
For $u_d$, the tower property instead decomposes $\|u\|_{L_2}^2$ over the two
treatment arms, and discarding the $d' \neq d$ term gives
\begin{align*}
    \|u\|_{L_2}^2
    &= \E\Bigl[\textstyle\sum_{d' \in \{0,1\}}
        \P(D_i {=} d' \mid X_i) \,
        \E\bigl[u(Y_i,X_i)^2 \mid \mathcal{I}^{c,1}, X_i, D_i {=} d'\bigr]
        \Bigm| \mathcal{I}^{c,1}\Bigr]
    \\ \; \geq  \; 
    &\E\Bigl[\P(D_i {=} d \mid X_i) \,
        \E\bigl[u(Y_i,X_i)^2 \mid \mathcal{I}^{c,1}, X_i, D_i {=} d\bigr]
        \Bigm| \mathcal{I}^{c,1}\Bigr],
\end{align*}
where $\P(D_i {=} d \mid \mathcal{I}^{c,1}, X_i) = \P(D_i {=} d \mid X_i)$ because
$i \in \mathcal{I}$ is independent of the auxiliary fold $\mathcal{I}^{c,1}$.
Under \autoref{asm:overlap} we have $\eta \leq e_0(X_i) \leq 1 - \eta$, hence
$\P(D_i {=} d \mid X_i) \geq \eta$ for both $d \in \{0,1\}$, and so
\begin{align*}
    \|u\|_{L_2}^2
    \geq \eta \, \E\Bigl[\E\bigl[u(Y_i,X_i)^2 \mid \mathcal{I}^{c,1}, X_i, D_i {=} d\bigr]
        \Bigm| \mathcal{I}^{c,1}\Bigr]
    \geq \eta \, \E\bigl[u_d(X_i)^2 \mid \mathcal{I}^{c,1}\bigr]
    = \eta \, \|u_d\|_{L_2}^2 ,
\end{align*}
where the second inequality is the bound due to Jensen's inequality above. Rearranging,
\begin{align*}
    \|u_d\|_{L_2} \leq \eta^{-1/2} \|u\|_{L_2},
    \qquad
    \|u_m\|_{L_2} \leq \|u\|_{L_2} .
\end{align*}

For $\aipw$ and $\dr$, $\Delta^{k}_{i,\hat{f}} {-} \Delta^{k}_{i,\tilde{f}_0}$ is the conditionally centered version of the score difference $\psi^{k}_{i,\hat{f}} {-} \psi^{k}_{i,\tilde{f}_0}$, so its conditional second moment is bounded by that of the score difference itself.
For the score differences, the triangle inequality in $L_2$ together with $\eta \leq e_0(X_i) \leq 1 - \eta$ yields
\begin{align*}
    \bigl\|\psi^{\aipw}_{i,\hat f} {-} \psi^{\aipw}_{i,\tilde f_0}\bigr\|_{L_2}
    &= \Bigl\|
        u_1 - u_0
        + \tfrac{D_i \, (u {-} u_1)}{e_0(X_i)}
        - \tfrac{(1 {-} D_i)(u {-} u_0)}{1 {-} e_0(X_i)}
    \Bigr\|_{L_2}
    \\
    &\leq 2 \left(\frac{1}{\eta^{1/2}} + \frac{1}{\eta} + \frac{1}{\eta^{3/2}}\right) \|u\|_{L_2}
    \leq \frac{6}{\eta^{3/2}} \, \|u\|_{L_2},
    \\
    \bigl\|\psi^{\dr}_{i,\hat f} {-} \psi^{\dr}_{i,\tilde f_0}\bigr\|_{L_2}
    &= \Bigl\|
        \tfrac{D_i - e_0(X_i)}{e_0(X_i)(1 - e_0(X_i))} \, (u {-} u_m)
    \Bigr\|_{L_2}
    \leq \frac{1}{\eta} \bigl(\|u\|_{L_2} + \|u_m\|_{L_2}\bigr)
    \leq \frac{2}{\eta} \, \|u\|_{L_2},
\end{align*}
using $\eta \leq 1$ and $\bigl|\tfrac{D_i - e_0(X_i)}{e_0(X_i)(1 - e_0(X_i))}\bigr| \leq \tfrac{1}{\eta}$; squaring gives $C_{\aipw}$ and $C_{\dr}$.
For $\drs$,
\begin{align*}
    \Delta^{\drs}_{i,\hat{f}} - \Delta^{\drs}_{i,\tilde{f}_0}
    = \bigl(D_i {-} e_0(X_i)\bigr)(u {-} u_m) - \bigl(\theta_{\hat f}^\star {-} \theta_{\tilde f_0}^\star\bigr)\bigl(D_i {-} e_0(X_i)\bigr)^2,
\end{align*}
where $|D_i - e_0(X_i)| \leq 1$ and, by the Cauchy--Schwarz inequality,
\begin{align*}
    \bigl|\theta_{\hat f}^\star - \theta_{\tilde f_0}^\star\bigr|
    = \frac{\bigl|\E\bigl[(D_i {-} e_0(X_i))(u {-} u_m) \,\big|\, \mathcal{I}^{c,1}\bigr]\bigr|}{\E[e_0(X_i)(1 {-} e_0(X_i))]}
    \leq \frac{2 \, \|u\|_{L_2}}{\E[e_0(X_i)(1 {-} e_0(X_i))]},
\end{align*}
so that
$\bigl\|\Delta^{\drs}_{i,\hat{f}} - \Delta^{\drs}_{i,\tilde{f}_0}\bigr\|_{L_2}
\leq 2 \left(1 + \frac{1}{\E[e_0(X_i)(1 {-} e_0(X_i))]}\right) \|u\|_{L_2}$,
and squaring gives $C_{\drs}$.

Since $\|\hat{f} - \tilde{f}_0\|_{L_2} = o_p(1)$,
it follows from \eqref{eq:Ckbound} that
\begin{align*}
    \frac{1}{\sqrt n}\sum_{i \in \mathcal{I}} \Delta^{k}_{i,\hat{f}}
    =
    \frac{1}{\sqrt n}\sum_{i \in \mathcal{I}} \Delta^{k}_{i,\tilde{f}_0} + o_p(1)
\end{align*}
unconditionally by Markov's inequality.

Since also $\frac{1}{\sqrt n}\sum_{i \in \mathcal{I}} \Delta^{k}_{i,\tilde{f}_0} \xrightarrow{d} \N(0,\Var(\Delta_{i,\tilde{f}_0}^k))$ by the Central Limit Theorem,
we obtain that
\begin{align}
    \sqrt n\bigl(\hat\theta^{\aipw}_{\hat f}{-}\theta_{\hat f}\bigr)
    &\xrightarrow{d} \N\bigl(0,\, \Var(\psi^{\aipw}_{i,\tilde{f}_0})\bigr),
    \notag
    \\
    \sqrt n\bigl(\hat\theta^{\dr}_{\hat f}{-}\theta_{\hat f}\bigr)
    &\xrightarrow{d} \N\bigl(0,\, \Var(\psi^{\dr}_{i,\tilde{f}_0})\bigr),
    \label{eq:oracleclt}
    \\
    \sqrt n\bigl(\hat\theta^{\drs,1}_{\hat f}{-}\hat\theta^{\drs,2}_{\hat f}\,\theta_{\hat f}^\star\bigr)
    &\xrightarrow{d} \N\bigl(0,\, \Var\bigl(\psi^{\drs,1}_{i,\tilde{f}_0} - \theta_{\tilde{f}_0}^\star\,\psi^{\drs,2}_{i,\tilde{f}_0}\bigr)\bigr).
    \notag
\end{align}
Note that, if $\tilde{f}_0 = f_0$ and the null hypothesis is true, then we have that all of these variances are zero.
In this case, the result still applies, with convergence to zero in probability.
Conversely, for misspecified $\tilde{f}_0 \neq f_0$, these variances are generally positive even under the null hypothesis, in which case \eqref{eq:oracleclt} is an ordinary non-degenerate central limit theorem; no step of the argument requires the variances to vanish under the null.

\paragraph{Part 3: vanishing differences between feasible and oracle estimator.}
We now show that the remaining terms on the right-hand side of \eqref{eq:tripledecomp} vanish in probability.
Write
\begin{align*}
    \sqrt{n}\bigl(\hat\theta^{k} - \hat\theta^{k}_{\hat f}\bigr)
    = \frac{1}{\sqrt n}\sum_{i \in \mathcal{I}} \nu^{k}_{i},
    \qquad
    \nu^{k}_{i} := \hat{\psi}^{k}_{i} - \psi^{k}_{i,\hat{f}},
    \qquad
    k \in \{\aipw,\, \dr,\, (\drs,1),\, (\drs,2)\}.
\end{align*}
Using $\hat{e} = \hat{e}^{\mathcal{I}^{c,2}}$, $\hat{\mu}_d = \hat{\mu}^{\mathcal{I}^{c,2}}_{d}$, $\hat{m} = \hat{m}^{\mathcal{I}^{c,2}}$, we find
\begin{align*}
    \nu^{\aipw}_{i}
    &=
     \bigl(\hat{\mu}_{1}(X_i) - \mu_{1,\hat f}(X_i)\bigr)
     + D_i \left(
        \frac{\hat f(Y_i,X_i) - \hat{\mu}_{1}(X_i)}{\hat{e}(X_i)}
      - \frac{\hat f(Y_i,X_i) - \mu_{1,\hat f}(X_i)}{e_0(X_i)}
     \right) \\
    &\quad
     - \bigl(\hat{\mu}_{0}(X_i) - \mu_{0,\hat f}(X_i)\bigr)
     - (1-D_i) \left(
        \frac{\hat f(Y_i,X_i) - \hat{\mu}_{0}(X_i)}{1 - \hat{e}(X_i)}
      - \frac{\hat f(Y_i,X_i) - \mu_{0,\hat f}(X_i)}{1 - e_0(X_i)}
     \right), \\
    \nu^{\dr}_{i}
    &= \frac{D_i - \hat{e}(X_i)}{\hat{e}(X_i)\bigl(1 - \hat{e}(X_i)\bigr)}
       \bigl(\hat f(Y_i,X_i) - \hat{m}(X_i)\bigr)
     - \frac{D_i - e_0(X_i)}{e_0(X_i)\bigl(1 - e_0(X_i)\bigr)}
       \bigl(\hat f(Y_i,X_i) - m_{\hat f}(X_i)\bigr), \\
    \nu^{\drs,1}_{i}
    &= \bigl(D_i - \hat{e}(X_i)\bigr)\bigl(\hat f(Y_i,X_i) - \hat{m}(X_i)\bigr)
     - \bigl(D_i - e_0(X_i)\bigr)\bigl(\hat f(Y_i,X_i) - m_{\hat f}(X_i)\bigr), \\
    \nu^{\drs,2}_{i}
    &= \bigl(D_i - \hat{e}(X_i)\bigr)^2 - \bigl(D_i - e_0(X_i)\bigr)^2.
\end{align*}
Now define $\tilde{D}_i := D_i - e_0(X_i)$, $\tilde{Z}_i := \hat{f}(Y_i, X_i) - m_{\hat f}(X_i)$, and $\tilde{Z}_{di} := \hat{f}(Y_i, X_i) - \mu_{d,\hat f}(X_i)$ for $d \in \{0,1\}$.
The strategy \citep[analogous to Theorem 3.2 of][]{wager2024causal} is to decompose the summand into terms that separate the outcome model error from the propensity score error.

To separate the two error sources, we add and subtract in each difference the score evaluated at the estimated outcome models and the true propensity score $e_0$. For candidate outcome models $(m_1, m_0, m)$ and a candidate propensity score $e$, write
\begin{align*}
    F^{\aipw}_i(m_1, m_0, e) &:= m_1(X_i) - m_0(X_i)
    + \frac{D_i \bigl(\hat f(Y_i,X_i) - m_1(X_i)\bigr)}{e(X_i)}
    - \frac{(1{-}D_i)\bigl(\hat f(Y_i,X_i) - m_0(X_i)\bigr)}{1 - e(X_i)},
    \\
    F^{\dr}_i(m, e) &:= \frac{D_i - e(X_i)}{e(X_i)\bigl(1 - e(X_i)\bigr)}\bigl(\hat f(Y_i,X_i) - m(X_i)\bigr),
    \\
    F^{\drs,1}_i(m, e) &:= \bigl(D_i - e(X_i)\bigr)\bigl(\hat f(Y_i,X_i) - m(X_i)\bigr),
\end{align*}
so that
\[
    \nu^{\aipw}_{i}
    = \underbrace{F^{\aipw}_i(\hat\mu_1, \hat\mu_0, \hat e) - F^{\aipw}_i(\hat\mu_1, \hat\mu_0, e_0)}_{\text{only the propensity score changes}}
    + \underbrace{F^{\aipw}_i(\hat\mu_1, \hat\mu_0, e_0) - F^{\aipw}_i(\mu_{1,\hat f}, \mu_{0,\hat f}, e_0)}_{\text{only the outcome models change}},
\]
and analogously for $\nu^{\dr}_i$ and $\nu^{\drs,1}_i$ with $(\hat m, m_{\hat f})$ in place of the outcome models.
The outcome-model brackets simplify directly: using $1 - \tfrac{D_i}{e_0(X_i)} = -\tfrac{\tilde{D}_i}{e_0(X_i)}$ and $1 - \tfrac{1-D_i}{1-e_0(X_i)} = \tfrac{\tilde{D}_i}{1-e_0(X_i)}$, the $\aipw$ bracket equals $-\tfrac{\tilde{D}_i}{e_0(X_i)}\bigl(\hat{\mu}_1(X_i) - \mu_{1,\hat f}(X_i)\bigr) - \tfrac{\tilde{D}_i}{1-e_0(X_i)}\bigl(\hat{\mu}_0(X_i) - \mu_{0,\hat f}(X_i)\bigr)$, and the $\dr$ and $\drs$ brackets equal $-\tfrac{\tilde{D}_i}{e_0(X_i)(1-e_0(X_i))}\bigl(\hat{m}(X_i) - m_{\hat f}(X_i)\bigr)$ and $-\tilde{D}_i\bigl(\hat{m}(X_i) - m_{\hat f}(X_i)\bigr)$, respectively.
The propensity brackets contain the estimated residuals $\hat f - \hat{\mu}_d$ and $\hat f - \hat{m}$ multiplied by differences of the (inverse) propensity scores (for $\dr$, expand $\tfrac{D_i - e(X_i)}{e(X_i)(1-e(X_i))} = \tfrac{D_i}{e(X_i)} - \tfrac{1-D_i}{1-e(X_i)}$); substituting $\hat f - \hat{\mu}_d = \tilde{Z}_{di} - (\hat{\mu}_d - \mu_{d,\hat f})$ and $\hat f - \hat{m} = \tilde{Z}_i - (\hat{m} - m_{\hat f})$ (for $\dr$, on $\{D_i = d\}$, $\hat f - \hat{m} = \tilde{Z}_{di} + (\mu_{d,\hat f} - m_{\hat f}) - (\hat{m} - m_{\hat f})$, where $\mu_{1,\hat f}(X_i) - m_{\hat f}(X_i) = \bigl(1{-}e_0(X_i)\bigr)\theta_{\hat f}(X_i)$ and $\mu_{0,\hat f}(X_i) - m_{\hat f}(X_i) = -e_0(X_i)\,\theta_{\hat f}(X_i)$) splits each propensity bracket into terms that are first-order in a single estimation error and products of two estimation errors.
Finally, $\nu^{\drs,2}_i = -2\,\tilde{D}_i\bigl(\hat{e}(X_i) - e_0(X_i)\bigr) + \bigl(\hat{e}(X_i) - e_0(X_i)\bigr)^2$ directly.
Collecting the terms that are first-order in one estimation error into $R^{k}_{1i}$ (with the terms proportional to $\theta_{\hat f}(X_i)$ collected into $R^{\dr}_{3i}$ for $\dr$) and the products of two estimation errors into $R^{k}_{2i}$ yields
{\allowdisplaybreaks
\begin{align*}
    \nu^{\aipw}_{i}
    &= R_{1i}^{\aipw} + R_{2i}^{\aipw}
    \\
    R_{1i}^{\aipw} &=
    - \tfrac{\tilde{D}_i}{e_0(X_i)} \bigl(\hat{\mu}_1(X_i) {-} \mu_{1,\hat f}(X_i)\bigr)
       - \tfrac{\tilde{D}_i}{1{-}e_0(X_i)} \bigl(\hat{\mu}_0(X_i) {-} \mu_{0,\hat f}(X_i)\bigr)
       \\
    &\quad\quad\quad    + D_i \, \tilde{Z}_{1i} \bigl(\tfrac{1}{\hat{e}(X_i)} {-} \tfrac{1}{e_0(X_i)}\bigr)
       - (1 {-} D_i) \, \tilde{Z}_{0i} \bigl(\tfrac{1}{1{-}\hat{e}(X_i)} {-} \tfrac{1}{1{-}e_0(X_i)}\bigr)
       \\
    R_{2i}^{\aipw} &=
       - D_i \bigl(\hat{\mu}_1(X_i) {-} \mu_{1,\hat f}(X_i)\bigr) \bigl(\tfrac{1}{\hat{e}(X_i)} {-} \tfrac{1}{e_0(X_i)}\bigr)
       + (1 {-} D_i) \bigl(\hat{\mu}_0(X_i) {-} \mu_{0,\hat f}(X_i)\bigr) \bigl(\tfrac{1}{1{-}\hat{e}(X_i)} {-} \tfrac{1}{1{-}e_0(X_i)}\bigr),
    \\[1em]
    \nu^{\dr}_{i}
    &= R_{1i}^{\dr} + R_{2i}^{\dr} + R_{3i}^{\dr}
    \\
    R_{1i}^{\dr} &=
    D_i \, \tilde{Z}_{1i} \bigl(\tfrac{1}{\hat{e}(X_i)} {-} \tfrac{1}{e_0(X_i)}\bigr)
     - (1 {-} D_i) \, \tilde{Z}_{0i} \bigl(\tfrac{1}{1{-}\hat{e}(X_i)} {-} \tfrac{1}{1{-}e_0(X_i)}\bigr)
     - \tfrac{\tilde{D}_i}{e_0(X_i)(1{-}e_0(X_i))} \bigl(\hat{m}(X_i) {-} m_{\hat f}(X_i)\bigr)
    \\
    R_{2i}^{\dr} &=
     - \Bigl[
        D_i \bigl(\tfrac{1}{\hat{e}(X_i)} {-} \tfrac{1}{e_0(X_i)}\bigr)
        - (1 {-} D_i) \bigl(\tfrac{1}{1{-}\hat{e}(X_i)} {-} \tfrac{1}{1{-}e_0(X_i)}\bigr)
     \Bigr] \bigl(\hat{m}(X_i) {-} m_{\hat f}(X_i)\bigr)
    \\
    R_{3i}^{\dr} &=
     \theta_{\hat f}(X_i) \Bigl[
        D_i \bigl(1 {-} e_0(X_i)\bigr) \bigl(\tfrac{1}{\hat{e}(X_i)} {-} \tfrac{1}{e_0(X_i)}\bigr)
        + (1 {-} D_i) \, e_0(X_i) \bigl(\tfrac{1}{1{-}\hat{e}(X_i)} {-} \tfrac{1}{1{-}e_0(X_i)}\bigr)
     \Bigr],
    \\[1em]
    \nu^{\drs,1}_{i}
    &= R_{1i}^{\drs,1} + R_{2i}^{\drs,1}
    \\
    R_{1i}^{\drs,1} &=
    - \tilde{D}_i \bigl(\hat{m}(X_i) {-} m_{\hat f}(X_i)\bigr)
       - \tilde{Z}_i \bigl(\hat{e}(X_i) {-} e_0(X_i)\bigr)
    \\
    R_{2i}^{\drs,1} &=
       \bigl(\hat{e}(X_i) {-} e_0(X_i)\bigr) \bigl(\hat{m}(X_i) {-} m_{\hat f}(X_i)\bigr),
    \\[1em]
    \nu^{\drs,2}_{i}
    &= R_{1i}^{\drs,2} + R_{2i}^{\drs,2}
    \\
    R_{1i}^{\drs,2} &=
    - 2 \, \tilde{D}_i \bigl(\hat{e}(X_i) {-} e_0(X_i)\bigr)
    \\
    R_{2i}^{\drs,2} &=
       \bigl(\hat{e}(X_i) {-} e_0(X_i)\bigr)^2.
\end{align*}}
In each case, the leading products combine a factor that is mean-zero conditional on $X_i$ and the auxiliary samples and does not involve the estimated nuisances ($\tilde{D}_i$, $D_i \tilde{Z}_{1i}$, $(1 - D_i) \tilde{Z}_{0i}$, $\tilde{Z}_i$) with a nuisance estimation error that depends on $X_i$ and the auxiliary samples only, while the trailing terms are products of two estimation errors.
For $\dr$, the additional component $R_{3i}^{\dr}$ is first-order in the propensity estimation error but proportional to $\theta_{\hat f}(X_i)$, and therefore vanishes under the null.

Writing
\begin{align*}
    R_1^k = \frac{1}{\sqrt{n}} \sum_{i \in \mathcal{I}} R_{1i}^{k},
    \qquad
    R_2^k = \frac{1}{\sqrt{n}} \sum_{i \in \mathcal{I}} R_{2i}^{k},
    \qquad
    R_3^{\dr} = \frac{1}{\sqrt{n}} \sum_{i \in \mathcal{I}} R_{3i}^{\dr},
\end{align*}
we now study each sum separately.

First, the $R_{1i}^k$ are mean-zero and uncorrelated with $R_{1j}^k$ for $j \neq i$, conditional on $\mathcal{I}^{c, 1},\mathcal{I}^{c, 2}, X_i, X_j$.
As a consequence,
\begin{align*}
    \E[(R_{1}^k)^2|\mathcal{I}^{c, 1},\mathcal{I}^{c, 2}]
    = \frac{1}{n} \sum_{i \in \mathcal{I}} \E[(R_{1i}^k)^2|\mathcal{I}^{c, 1},\mathcal{I}^{c, 2}]
    = \E[(R_{1i}^k)^2|\mathcal{I}^{c, 1},\mathcal{I}^{c, 2}].
\end{align*}
In particular, for $\|\hat{f}\|_{L_\infty} \leq C$,
\begin{align*}
    \E[(R_{1}^{\aipw})^2|\mathcal{I}^{c, 1},\mathcal{I}^{c, 2}]^{1/2}
    &\leq \tfrac{1}{\eta} \bigl\|\hat{\mu}_1 {-} \mu_{1,\hat f}\bigr\|_{L_2}
        + \tfrac{1}{\eta} \bigl\|\hat{\mu}_0 {-} \mu_{0,\hat f}\bigr\|_{L_2}
        + 2 C \, \bigl\|\tfrac{1}{\hat{e}} {-} \tfrac{1}{e_0}\bigr\|_{L_2}
        + 2 C \, \bigl\|\tfrac{1}{1{-}\hat{e}} {-} \tfrac{1}{1{-}e_0}\bigr\|_{L_2}, \\
    \E[(R_{1}^{\dr})^2|\mathcal{I}^{c, 1},\mathcal{I}^{c, 2}]^{1/2}
    &\leq 2 C \, \bigl\|\tfrac{1}{\hat{e}} {-} \tfrac{1}{e_0}\bigr\|_{L_2}
        + 2 C \, \bigl\|\tfrac{1}{1{-}\hat{e}} {-} \tfrac{1}{1{-}e_0}\bigr\|_{L_2}
        + \tfrac{1}{\eta} \bigl\|\hat{m} {-} m_{\hat f}\bigr\|_{L_2}, \\
    \E[(R_{1}^{\drs,1})^2|\mathcal{I}^{c, 1},\mathcal{I}^{c, 2}]^{1/2}
    &\leq \bigl\|\hat{m} {-} m_{\hat f}\bigr\|_{L_2}
        + 2 C \, \bigl\|\hat{e} {-} e_0\bigr\|_{L_2}, \\
    \E[(R_{1}^{\drs,2})^2|\mathcal{I}^{c, 1},\mathcal{I}^{c, 2}]^{1/2}
    &\leq 2 \, \bigl\|\hat{e} {-} e_0\bigr\|_{L_2},
\end{align*}
by the triangle inequality in $L_2$, using that $|\tilde{D}_i| \leq 1$, that $|D_i \tilde{Z}_{1i}|, |(1 {-} D_i) \tilde{Z}_{0i}|, |\tilde{Z}_i| \leq 2 C$, and that $\eta \leq e_0(X_i) \leq 1 - \eta$ (\autoref{asm:overlap}) for the $\aipw$ and $\dr$ bounds.
As a result, by Markov's inequality, $R_{1}^k \xrightarrow{p} 0$ unconditionally.

Second, 
$\E[|R_{2}^k| \, |\mathcal{I}^{c, 1},\mathcal{I}^{c, 2}] \leq \tfrac{1}{\sqrt{n}} \sum_{i \in \mathcal{I}} \E[|R_{2i}^k| \, |\mathcal{I}^{c, 1},\mathcal{I}^{c, 2}] = \sqrt{n} \, \E[|R_{2i}^k| \, |\mathcal{I}^{c, 1},\mathcal{I}^{c, 2}]$,
and so, by the Cauchy--Schwarz inequality,
\begin{align*}
    \E[|R_{2}^{\aipw}| \, |\mathcal{I}^{c, 1},\mathcal{I}^{c, 2}]
    &\leq \sqrt{n} \left(
        \bigl\|\hat{\mu}_1 {-} \mu_{1,\hat f}\bigr\|_{L_2} \bigl\|\tfrac{1}{\hat{e}} {-} \tfrac{1}{e_0}\bigr\|_{L_2}
        + \bigl\|\hat{\mu}_0 {-} \mu_{0,\hat f}\bigr\|_{L_2} \bigl\|\tfrac{1}{1{-}\hat{e}} {-} \tfrac{1}{1{-}e_0}\bigr\|_{L_2}
    \right), \\
    \E[|R_{2}^{\dr}| \, |\mathcal{I}^{c, 1},\mathcal{I}^{c, 2}]
    &\leq \sqrt{n} \Bigl(\bigl\|\tfrac{1}{\hat{e}} {-} \tfrac{1}{e_0}\bigr\|_{L_2}
        + \bigl\|\tfrac{1}{1{-}\hat{e}} {-} \tfrac{1}{1{-}e_0}\bigr\|_{L_2}\Bigr) \bigl\|\hat{m} {-} m_{\hat f}\bigr\|_{L_2}, \\
    \E[|R_{2}^{\drs,1}| \, |\mathcal{I}^{c, 1},\mathcal{I}^{c, 2}]
    &\leq \sqrt{n} \, \bigl\|\hat{e} {-} e_0\bigr\|_{L_2} \, \bigl\|\hat{m} {-} m_{\hat f}\bigr\|_{L_2}, \\
    \E[|R_{2}^{\drs,2}| \, |\mathcal{I}^{c, 1},\mathcal{I}^{c, 2}]
    &\leq \sqrt{n} \, \bigl\|\hat{e} {-} e_0\bigr\|_{L_2}^2.
\end{align*}
Since all nuisance components are $o_p(n^{-1/4})$ by \autoref{asm:nuisance_rates}, by Markov's inequality we also have that $R_{2}^k \xrightarrow{p} 0$ unconditionally.

In addition, $R_{3i}^{\dr} = 0$ a.s.\ under the null hypothesis (since $\theta_{\hat{f}}(X_i) = 0$ a.s.), so $R_3^{\dr} = 0$.

The key to these bounds is the orthogonality of the $\aipw$ and $\drs$ components, conditional on $\mathcal{I}^{c,1}$:
since the expectation of the first derivative of the increments $\hat \psi_i^k$ with respect to the nuisance functions is zero around their true value, moderate deviations in estimating those nuisance components do not have a first-order effect, but only a second-order effect.
For $\dr$, this orthogonality generally only holds under the null, and it is possible that $\sqrt{n} \bigl(\hat\theta^{\dr} - \hat\theta^{\dr}_{\hat f}\bigr)$ does not vanish asymptotically when the null does not hold.

Taking all pieces together, we conclude that under the conditions of the theorem
\begin{align}
    \sqrt{n} \bigl(\hat\theta^{k} {-} \hat\theta^{k}_{\hat f}\bigr)
    = R_1^{k} + R_2^{k}
    \xrightarrow{p} 0,
    \qquad
    k \in \{\aipw,\, \dr,\, (\drs,1),\, (\drs,2)\}.
    \label{eq:remaindersvanish}
\end{align}

\paragraph{Part 4: overall asymptotic normality.}

For $\aipw$, the result follows directly by combining \eqref{eq:tripledecomp}, \eqref{eq:oracleclt}, and \eqref{eq:remaindersvanish}.

For $\drs$, note that $\theta_{\hat f}^\star$ is bounded, so from combining \eqref{eq:tripledecomp}, \eqref{eq:oracleclt}, and \eqref{eq:remaindersvanish} we obtain
\begin{align*}
    \sqrt{n} \bigl(\hat\theta^{\drs,1} {-} \hat\theta^{\drs,2} \, \theta_{\hat f}^\star\bigr)
    \xrightarrow{d} \N\bigl(0,\, \Var\bigl(\psi^{\drs,1}_{i,\tilde{f}_0} - \theta_{\tilde{f}_0}^\star\,\psi^{\drs,2}_{i,\tilde{f}_0}\bigr)\bigr).
\end{align*}
In addition, by \eqref{eq:remaindersvanish}, $\hat\theta^{\drs,2} = \hat\theta^{\drs,2}_{\hat f} + o_p(1)$
and
$\hat\theta^{\drs,2}_{\hat f} \xrightarrow{p} \E[e_0(X_i)(1 {-} e_0(X_i))]$ by the Law of Large Numbers.
Hence, $\hat\theta^{\drs,2} \xrightarrow{p} \E[e_0(X_i)(1 {-} e_0(X_i))]$, which is positive since $\P(e_0(X_i) \in (0,1)) > 0$, and by Slutsky we have that
\begin{align*}
    \sqrt{n} \bigl(\hat\theta^{\drs} {-} \theta_{\hat f}^\star\bigr)
    = \bigl(\hat\theta^{\drs,2}\bigr)^{-1} \sqrt{n} \bigl(\hat\theta^{\drs,1} {-} \hat\theta^{\drs,2} \, \theta_{\hat f}^\star\bigr)
    \xrightarrow{d} \N\left(0,\, \frac{\Var\bigl(\psi^{\drs,1}_{i,\tilde{f}_0} - \theta_{\tilde{f}_0}^\star\,\psi^{\drs,2}_{i,\tilde{f}_0}\bigr)}{\bigl(\E[e_0(X_i)(1 {-} e_0(X_i))]\bigr)^2}\right).
\end{align*}

For $\dr$ under the null, note that the variance in \eqref{eq:oracleclt} equals $\Var(\psi^{\dr}_{i,\tilde{f}_0}) = 0$ if $\tilde{f}_0 = f_0$. However, all results apply even in the case when the variance is zero (in which case \eqref{eq:oracleclt} simply yields convergence in probability to zero), and the result extends to other $\tilde{f}_0$ that may yield asymptotic normality with non-trivial variance. Combining \eqref{eq:tripledecomp}, \eqref{eq:oracleclt}, and \eqref{eq:remaindersvanish} (where $R_3^{\dr} = 0$ under the null) thus yields $\sqrt{n}\bigl(\hat\theta^{\dr} {-} \theta_{\hat f}\bigr) \xrightarrow{p} 0$.
\end{proof}

\begin{remark}[Conditional Neyman orthogonality of the $\aipw$ and $\drs$ scores, as well as the $\dr$ score under the null]\label{rem:orthogonality_scores}
The scores corresponding to the $\aipw$ and $\drs$ test statistics are Neyman orthogonal with respect to the propensity score and outcome models, conditional on the predicted treatment probability.
The same is true for the $\dr$ test statistic under the null.
\end{remark}

\begin{proof}

For realizations $(Y,X,D)$, a parameter value $\theta$, and conformable functions $(f, e, m_1, m_0, m)$ define
the score functions
\begin{align*}
    \psi^{\aipw}_0(Y,X,D; \theta, f, e, m_1, m_0)
    &= m_1(X) {-} m_0(X)
    \\
    &\quad
    + \frac{D \, \bigl(f(Y,X) {-} m_1(X)\bigr)}{e(X)}
    - \frac{(1{-}D) \bigl(f(Y,X) {-} m_0(X)\bigr)}{1 {-} e(X)}
    - \theta, \\
    \psi^{\dr}_0(Y,X,D; \theta, f, e, m)
    &= \frac{D {-} e(X)}{e(X)\bigl(1 {-} e(X)\bigr)} \bigl(f(Y,X) {-} m(X)\bigr) - \theta, \\
    \psi^{\drs}_0(Y,X,D; \theta, f, e, m)
    &= \bigl(D {-} e(X)\bigr)\bigl(f(Y,X) {-} m(X)\bigr) - \theta \, \bigl(D {-} e(X)\bigr)^2.
\end{align*}
These are the score functions corresponding to the test statistics $\hat\theta^{\aipw}$, $\hat\theta^{\dr}$, and $\hat\theta^{\drs}$.

All three score functions only depend on the nuisance parameters through their fitted values. Specifically:
\begin{align*}
    \psi^{\aipw}_0(Y,X,D; \theta, f, e, m_1, m_0)
    &= \psi^{\aipw}\bigl(D; \theta, f(Y,X), e(X), m_1(X), m_0(X)\bigr),
    \\
    \psi^{\aipw}(D; \theta, F, E, M_1, M_0)
    &:= M_1 {-} M_0 + \frac{D \, (F {-} M_1)}{E} - \frac{(1{-}D) (F {-} M_0)}{1 {-} E} - \theta,
    \\[0.5em]
    \psi^{\dr}_0(Y,X,D; \theta, f, e, m)
    &= \psi^{\dr}\bigl(D; \theta, f(Y,X), e(X), m(X)\bigr),
    \\
    \psi^{\dr}(D; \theta, F, E, M)
    &:= \frac{D {-} E}{E (1 {-} E)} (F {-} M) - \theta,
    \\[0.5em]
    \psi^{\drs}_0(Y,X,D; \theta, f, e, m)
    &= \psi^{\drs}\bigl(D; \theta, f(Y,X), e(X), m(X)\bigr),
    \\
    \psi^{\drs}(D; \theta, F, E, M)
    &:= (D {-} E)(F {-} M) - \theta \, (D {-} E)^2.
\end{align*}

Conditional Neyman orthogonality corresponds to the Gateaux derivative of $\E[\psi^{k}_0(\ldots) \,|\, \mathcal{I}^{c,1}]$ with respect to $(e, m_1, m_0, m)$ at their oracle values $(e_0, \mu_{1,\hat f}, \mu_{0,\hat f}, m_{\hat f})$ being zero, given $\mathcal{I}^{c,1}$.
Here, since the scores only depend on the fitted values, Neyman orthogonality requires that, a.s.,

\begin{align*}
    \E\left[\frac{\partial \psi^{\aipw}(D_i; \theta_{\hat f}, \hat{f}(Y_i, X_i), E, M_1, M_0)}{\partial (E, M_1, M_0)} \Bigg|_{\begin{subarray}{l} E = e_0(X_i) \\ M_1 = \mu_{1,\hat f}(X_i) \\ M_0 = \mu_{0,\hat f}(X_i) \end{subarray}} \,\middle|\, \mathcal{I}^{c,1}, X_i\right] &= 0,
    \\[0.5em]
    \E\left[\frac{\partial \psi^{\dr}(D_i; \theta_{\hat f}, \hat{f}(Y_i, X_i), E, M)}{\partial (E, M)} \Bigg|_{\begin{subarray}{l} E = e_0(X_i) \\ M = m_{\hat f}(X_i) \end{subarray}} \,\middle|\, \mathcal{I}^{c,1}, X_i\right] &= 0,
    \\[0.5em]
    \E\left[\frac{\partial \psi^{\drs}(D_i; \theta_{\hat f}^\star, \hat{f}(Y_i, X_i), E, M)}{\partial (E, M)} \Bigg|_{\begin{subarray}{l} E = e_0(X_i) \\ M = m_{\hat f}(X_i) \end{subarray}} \,\middle|\, \mathcal{I}^{c,1}, X_i\right] &= 0.
\end{align*}

The derivatives evaluate as
\begin{align*}
    \frac{\partial \psi^{\aipw}}{\partial E}
    &= - \frac{D \, (F {-} M_1)}{E^2} - \frac{(1{-}D)(F {-} M_0)}{(1{-}E)^2},
    &
    \frac{\partial \psi^{\aipw}}{\partial M_1}
    &= 1 - \frac{D}{E},
    &
    \frac{\partial \psi^{\aipw}}{\partial M_0}
    &= \frac{1{-}D}{1{-}E} - 1,
    \\
    \frac{\partial \psi^{\dr}}{\partial E}
    &= - \left(\frac{D}{E^2} + \frac{1{-}D}{(1{-}E)^2}\right)(F {-} M),
    &
    \frac{\partial \psi^{\dr}}{\partial M}
    &= - \frac{D {-} E}{E(1{-}E)},
    \\
    \frac{\partial \psi^{\drs}}{\partial E}
    &= - (F {-} M) + 2 \theta \, (D {-} E),
    &
    \frac{\partial \psi^{\drs}}{\partial M}
    &= - (D {-} E).
\end{align*}
The conditional expectations are zero for $\aipw$ and $\drs$, and for $\dr$ under the null.
\end{proof}

\subsection{Proof of \autoref{cor:asymptotics_test}}
\label{app:proof_corollary}

\begin{proof}[Proof of \autoref{cor:asymptotics_test}]
    We leverage the results from the proof of \autoref{thm:asymptotics_orthogonal} in order to describe the asymptotic distribution of the three test statistics.
    Throughout, we use the notation from that proof, including
    the abbreviations $\hat{e} = \hat{e}^{\mathcal{I}^{c,2}}$, $\hat{\mu}_d = \hat{\mu}^{\mathcal{I}^{c,2}}_{d}$, and $\hat{m} = \hat{m}^{\mathcal{I}^{c,2}}$.

    \paragraph{Part 1: convergence of estimates in probability.}
    By \autoref{thm:asymptotics_orthogonal}, $\hat\theta^{\aipw} - \theta_{\hat f} \xrightarrow{p} 0$ and $\hat\theta^{\drs} - \theta_{\hat f}^\star \xrightarrow{p} 0$.
    Furthermore, in the proof of \autoref{thm:asymptotics_orthogonal},
    \begin{align*}
        \E[|n^{-1/2} R_3^{\dr}| \, |\mathcal{I}^{c, 1},\mathcal{I}^{c, 2}]
        &\leq \frac{1}{n} \sum_{i \in \mathcal{I}} \E[|R_{3i}^{\dr}| \, |\mathcal{I}^{c, 1},\mathcal{I}^{c, 2}]
        = \E[|R_{3i}^{\dr}| \, |\mathcal{I}^{c, 1},\mathcal{I}^{c, 2}]
        \\
        &\leq 2 C \left(
            \bigl\|\tfrac{1}{\hat{e}} {-} \tfrac{1}{e_0}\bigr\|_{L_2}
            + \bigl\|\tfrac{1}{1{-}\hat{e}} {-} \tfrac{1}{1{-}e_0}\bigr\|_{L_2}
        \right)
        \xrightarrow{p} 0
    \end{align*}
    by \autoref{asm:nuisance_rates},
    so $n^{-1/2} R_3^{\dr} \xrightarrow{p} 0$ by Markov's inequality
    and
    \begin{align*}
        \hat\theta^{\dr} - \theta_{\hat f}
        =
        n^{-1/2} \left(
            \sqrt{n} \bigl(\hat\theta^{\dr}_{\hat f} {-} \theta_{\hat f}\bigr)
            + R_1^{\dr} + R_2^{\dr}\right) + n^{-1/2} R_3^{\dr}
        \xrightarrow{p} 0,
    \end{align*}
since $\sqrt{n} \bigl(\hat\theta^{\dr}_{\hat f} {-} \theta_{\hat f}\bigr) = O_p(1)$ by \eqref{eq:oracleclt} and $R_1^{\dr}, R_2^{\dr} \xrightarrow{p} 0$ as in the proof of \autoref{thm:asymptotics_orthogonal}.
    Also,
    \begin{equation}
    \label{eq:targetconvergence}
    \begin{aligned}
        \bigl|\theta_{\hat f} - \theta_{\tilde{f}_0}\bigr|
        &\leq \bigl\|\mu_{1,\hat f} {-} \mu_{1,\tilde f_0}\bigr\|_{L_2}
        + \bigl\|\mu_{0,\hat f} {-} \mu_{0,\tilde f_0}\bigr\|_{L_2}
        \leq \frac{2}{\sqrt{\eta}} \, \bigl\|\hat{f} - \tilde{f}_0\bigr\|_{L_2}
        \xrightarrow{p} 0,
        \\
        \bigl|\theta^\star_{\hat f} - \theta^\star_{\tilde{f}_0}\bigr|
        &\leq \frac{2 \, \bigl\|\hat{f} - \tilde{f}_0\bigr\|_{L_2}}{\E[e_0(X_i)(1 {-} e_0(X_i))]}
        \xrightarrow{p} 0,
    \end{aligned}
    \end{equation}
    by Jensen's inequality and the Cauchy--Schwarz inequality, respectively.
    Taken together,
    \begin{align}
        \hat\theta^{\aipw} \xrightarrow{p} \theta_{\tilde{f}_0},
        \qquad
        \hat\theta^{\dr} \xrightarrow{p} \theta_{\tilde{f}_0},
        \qquad
        \hat\theta^{\drs} \xrightarrow{p} \theta^\star_{\tilde{f}_0}.
        \label{eq:estimandconvergence}
    \end{align}

    \paragraph{Part 2: convergence of variance estimates.}

    We next show that the variance estimators from \autoref{alg:aipw_dr} are consistent for the asymptotic variances from \autoref{thm:asymptotics_orthogonal}.
    Recall that
    \begin{align*}
        \hat{\sigma}^2_{\aipw}
        &= \frac{1}{n}\sum_{i\in\mathcal{I}}(\hat{\psi}_i^{\aipw} - \hat{\theta}^{\aipw})^2,
        \qquad
        \hat{\sigma}^2_{\dr}
        = \frac{1}{n}\sum_{i\in\mathcal{I}}(\hat{\psi}_i^{\dr} - \hat{\theta}^{\dr})^2,
        \\
        \hat{\sigma}^2_{\drs}
        &= \frac{\frac{1}{n}\sum_{i\in\mathcal{I}} (\hat{\psi}_i^{\drs,1} - \hat{\theta}^{\drs} \hat{\psi}_i^{\drs,2})^2}{(\frac{1}{n}\sum_{i\in\mathcal{I}} \hat{\psi}_i^{\drs,2})^2}.
    \end{align*}
    We have, for $k, l \in \{\aipw,\, \dr,\, (\drs,1),\, (\drs,2)\}$, that
    \begin{align}
        &\E\left[\left|
            \frac{1}{n}\sum_{i \in \mathcal{I}} \hat{\psi}^{k}_i \hat{\psi}^{l}_i
            - \frac{1}{n}\sum_{i \in \mathcal{I}} \psi^{k}_{i,\tilde{f}_0} \psi^{l}_{i,\tilde{f}_0}
        \right| \,\middle|\, \mathcal{I}^{c, 1},\mathcal{I}^{c, 2}\right]
        \notag
        \\
        &\leq
        \bigl\|\hat{\psi}^{k}_i {-} \psi^{k}_{i,\tilde{f}_0}\bigr\|_{L_2} \bigl\|\hat{\psi}^{l}_i {-} \psi^{l}_{i,\tilde{f}_0}\bigr\|_{L_2}
        + \bigl\|\hat{\psi}^{k}_i {-} \psi^{k}_{i,\tilde{f}_0}\bigr\|_{L_2} \bigl\|\psi^{l}_{i,\tilde{f}_0}\bigr\|_{L_2}
        + \bigl\|\psi^{k}_{i,\tilde{f}_0}\bigr\|_{L_2} \bigl\|\hat{\psi}^{l}_i {-} \psi^{l}_{i,\tilde{f}_0}\bigr\|_{L_2}
        \label{eq:CSbound}
    \end{align}
    by the Cauchy--Schwarz inequality.
    Using the decomposition in the proof of \autoref{thm:asymptotics_orthogonal},
    \begin{align*}
        \bigl\|\hat{\psi}^{k}_i {-} \psi^{k}_{i,\tilde{f}_0}\bigr\|_{L_2}
        \leq \bigl\|\nu^{k}_{i}\bigr\|_{L_2} + \bigl\|\psi^{k}_{i,\hat{f}} {-} \psi^{k}_{i,\tilde{f}_0}\bigr\|_{L_2},
    \end{align*}
    where $\bigl\|\psi^{k}_{i,\hat{f}} {-} \psi^{k}_{i,\tilde{f}_0}\bigr\|_{L_2} \xrightarrow{p} 0$ as in that proof,
    and $\bigl\|\nu^{k}_{i}\bigr\|_{L_2} \leq \bigl\|R^{k}_{1i}\bigr\|_{L_2} + \bigl\|R^{k}_{2i}\bigr\|_{L_2}$ (plus $\bigl\|R^{\dr}_{3i}\bigr\|_{L_2}$ for $k = \dr$), all of which $\xrightarrow{p} 0$:
    $R^{k}_{1i}$ and $R^{\dr}_{3i}$ are bounded in $L_2$ by nuisance-error norms as in the proof of \autoref{thm:asymptotics_orthogonal},
    while the $L_\infty$ bounds on $\hat{e}, \hat{\mu}_1, \hat{\mu}_0, \hat{m}$, and $\hat{f}$ (and hence on $\mu_{d,\hat f}$ and $m_{\hat f}$) imply that, for a constant $C'$,
    \begin{align*}
        \bigl\|R^{\aipw}_{2i}\bigr\|_{L_2}, \bigl\|R^{\dr}_{2i}\bigr\|_{L_2}
        &\leq C' \left(
            \bigl\|\tfrac{1}{\hat{e}} {-} \tfrac{1}{e_0}\bigr\|_{L_2}
            + \bigl\|\tfrac{1}{1{-}\hat{e}} {-} \tfrac{1}{1{-}e_0}\bigr\|_{L_2}
        \right),
        \\
        \bigl\|R^{\drs,1}_{2i}\bigr\|_{L_2}, \bigl\|R^{\drs,2}_{2i}\bigr\|_{L_2}
        &\leq C' \, \bigl\|\hat{e} {-} e_0\bigr\|_{L_2},
    \end{align*}
    which vanish in probability by \autoref{asm:nuisance_rates}.
    Since the $\psi^{l}_{i,\tilde{f}_0}$ are bounded, all terms in \eqref{eq:CSbound} converge to zero in probability.
    Hence, by Markov's inequality and the Law of Large Numbers, we obtain
    \begin{align*}
        \frac{1}{n}\sum_{i \in \mathcal{I}} \hat{\psi}^{k}_i \hat{\psi}^{l}_i
        =
        \frac{1}{n}\sum_{i \in \mathcal{I}} \psi^{k}_{i,\tilde{f}_0} \psi^{l}_{i,\tilde{f}_0} + o_p(1)
        \xrightarrow{p}
        \E[\psi^{k}_{i,\tilde{f}_0} \psi^{l}_{i,\tilde{f}_0}].
    \end{align*}
    Resolving the squares, the variance estimators equal
    \begin{align*}
        \hat{\sigma}^2_{\aipw}
        &= \frac{1}{n}\sum_{i\in\mathcal{I}} (\hat{\psi}_i^{\aipw})^2 - (\hat{\theta}^{\aipw})^2,
        \qquad
        \hat{\sigma}^2_{\dr}
        = \frac{1}{n}\sum_{i\in\mathcal{I}} (\hat{\psi}_i^{\dr})^2 - (\hat{\theta}^{\dr})^2,
        \\
        \hat{\sigma}^2_{\drs}
        &= \frac{
            \frac{1}{n}\sum_{i\in\mathcal{I}} (\hat{\psi}_i^{\drs,1})^2
            - 2 \, \hat{\theta}^{\drs} \, \frac{1}{n}\sum_{i\in\mathcal{I}} \hat{\psi}_i^{\drs,1} \hat{\psi}_i^{\drs,2}
            + (\hat{\theta}^{\drs})^2 \, \frac{1}{n}\sum_{i\in\mathcal{I}} (\hat{\psi}_i^{\drs,2})^2
        }{\bigl(\frac{1}{n}\sum_{i\in\mathcal{I}} \hat{\psi}_i^{\drs,2}\bigr)^2},
    \end{align*}
    so with \eqref{eq:estimandconvergence}, $\frac{1}{n}\sum_{i\in\mathcal{I}} \hat{\psi}_i^{\drs,2} = \hat\theta^{\drs,2} \xrightarrow{p} \E[e_0(X_i)(1 {-} e_0(X_i))]$ (positive since $\P(e_0(X_i) \in (0,1)) > 0$) from the proof of \autoref{thm:asymptotics_orthogonal}, and the continuous mapping theorem,
    \begin{align}
        \hat{\sigma}^2_{\aipw}
        &\xrightarrow{p} \E[(\psi^{\aipw}_{i,\tilde{f}_0})^2] - \theta_{\tilde{f}_0}^2
        = \Var(\psi^{\aipw}_{i,\tilde{f}_0}),
        \qquad
        \hat{\sigma}^2_{\dr}
        \xrightarrow{p} \E[(\psi^{\dr}_{i,\tilde{f}_0})^2] - \theta_{\tilde{f}_0}^2
        = \Var(\psi^{\dr}_{i,\tilde{f}_0}),
        \notag
        \\
        \hat{\sigma}^2_{\drs}
        &\xrightarrow{p}
        \frac{\E[(\psi^{\drs,1}_{i,\tilde{f}_0} - \theta^\star_{\tilde{f}_0} \psi^{\drs,2}_{i,\tilde{f}_0})^2]}{\bigl(\E[e_0(X_i)(1 {-} e_0(X_i))]\bigr)^2}
        =
        \frac{\Var\bigl(\psi^{\drs,1}_{i,\tilde{f}_0} - \theta^\star_{\tilde{f}_0} \psi^{\drs,2}_{i,\tilde{f}_0}\bigr)}{\bigl(\E[e_0(X_i)(1 {-} e_0(X_i))]\bigr)^2},
        \label{eq:varianceconvergence}
    \end{align}
    which are the asymptotic variances from \eqref{eq:oracleclt}, with the $\drs$ normalization from \autoref{thm:asymptotics_orthogonal}.

    \paragraph{Part 3: CLT for the test statistics.}
    Write $\theta^{\aipw}_{\hat f} := \theta_{\hat f}$, $\theta^{\dr}_{\hat f} := \theta_{\hat f}$, and $\theta^{\drs}_{\hat f} := \theta^\star_{\hat f}$.
    We have that
    \begin{align*}
        T_n^{k} - \sqrt{n} \frac{\theta^{k}_{\hat f}}{\sqrt{\hat{\sigma}^2_{k} + \delta^2}}
        = \frac{\sqrt{n}\bigl(\hat{\theta}^{k} {-} \theta^{k}_{\hat f}\bigr) + \delta\,Z_k}{\sqrt{\hat{\sigma}^2_{k} + \delta^2}},
        \qquad
        k \in \{\aipw,\, \dr,\, \drs\},
    \end{align*}
    where $Z_k \sim \N(0,1)$ is drawn independently of the data.
    Write also $\sigma^2_{k}$ for the respective probability limit from \eqref{eq:varianceconvergence}.
    Under the assumptions of the corollary, we have that
    \begin{align*}
        \sqrt{n}\bigl(\hat{\theta}^{k} {-} \theta^{k}_{\hat f}\bigr) + \delta\,Z_k
        \xrightarrow{d} \N(0,\, \sigma^2_{k} + \delta^2),
        \qquad
        k \in \{\aipw,\, \drs\},
    \end{align*}
    by \autoref{thm:asymptotics_orthogonal} and the independence of $Z_k$ from the data,
    which together with \eqref{eq:varianceconvergence} and Slutsky (using $\sigma^2_{k} + \delta^2 \geq \delta^2 > 0$) yields
    \begin{align*}
        T_n^{k} - \sqrt{n} \frac{\theta^{k}_{\hat f}}{\sqrt{\hat{\sigma}^2_{k} + \delta^2}}
        =
        \frac{\sqrt{n}\bigl(\hat{\theta}^{k} {-} \theta^{k}_{\hat f}\bigr) + \delta\,Z_k}{\sqrt{\hat{\sigma}^2_{k} + \delta^2}}
        \xrightarrow{d} \N(0,\, 1),
        \qquad
        k \in \{\aipw,\, \drs\}.
    \end{align*}
    Furthermore, the same holds true for $k = \dr$ under the null by \autoref{thm:asymptotics_orthogonal}.
    Since also $\theta_{\hat f}(X_i) = 0$ a.s.\ and thus $\theta^{k}_{\hat f} = 0$ under the null,
    we have that $T_n^{k} \xrightarrow{d} \N(0,1)$ under the null for all $k \in \{\aipw,\, \dr,\, \drs\}$.
    Note that this argument uses only that $\sigma_k^2 + \delta^2 \geq \delta^2 > 0$ and that $\theta^{k}_{\hat f} = 0$ under the null, which holds for any prediction function $\hat{f}$; in particular, the conclusion does not rely on the degeneracy of the asymptotic variance that arises under the well-specified null.

    \paragraph{Part 4: asymptotic size and power.}
    Under the null, we obtain from $T_n^{k} \xrightarrow{d} \N(0,1)$ that $\P(T_n^{k} > z_{1-\alpha}) \rightarrow \alpha$, so the tests have asymptotic size $\alpha$.

    Under the alternative,
    we consider two cases.
    First, if $\tilde{f}_0$ is well-specified, that is, $\tilde{f}_0 = f_0$,
    then
    $\theta_{f_0} = \theta_0 > 0$ by \autoref{thm:one_sided_aipw}
    and $\theta^\star_{f_0} = \theta^\star_0 > 0$ by \autoref{thm:one_sided_drs}.
    In this case, by \eqref{eq:targetconvergence},
    $\theta_{\hat f} \xrightarrow{p} \theta_0 > 0$ and $\theta^\star_{\hat f} \xrightarrow{p} \theta^\star_0 > 0$.
    Writing $\theta^{\aipw}_0 := \theta_0, \theta^{\dr}_0 := \theta_0, \theta^{\drs}_0 := \theta^\star_0$,
    by the continuous mapping theorem we have
    \begin{align*}
        \frac{1}{\sqrt{n}} T_n^k
        = \frac{\hat{\theta}^{k} + \delta\,Z_k / \sqrt{n}}{\sqrt{\hat{\sigma}^2_{k} + \delta^2}}
        \xrightarrow{p} \frac{\theta^{k}_0}{\sqrt{\sigma^2_{k} + \delta^2}} > 0,
        \qquad
        k \in \{\aipw,\, \dr,\, \drs\}.
    \end{align*}
    As a consequence,
    $\P(T_n^k > z_{1-\alpha}) = \P(T_n^k / \sqrt{n}> z_{1-\alpha} / \sqrt{n}) \rightarrow 1$.
    Hence, the tests are consistent (that is, have power converging to one) under the alternative.

    Second, $\tilde{f}_0$ may be misspecified ($\tilde{f}_0 \neq f_0$), in which case $\theta_{\tilde{f}_0}$ and $\theta^\star_{\tilde{f}_0}$ are not guaranteed to be positive any more, and we may lose power. In this case, consistency only holds when $\theta_{\tilde{f}_0} > 0$ and $\theta^\star_{\tilde{f}_0} > 0$, respectively.

\end{proof}

\section{Joint Wald Test}
\label{app:wald}

Let $i = 1,\dots, N$ index individuals and $m = 1,\dots,M$ index outcomes.
For each outcome $m$ we specify the linear equation
\begin{equation*}
  Y_{im} = \alpha_m + \tau_m\,D_i + X_i^\top\beta_m + \varepsilon_{im},
  \qquad
  \mathbb E[\varepsilon_{im}|X_i,D_i]=0,
\end{equation*}
where $D_i\in\{0,1\}$ is the treatment indicator, $X_i$ is a \(K\)-dimensional vector of covariates as before, and the error terms $\varepsilon_i=(\varepsilon_{i1},\dots,\varepsilon_{iM})^\top$ form a mean-zero vector with covariance matrix $\Sigma=\mathbb E[\varepsilon_i\varepsilon_i^\top]$ that is identical across $i$ but not assumed diagonal, i.e. the system is seemingly unrelated. \(\alpha_m\) and \(\tau_m\) represent intercept and treatment effect respectively for the \(m\)-th outcome. The coefficient of interest is
\[
  \boldsymbol{\tau}
  =
  (\tau_1,\dots,\tau_M)^\top \;\in\; \mathbb R^{M}.
\]
We test the joint null
\[
  H_0:\; \boldsymbol{\tau}=0
  \quad\text{versus}\quad
  H_1:\; \boldsymbol{\tau}\neq 0 .
\]

\noindent Let $\widehat{\boldsymbol{\tau}}
=(\hat\tau_1,\dots,\hat\tau_M)^\top$
be the feasible generalized least-squares estimates of treatment effects obtained from the
systemfit package in R \citep*{henningsen2008systemfit}. Let
$\widehat{\mathcal V}_{\tau}\in\mathbb R^{M\times M}$ be the
covariance matrix corresponding to
this vector. The Wald test statistic is
\begin{equation*}
  W
  \;=\;
  \widehat{\boldsymbol{\tau}}^\top
  \,\widehat{\mathcal V}_{\tau}^{-1}\,
  \widehat{\boldsymbol{\tau}}
  \;\;\in\;[0,\infty).
\end{equation*}
Under standard regularity conditions (full-rank design, $\Sigma$ positive definite), \(W\) converges to a chi-squared distribution when $N\to\infty$:
\[
  W \;\xrightarrow{d}\; \chi^2_{M}\quad\text{under }H_0 .
\]

\noindent For a prespecified significance level $\alpha\in(0,1)$, let
$\,\chi^2_{M,1-\alpha}\,$ denote the $1-\alpha$ upper quantile of
the $\chi^2_{M}$ distribution.
Reject $H_0$ whenever
\[
  W \;>\; \chi^2_{M,1-\alpha}.
\]

\clearpage

\section{Distributions of Test Statistics} \label{app:sims_distributions}

\begin{figure}[H]

\centering

\resizebox{0.85\textwidth}{!}{%
\begin{minipage}{\textwidth}
\centering

% --- Row 1 ---
\begin{subfigure}[t]{0.42\textwidth}
    \centering
    \includegraphics[width=\linewidth]{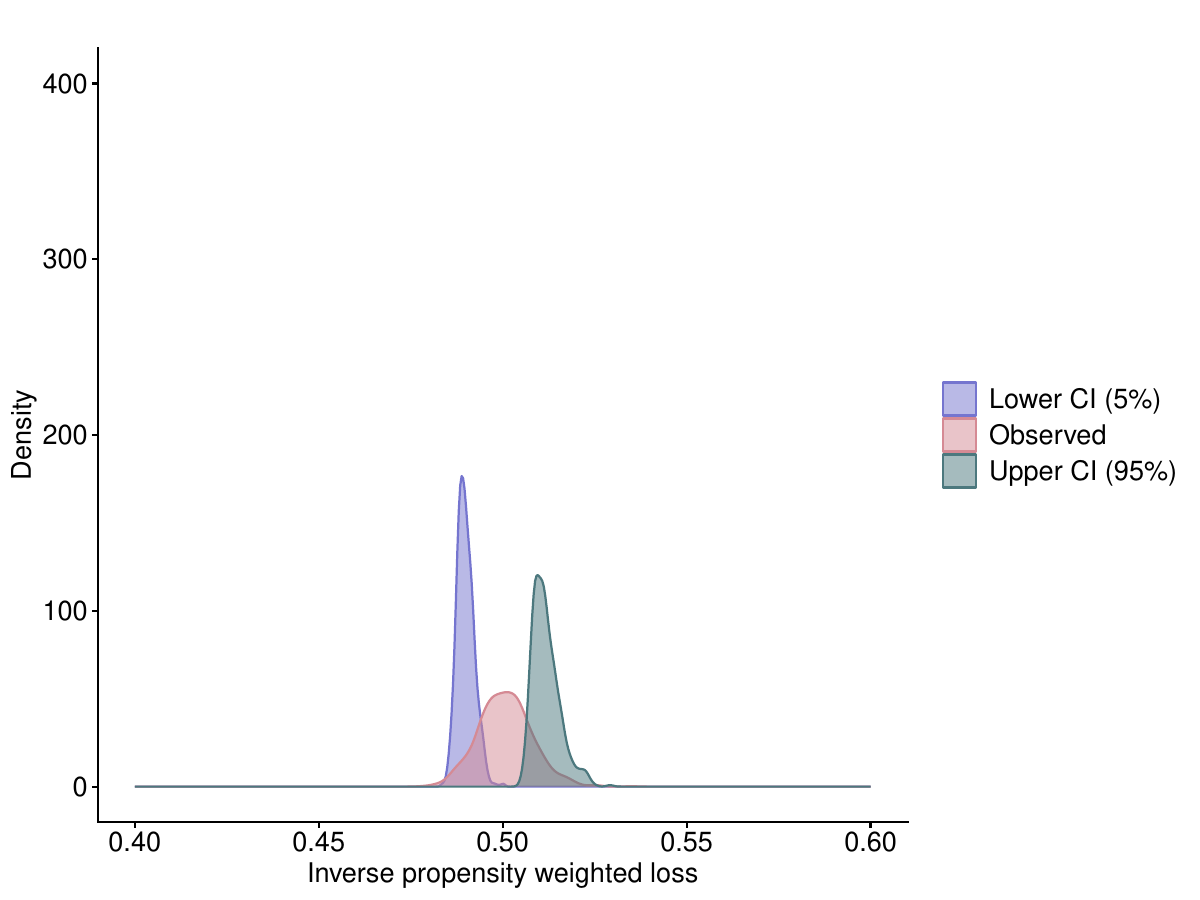}
    \caption{Full Heterogeneity}
\end{subfigure}
\hfill
\begin{subfigure}[t]{0.41\textwidth}
    \centering
    \includegraphics[width=\linewidth]{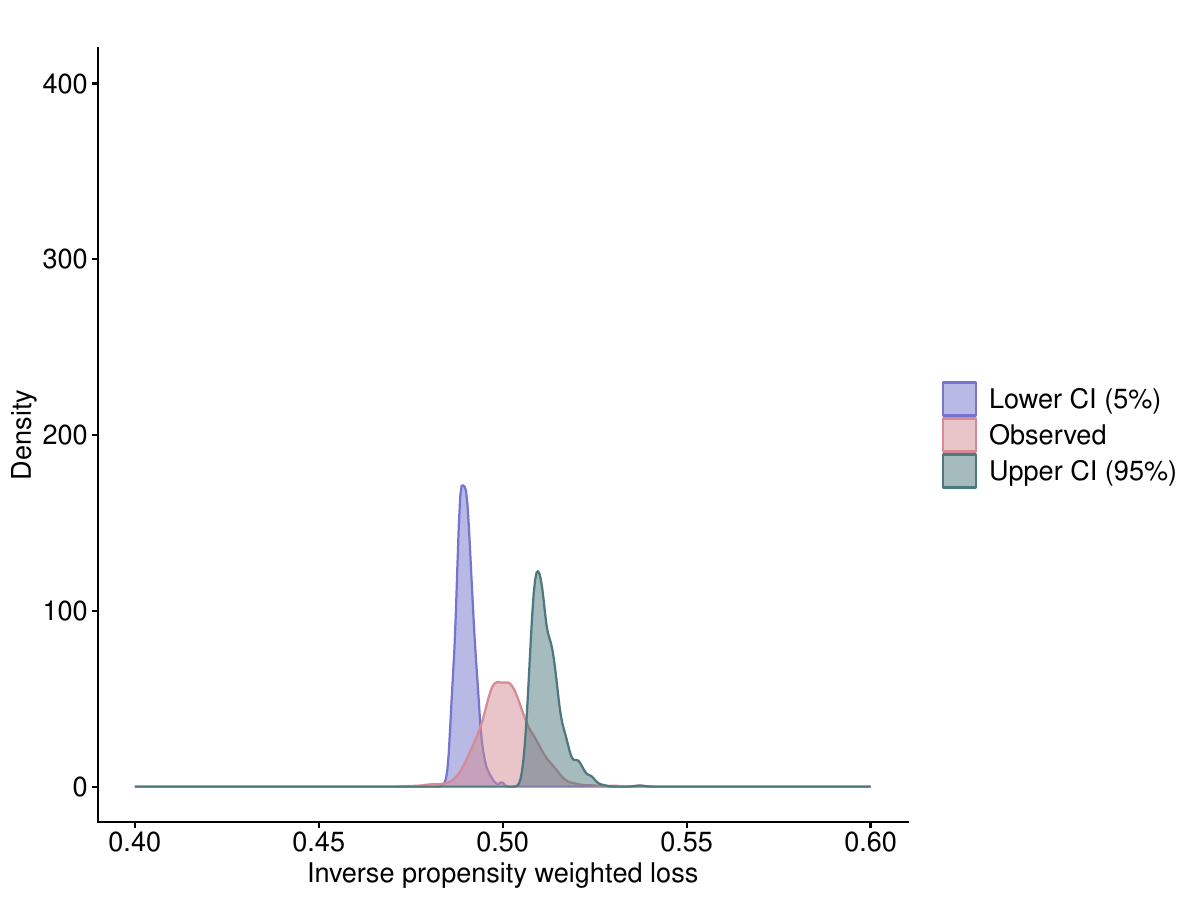}
    \caption{Interactions}
\end{subfigure}

\vspace{1em}

% --- Row 2 ---
\begin{subfigure}[t]{0.41\textwidth}
    \centering
    \includegraphics[width=\linewidth]{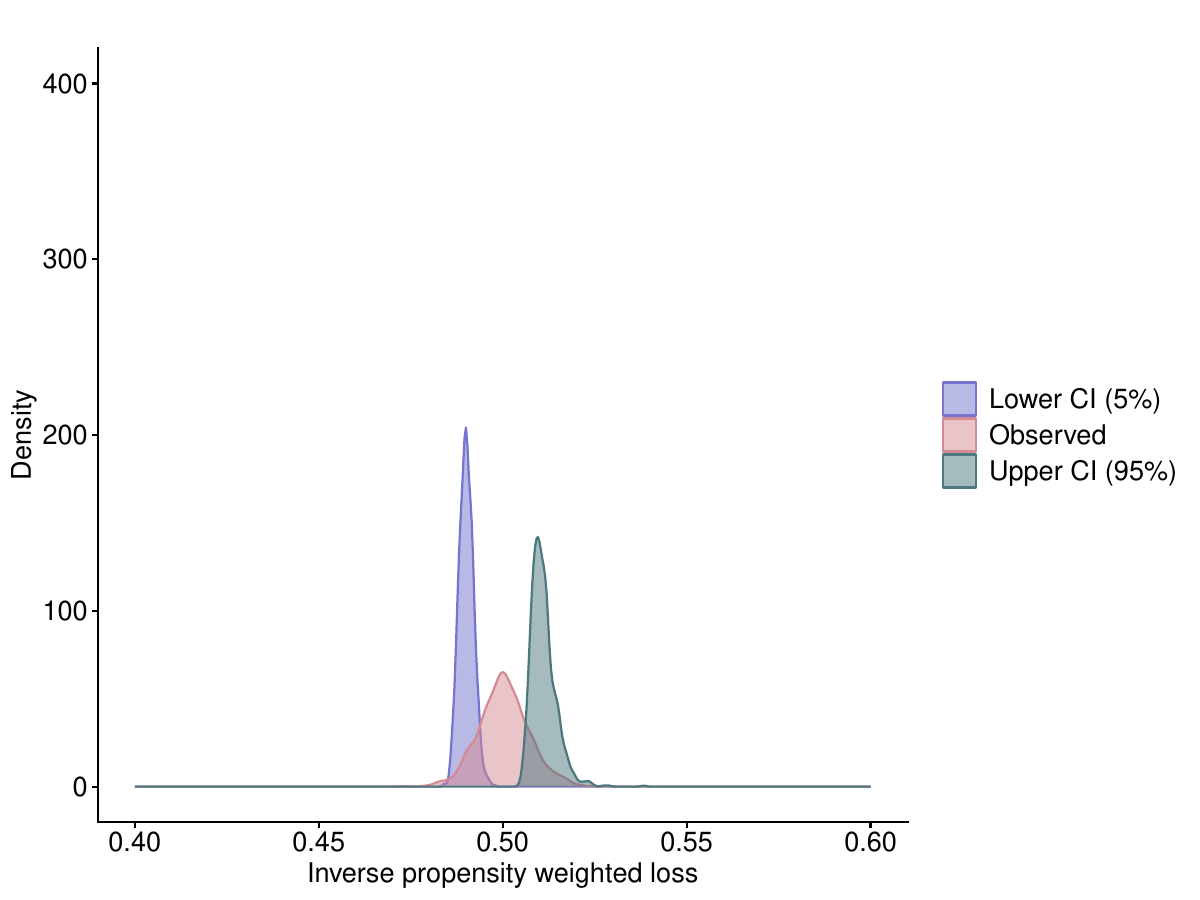}
    \caption{Mixtures}
\end{subfigure}
\hfill
\begin{subfigure}[t]{0.41\textwidth}
    \centering
    \includegraphics[width=\linewidth]{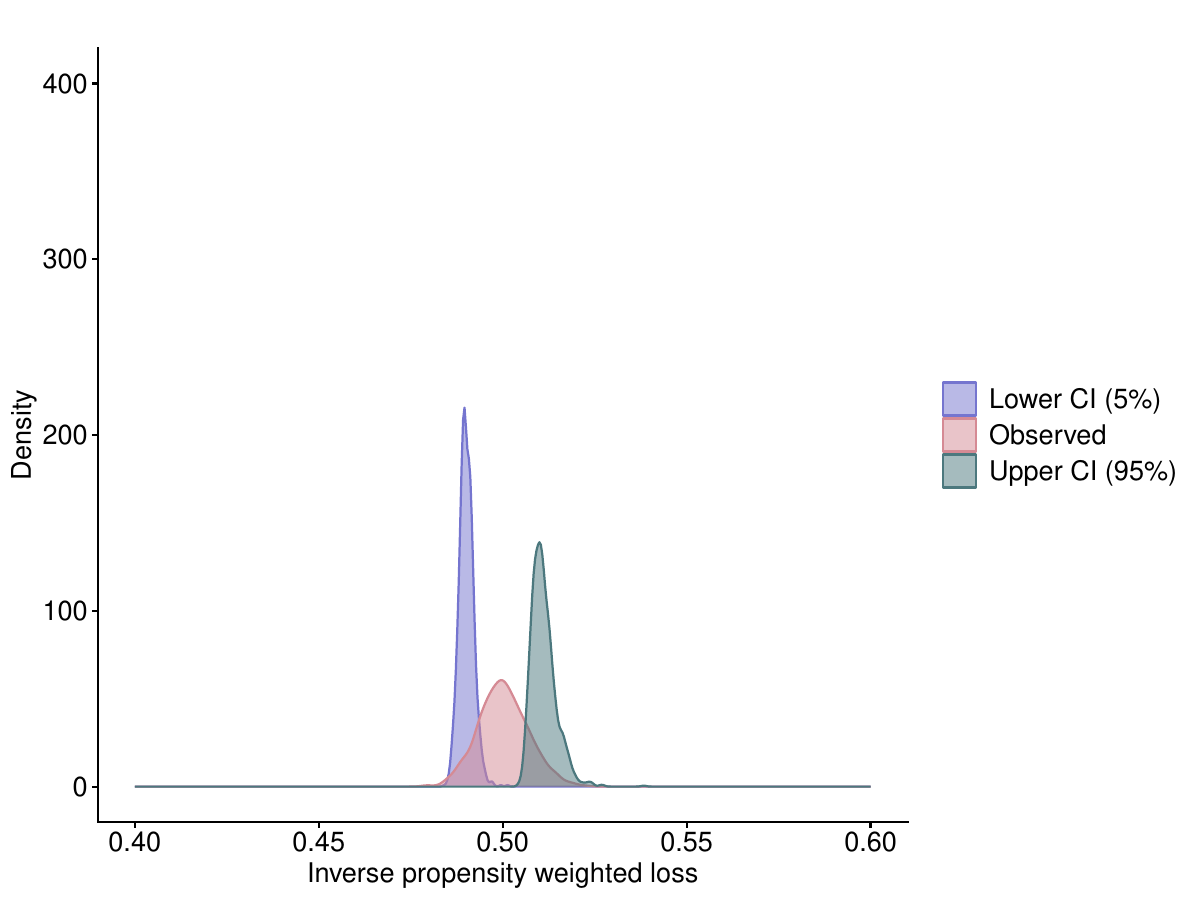}
    \caption{No Baseline}
\end{subfigure}

\vspace{1em}

% --- Row 3: Mixed, centered ---
\begin{subfigure}[t]{0.41\textwidth}
    \centering
    \includegraphics[width=\linewidth]{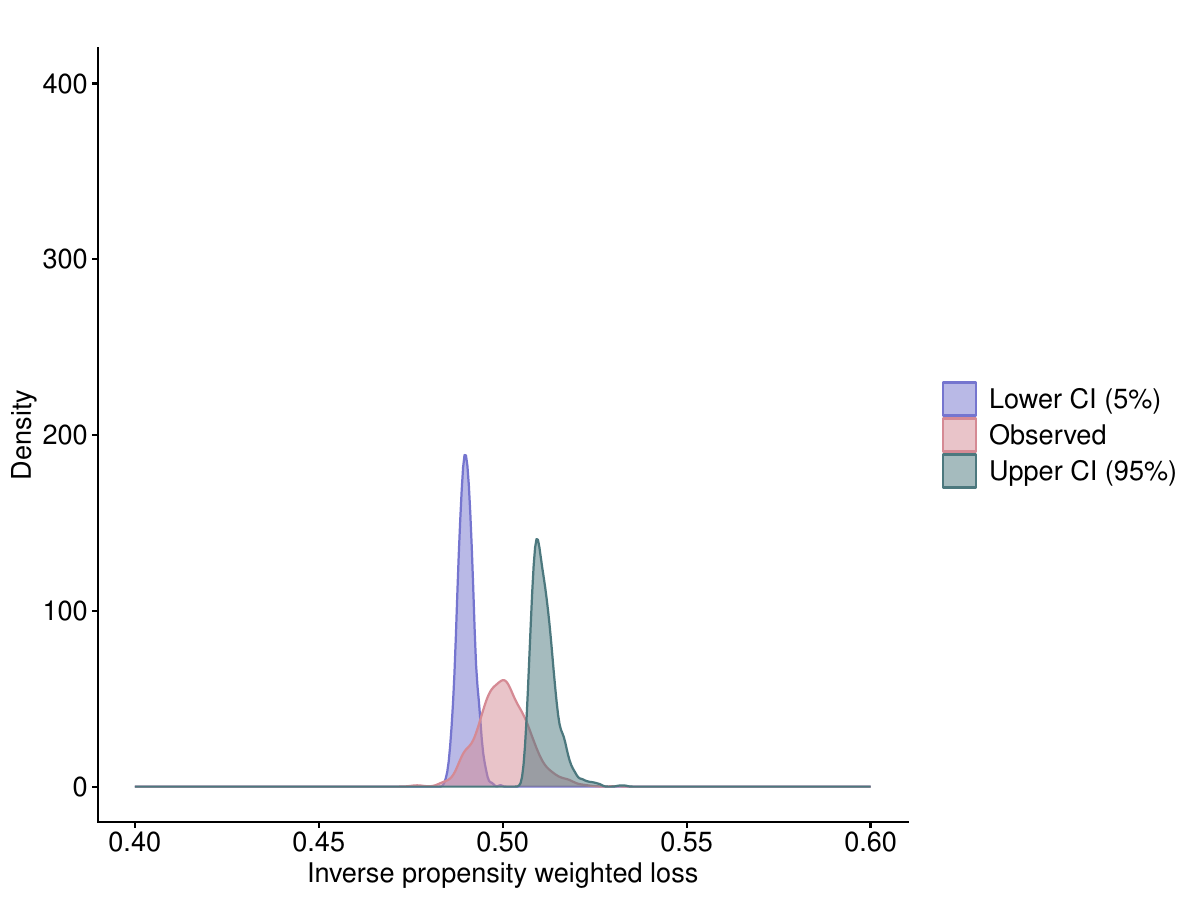}
    \caption{Mixed}
\end{subfigure}

\end{minipage}%
}

\caption{The observed IPW loss-based test statistic and the reference distribution under the null hypothesis (no treatment effect). Observed is the density of the realized test statistic across replications. Lower CI (5\%) and Upper CI (95\%) are the densities of the 5th and 95th percentiles of each replication's own permutation null (\autoref{alg:crt}). Results are based on 1,000 Monte Carlo replications, each with a sample of 1,000 observations and five outcome variables. }
\label{fig:ipw_loss_null}
\end{figure}

\begin{figure}[H]
\centering
\begin{subfigure}[t]{0.45\textwidth}
    \centering
    \includegraphics[width=\linewidth]{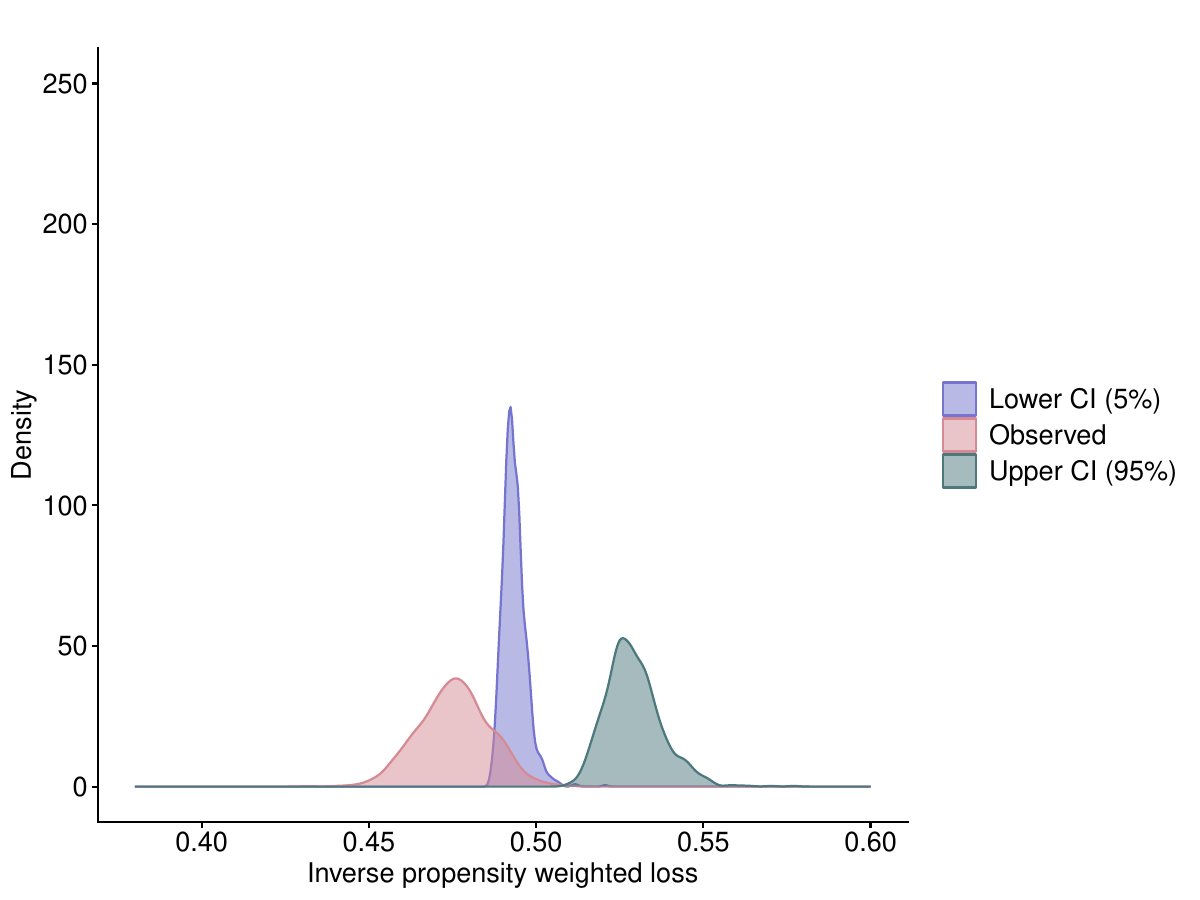}
    \caption{Full Heterogeneity}
\end{subfigure}
\hfill
\begin{subfigure}[t]{0.45\textwidth}
    \centering
    \includegraphics[width=\linewidth]{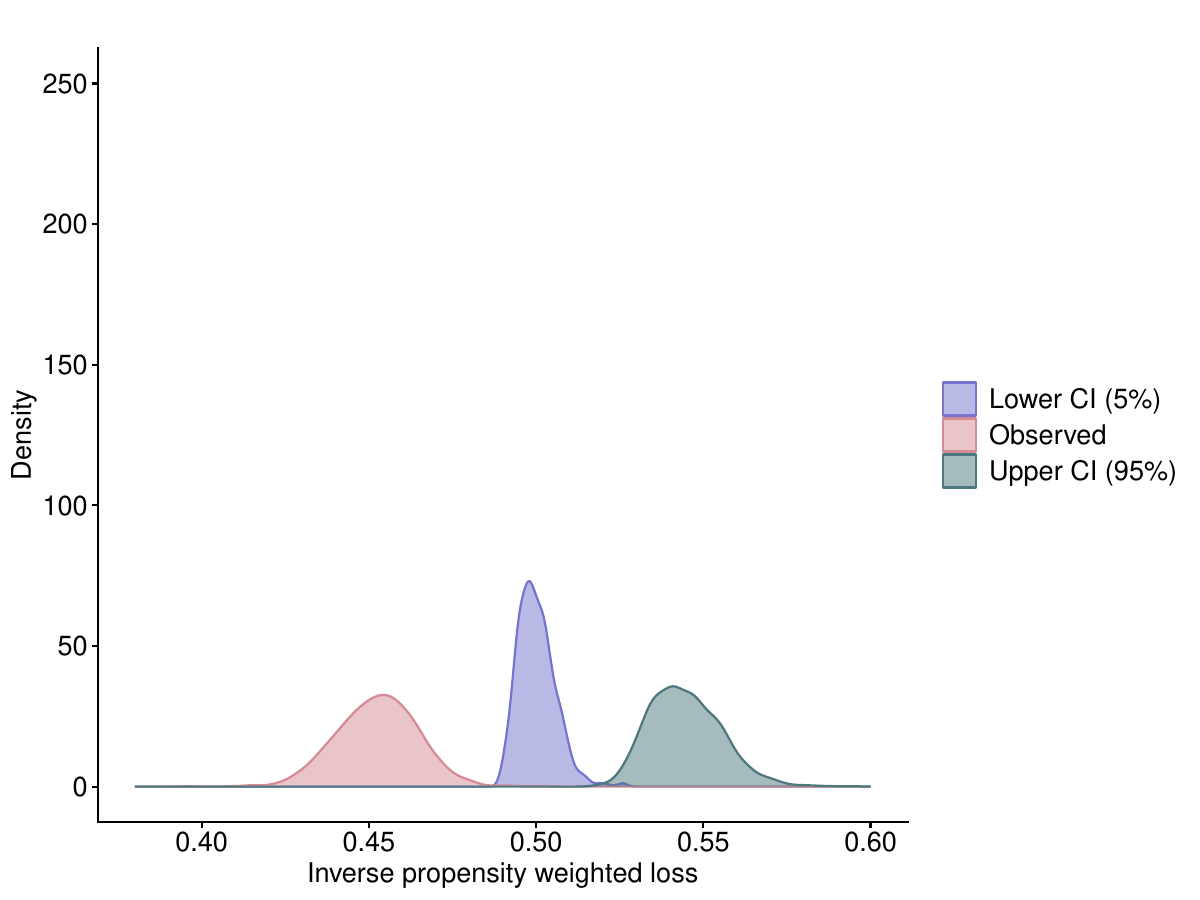}
    \caption{Interactions}
\end{subfigure}

\vspace{1em}

\begin{subfigure}[t]{0.45\textwidth}
    \centering
    \includegraphics[width=\linewidth]{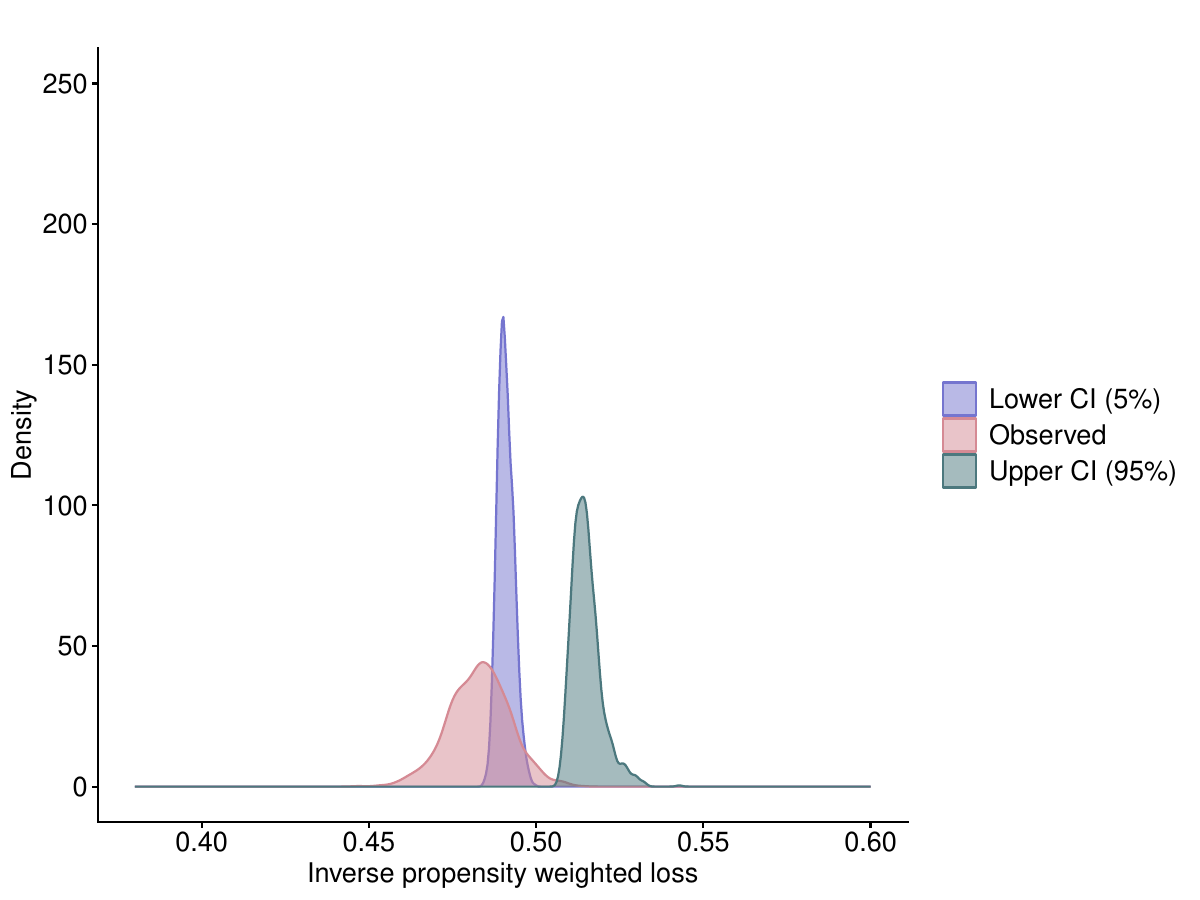}
    \caption{Mixtures}
\end{subfigure}
\hfill
\begin{subfigure}[t]{0.45\textwidth}
    \centering
    \includegraphics[width=\linewidth]{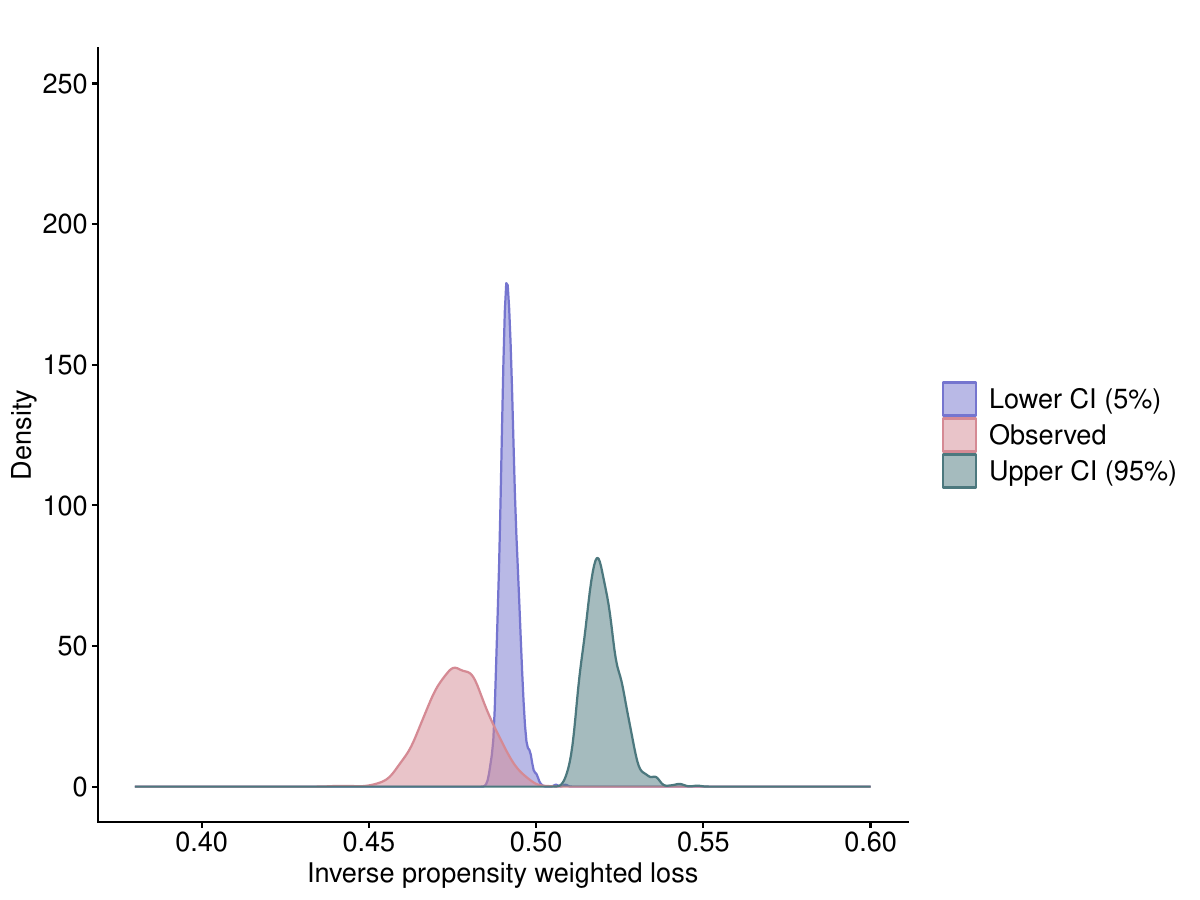}
    \caption{No Baseline}
\end{subfigure}

\vspace{1em}

\begin{subfigure}[t]{0.45\textwidth}
    \centering
    \includegraphics[width=\linewidth]{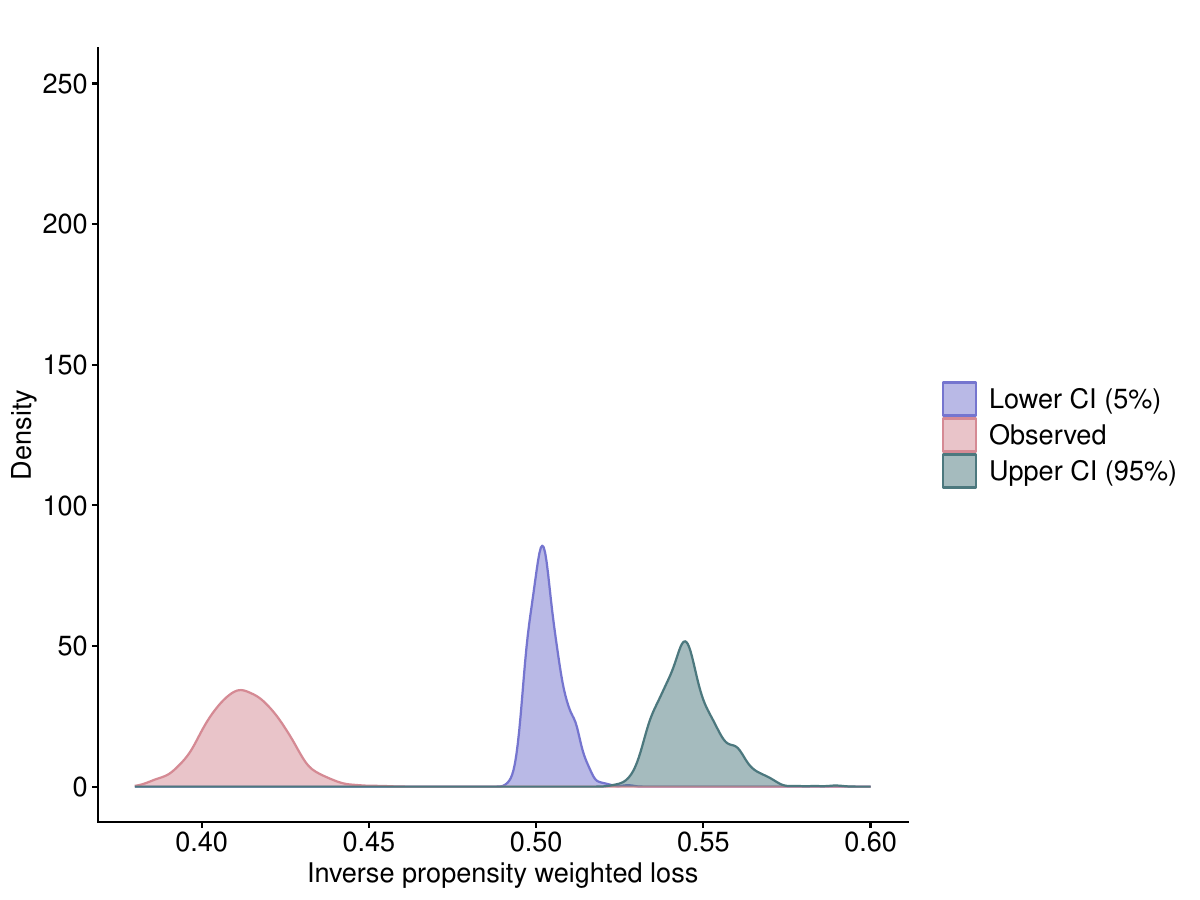}
    \caption{Mixed}
\end{subfigure}

\caption{
The observed IPW loss-based test statistic and the reference distribution under the alternative hypothesis (weak treatment effects). Observed is the density of the realized test statistic across replications. Lower CI (5\%) and Upper CI (95\%) are the densities of the 5th and 95th percentiles of each replication's own permutation null (\autoref{alg:crt}). Results are based on 1,000 Monte Carlo replications, each with a sample of 1,000 observations and five outcome variables. 
}
\label{fig:ipw_loss_alternative}
\end{figure}

\begin{figure}[H]
\centering

% --- Row 1 ---
\begin{subfigure}[t]{0.45\textwidth}
    \centering
    \includegraphics[width=\linewidth]{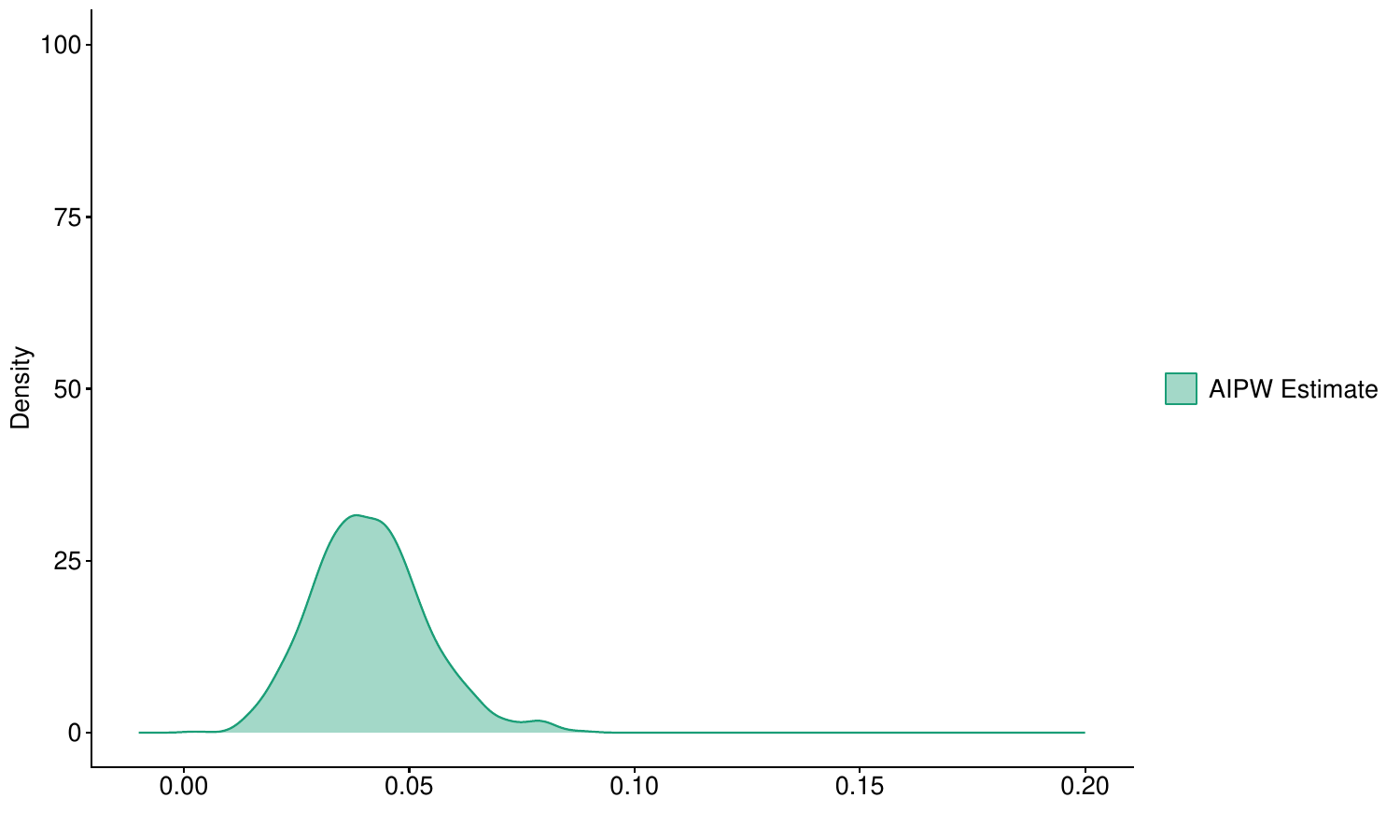}
    \caption{Full Heterogeneity}
\end{subfigure}
\hfill
\begin{subfigure}[t]{0.45\textwidth}
    \centering
    \includegraphics[width=\linewidth]{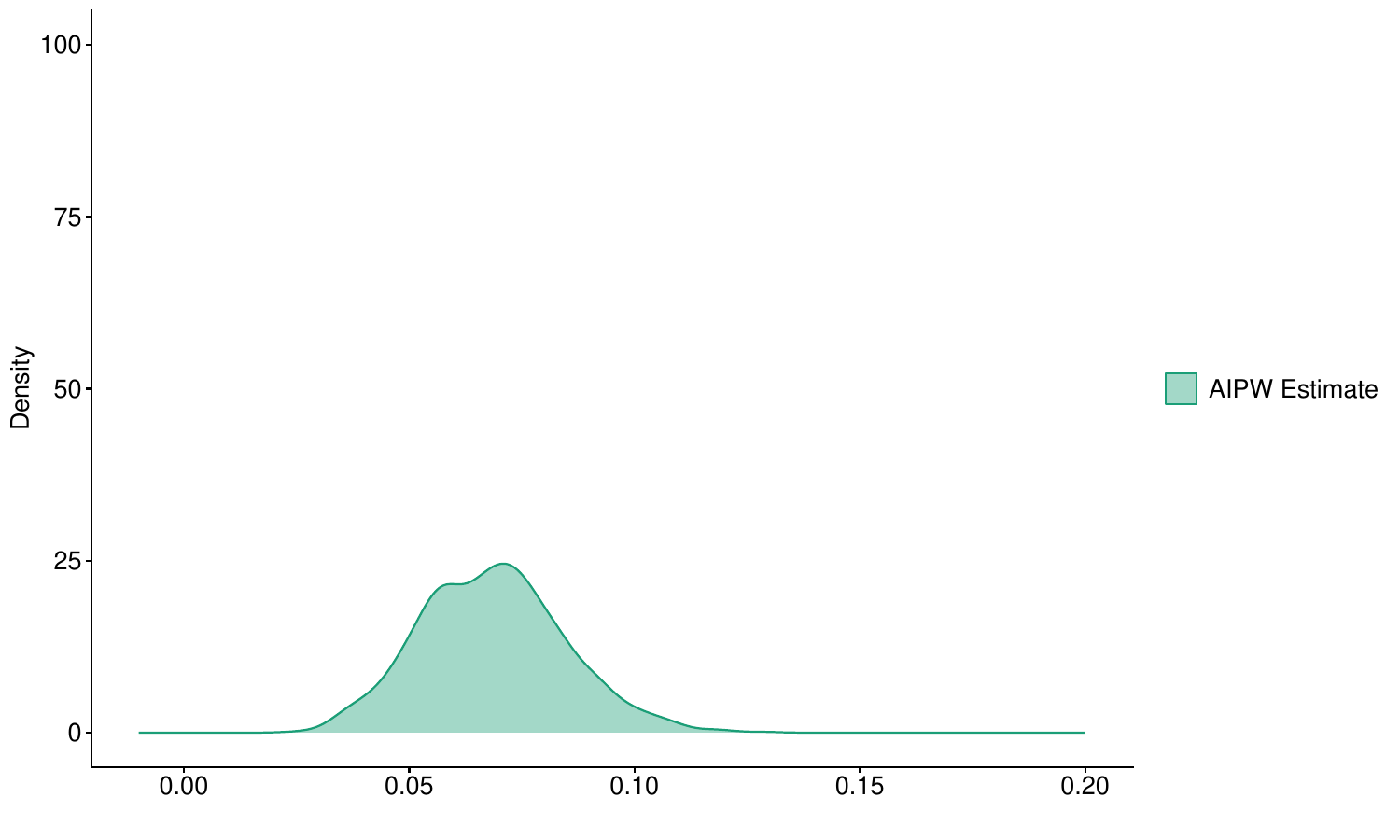}
    \caption{Interactions}
\end{subfigure}

\vspace{1em}

% --- Row 2 ---
\begin{subfigure}[t]{0.45\textwidth}
    \centering
    \includegraphics[width=\linewidth]{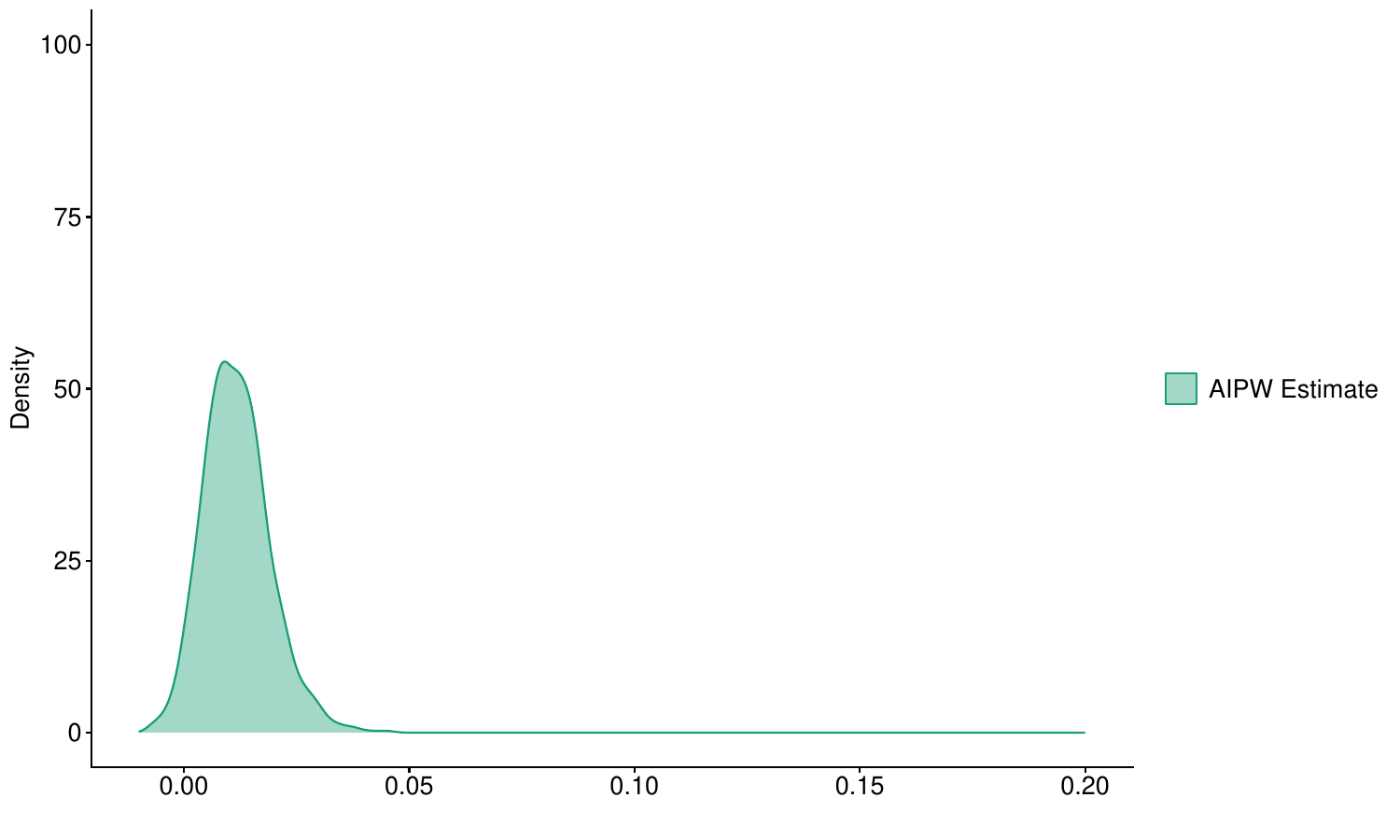}
    \caption{Mixtures}
\end{subfigure}
\hfill
\begin{subfigure}[t]{0.45\textwidth}
    \centering
    \includegraphics[width=\linewidth]{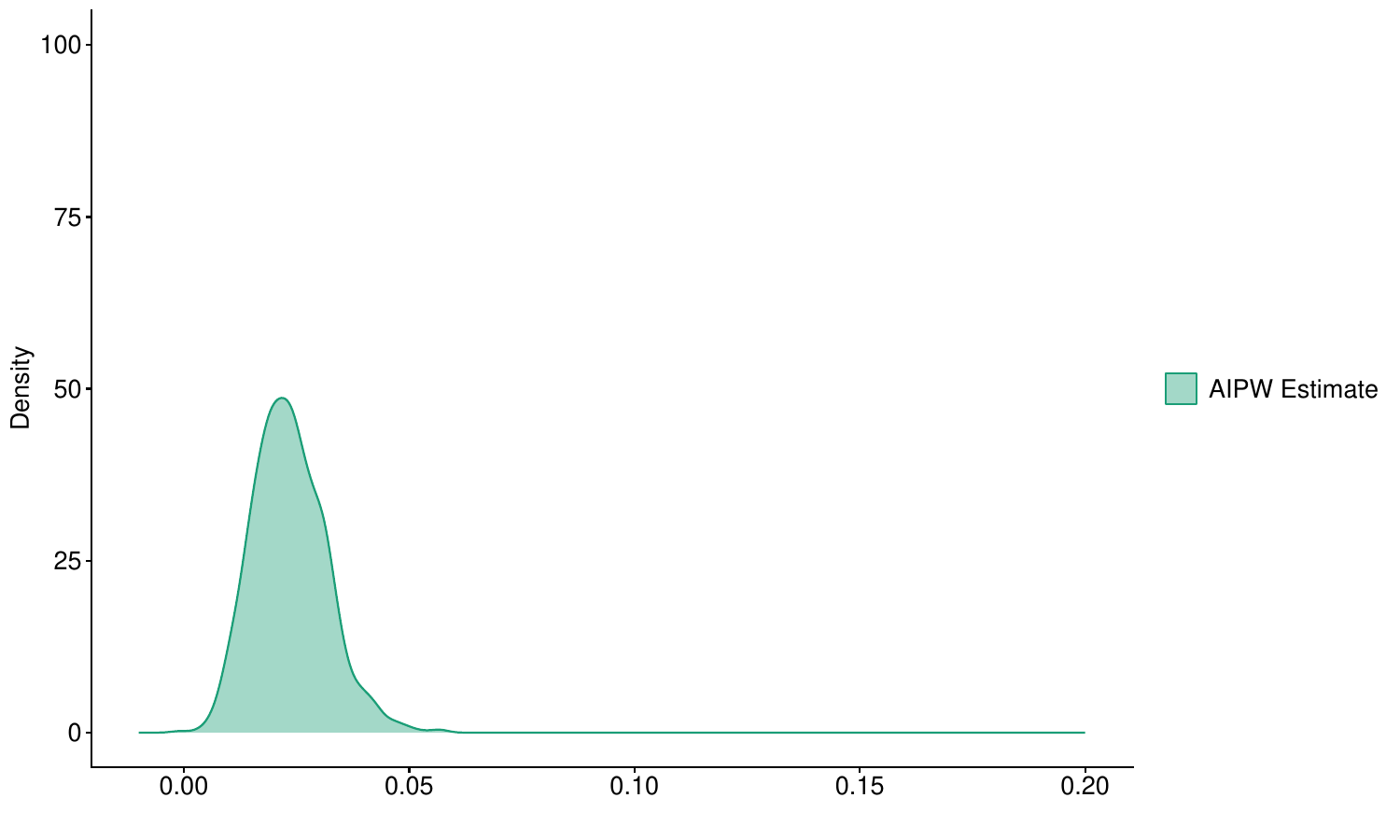}
    \caption{No Baseline}
\end{subfigure}

\vspace{1em}

% --- Row 3: Mixed, centered ---
\begin{subfigure}[t]{0.45\textwidth}
    \centering
    \includegraphics[width=\linewidth]{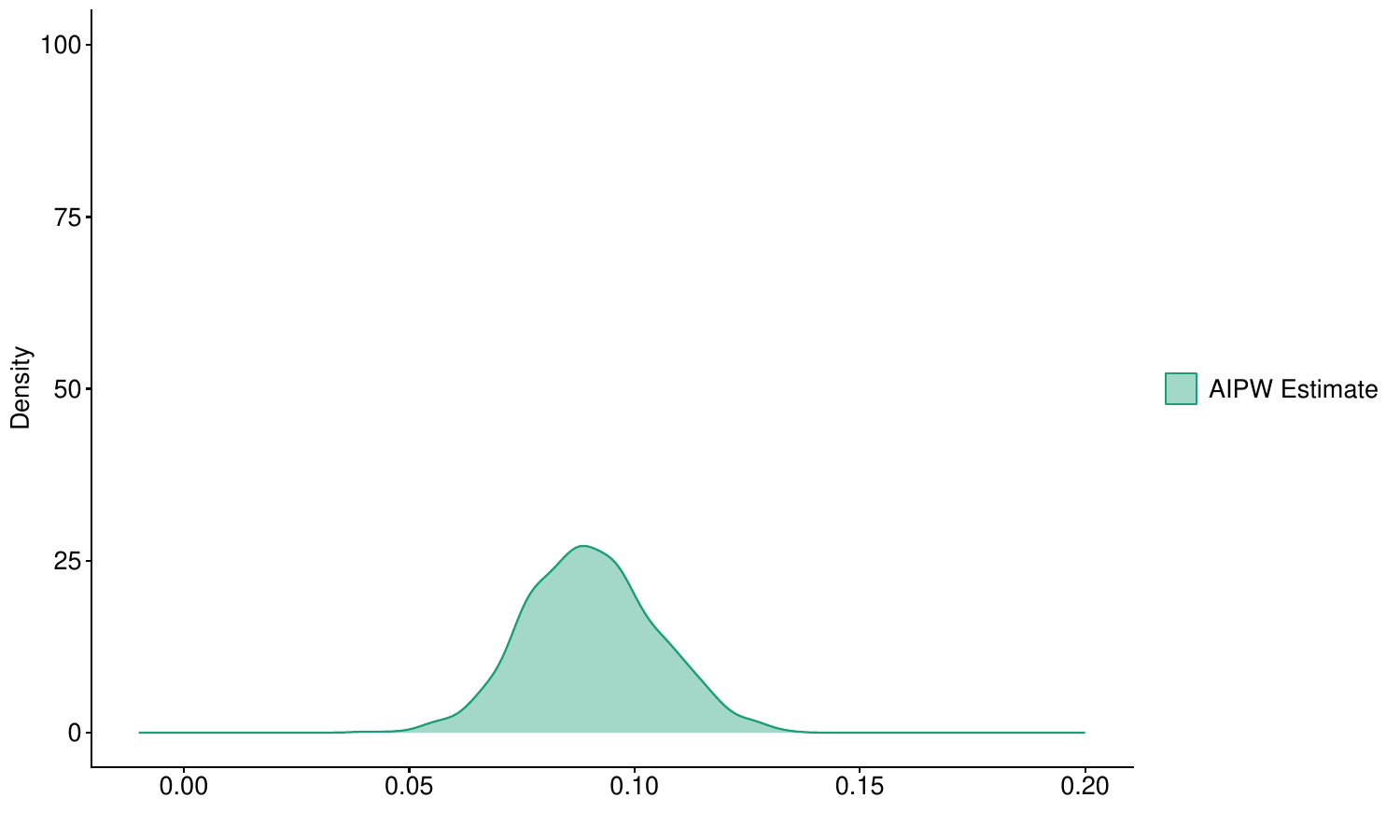}
    \caption{Mixed}
\end{subfigure}

\caption{Distributions of the estimated $\aipw$ test statistic under the alternative hypothesis (weak treatment effects). Results are based on 1,000 Monte Carlo replications, each with a sample of 1,000 observations and five outcome variables.}
\label{fig:aipw_density}
\end{figure}

\begin{figure}[H]
\centering

% --- Row 1 ---
\begin{subfigure}[t]{0.45\textwidth}
    \centering
    \includegraphics[width=\linewidth]{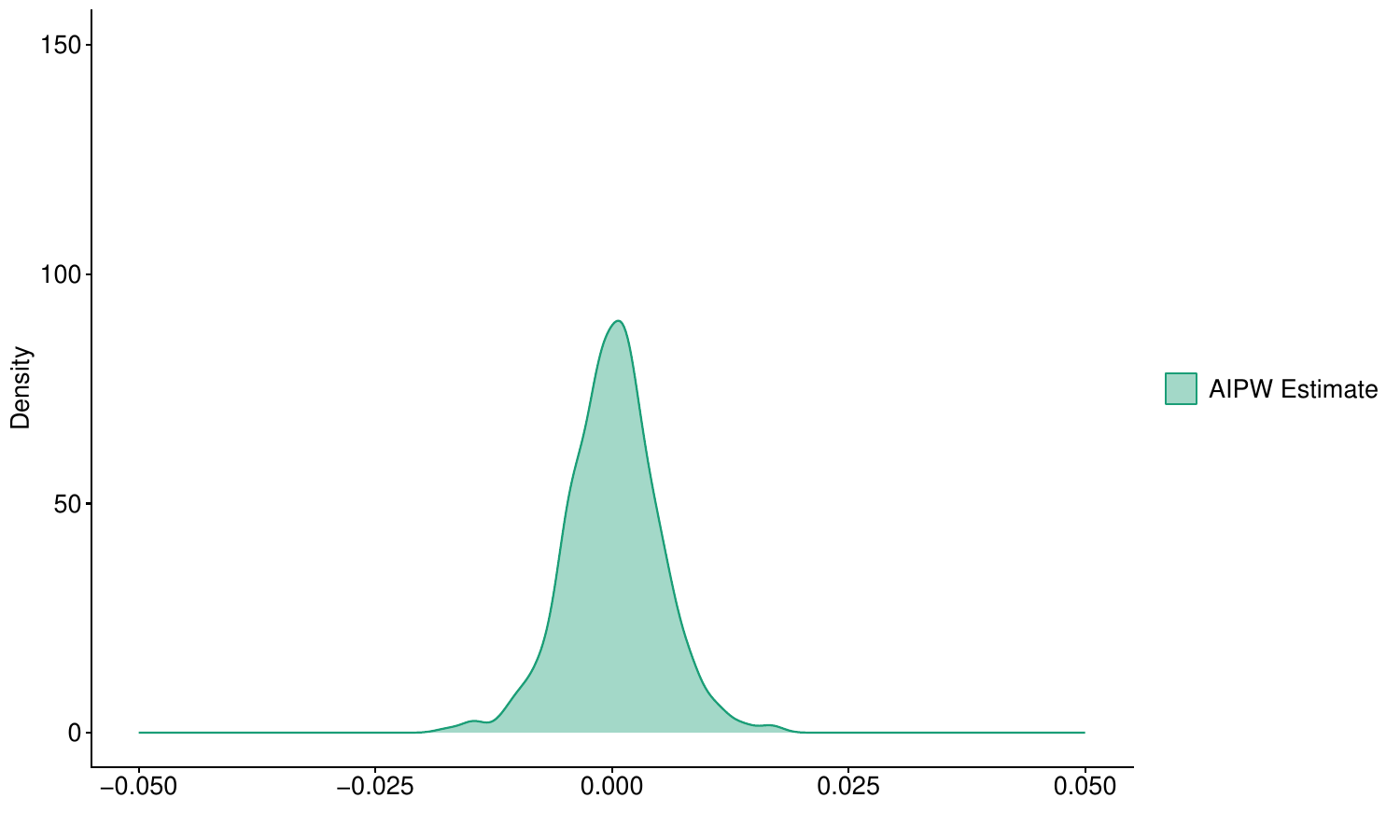}
    \caption{Full Heterogeneity}
\end{subfigure}
\hfill
\begin{subfigure}[t]{0.45\textwidth}
    \centering
    \includegraphics[width=\linewidth]{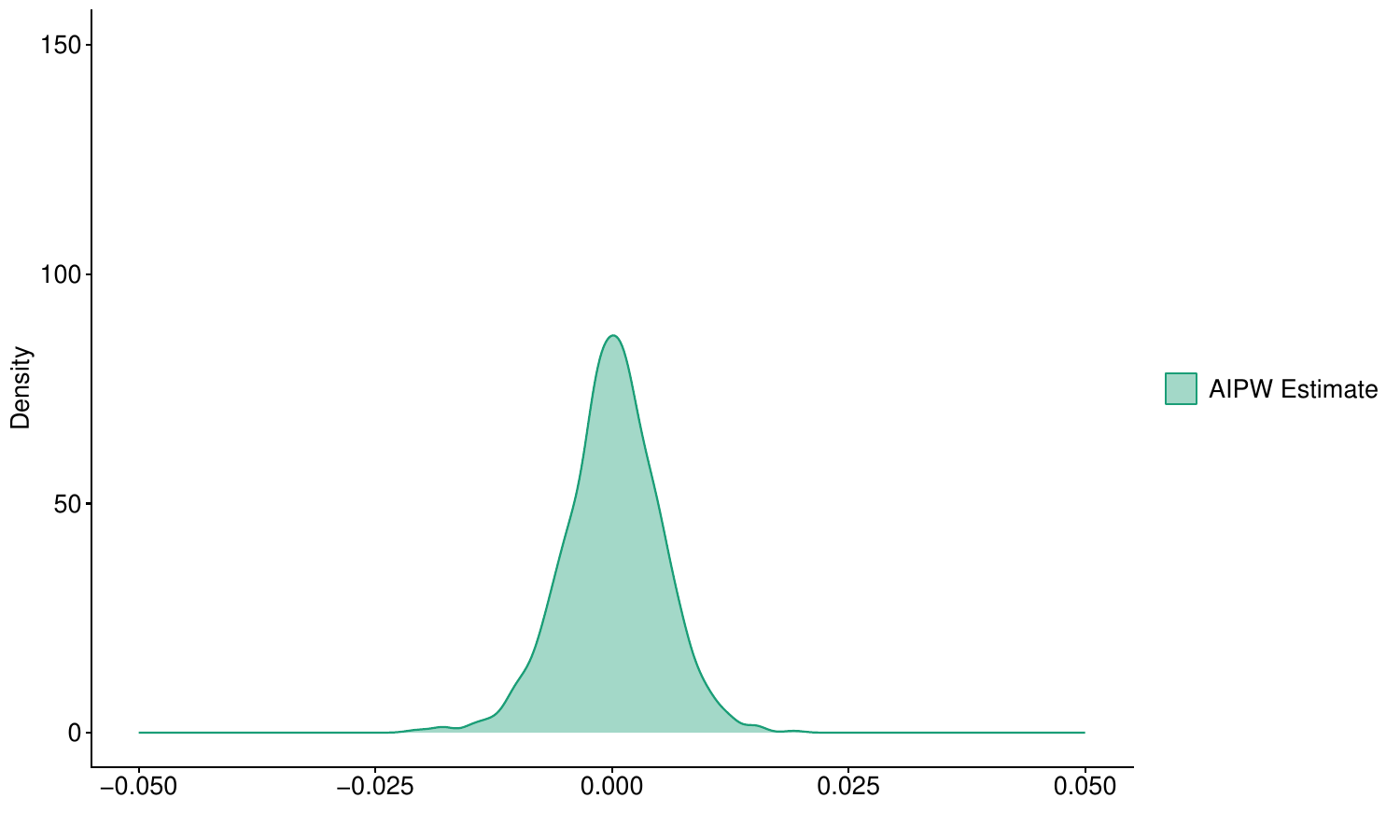}
    \caption{Interactions}
\end{subfigure}

\vspace{1em}

% --- Row 2 ---
\begin{subfigure}[t]{0.45\textwidth}
    \centering
    \includegraphics[width=\linewidth]{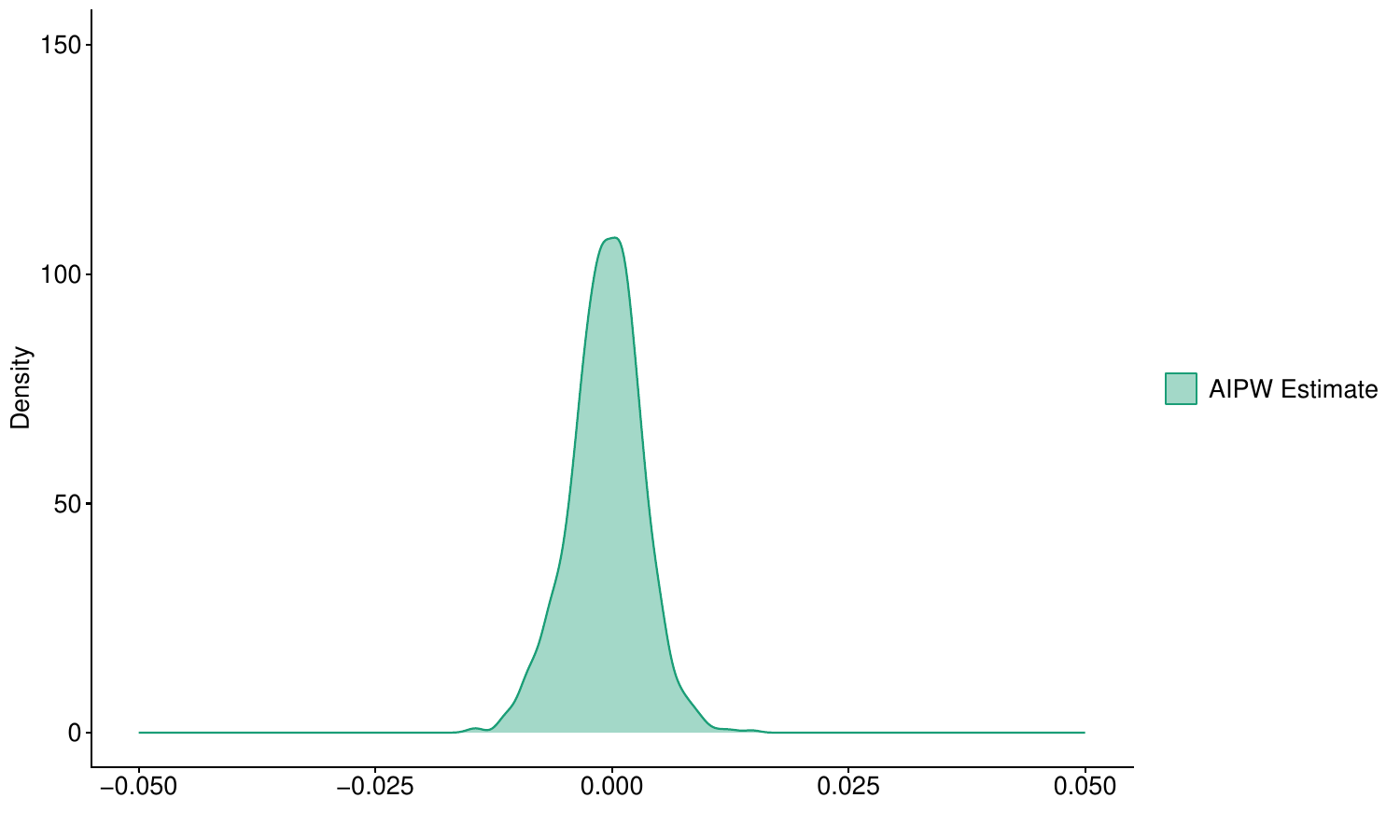}
    \caption{Mixtures}
\end{subfigure}
\hfill
\begin{subfigure}[t]{0.45\textwidth}
    \centering
    \includegraphics[width=\linewidth]{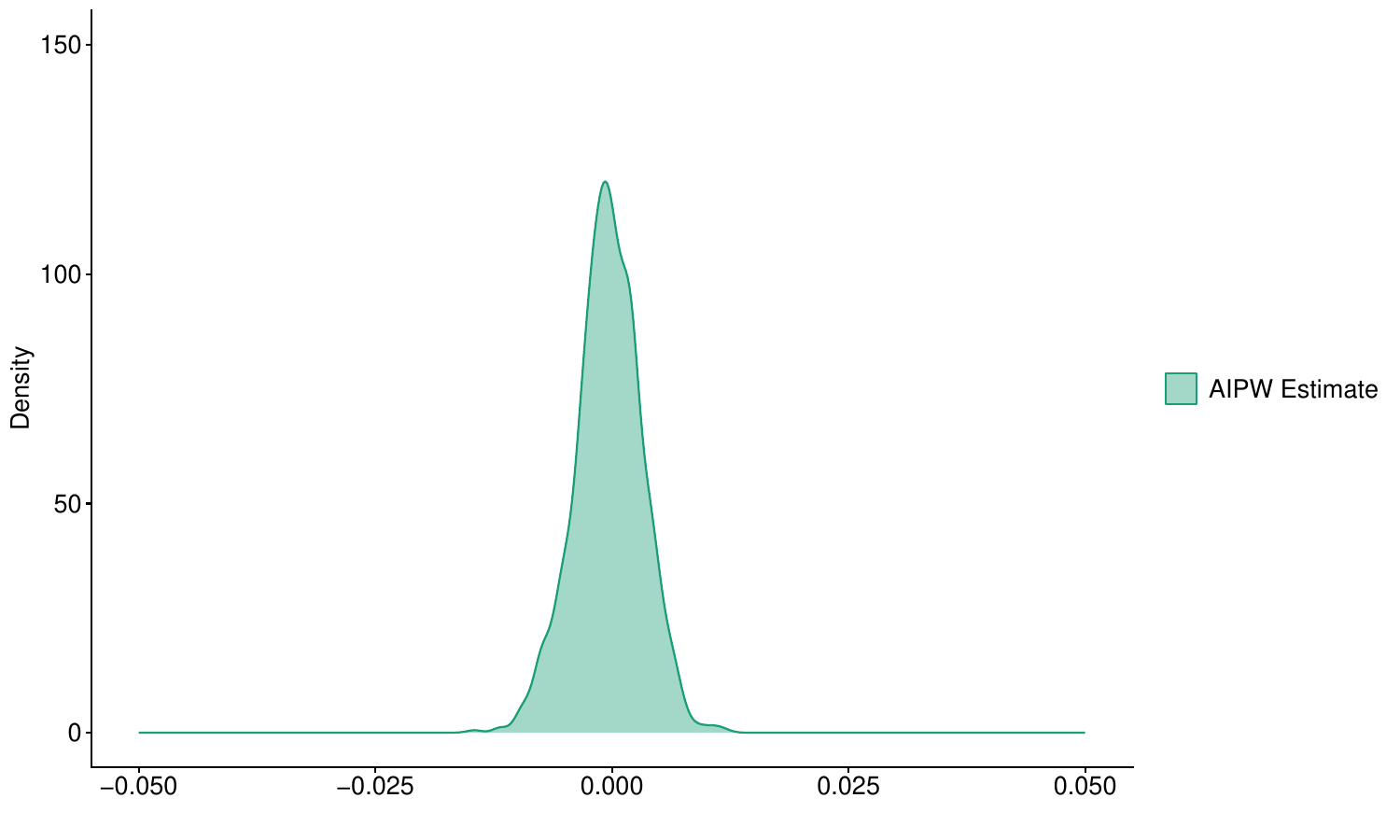}
    \caption{No Baseline}
\end{subfigure}

\vspace{1em}

% --- Row 3: Mixed, centered ---
\begin{subfigure}[t]{0.45\textwidth}
    \centering
    \includegraphics[width=\linewidth]{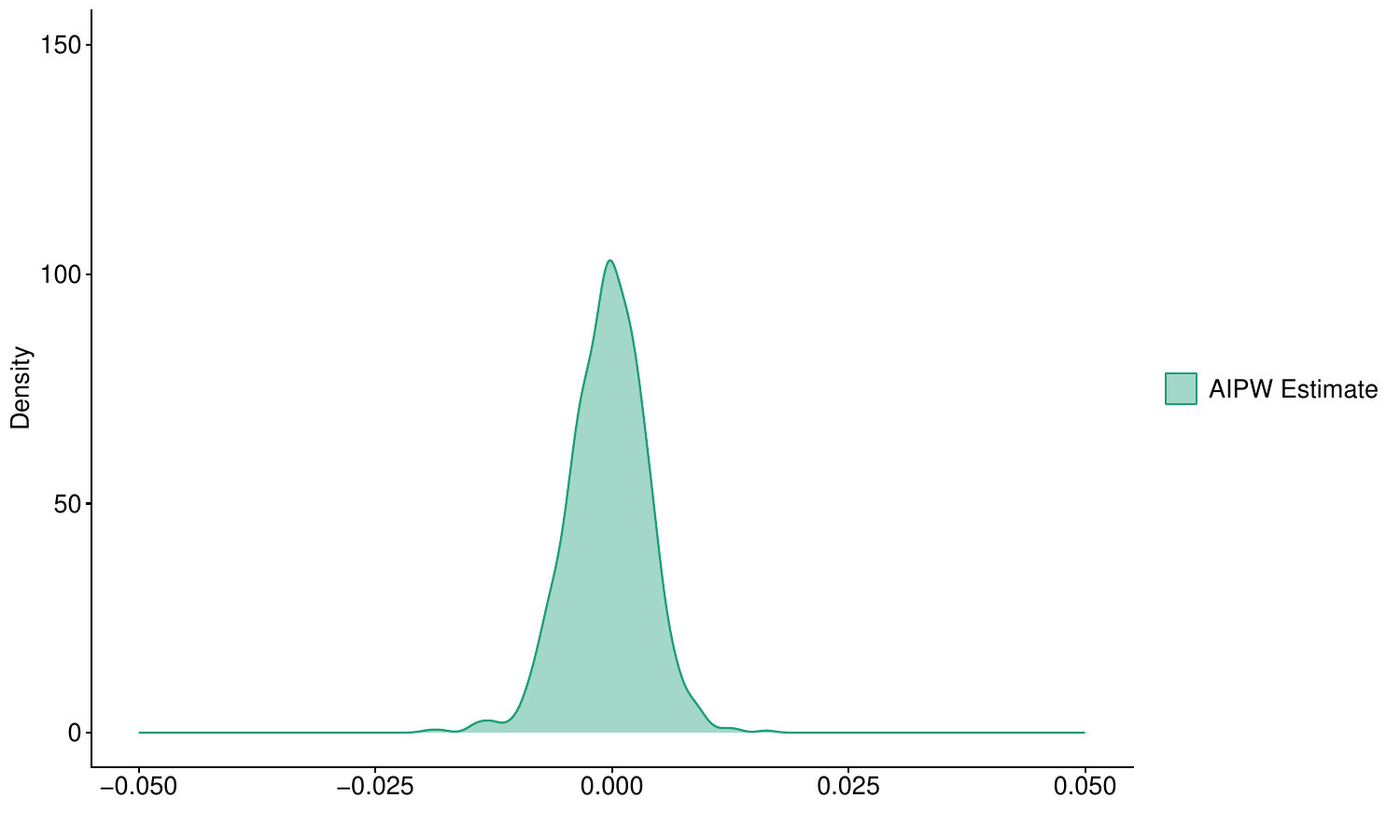}
    \caption{Mixed}
\end{subfigure}

\caption{Distributions of the estimated $\aipw$ test statistic under the null hypothesis. Results are based on 1,000 Monte Carlo replications, each with a sample of 1,000 observations and five outcome variables.}
\label{fig:aipw_density_null}
\end{figure}

\begin{figure}[H]
\centering
% --- Row 1 ---
\begin{subfigure}[t]{0.45\textwidth}
    \centering
    \includegraphics[width=\linewidth]{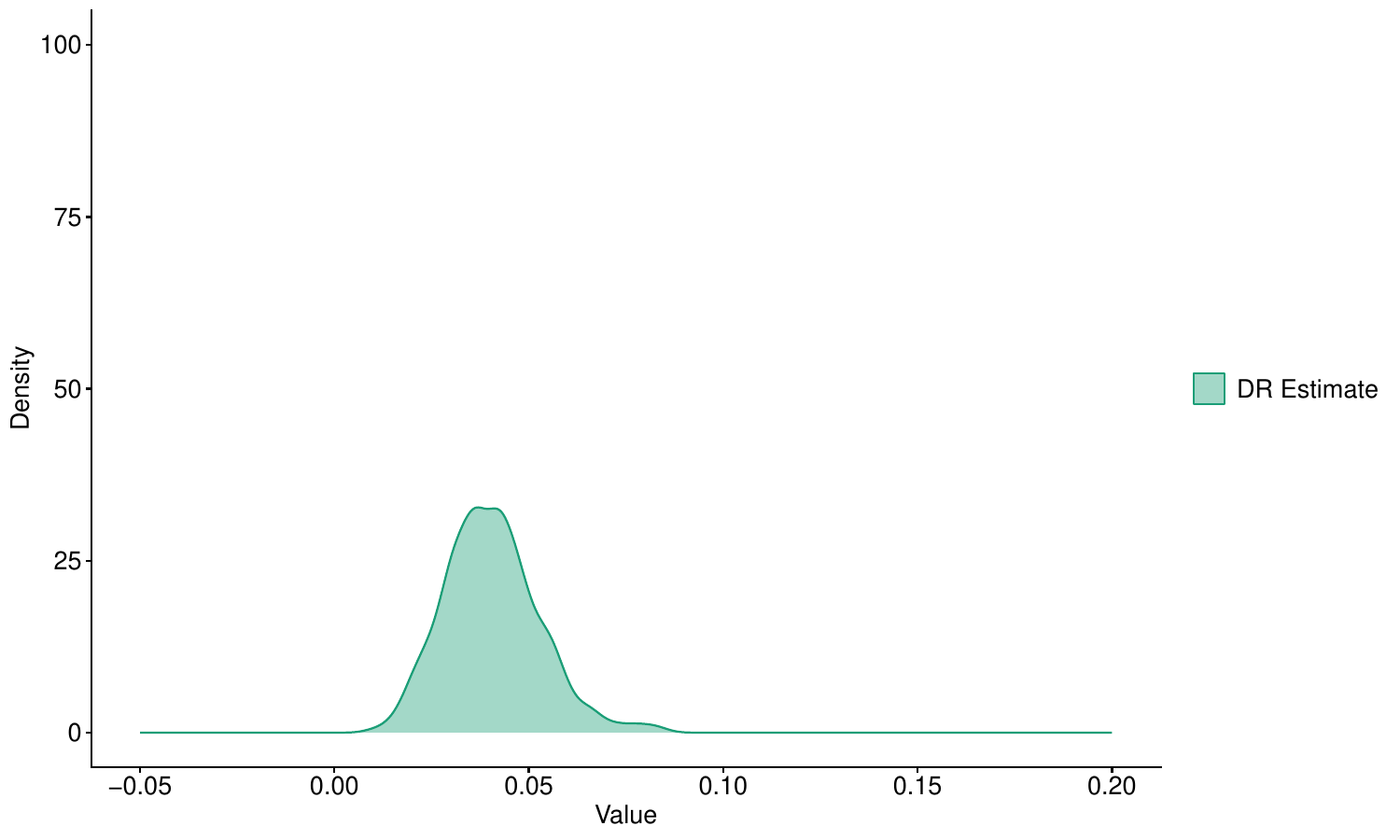}
    \caption{Full Heterogeneity}
\end{subfigure}
\hfill
\begin{subfigure}[t]{0.45\textwidth}
    \centering
    \includegraphics[width=\linewidth]{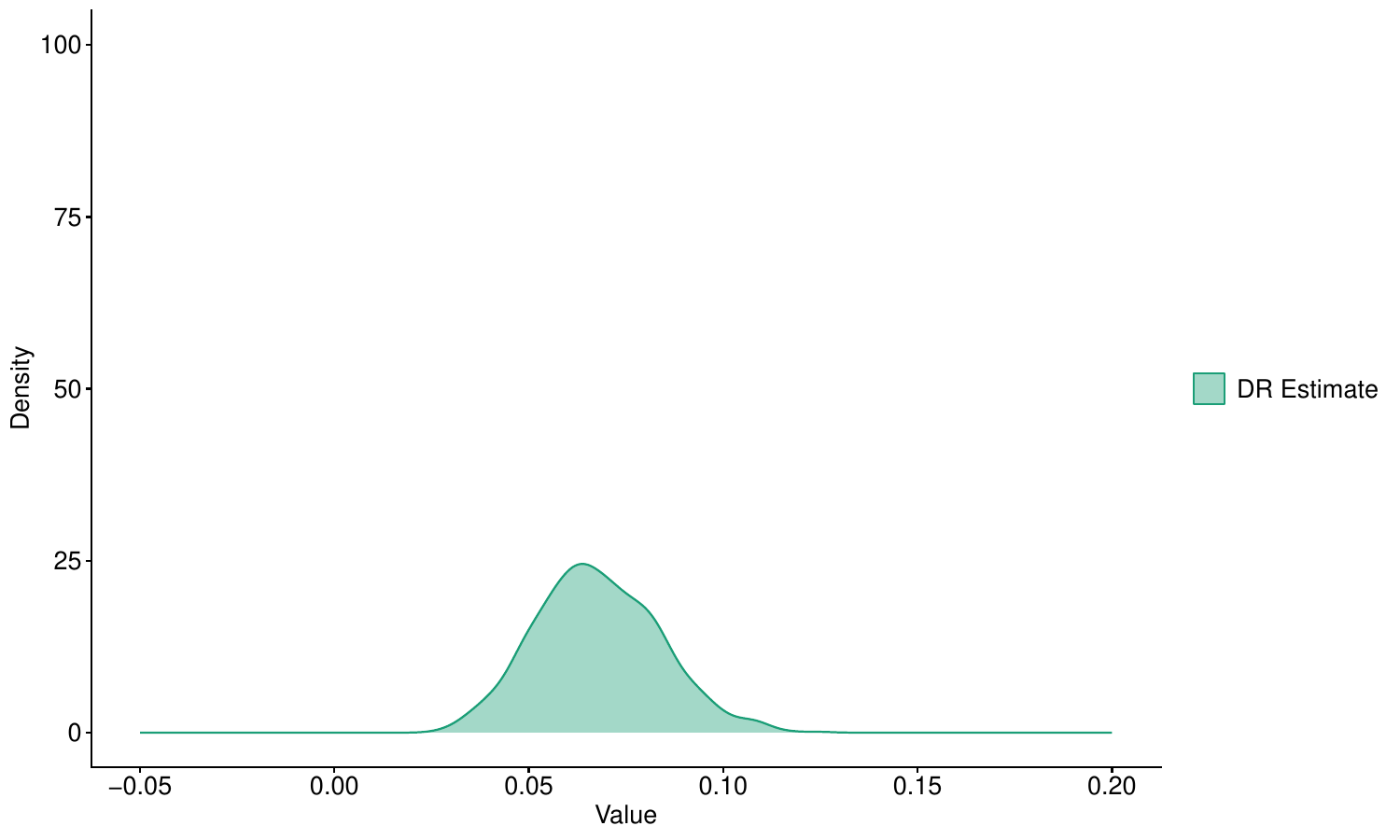}
    \caption{Interactions}
\end{subfigure}

\vspace{1em}

% --- Row 2 ---
\begin{subfigure}[t]{0.45\textwidth}
    \centering
    \includegraphics[width=\linewidth]{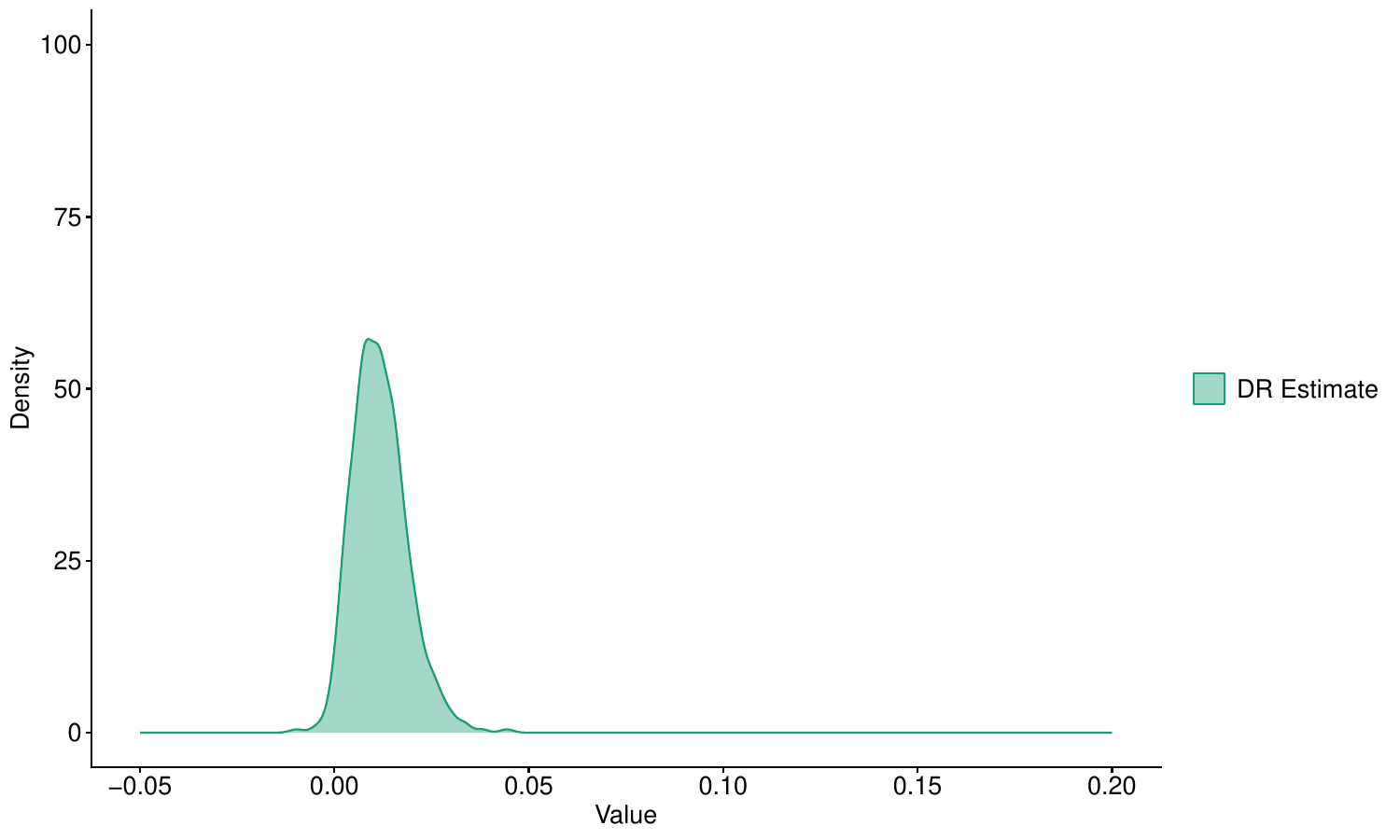}
    \caption{Mixtures}
\end{subfigure}
\hfill
\begin{subfigure}[t]{0.45\textwidth}
    \centering
    \includegraphics[width=\linewidth]{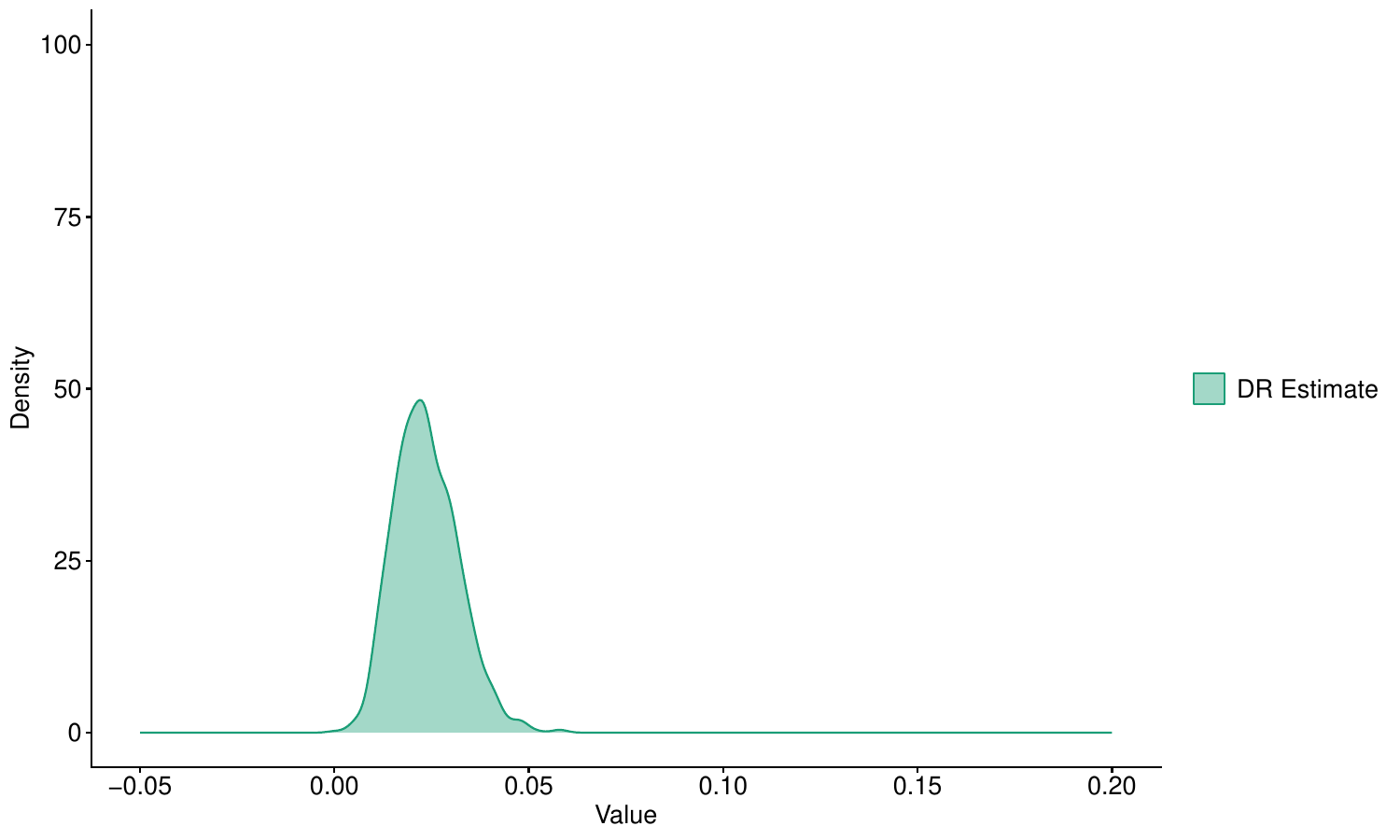}
    \caption{No Baseline}
\end{subfigure}

\vspace{1em}

% --- Row 3: Mixed, centered ---
\begin{subfigure}[t]{0.45\textwidth}
    \centering
    \includegraphics[width=\linewidth]{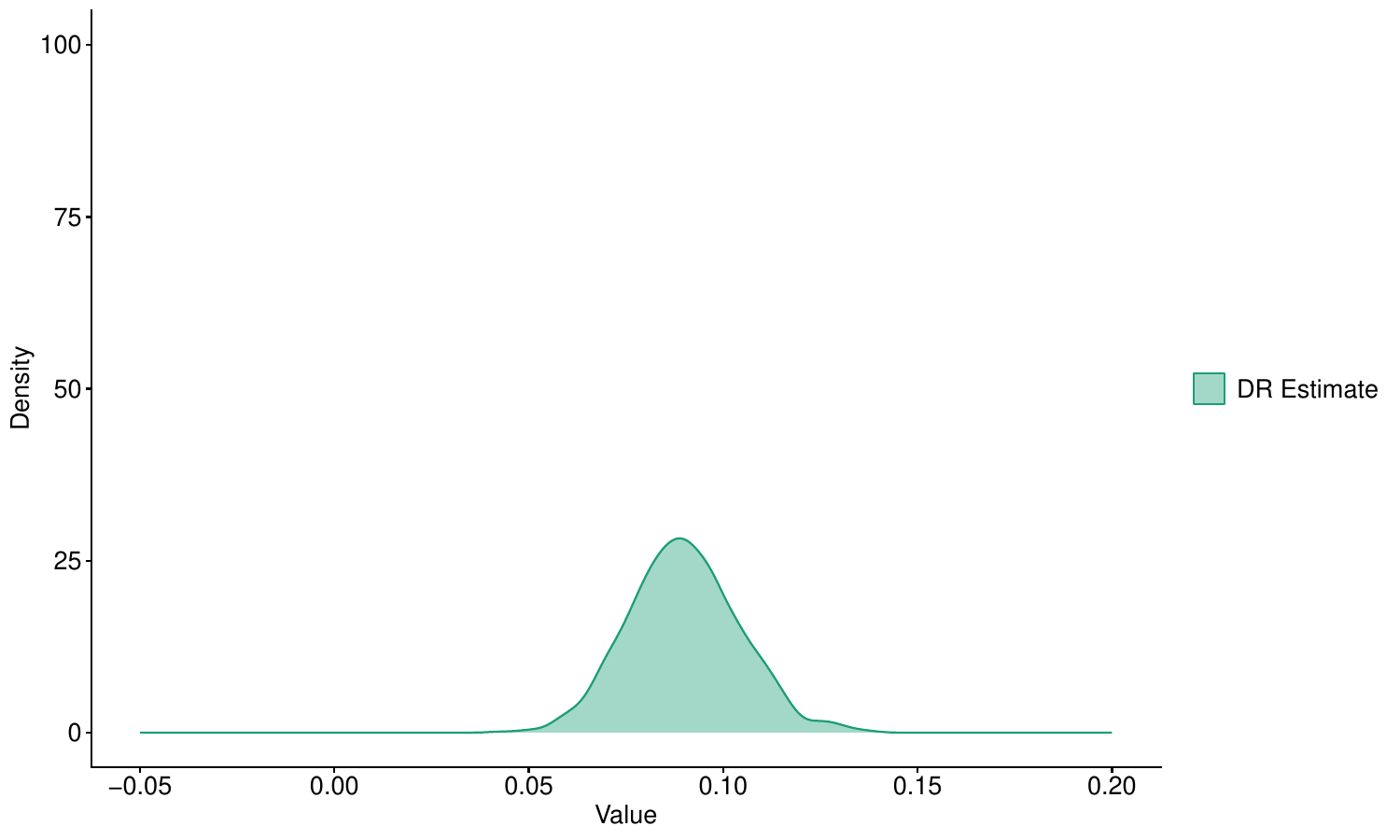}
    \caption{Mixed}
\end{subfigure}

\caption{Distributions of the estimated $\dr$ test statistic under the alternative hypothesis (weak treatment effects). Results are based on 1,000 Monte Carlo replications, each with a sample of 1,000 observations and five outcome variables.}
\label{fig:r_learner_density}
\end{figure}

\begin{figure}[H]
\centering

% --- Row 1 ---
\begin{subfigure}[t]{0.45\textwidth}
    \centering
    \includegraphics[width=\linewidth]{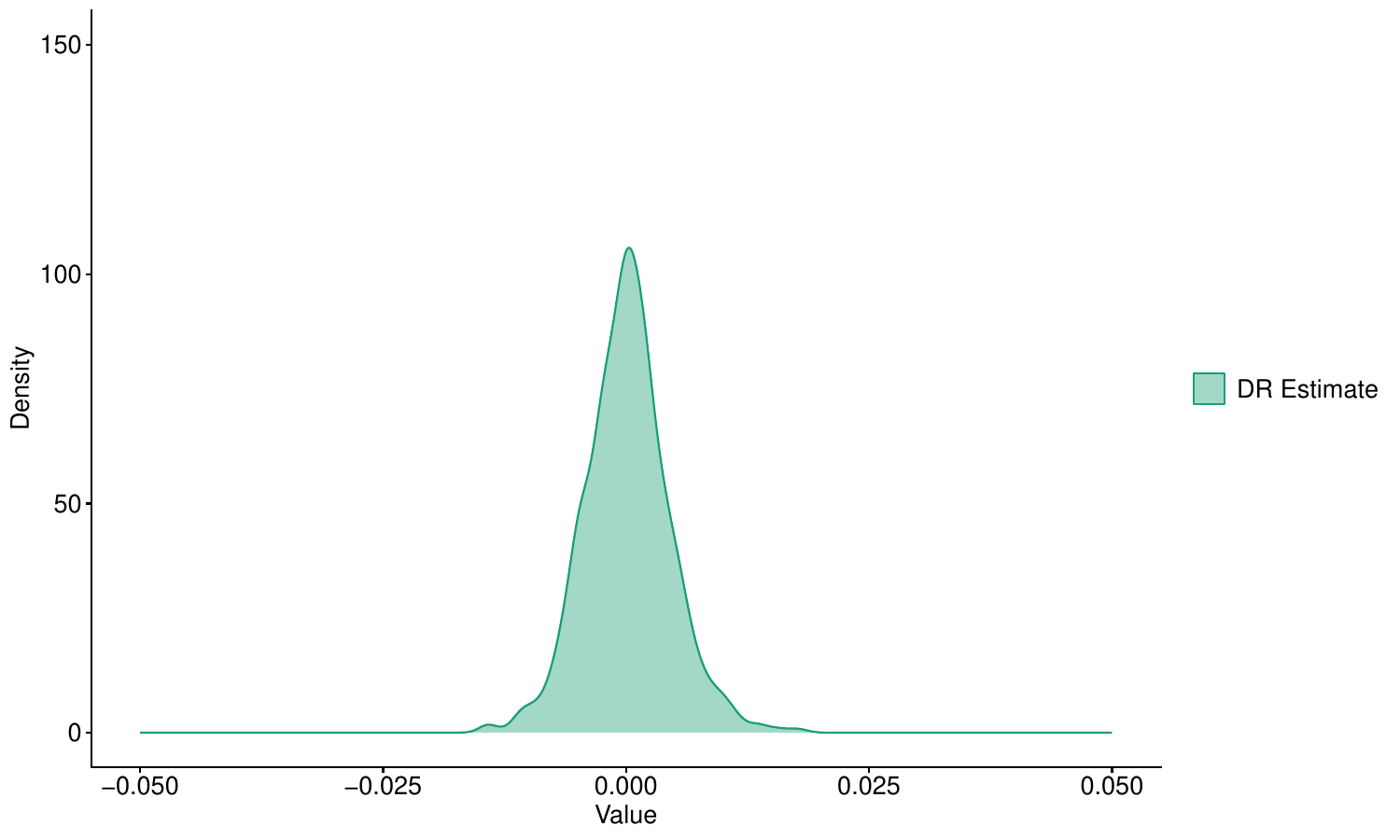}
    \caption{Full Heterogeneity}
\end{subfigure}
\hfill
\begin{subfigure}[t]{0.45\textwidth}
    \centering
    \includegraphics[width=\linewidth]{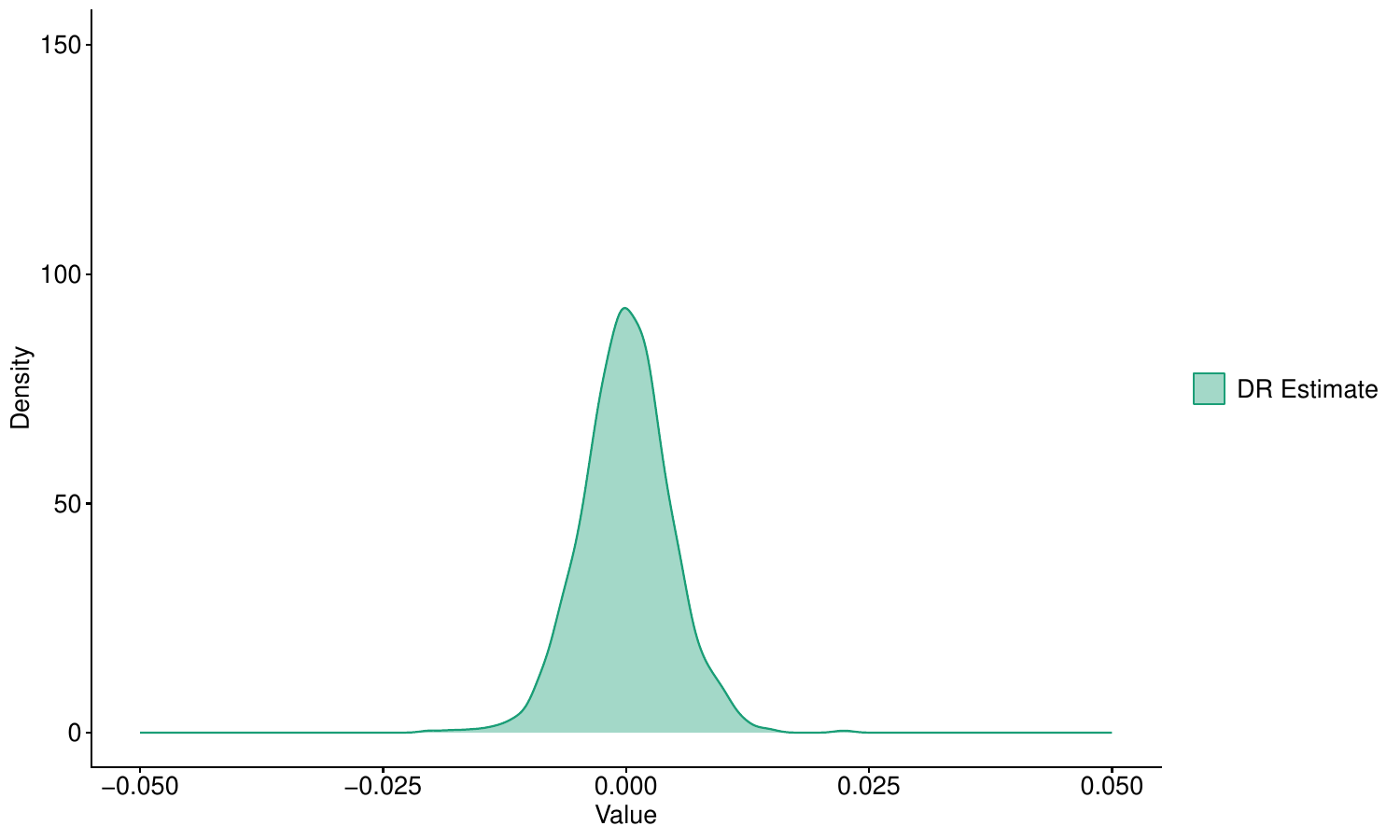}
    \caption{Interactions}
\end{subfigure}

\vspace{1em}

% --- Row 2 ---
\begin{subfigure}[t]{0.45\textwidth}
    \centering
    \includegraphics[width=\linewidth]{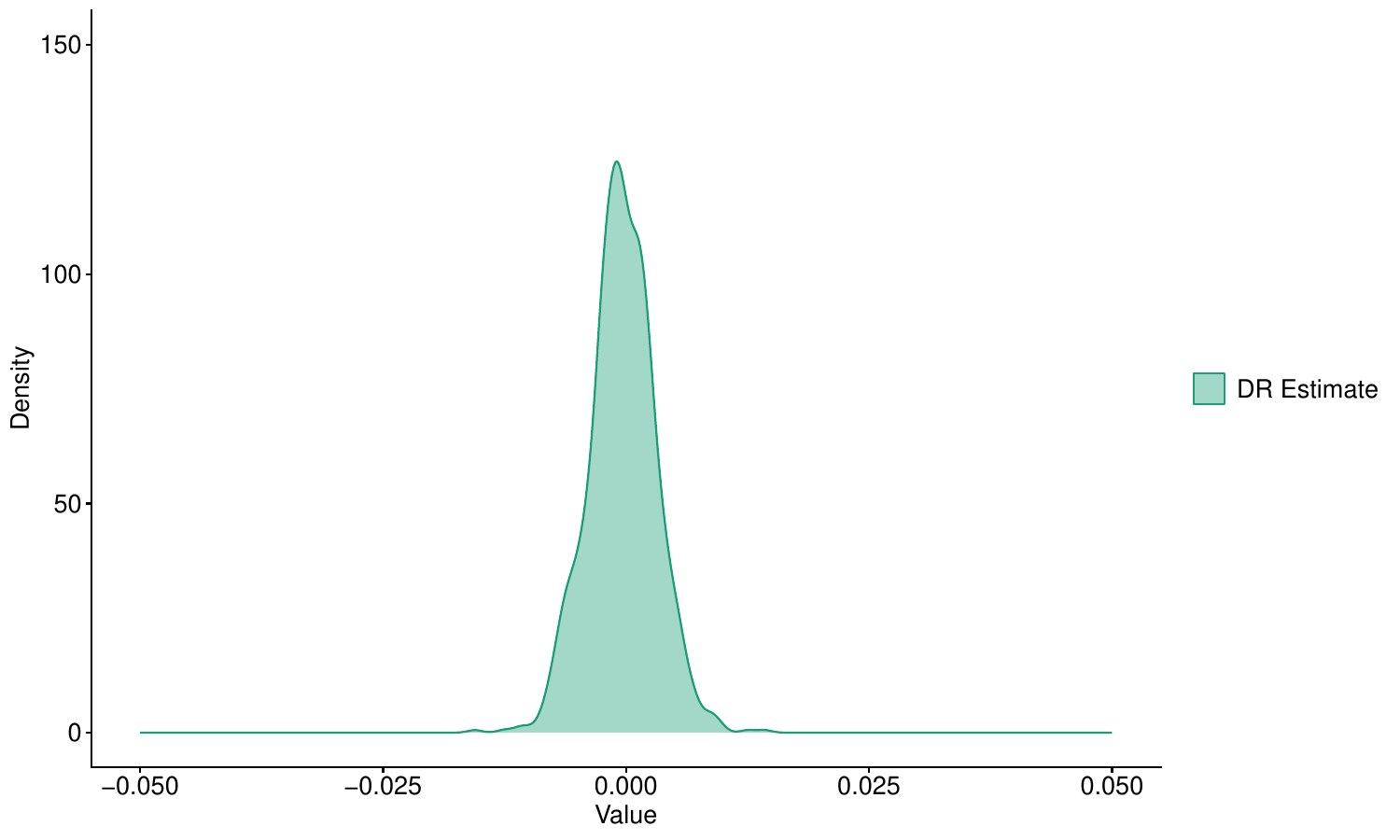}
    \caption{Mixtures}
\end{subfigure}
\hfill
\begin{subfigure}[t]{0.45\textwidth}
    \centering
    \includegraphics[width=\linewidth]{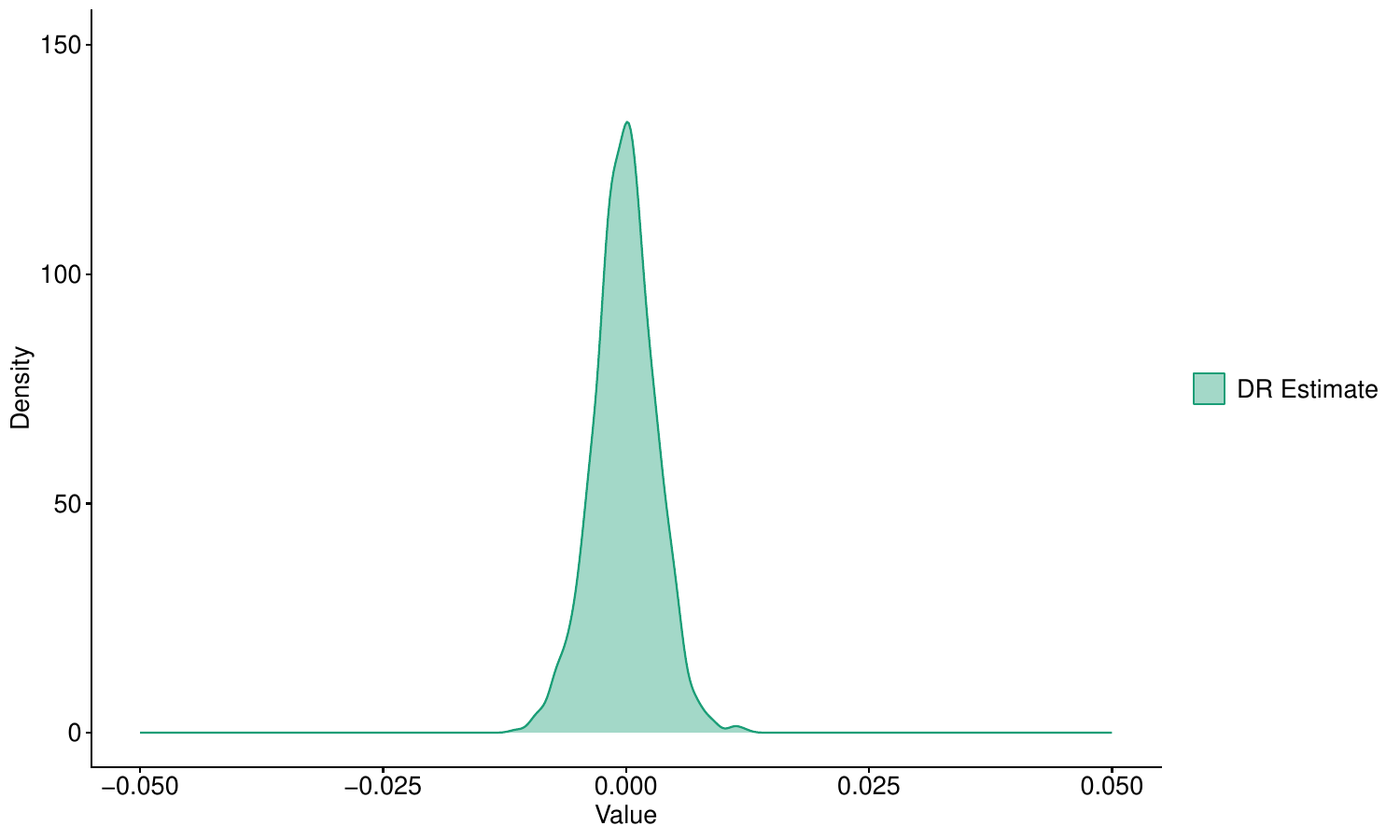}
    \caption{No Baseline}
\end{subfigure}

\vspace{1em}

% --- Row 3: Mixed, centered ---
\begin{subfigure}[t]{0.45\textwidth}
    \centering
    \includegraphics[width=\linewidth]{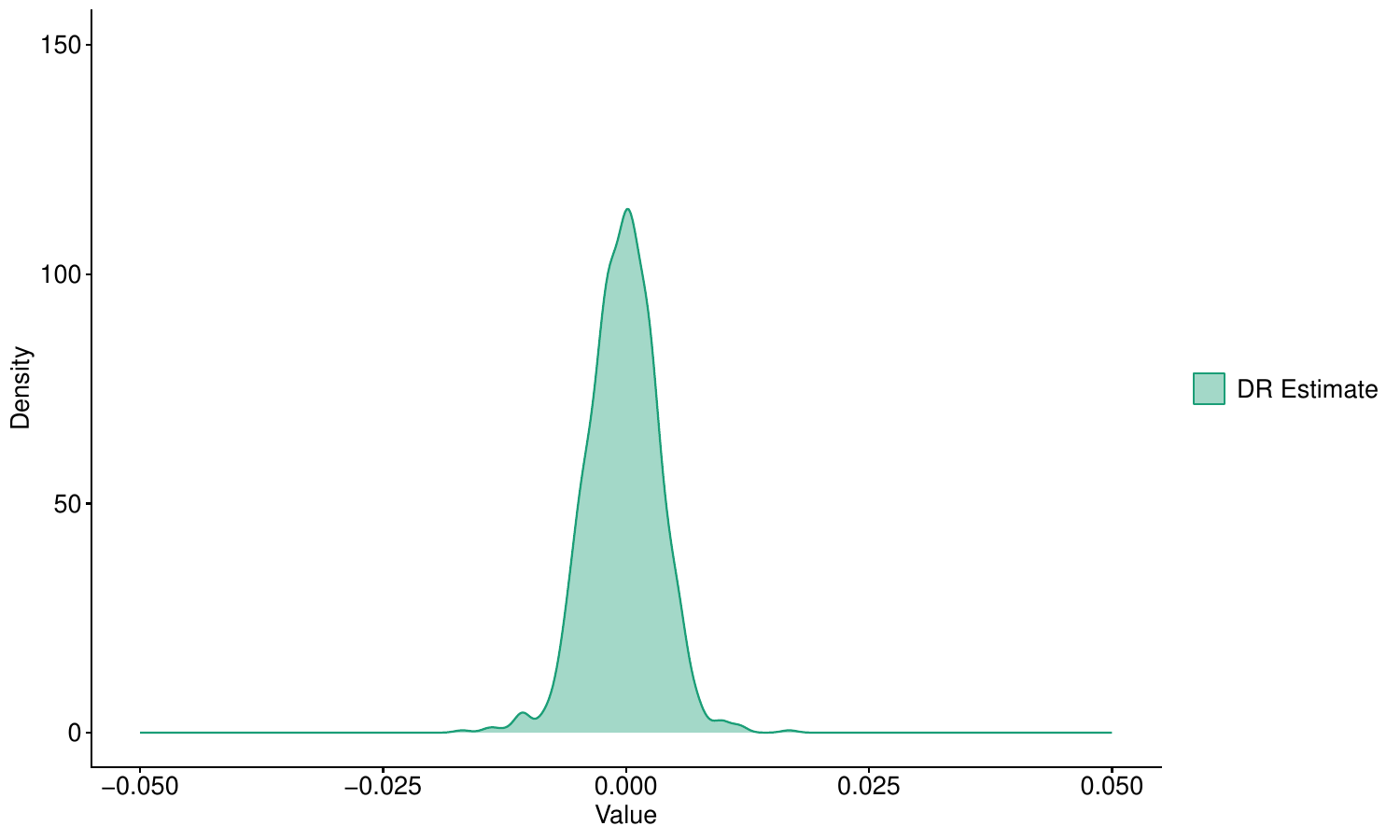}
    \caption{Mixed}
\end{subfigure}

\caption{Distributions of the estimated $\dr$ test statistic under the null hypothesis. Results are based on 1,000 Monte Carlo replications, each with a sample of 1,000 observations and five outcome variables.}
\label{fig:r_learner_density_null}
\end{figure}

\begin{figure}[H]
\centering

% --- Row 1 ---
\begin{subfigure}[t]{0.45\textwidth}
    \centering
    \includegraphics[width=\linewidth]{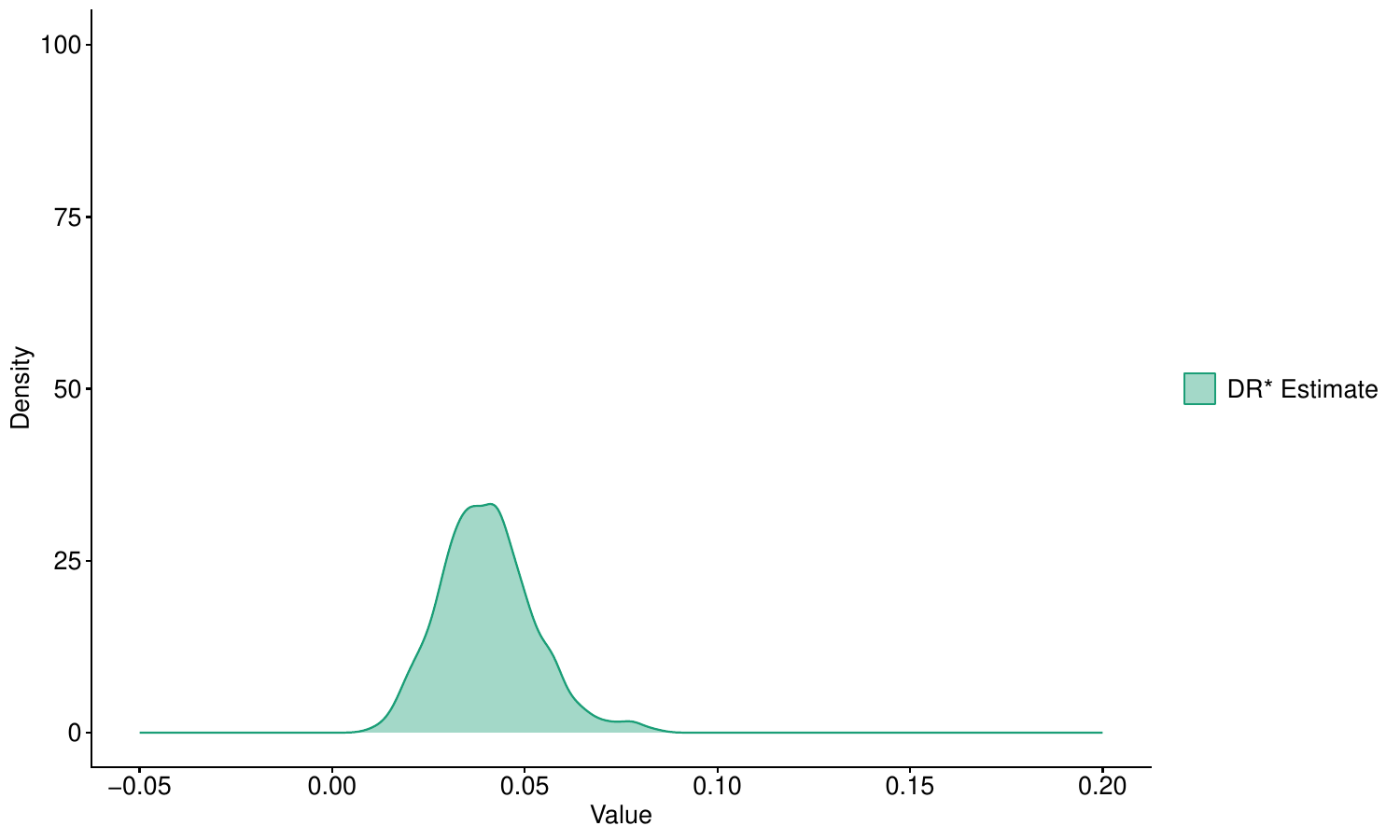}
    \caption{Full Heterogeneity}
\end{subfigure}
\hfill
\begin{subfigure}[t]{0.45\textwidth}
    \centering
    \includegraphics[width=\linewidth]{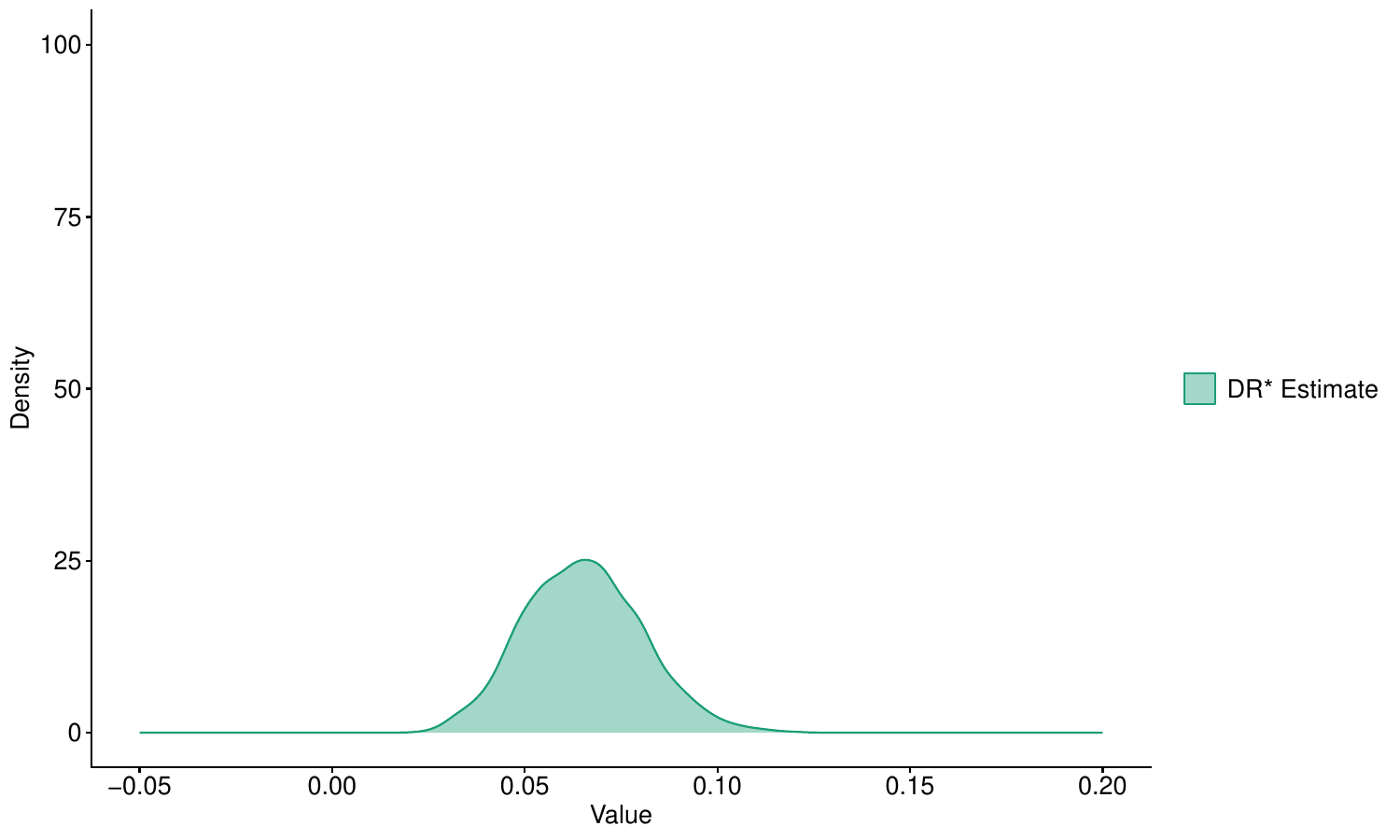}
    \caption{Interactions}
\end{subfigure}

\vspace{1em}

% --- Row 2 ---
\begin{subfigure}[t]{0.45\textwidth}
    \centering
    \includegraphics[width=\linewidth]{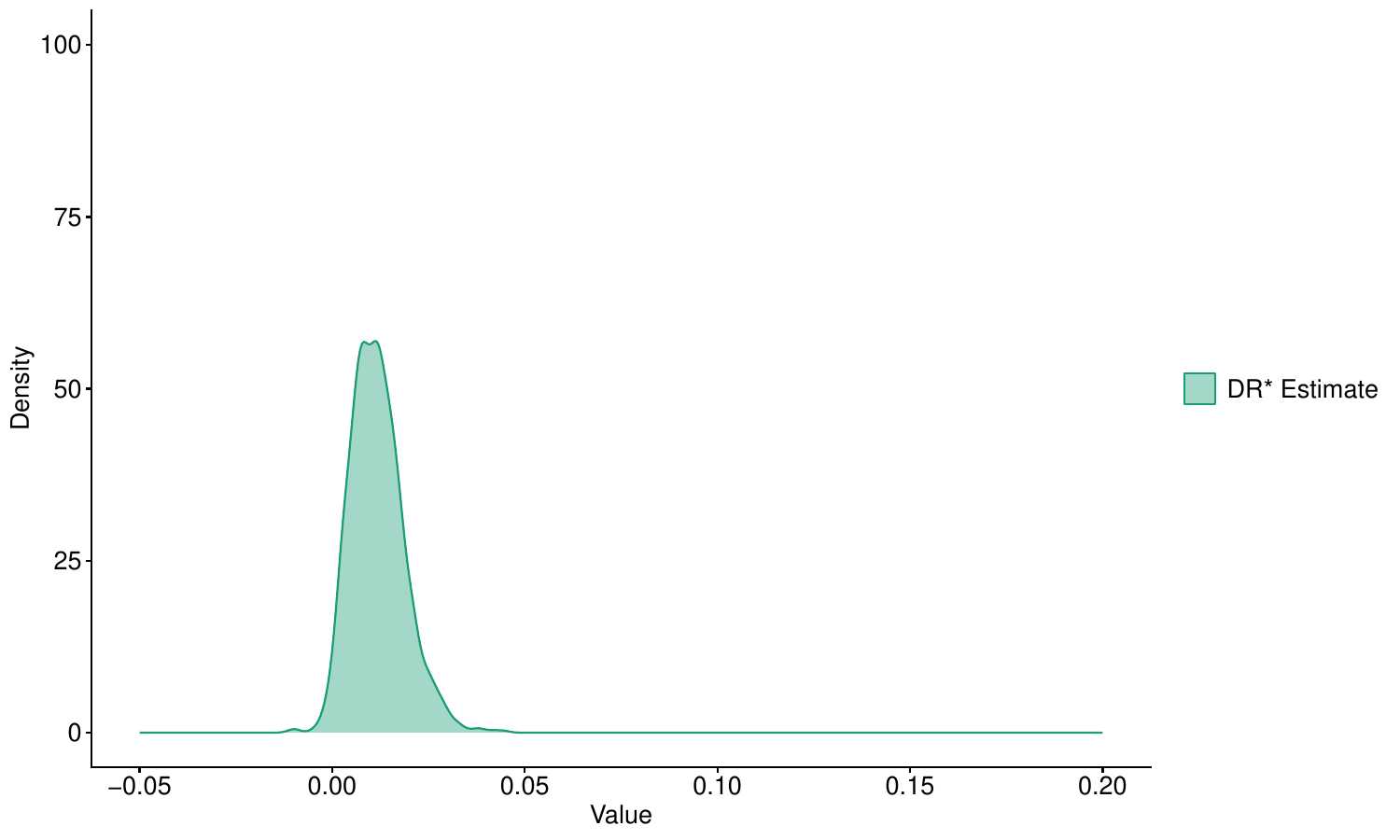}
    \caption{Mixtures}
\end{subfigure}
\hfill
\begin{subfigure}[t]{0.45\textwidth}
    \centering
    \includegraphics[width=\linewidth]{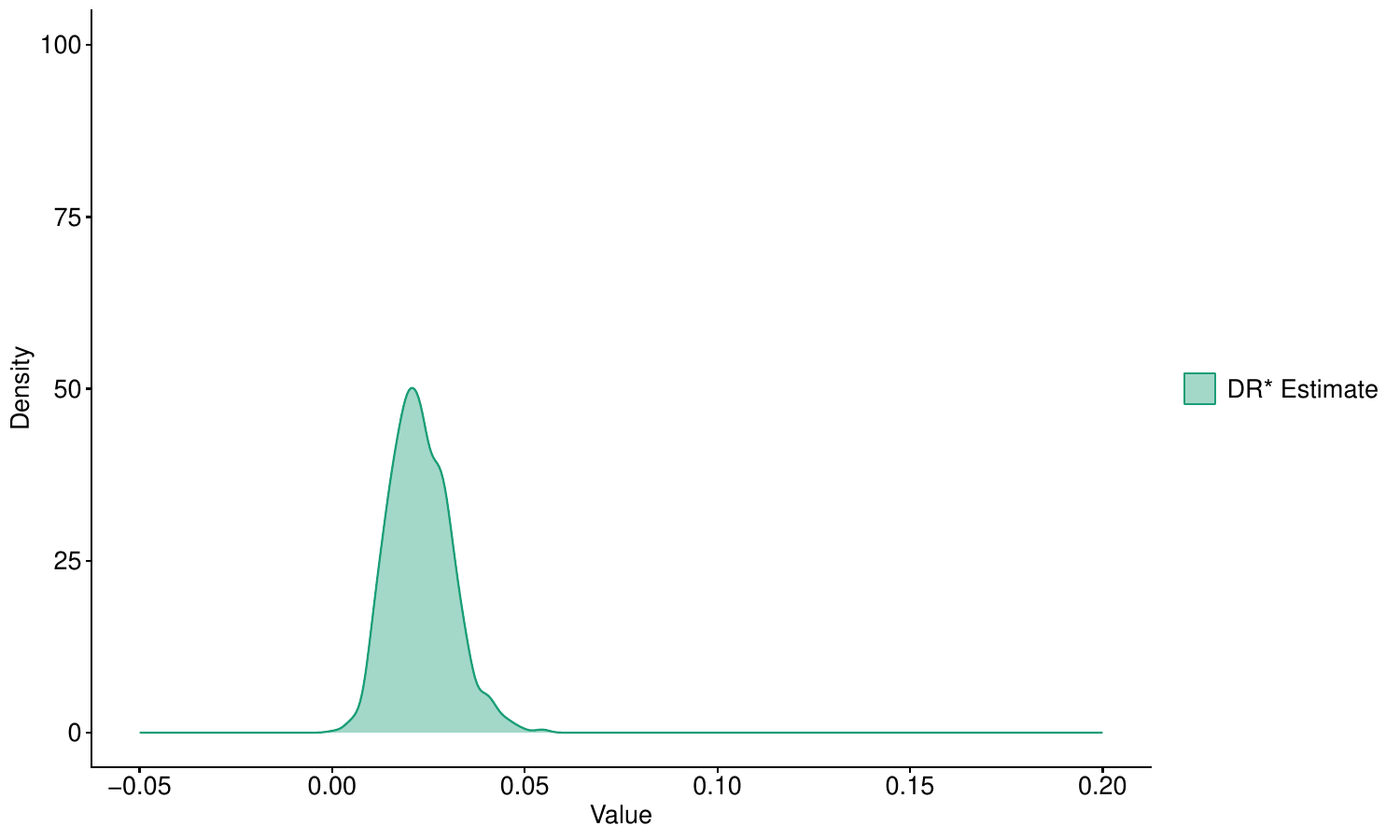}
    \caption{No Baseline}
\end{subfigure}

\vspace{1em}

% --- Row 3: Mixed, centered ---
\begin{subfigure}[t]{0.45\textwidth}
    \centering
    \includegraphics[width=\linewidth]{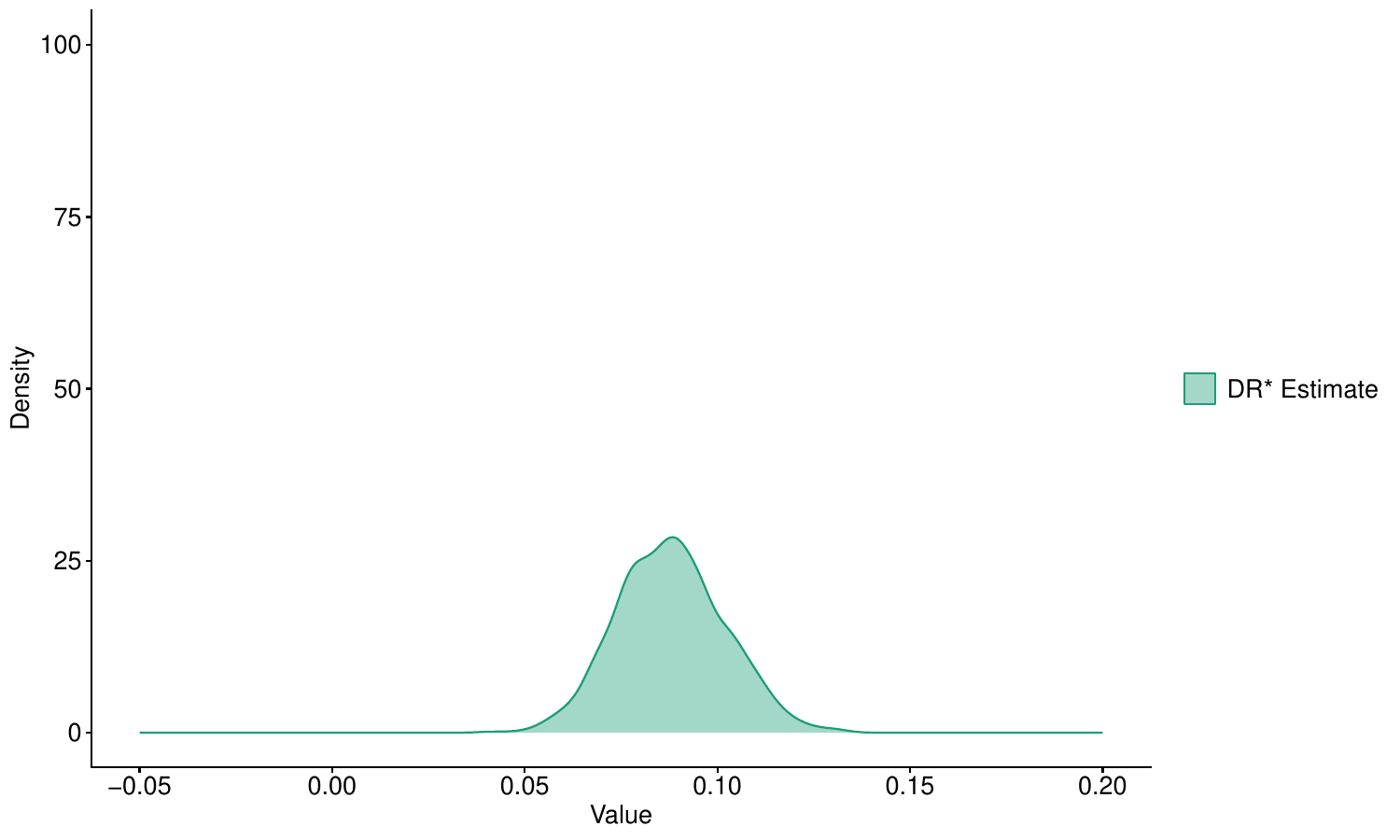}
    \caption{Mixed}
\end{subfigure}

\caption{Distributions of the estimated \(\drs\) test statistic under the alternative hypothesis (weak treatment effects). Results are based on 1,000 Monte Carlo replications, each with a sample of 1,000 observations and five outcome variables.}
\label{fig:dr_star_density_alternative}
\end{figure}

\begin{figure}[H]
\centering

% --- Row 1 ---
\begin{subfigure}[t]{0.45\textwidth}
    \centering
    \includegraphics[width=\linewidth]{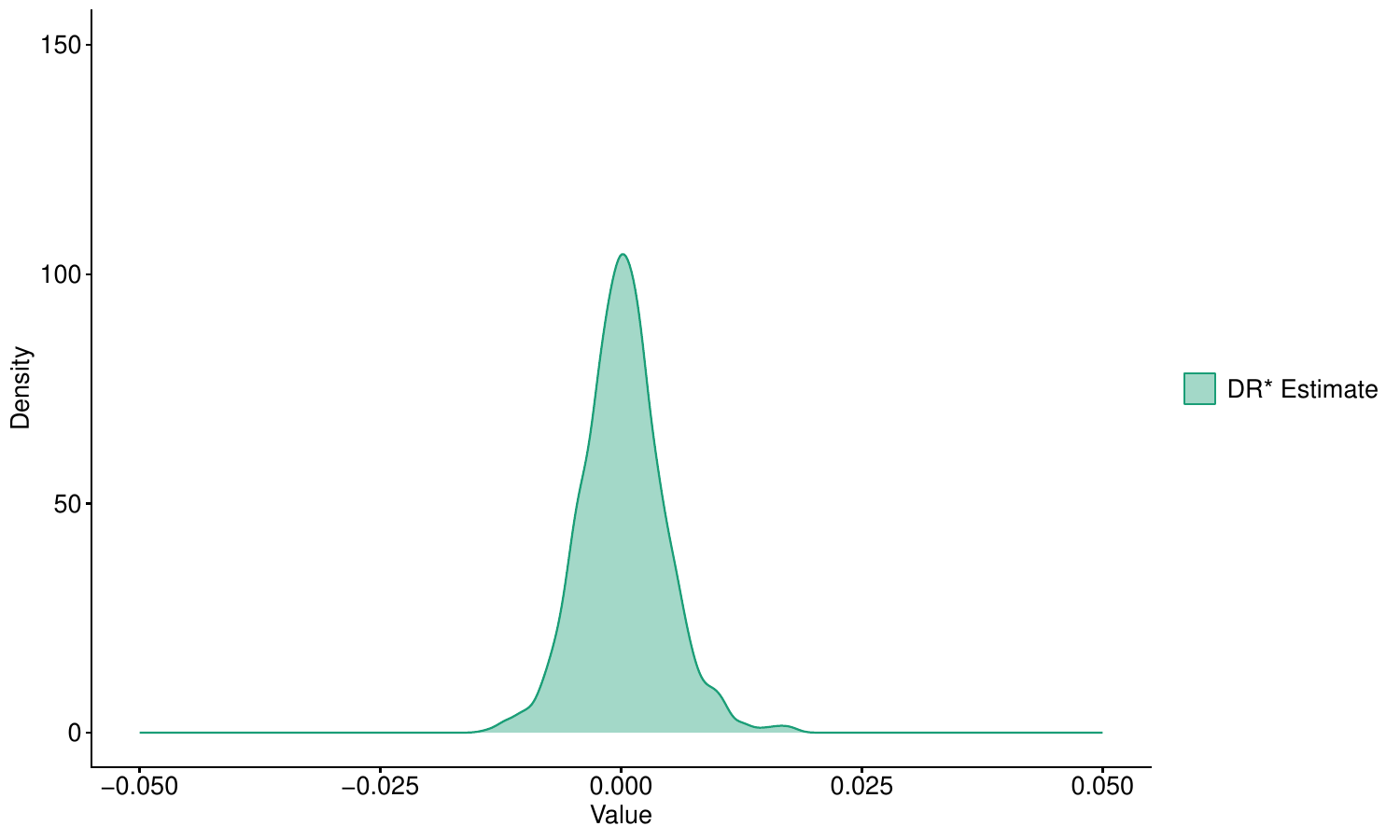}
    \caption{Full Heterogeneity}
\end{subfigure}
\hfill
\begin{subfigure}[t]{0.45\textwidth}
    \centering
    \includegraphics[width=\linewidth]{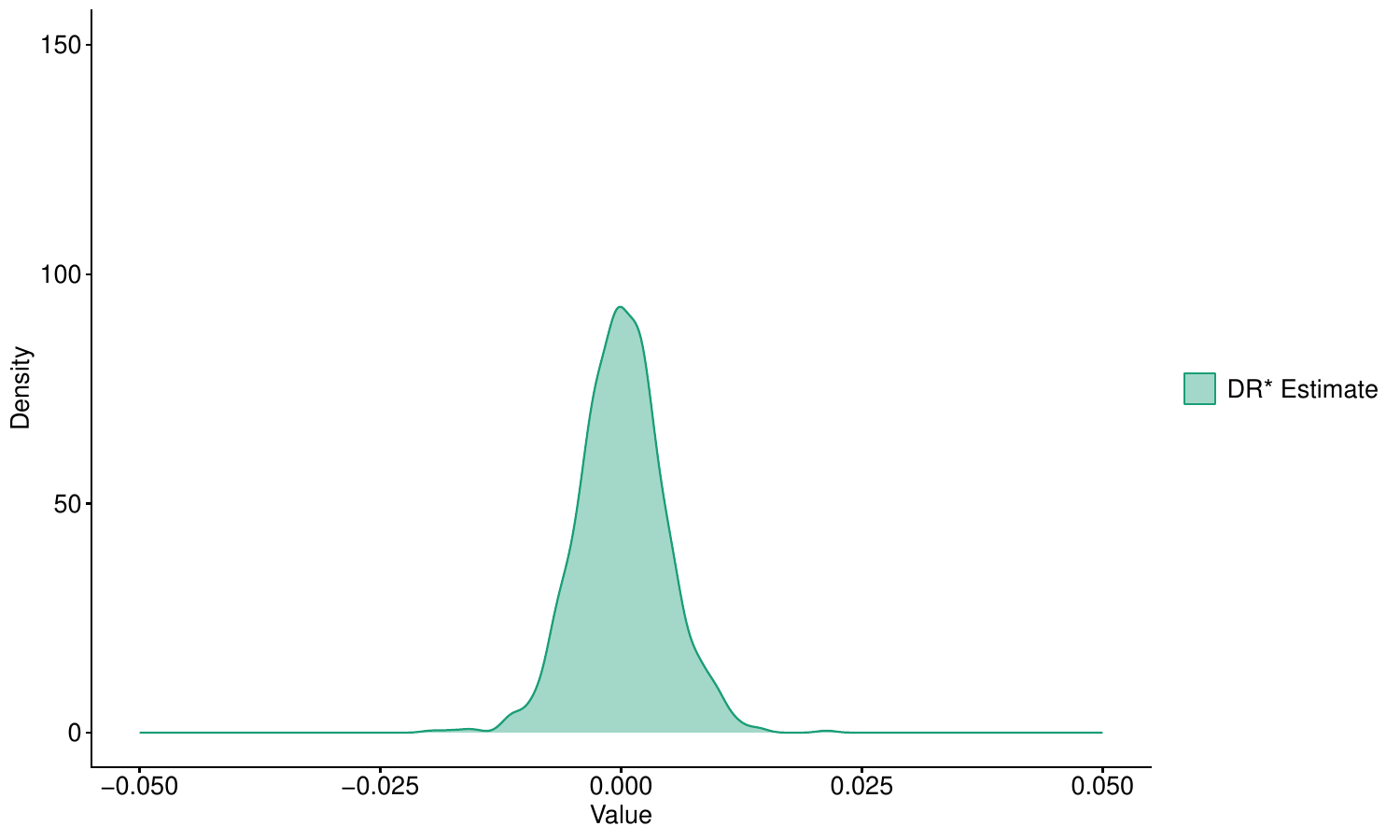}
    \caption{Interactions}
\end{subfigure}

\vspace{1em}

% --- Row 2 ---
\begin{subfigure}[t]{0.45\textwidth}
    \centering
    \includegraphics[width=\linewidth]{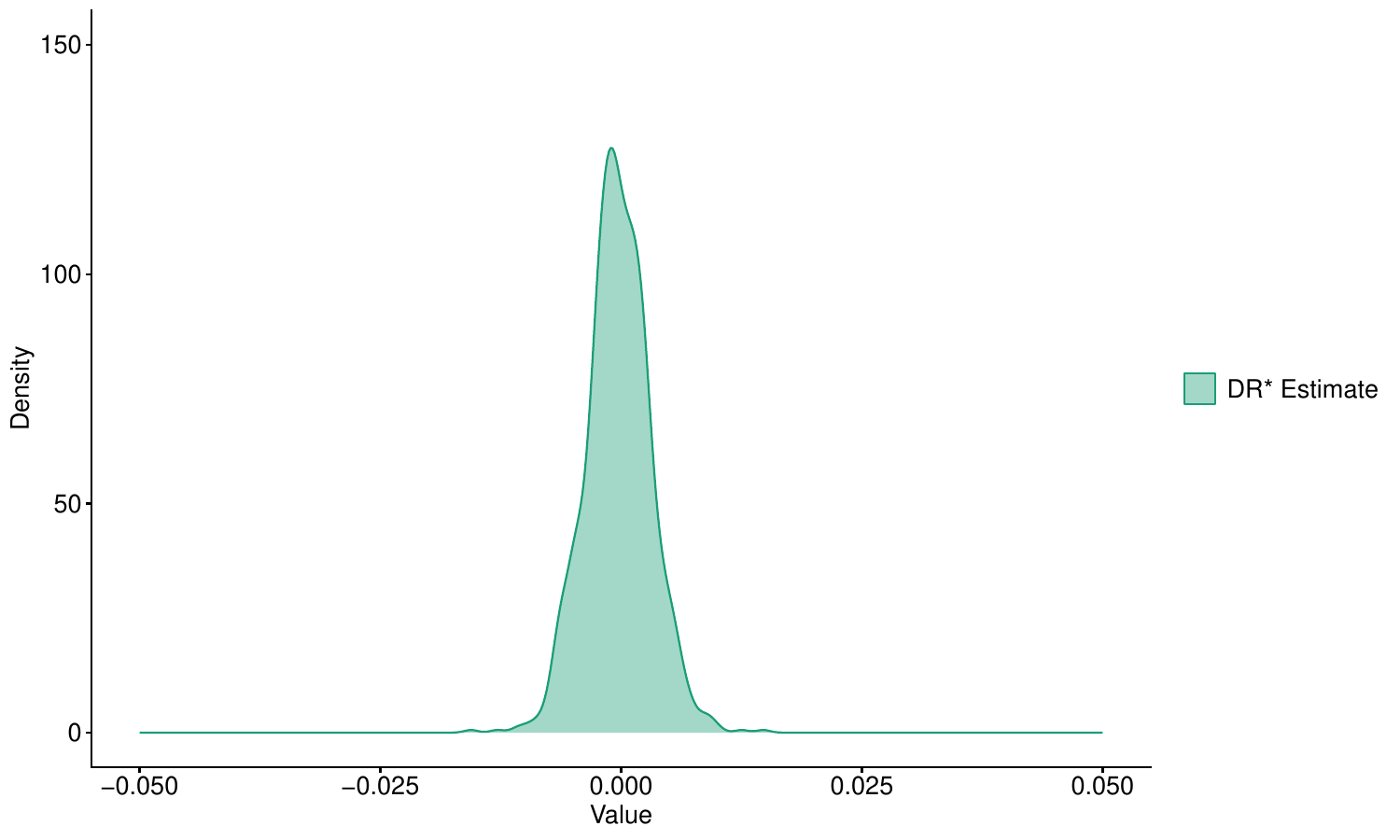}
    \caption{Mixtures}
\end{subfigure}
\hfill
\begin{subfigure}[t]{0.45\textwidth}
    \centering
    \includegraphics[width=\linewidth]{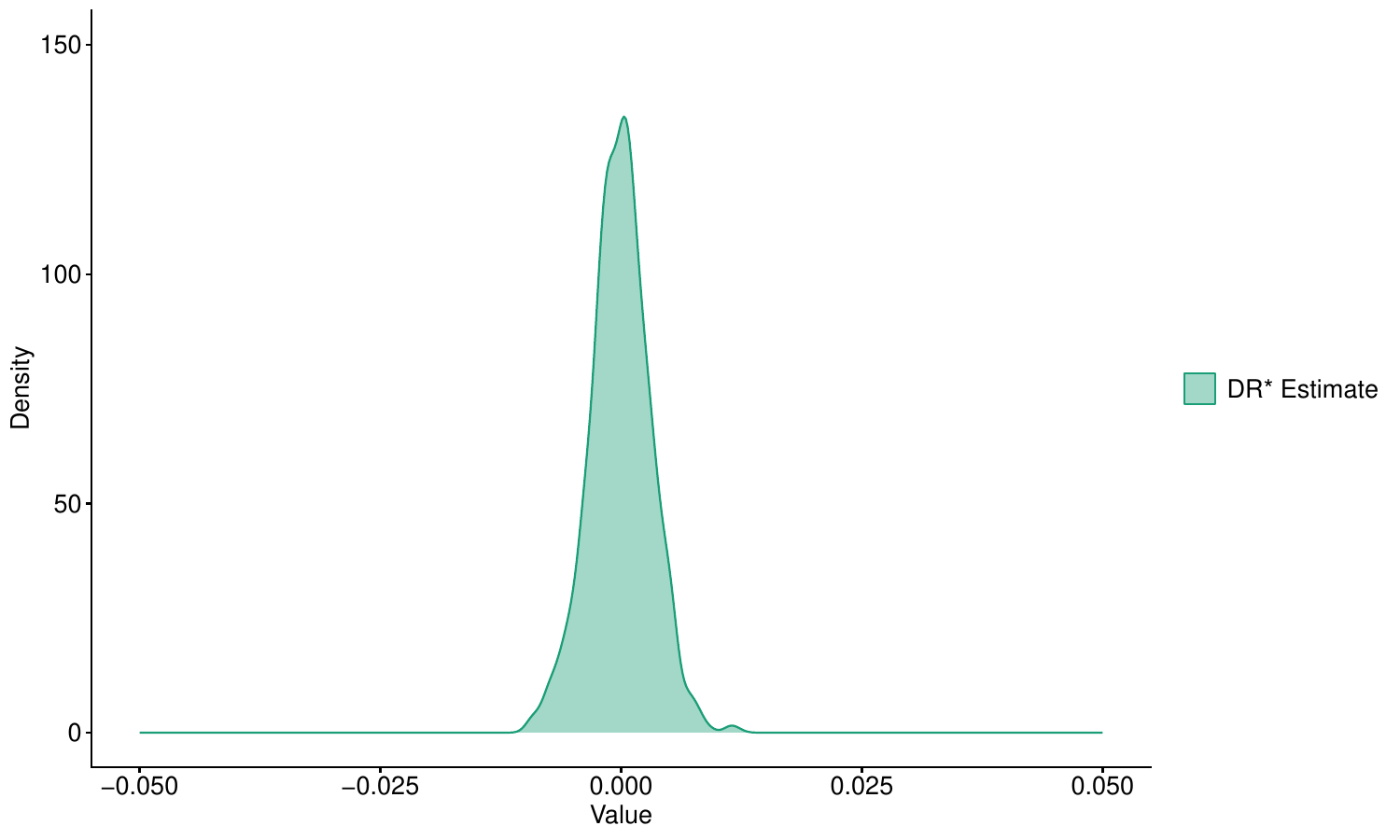}
    \caption{No Baseline}
\end{subfigure}

\vspace{1em}

% --- Row 3: Mixed, centered ---
\begin{subfigure}[t]{0.45\textwidth}
    \centering
    \includegraphics[width=\linewidth]{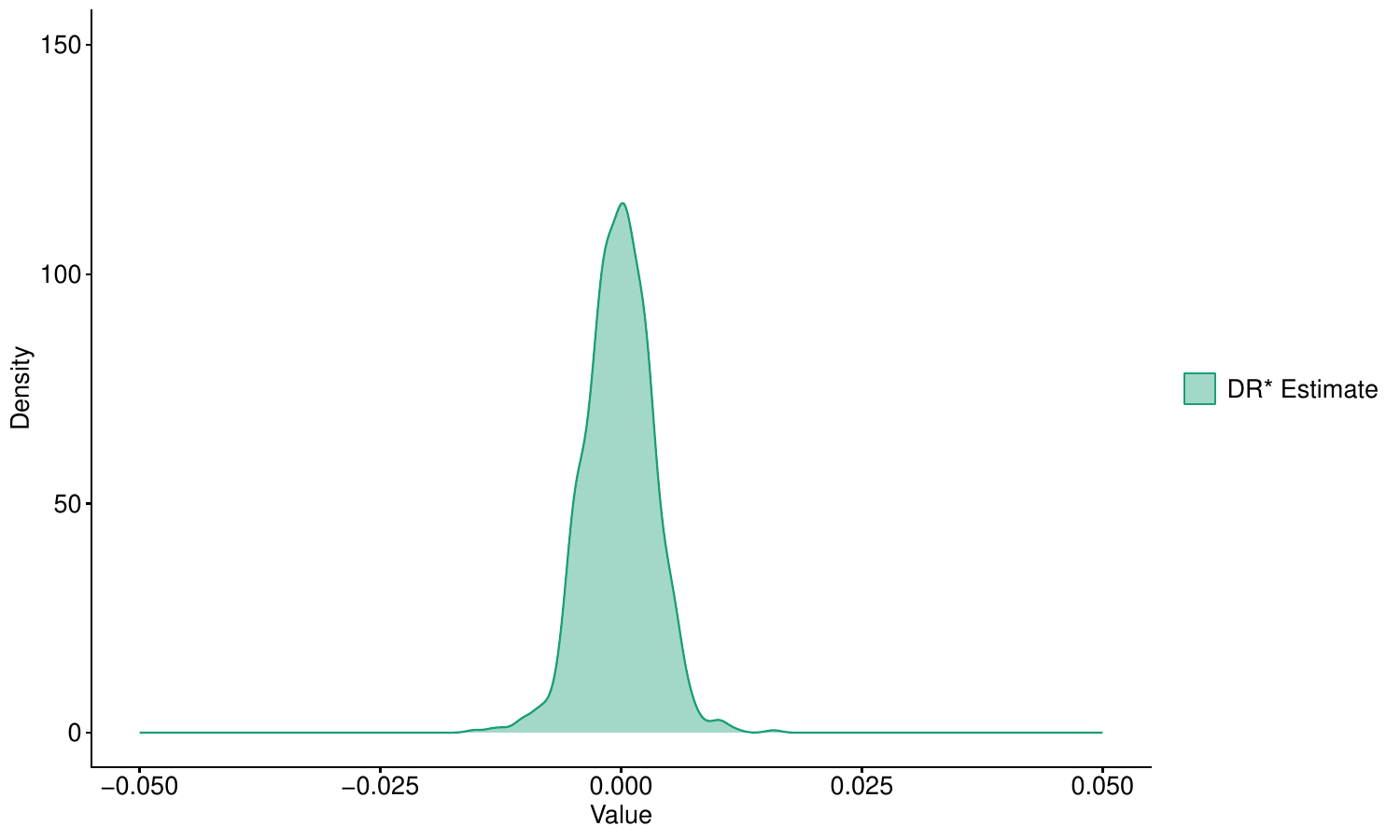}
    \caption{Mixed}
\end{subfigure}

\caption{Distributions of the estimated \(\drs\) test statistic under the null hypothesis. Results are based on 1,000 Monte Carlo replications, each with a sample of 1,000 observations and five outcome variables.}
\label{fig:dr_star_density_null}
\end{figure}

\clearpage

\section{Additional Simulations}\label{app:sims_additional}

We simulate \(N\) individuals, indexed by \(i=1,\dots,N\), each with an \(M\)-dimensional outcome vector \(Y_i = (Y_{i1},\dots,Y_{iM})^\top\). For each individual we generate covariates
\[
  \text{Age}_i \sim \text{Discrete Uniform}\{18,\dots,65\},\quad
  \text{Gender}_i \sim \text{Bernoulli}(0.5),
\]
\[
  \text{Edu}_i \sim \text{Categorical}(\{1,2,3,4\};\,0.3,0.4,0.2,0.1),\quad
  \text{Emp}_i \sim \text{Bernoulli}(0.8),
\]
where \(\text{Gender}_i = 1\) for males, \(\text{Edu}_i\) represents increasing education levels (High school, Bachelor's, Master's, PhD), and \(\text{Emp}_i = 1\) for the employed. Treatment is assigned observationally:
\begin{align*}
  \mathrm{logit}(\pi_i)
  &= \mathrm{logit}(0.5) + 0.01\,(\text{Age}_i - \overline{\text{Age}})
     + 0.1\,(\text{Gender}_i - \overline{\text{Gender}}) + \varepsilon_i,\\
  \varepsilon_i &\sim \N(0, 0.005^2),\qquad
  D_i \sim \text{Bernoulli}(\pi_i).
\end{align*}
The baseline outcome vector is
\[
  U_i = (U_{i1},\dots,U_{iM})^\top \sim \N(0,\Sigma),
\]
where \(\Sigma\) has unit variances and off-diagonal correlations decaying with \(|m-m'|\). The observed outcomes \(Y_{im}\) are obtained from \(U_{im}\) via the transformations below. Under the null we set \(\tau = 0\) and under the alternative \(\tau = 0.5\), except for the Heavy Tails and Mixtures designs, whose null outcome equations are stated separately below.

\begin{description}[style=nextline, font=\normalfont]

\item[$\circ$] Heavy Tails:
Under the alternative,
\[
  Y_{im} = U_{im} + (D_i\,\tau + 0.5)\,T_i + 0.5\,Z_i,
  \qquad T_i \sim t_{1},\ Z_i \sim \N(-2,1).
\]
Under the null,
\[
  Y_{im} = U_{im}\,\tilde T_i + 1,
  \qquad \tilde T_i \sim t_{2},
\]
where $t_1$ and $t_2$ denote the Student's $t$ distribution with 1 and 2 degrees of freedom, respectively. 
\item[$\circ$] Mixtures:
Draw
\[
  I_i \sim \text{Bernoulli}(0.5),\quad
  Z_i^- \sim \N(-2,1),\quad
  Z_i^+ \sim \N(10,1),\quad
  W_i^- \sim \N(-2,1),
\]
independently, and set
\(M_i = I_i Z_i^- + (1 - I_i)Z_i^+\). Under the alternative,
\[
  Y_{im} = U_{im}\,M_i\,D_i\,\tau + 0.5\,I_i\,W_i^-.
\]
Under the null,
\[
  Y_{im} = 0.1\,U_{im}\,M_i + 0.5\,I_i\,W_i^-.
\]

\item[$\circ$] Mixtures Separable:
Define
\[
  G_i = \mathbf{1}\{\text{Edu}_i \le 2\} \cdot 1.5 - \mathbf{1}\{\text{Edu}_i > 2\}\cdot 1.5,
\]
and draw independently
\[
  I_i \sim \text{Bernoulli}(0.5),\quad
  \eta_i^- \sim \N(-2,1),\quad
  \eta_i^+ \sim \N(2,1),\qquad
  \eta_i = I_i\,\eta_i^- + (1 - I_i)\,\eta_i^+.
\]
Then
\[
  Y_{im} = U_{im} + D_i\,\tau\,G_i + \eta_i.
\]

\item[$\circ$] Normal:
\[
  Y_{im} = U_{im} + D_i\,\tau + 0.5.
\]

\item[$\circ$] Outliers:
Draw
\[
  C_i \sim \text{Bernoulli}(0.05),\qquad v_i \sim \N(0,1),
\]
set \(\tilde U_{im} = (1 + 9C_i)\,U_{im}\), and let
\[
  Y_{im} = \tilde U_{im} + D_i\,\tau + 0.5 + (1 + 9C_i)\,v_i.
\]
\end{description}

\autoref{tab:strong_effects_app}
shows that the proposed tests achieve consistently high power across a wide range of outcome distributions. Under Normal and Mixtures designs, all four test statistics (IPW, $\aipw$, $\dr$, $\drs$) reach power near 100\% even with as few as 500 observations, while the Wald test remains below 20\% in designs with Mixtures and Heavy Tails. Across all designs, performance converges to one as the sample size or outcome dimension increases, confirming that the proposed procedures deliver reliable power even in challenging scenarios. \autoref{tab:null_effects_app} confirms that the Type I error rate is around 5\% for the proposed tests, while the Wald test does not reliably control size.

\begin{table}[H]
\centering
\caption{\label{tab:strong_effects_app}Power of the conditional independence tests. The table reports rejection probabilities under the alternative hypothesis for test statistics based on inverse propensity weighting (IPW, \autoref{alg:crt}), augmented inverse propensity weighting (AIPW, \autoref{alg:aipw_dr}), double residualization (DR / \(\drs\), \autoref{alg:aipw_dr}), and a joint Wald test (\autoref{app:wald}). Each cell reports the average over 1,000 Monte Carlo replications. }
\centering
\resizebox{\ifdim\width>\linewidth\linewidth\else\width\fi}{!}{
\fontsize{9}{11}\selectfont
\begin{tabular}[t]{r>{}rrr>{}rrr>{}rrr>{}rrr>{}rrr}
\toprule
\multicolumn{1}{c}{ } & \multicolumn{3}{c}{IPW} & \multicolumn{3}{c}{$\aipw$} & \multicolumn{3}{c}{$\dr$} & \multicolumn{3}{c}{\(\drs\)} & \multicolumn{3}{c}{Wald} \\
\cmidrule(l{3pt}r{3pt}){2-4} \cmidrule(l{3pt}r{3pt}){5-7} \cmidrule(l{3pt}r{3pt}){8-10} \cmidrule(l{3pt}r{3pt}){11-13} \cmidrule(l{3pt}r{3pt}){14-16}
\multicolumn{1}{c}{ } & \multicolumn{15}{c}{Number of outcomes} \\
\cmidrule(l{3pt}r{3pt}){2-16}
\multirow{2}{*}{\raisebox{2ex}[0pt][0pt]{\makecell[l]{Sample size}}} & 5 & 10 & 30 & 5 & 10 & 30 & 5 & 10 & 30 & 5 & 10 & 30 & 5 & 10 & 30\\
\midrule
\addlinespace[0.8em]
\multicolumn{16}{l}{Heavy Tails}\\
\hspace{1em}500 & 0.572 & 0.613 & 0.691 & 0.428 & 0.488 & 0.576 & 0.454 & 0.510 & 0.592 & 0.457 & 0.517 & 0.598 & 0.049 & 0.062 & 0.094\\
\hspace{1em}1000 & 0.890 & 0.918 & 0.954 & 0.754 & 0.810 & 0.879 & 0.772 & 0.838 & 0.898 & 0.771 & 0.836 & 0.899 & 0.047 & 0.039 & 0.065\\
\hspace{1em}2000 & 0.999 & 1.000 & 1.000 & 0.975 & 0.985 & 0.997 & 0.977 & 0.987 & 0.996 & 0.979 & 0.987 & 0.996 & 0.048 & 0.040 & 0.062\\
\midrule
\addlinespace[0.8em]
\multicolumn{16}{l}{Mixtures}\\
\hspace{1em}500 & 1.000 & 1.000 & 1.000 & 1.000 & 1.000 & 1.000 & 1.000 & 1.000 & 1.000 & 1.000 & 1.000 & 1.000 & 0.055 & 0.076 & 0.178\\
\hspace{1em}1000 & 1.000 & 1.000 & 1.000 & 1.000 & 1.000 & 1.000 & 1.000 & 1.000 & 1.000 & 1.000 & 1.000 & 1.000 & 0.061 & 0.055 & 0.078\\
\hspace{1em}2000 & 1.000 & 1.000 & 1.000 & 1.000 & 1.000 & 1.000 & 1.000 & 1.000 & 1.000 & 1.000 & 1.000 & 1.000 & 0.047 & 0.050 & 0.080\\
\midrule
\addlinespace[0.8em]
\multicolumn{16}{l}{Mixtures Separable}\\
\hspace{1em}500 & 0.250 & 0.292 & 0.296 & 0.153 & 0.190 & 0.242 & 0.187 & 0.221 & 0.262 & 0.191 & 0.219 & 0.261 & 0.158 & 0.144 & 0.130\\
\hspace{1em}1000 & 0.595 & 0.642 & 0.675 & 0.388 & 0.473 & 0.558 & 0.431 & 0.501 & 0.575 & 0.435 & 0.505 & 0.578 & 0.283 & 0.241 & 0.187\\
\hspace{1em}2000 & 0.979 & 0.977 & 0.988 & 0.853 & 0.897 & 0.952 & 0.875 & 0.910 & 0.957 & 0.876 & 0.909 & 0.958 & 0.553 & 0.425 & 0.297\\
\midrule
\addlinespace[0.8em]
\multicolumn{16}{l}{Normal}\\
\hspace{1em}500 & 0.997 & 1.000 & 1.000 & 0.981 & 0.998 & 1.000 & 0.989 & 1.000 & 1.000 & 0.989 & 1.000 & 1.000 & 1.000 & 1.000 & 1.000\\
\hspace{1em}1000 & 1.000 & 1.000 & 1.000 & 1.000 & 1.000 & 1.000 & 1.000 & 1.000 & 1.000 & 1.000 & 1.000 & 1.000 & 1.000 & 1.000 & 1.000\\
\hspace{1em}2000 & 1.000 & 1.000 & 1.000 & 1.000 & 1.000 & 1.000 & 1.000 & 1.000 & 1.000 & 1.000 & 1.000 & 1.000 & 1.000 & 1.000 & 1.000\\
\midrule
\addlinespace[0.8em]
\multicolumn{16}{l}{Outliers}\\
\hspace{1em}500 & 0.714 & 0.789 & 0.875 & 0.597 & 0.685 & 0.816 & 0.622 & 0.710 & 0.831 & 0.621 & 0.715 & 0.835 & 0.319 & 0.324 & 0.486\\
\hspace{1em}1000 & 0.967 & 0.981 & 0.990 & 0.882 & 0.938 & 0.967 & 0.899 & 0.946 & 0.971 & 0.902 & 0.948 & 0.970 & 0.527 & 0.513 & 0.561\\
\hspace{1em}2000 & 1.000 & 1.000 & 1.000 & 0.993 & 1.000 & 1.000 & 0.994 & 1.000 & 1.000 & 0.994 & 1.000 & 1.000 & 0.850 & 0.809 & 0.784\\
\bottomrule
\end{tabular}}
\end{table}

\begin{table}[H]
\centering
\caption{\label{tab:null_effects_app} Size of the conditional independence tests. The table reports rejection probabilities under the null hypothesis for test statistics based on inverse propensity weighting (IPW, \autoref{alg:crt}), augmented inverse propensity weighting (AIPW, \autoref{alg:aipw_dr}), double residualization (DR / \(\drs\), \autoref{alg:aipw_dr}), and a joint Wald test (\autoref{app:wald}). Each cell reports the average over 1,000 Monte Carlo replications. }
\centering
\resizebox{\ifdim\width>\linewidth\linewidth\else\width\fi}{!}{
\fontsize{9}{11}\selectfont
\begin{tabular}[t]{r>{}rrr>{}rrr>{}rrr>{}rrr>{}rrr}
\toprule
\multicolumn{1}{c}{ } & \multicolumn{3}{c}{IPW} & \multicolumn{3}{c}{$\aipw$} & \multicolumn{3}{c}{$\dr$} & \multicolumn{3}{c}{\(\drs\)} & \multicolumn{3}{c}{Wald} \\
\cmidrule(l{3pt}r{3pt}){2-4} \cmidrule(l{3pt}r{3pt}){5-7} \cmidrule(l{3pt}r{3pt}){8-10} \cmidrule(l{3pt}r{3pt}){11-13} \cmidrule(l{3pt}r{3pt}){14-16}
\multicolumn{1}{c}{ } & \multicolumn{15}{c}{Number of outcomes} \\
\cmidrule(l{3pt}r{3pt}){2-16}
\multirow{2}{*}{\raisebox{2ex}[0pt][0pt]{\makecell[l]{Sample size}}} & 5 & 10 & 30 & 5 & 10 & 30 & 5 & 10 & 30 & 5 & 10 & 30 & 5 & 10 & 30\\
\midrule
\addlinespace[0.8em]
\multicolumn{16}{l}{Heavy Tails}\\
\hspace{1em}500 & 0.049 & 0.047 & 0.074 & 0.030 & 0.041 & 0.053 & 0.035 & 0.050 & 0.060 & 0.036 & 0.050 & 0.060 & 0.044 & 0.042 & 0.069\\
\hspace{1em}1000 & 0.059 & 0.045 & 0.043 & 0.031 & 0.034 & 0.042 & 0.033 & 0.045 & 0.050 & 0.032 & 0.043 & 0.049 & 0.040 & 0.035 & 0.052\\
\hspace{1em}2000 & 0.060 & 0.046 & 0.048 & 0.045 & 0.035 & 0.044 & 0.051 & 0.042 & 0.053 & 0.055 & 0.041 & 0.053 & 0.040 & 0.040 & 0.042\\
\midrule
\addlinespace[0.8em]
\multicolumn{16}{l}{Mixtures}\\
\hspace{1em}500 & 0.043 & 0.051 & 0.043 & 0.032 & 0.048 & 0.039 & 0.042 & 0.051 & 0.044 & 0.043 & 0.051 & 0.046 & 0.061 & 0.053 & 0.107\\
\hspace{1em}1000 & 0.059 & 0.058 & 0.053 & 0.042 & 0.041 & 0.055 & 0.045 & 0.045 & 0.061 & 0.046 & 0.047 & 0.061 & 0.057 & 0.048 & 0.073\\
\hspace{1em}2000 & 0.047 & 0.039 & 0.061 & 0.037 & 0.038 & 0.041 & 0.047 & 0.046 & 0.051 & 0.049 & 0.044 & 0.050 & 0.049 & 0.053 & 0.070\\
\midrule
\addlinespace[0.8em]
\multicolumn{16}{l}{Mixtures Separable}\\
\hspace{1em}500 & 0.049 & 0.066 & 0.054 & 0.031 & 0.052 & 0.045 & 0.038 & 0.051 & 0.051 & 0.039 & 0.051 & 0.052 & 0.079 & 0.091 & 0.105\\
\hspace{1em}1000 & 0.059 & 0.052 & 0.051 & 0.030 & 0.056 & 0.047 & 0.034 & 0.060 & 0.050 & 0.036 & 0.062 & 0.050 & 0.114 & 0.093 & 0.090\\
\hspace{1em}2000 & 0.049 & 0.057 & 0.052 & 0.034 & 0.031 & 0.049 & 0.042 & 0.038 & 0.049 & 0.043 & 0.039 & 0.050 & 0.179 & 0.146 & 0.115\\
\midrule
\addlinespace[0.8em]
\multicolumn{16}{l}{Normal}\\
\hspace{1em}500 & 0.049 & 0.059 & 0.052 & 0.043 & 0.049 & 0.040 & 0.048 & 0.054 & 0.047 & 0.051 & 0.055 & 0.050 & 0.293 & 0.299 & 0.311\\
\hspace{1em}1000 & 0.044 & 0.045 & 0.042 & 0.037 & 0.035 & 0.049 & 0.047 & 0.041 & 0.054 & 0.048 & 0.041 & 0.058 & 0.290 & 0.287 & 0.303\\
\hspace{1em}2000 & 0.052 & 0.059 & 0.051 & 0.026 & 0.038 & 0.048 & 0.030 & 0.048 & 0.049 & 0.031 & 0.050 & 0.050 & 0.288 & 0.290 & 0.289\\
\midrule
\addlinespace[0.8em]
\multicolumn{16}{l}{Outliers}\\
\hspace{1em}500 & 0.058 & 0.057 & 0.045 & 0.037 & 0.047 & 0.049 & 0.043 & 0.050 & 0.050 & 0.041 & 0.049 & 0.052 & 0.118 & 0.109 & 0.165\\
\hspace{1em}1000 & 0.065 & 0.054 & 0.044 & 0.051 & 0.041 & 0.046 & 0.054 & 0.046 & 0.050 & 0.054 & 0.044 & 0.050 & 0.167 & 0.162 & 0.172\\
\hspace{1em}2000 & 0.038 & 0.042 & 0.036 & 0.030 & 0.038 & 0.038 & 0.039 & 0.040 & 0.039 & 0.041 & 0.040 & 0.042 & 0.245 & 0.237 & 0.230\\
\bottomrule
\end{tabular}}
\end{table}

\section{Small Sample Performance}\label{app:small_sample}

We benchmark the proposed procedures against the kernel conditional independence (KCI) test of \citet*{zhang2012kernel} and the generalized covariance measure (GCM) of \citet*{shah2020hardness} based on the simulated experiments described in \autoref{sec:simulations} where the effects decay across outcomes as defined by \eqref{eq:dgp_decay} and \eqref{eq:dgp_mixed}. \autoref{tab:power_small_sample} illustrates power in small samples. 

Under the weak alternative $\rho = 0.2$ and small samples ($N \in \{150, 200\}$),
the competitors (KCI, GCM, joint Wald) dominate the proposed $\dr$,
$\aipw$, and $\drs$ tests in the Full Heterogeneity, Mixed Heterogeneity, and Interactions designs, 
where $\E[\tau(X_i)] \neq 0$ generates a detectable shift in the marginal mean. The ordering reverses in the Mixtures and No Baseline designs, where $\E[\tau(X_i)] = 0$ but distributional effects remain. 
The two families of procedures are
therefore complementary: $\dr$, $\aipw$, and $\drs$ are sensitive to any departure
from conditional independence and dominate when the treatment effect averages to
zero, while KCI, GCM, and the joint Wald tests dominate when $\E[\tau(X_i)] \neq 0$. \autoref{tab:size_small} shows that under the null hypothesis, $\dr$/$\drs$ and $\aipw$ have higher rejection probabilities with small samples when compared to KCI and GCM conditional independence tests and the joint Wald test. Because the nuisance functions are estimated with flexible machine-learning methods, their estimates are noisy in small samples; this additional variability inflates the variance of the test statistic and distorts its size.

\begin{table}[H]
\centering
\caption{\label{tab:power_small_sample}Power of the conditional independence tests in small samples. The table reports rejection probabilities under the alternative hypothesis (weak treatment effects with \(\rho = 0.2\)) for the double residualized (DR/\(\drs\)), augmented inverse propensity weighted (AIPW), kernel conditional independence (KCI, \citealp{zhang2012kernel}), generalized covariance measure (GCM, \citealp{shah2020hardness}), and joint Wald tests. Each cell reports the average over 100 Monte Carlo replications. }
\centering
\resizebox{\ifdim\width>\linewidth\linewidth\else\width\fi}{!}{
\fontsize{9}{11}\selectfont
\begin{tabular}[t]{l>{}rrr>{}rrr>{}rrr>{}rrr>{}rrr>{}rrr}
\toprule
\multicolumn{1}{c}{ } & \multicolumn{3}{c}{$\dr$} & \multicolumn{3}{c}{$\aipw$} & \multicolumn{3}{c}{$\drs$} & \multicolumn{3}{c}{KCI} & \multicolumn{3}{c}{GCM} & \multicolumn{3}{c}{Wald} \\
\cmidrule(l{3pt}r{3pt}){2-4} \cmidrule(l{3pt}r{3pt}){5-7} \cmidrule(l{3pt}r{3pt}){8-10} \cmidrule(l{3pt}r{3pt}){11-13} \cmidrule(l{3pt}r{3pt}){14-16} \cmidrule(l{3pt}r{3pt}){17-19}
\multicolumn{1}{c}{ } & \multicolumn{18}{c}{Number of outcomes} \\
\cmidrule(l{3pt}r{3pt}){2-19}
Sample size & 1 & 2 & 3 & 1 & 2 & 3 & 1 & 2 & 3 & 1 & 2 & 3 & 1 & 2 & 3 & 1 & 2 & 3\\
\midrule
\addlinespace[0.8em]
\multicolumn{19}{l}{Full Heterogeneity}\\
\hspace{1em}150 & 0.36 & 0.43 & 0.32 & 0.31 & 0.39 & 0.37 & 0.37 & 0.43 & 0.37 & 0.64 & 0.71 & 0.62 & 0.68 & 0.73 & 0.68 & 0.73 & 0.72 & 0.64\\
\hspace{1em}200 & 0.51 & 0.53 & 0.59 & 0.42 & 0.51 & 0.52 & 0.51 & 0.54 & 0.58 & 0.75 & 0.78 & 0.77 & 0.83 & 0.86 & 0.87 & 0.86 & 0.84 & 0.86\\
\midrule
\addlinespace[0.8em]
\multicolumn{19}{l}{No Baseline}\\
\hspace{1em}150 & 0.10 & 0.12 & 0.12 & 0.12 & 0.16 & 0.15 & 0.10 & 0.12 & 0.11 & 0.08 & 0.11 & 0.18 & 0.04 & 0.06 & 0.02 & 0.04 & 0.08 & 0.07\\
\hspace{1em}200 & 0.19 & 0.14 & 0.19 & 0.14 & 0.17 & 0.21 & 0.18 & 0.14 & 0.19 & 0.11 & 0.16 & 0.22 & 0.03 & 0.04 & 0.08 & 0.03 & 0.03 & 0.07\\
\midrule
\addlinespace[0.8em]
\multicolumn{19}{l}{Mixtures}\\
\hspace{1em}150 & 0.17 & 0.16 & 0.18 & 0.18 & 0.15 & 0.15 & 0.17 & 0.16 & 0.17 & 0.04 & 0.10 & 0.08 & 0.05 & 0.03 & 0.06 & 0.03 & 0.06 & 0.06\\
\hspace{1em}200 & 0.13 & 0.12 & 0.19 & 0.12 & 0.10 & 0.21 & 0.13 & 0.11 & 0.19 & 0.10 & 0.15 & 0.14 & 0.04 & 0.05 & 0.06 & 0.04 & 0.06 & 0.05\\
\midrule
\addlinespace[0.8em]
\multicolumn{19}{l}{Interactions}\\
\hspace{1em}150 & 0.55 & 0.67 & 0.58 & 0.41 & 0.62 & 0.57 & 0.56 & 0.68 & 0.59 & 0.96 & 0.86 & 0.91 & 0.96 & 0.91 & 0.95 & 0.97 & 0.91 & 0.94\\
\hspace{1em}200 & 0.75 & 0.81 & 0.72 & 0.71 & 0.67 & 0.64 & 0.75 & 0.81 & 0.70 & 0.97 & 0.98 & 0.94 & 0.99 & 1.00 & 0.98 & 0.99 & 0.99 & 0.96\\
\midrule
\addlinespace[0.8em]
\multicolumn{19}{l}{Mixed Heterogeneity}\\
\hspace{1em}150 & 0.49 & 0.39 & 0.39 & 0.43 & 0.36 & 0.35 & 0.47 & 0.38 & 0.39 & 0.72 & 0.74 & 0.92 & 0.74 & 0.73 & 0.60 & 0.78 & 0.73 & 0.69\\
\hspace{1em}200 & 0.47 & 0.51 & 0.52 & 0.45 & 0.44 & 0.46 & 0.47 & 0.52 & 0.53 & 0.81 & 0.86 & 0.98 & 0.87 & 0.85 & 0.85 & 0.90 & 0.89 & 0.88\\
\bottomrule
\end{tabular}}
\end{table}

\begin{table}[H]
\centering
\caption{\label{tab:size_small} Size of the conditional independence tests in small samples. The table reports rejection probabilities under the null hypothesis for the double residualized (DR/\(\drs\)), augmented inverse propensity weighted (AIPW), kernel conditional independence (KCI, \citealp{zhang2012kernel}), generalized covariance measure (GCM, \citealp{shah2020hardness}), and joint Wald tests. Each cell reports the average over 100 Monte Carlo replications.}
\centering
\resizebox{\ifdim\width>\linewidth\linewidth\else\width\fi}{!}{
\fontsize{9}{11}\selectfont
\begin{tabular}[t]{l>{}rrr>{}rrr>{}rrr>{}rrr>{}rrr>{}rrr}
\toprule
\multicolumn{1}{c}{ } & \multicolumn{3}{c}{$\dr$} & \multicolumn{3}{c}{\(\aipw\)} & \multicolumn{3}{c}{$\drs$} & \multicolumn{3}{c}{KCI} & \multicolumn{3}{c}{GCM} & \multicolumn{3}{c}{Wald} \\
\cmidrule(l{3pt}r{3pt}){2-4} \cmidrule(l{3pt}r{3pt}){5-7} \cmidrule(l{3pt}r{3pt}){8-10} \cmidrule(l{3pt}r{3pt}){11-13} \cmidrule(l{3pt}r{3pt}){14-16} \cmidrule(l{3pt}r{3pt}){17-19}
\multicolumn{1}{c}{ } & \multicolumn{18}{c}{Number of outcomes} \\
\cmidrule(l{3pt}r{3pt}){2-19}
Sample size & 1 & 2 & 3 & 1 & 2 & 3 & 1 & 2 & 3 & 1 & 2 & 3 & 1 & 2 & 3 & 1 & 2 & 3\\
\midrule
\addlinespace[0.8em]
\multicolumn{19}{l}{Full Heterogeneity}\\
\hspace{1em}150 & 0.12 & 0.12 & 0.08 & 0.11 & 0.12 & 0.11 & 0.13 & 0.12 & 0.08 & 0.05 & 0.06 & 0.06 & 0.06 & 0.13 & 0.07 & 0.06 & 0.05 & 0.10\\
\hspace{1em}200 & 0.16 & 0.15 & 0.12 & 0.15 & 0.20 & 0.15 & 0.15 & 0.15 & 0.13 & 0.06 & 0.05 & 0.05 & 0.05 & 0.07 & 0.09 & 0.06 & 0.08 & 0.05\\
\midrule
\addlinespace[0.8em]
\multicolumn{19}{l}{No Baseline}\\
\hspace{1em}150 & 0.15 & 0.14 & 0.11 & 0.14 & 0.13 & 0.15 & 0.15 & 0.14 & 0.11 & 0.04 & 0.05 & 0.07 & 0.06 & 0.04 & 0.05 & 0.06 & 0.07 & 0.09\\
\hspace{1em}200 & 0.13 & 0.12 & 0.10 & 0.13 & 0.11 & 0.14 & 0.11 & 0.13 & 0.09 & 0.04 & 0.03 & 0.08 & 0.04 & 0.04 & 0.09 & 0.03 & 0.04 & 0.08\\
\midrule
\addlinespace[0.8em]
\multicolumn{19}{l}{Mixtures}\\
\hspace{1em}150 & 0.12 & 0.11 & 0.15 & 0.15 & 0.12 & 0.14 & 0.12 & 0.10 & 0.15 & 0.04 & 0.04 & 0.05 & 0.04 & 0.03 & 0.05 & 0.04 & 0.03 & 0.07\\
\hspace{1em}200 & 0.09 & 0.07 & 0.09 & 0.08 & 0.09 & 0.12 & 0.08 & 0.07 & 0.09 & 0.05 & 0.06 & 0.03 & 0.06 & 0.03 & 0.04 & 0.05 & 0.04 & 0.05\\
\midrule
\addlinespace[0.8em]
\multicolumn{19}{l}{Interactions}\\
\hspace{1em}150 & 0.09 & 0.19 & 0.17 & 0.11 & 0.21 & 0.23 & 0.09 & 0.19 & 0.17 & 0.04 & 0.05 & 0.04 & 0.04 & 0.04 & 0.00 & 0.03 & 0.07 & 0.03\\
\hspace{1em}200 & 0.12 & 0.13 & 0.17 & 0.16 & 0.14 & 0.11 & 0.11 & 0.12 & 0.17 & 0.04 & 0.08 & 0.06 & 0.04 & 0.06 & 0.06 & 0.03 & 0.05 & 0.07\\
\midrule
\addlinespace[0.8em]
\multicolumn{19}{l}{Mixed}\\
\hspace{1em}150 & 0.14 & 0.10 & 0.10 & 0.18 & 0.13 & 0.13 & 0.14 & 0.10 & 0.10 & 0.03 & 0.04 & 0.06 & 0.08 & 0.03 & 0.08 & 0.04 & 0.03 & 0.08\\
\hspace{1em}200 & 0.09 & 0.17 & 0.07 & 0.07 & 0.19 & 0.13 & 0.10 & 0.17 & 0.08 & 0.06 & 0.04 & 0.03 & 0.06 & 0.08 & 0.04 & 0.03 & 0.07 & 0.05\\
\bottomrule
\bottomrule
\end{tabular}}
\end{table}

\end{appendix}

\end{document}